\documentclass[journal]{vgtc}                % final (journal style)

\ifpdf%                                % if we use pdflatex
  \pdfoutput=1\relax                   % create PDFs from pdfLaTeX
  \usepackage{graphicx}                % allow us to embed graphics files
  \DeclareGraphicsExtensions{.pdf,.png,.jpg,.jpeg} % for pdflatex we expect .pdf, .png, or .jpg files
\else%                                 % else we use pure latex
  \usepackage{graphicx}                % allow us to embed graphics files
  \DeclareGraphicsExtensions{.eps}     % for pure latex we expect eps files
\fi%

\graphicspath{{figures/}{pictures/}{images/}{./}} % where to search for the images

\usepackage{microtype}                 % use micro-typography (slightly more compact, better to read)
\PassOptionsToPackage{warn}{textcomp}  % to address font issues with \textrightarrow
\usepackage{textcomp}                  % use better special symbols
\usepackage{mathptmx}                  % use matching math font
\usepackage{times}                     % we use Times as the main font
\usepackage{cite}                      % needed to automatically sort the references
\usepackage{tabu}                      % only used for the table example
\usepackage{booktabs}                  % only used for the table example
\usepackage{amsfonts}
\usepackage{adjustbox}
\usepackage{amsmath}
\usepackage{amssymb}
\usepackage{subfigure}
\usepackage{float}
\usepackage{xcolor}
\usepackage{xspace}
\usepackage{enumitem}
\usepackage{comment}
\usepackage{multirow}
\usepackage{setspace}

\newcommand{\Xspace}        {{\mathbb X}}

\newcommand{\Hgroup}        {{\mathsf H}}

\newcommand{\Rips}{{R}}

\makeatletter
\newcommand{\thickhline}{%
    \noalign {\ifnum 0=`}\fi \hrule height 1pt
    \futurelet \reserved@a \@xhline
}
\makeatother

\definecolor{colorAnger}{rgb}{0.55, 0.34, 0.29}
\definecolor{colorDisgust}{rgb}{0.58, 0.40, 0.74}
\definecolor{colorFear}{rgb}{0.84, 0.15, 0.15}
\definecolor{colorHappy}{rgb}{0.17, 0.63, 0.17}
\definecolor{colorSad}{rgb}{1.0, 0.50, 0.05}
\definecolor{colorSurprise}{rgb}{0.12, 0.47, 0.70}

\newcommand{\anger}{\textcolor{colorAnger}{\textit{anger}}\xspace}
\newcommand{\disgust}{\textcolor{colorDisgust}{\textit{disgust}}\xspace}
\newcommand{\fear}{\textcolor{colorFear}{\textit{fear}}\xspace}
\newcommand{\happiness}{\textcolor{colorHappy}{\textit{happiness}}\xspace}
\newcommand{\sadness}{\textcolor{colorSad}{\textit{sadness}}\xspace}
\newcommand{\surprise}{\textcolor{colorSurprise}{\textit{surprise}}\xspace}

\newcommand{\Anger}{\textcolor{colorAnger}{\textit{Anger}}\xspace}
\newcommand{\Disgust}{\textcolor{colorDisgust}{\textit{Disgust}}\xspace}
\newcommand{\Fear}{\textcolor{colorFear}{\textit{Fear}}\xspace}
\newcommand{\Happiness}{\textcolor{colorHappy}{\textit{Happiness}}\xspace}
\newcommand{\Sadness}{\textcolor{colorSad}{\textit{Sadness}}\xspace}
\newcommand{\Surprise}{\textcolor{colorSurprise}{\textit{Surprise}}\xspace}

\newcommand{\para}[1]{\vspace{5pt}\noindent\textbf{#1} \ \ }

\makeatletter
\newcommand{\labeltext}[3][]{%
    \@bsphack%
    \csname phantomsection\endcsname% in case hyperref is used
    \def\tst{#1}%
    \def\labelmarkup{\textbf}% How to markup the label itself
    \def\refmarkup{}%
    \ifx\tst\empty\def\@currentlabel{\refmarkup{#2}}{\label{#3}}%
    \else\def\@currentlabel{\refmarkup{#1}}{\label{#3}}\fi%
    \@esphack%
    \labelmarkup{#2}% visible printed text.
}
\makeatother

\makeatletter
\let\orgdescriptionlabel\descriptionlabel
\renewcommand*{\descriptionlabel}[1]{%
  \let\orglabel\label
  \let\label\@gobble
  \phantomsection
  \edef\@currentlabel{#1}%
  \let\label\orglabel
  \orgdescriptionlabel{#1}%
}
\makeatother

\title{AffectiveTDA: Using Topological Data Analysis to \\ Improve Analysis and Explainability in Affective Computing}

\author{Hamza Elhamdadi, Shaun Canavan, and Paul Rosen}
\authorfooter{
\item
 Hamza Elhamdadi is with the University of Massachusetts Amherst. E-mail: helhamdadi@umass.edu.
\item
 Shaun Canavan is with the University of South Florida. E-mail: scanavan@usf.edu.
\item
 Paul Rosen is with the University of South Florida. E-mail: prosen@usf.edu.
}

\shortauthortitle{Elhamdadi \MakeLowercase{\textit{et al.}}: AffectiveTDA: Topology for Explainability in Affective Computing}
\abstract{%
  We present an approach utilizing Topological Data Analysis to study the structure of face poses used in affective computing, i.e., the process of recognizing human emotion. The approach uses a conditional comparison of different emotions, both respective and irrespective of time, with multiple topological distance metrics, dimension reduction techniques, and face subsections (e.g., eyes, nose, mouth, etc.). The results confirm that our topology-based approach captures known patterns, distinctions between emotions, and distinctions between individuals, which is an important step towards more robust and explainable emotion recognition by machines.
} % end of abstract


\teaser{
    \centering
    \fbox{\includegraphics[trim=0 5pt 1pt 1pt, clip, width=0.825\linewidth]{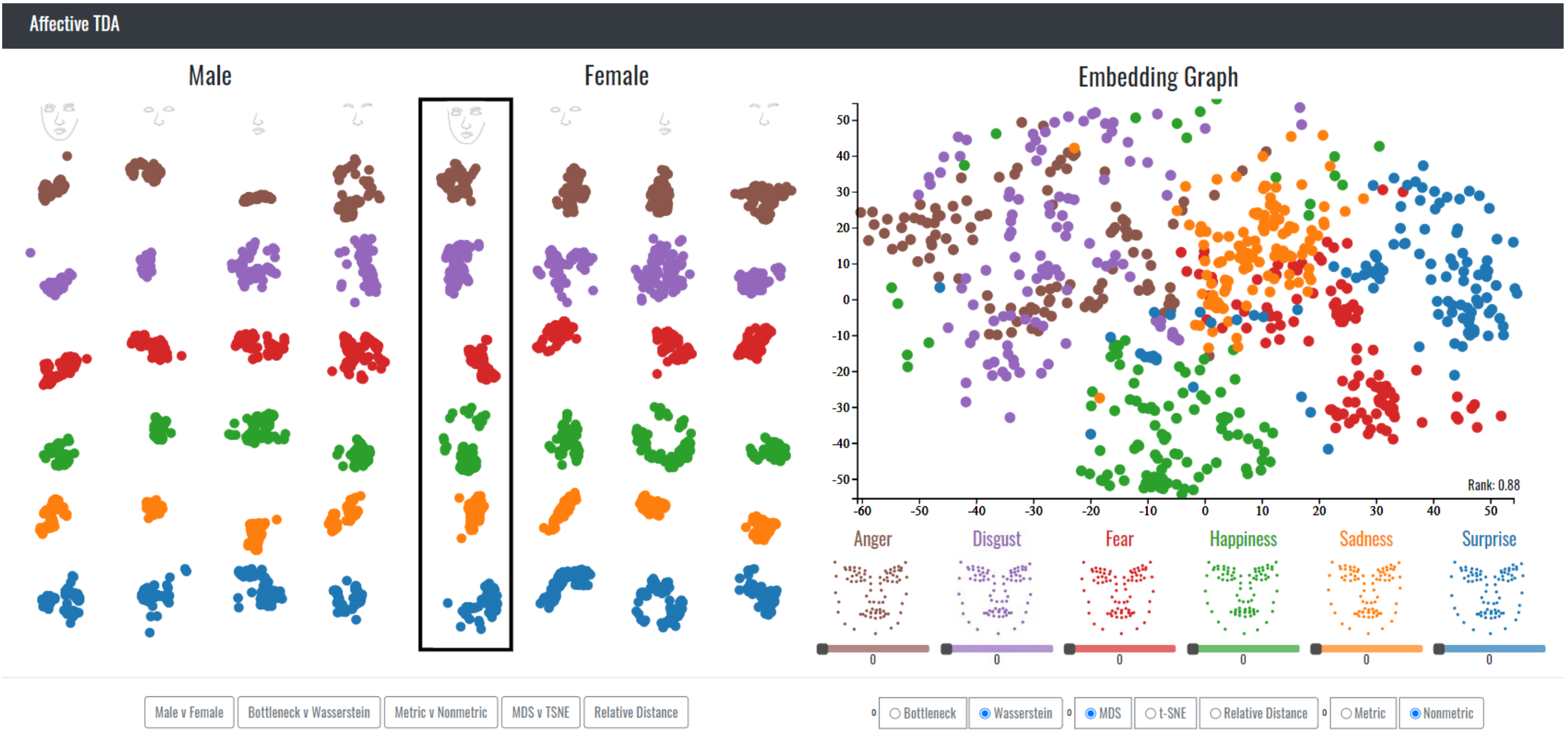}}
    \caption{Our affective computing visualization provides numerous options for comparing and contrasting the data. For example, the small multiples view (left) is comparing a male (left) subject to a female (right) subject considering different subsets of facial landmarks (columns), across emotions (rows) of \anger, \disgust, \fear, \happiness, \sadness, and \surprise. Each point represents one facial pose. The embedding graph (top right) compares all facial poses across all emotions, in this case showing the female full face using MDS on non-metric topology. The 3D landmarks (lower right) show a single facial pose per emotion. Settings are selected on the bottom.}
    \label{fig:teaser}
}

\vgtcinsertpkg

\begin{document}

\firstsection{Introduction}

\maketitle

\setstretch{0.965}

Affective computing, computer-based detection of human affects, has applications that span education (e.g., judging learners' confidence), healthcare (e.g., judging pain), and product marketing (e.g., measuring consumers' response to products). Early work in measuring affect began in the late 1960's spearheaded by Ekman and Friesen~\cite{ekman1969repertoire}. Their work culminated in a classification of six basic emotions: \anger, \disgust, \fear, \happiness, \sadness, and \surprise~\cite{ekman1987universals}, which were later expanded~\cite{ekman1999basic}. The field of affective computing has seen significant growth since the seminal work from Rosalind Picard~\cite{picard2000affective}. The vast majority of research in affective computing has been focused on machine learning algorithms trained on emotion data to classify affect. Like many machine learning solutions, these neural networks focused on classifying the input emotion and ignored data inspection and decision-making explainability.

To inspect the data, an effective visual representation of emotion data must address numerous challenges. First, the affect data are quite large, captured by multiple high-speed video cameras. Fortunately, previous affective computing research already partially addressed this issue by reducing the data to 83 landmark points tracked temporally. Nevertheless, the problem remains challenging because patterns in emotion occur over extended time periods, represented by a series of 83-landmark poses. Furthermore, patterns of interest may occur in different time sequences lasting for different lengths of time, making alignment and comparison non-trivial. Finally, changes in landmarks are simultaneously subtle and subject to noise from the extraction process, making them difficult to observe. 

This paper presents a visual analytics approach utilizing Topological Data Analysis (TDA) to examine emotion data respective and irrespective of time. By using TDA to address this problem, our approach can capture and track the topological ``shape'' of facial landmarks over time in a manner robust to noise~\cite{EdelsbrunnerHarer2010}. After analysis, the data are presented for investigation using familiar visualizations, e.g., timelines (see \autoref{fig:au_example} top) and scatterplots (see \autoref{fig:teaser} top right), and through landmark-based representations (see \autoref{fig:teaser} bottom right or \autoref{fig:representative_ex}). These interfaces enable tracking facial movement, comparing emotions, and comparing individuals, while also providing the ability to derive precise explanations for features identified in the data.

A natural question at this point would be, why is TDA well-suited to this problem? Our approach utilizes one of the foundational tools of TDA, namely persistent homology. There are four main advantages to this tool. (1)~Persistent homology has a solid mathematical grounding, and its output is explainable. (2)~Persistent homology extracts homology groups, which in our context are (connected) components and tunnels/cycles. These fundamental shapes match well with the shapes of a face. (3)~The homology groups are extracted at multiple scales, without the need to specify any thresholds or other parameters. This means that persistent homology captures all of the topological structures without any user intervention\footnote{Note that while persistent homology itself has no parameters, our pipeline does have some options available to users.}. (4)~Finally, it classifies features by their importance with a measure called persistence, which automatically differentiates topological signal from noise~\cite{cohen2007stability}.

\vspace{3pt}
\noindent The specific contributions of this paper are:
\vspace{-3pt}
\begin{enumerate}[noitemsep, itemsep=2pt]
    \item a mapping of affective computing data to TDA (see \autoref{sec:tda}), including a novel non-metric formulation of geometry for faster and more accurate topology extraction (see \autoref{sec:tda:nonmetric});
    \item a visual analytics interface that enables analyzing, comparing, and contrasting multiple data configurations (see \autoref{sec:vis});
    \item an evaluation that uses our methodology to explain features in data that were extracted by state-of-the-art emotion detection machine learning algorithms (see \autoref{sec:eval:AUs}); and
    \item an evaluation of the ability of TDA to differentiate emotions within the same individual (see \autoref{sec:eval:exp}) and differentiate multiple individuals showing the same emotion (see \autoref{sec:eval:ind}).
\end{enumerate} 

\vspace{-3pt}
\noindent Perhaps most importantly, our approach opens the door to explainability in a way that may help to unlock open questions in the affective computing community.

\setstretch{0.97}

\section{Background in Affective Computing}

Affective computing has applications in fields as varied as medicine~\cite{zamzmi2016approach}, entertainment~\cite{fleureau2012physiological}, and security~\cite{kim2020contverb}. Most notably, the expression recognition sub-field focuses on detecting subjects' affective states automatically.

\subsection{Expression Recognition}
\label{sec:expRecRW}

While successful 2D facial-expression image recognition exists~\cite{kalam2019facial, fan2020facial, li2020deep}, the approaches suffer from weaknesses, such as occlusion from, e.g., a rotating head. We focus our discussion instead on a few representative 3D facial recognition approaches. Zhen et al.~\cite{zhen2016muscular} developed a model that localized points within each muscular region of the face and extracted features that include coordinate, normal, and shape index~\cite{koenderink1992surface}. The features were then used to train a Support Vector Machine (SVM)~\cite{vapnik2013nature} to recognize expressions. Xue et al.~\cite{xue2015automatic} proposed a method for 4D (3D + time) expression recognition, which showed promise differentiating difficult emotions, such as \anger and \sadness. The method extracted local patch sequences from consecutive 3D video frames and represented them with a 3D discrete cosine transform. Then, a nearest-neighbor classifier was used to recognize the expressions. Hariri et al.~\cite{hariri20173d} proposed an approach to expression recognition using manifold-based classification. The approach sampled the face by extracting local geometry as covariance regions, which were used with an SVM to recognize expressions.

Some recent techniques showed that not all regions of the face carry the same importance in emotion recognition. Hernandez-Matamoros et al.~\cite{hernandez2016facial} found that segmenting the face based on the eyes and mouth resulted in improved expression recognition. Fabiano et al.~\cite{fabiano2020impact} further illustrated that different areas of the face carry different levels of importance for emotions, e.g., one subject \happiness had more important features on the right eye and eyebrow, while embarrassment had more on the left eye and eyebrow. We utilize this information in our visualization design by targeting specific subsets of facial features.

\subsection{Affective Computing in Visualization}

There has been limited work in the visualization community on affective computing; what exists has been primarily focused on \textit{visualizing affective states}, i.e., considering valence and arousal, not inspecting the landmarks used as input to affective computing algorithms. 

Early work on visualizing affective states concerns the glyph-based Self-Assessment Manikin (SAM), which measures pleasure, arousal, and dominance of a person's affective state~\cite{Bradley:1994}. Cernea et al.~\cite{Cernea:2013} later described guidelines for conveying the user emotion through the use of widgets that depict the affective states of valence and arousal. The widgets employed \textit{emotion scents}, hue-varied colormaps representing either valance or arousal, e.g., red and green represent negative and positive valance, respectively. Emotion-prints was an early system to provided real-time feedback of valance and arousal to users using touch-displays~\cite{cernea2015emotion}. More recently, Kovacevik et al.~\cite{Kovacevik:2020} employed ideas from SAM and {emotion scents} to create a glyph for simultaneous representation of valence and arousal. Their research focused on video game players' and developers' awareness of emotions elicited from a particular gaming experience. For visualizing affect over extended periods, AffectAura provided an interface that enabled users to visualize emotional states over time for the purpose of reflection~\cite{mcduff2012affectaura}.

There has also been some work visualizing the affective state of multiple individuals using, e.g., virtual agents in collaborative work~\cite{cernea2014group} or using a visual analytics interface to access the emotional state of students in a classroom~\cite{zeng2020emotioncues}. Qin et al.~\cite{qin2020heartbees} created HeartBees, which was an interface to demonstrate the affect of a crowd using physiological data. The interface used an abstract flocking behavior to demonstrate the collective emotional state.

In contrast to all of these prior approaches, our work focuses on using TDA and visualization to investigate the data used in classifying expression, i.e., the input data, not the emotional state itself. There has been some recent work that looked at the explainability of deep networks in expression recognition, e.g.,~\cite{otberdout2019automatic}. These approaches focus on visualizing heatmaps that highlight \textit{what} parts of the image most influenced decision making, not necessarily \textit{why}.

\subsection{Dataset}
\label{dataset}

To evaluate our approach, we use the BU4DFE 3D facial expression dataset~\cite{yin126high}, which has been extensively used for expression recognition~\cite{fabiano2019deformable, tornincasa20193d, patil2017emotion, chinaev2018mobileface}, 3D shape reconstruction~\cite{grasshof2020multilinear, gao20193d, liu20193d}, face tracking~\cite{pham2016robust, canavan2015landmark}, and face recognition~\cite{aouada20203d, jannat2020subject, lawrence20143d, sun20083d}. The dataset contains 101 subjects (58 female and 43 male) from multiple ethnicities, including Caucasian, African American, Asian, and Hispanic, with an age range of 18-45 years old. Each modality has the six basic emotions: \anger, \disgust, \fear, \happiness, \sadness, and \surprise. For each sequence, the expression is the result of gradually building from neutral to low then high intensity and back again. Each of the video sequences is 3-4 seconds in length.

\begin{figure*}[!t]
    \centering
    
    \hspace{20pt}
    \includegraphics[width=0.90\linewidth]{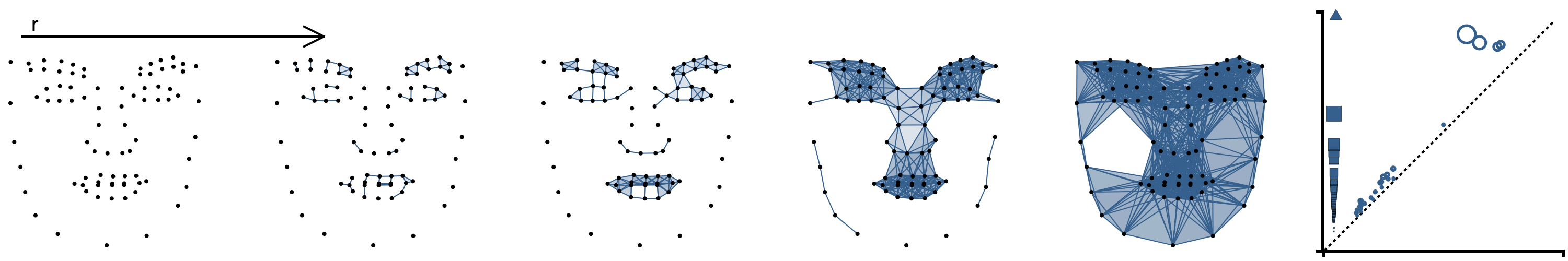}
    \begin{minipage}[t]{0\textwidth}\vspace{-45pt}\hspace{-475pt}\subfigure[\label{fig:ph:rips}]{}\end{minipage}%
    \begin{minipage}[t]{0\textwidth}\vspace{-25pt}\hspace{-15pt}\subfigure[\label{fig:ph:pd}]{}\end{minipage}
    \caption{An illustration of persistent homology on the 83 facial landmarks on the female subject F001. (a) The persistent homology is calculated by forming a Rips filtration and tracking/extracting the associated homology groups. Starting at $r=0$, if the pairwise distance between any two or three points is less than $r$, an edge or triangle is formed, respectively. As $r$ increases, components merge, and tunnels form and disappear. (b) The topology is visualized with a persistence diagram. Square points are $\Hgroup_0$ components (the triangle indicates a single infinite $\Hgroup_0$ component), and the hollow circles are the $\Hgroup_1$ tunnels. The horizontal position of points is their birth $r_{b_i}$ and their vertical position is their death $r_{d_i}$. Distance from the dotted diagonal, as well as object size, is proportional to its persistence.}
    \label{fig:ph}
\end{figure*}

The data are captured using the Di3D dynamic face capturing system~\cite{di3d}, which consists of three cameras, two to capture stereo and one to capture texture. Passive stereophotogrammetry is used on each pair of stereo images to create the 3D facial pose models with an RMS accuracy of $0.2$ mm. Each 3D model contains 83 facial landmarks (see \autoref{fig:ph}), which correspond to the key areas of the face that include the mouth, eyes, eyebrows, nose, and jawline. The landmarks are the result of using an active appearance model~\cite{cootes2001active} that detects the landmarks on the 2D texture images, which are aligned and projected into the corresponding 3D models.

\section{Overview of the Pipeline} 

TDA has received significant attention in the visualization community, e.g., \cite{tierny2017topology}. We utilize a foundational tool of TDA, persistent homology, which has been studied in graph analysis~\cite{rieck2017clique,suh2019persistent,hajij2018visual,hajij2020fast}, high-dimensional data analysis~\cite{wang2011branching}, and multivariate analysis~\cite{rieck2012multivariate}. We utilize persistent homology to capture the topology of the landmarks of each facial pose into a structure known as a persistence diagram. We then compare the topology of different subsets of facial poses to reveal their relationships. Our processing pipeline contains three main stages, which are fed into the visualization (see \autoref{sec:vis}). 
\begin{description}[noitemsep,itemsep=3pt]
    \item[Stage 1:] The first stage is extracting the topology of a single facial pose. We offer two variations, a Euclidean metric-based approach (see \autoref{sec:tda:ph}) and a novel non-metric-based approach (see \autoref{sec:tda:nonmetric}).
    \item[Stage 2:] Once the topologies of individual poses are extracted, we compare the topology to that of other poses to determine their pairwise dissimilarity (see \autoref{sec:tda:topoDist}).
    \item[Stage 3:] Finally, using the topological dissimilarity, we utilize a variety of dimension reduction techniques to highlight different aspects of the dissimilarity between groups of facial poses (see \autoref{sec:tda:dimRed}).
\end{description}

\section{Topological Data Analysis of Facial Landmarks}
\label{sec:tda}

We consider two variations for extracting the topology of facial poses, a Euclidean metric approach, followed by a novel non-metric variant.

\subsection{Euclidean Metric Persistent Homology on Landmarks}
\label{sec:tda:ph}

Homology deals with the topological features of a space. Given a topological space $\Xspace$, we are interested in extracting the $\Hgroup_0(\Xspace)$ and $\Hgroup_1(\Xspace)$ homology groups, which correspond to (connected) components and tunnels/cycles of $\Xspace$, respectively\footnote{We do not consider $\Hgroup_2$ (voids) because despite our data being 3D, it is nearly flat. Thus, voids rarely occur, and when they do, they have low persistence.}. In practice, there may not exist a single scale that captures the topological structures of the data. Instead, we use a multi-scale notion of homology, called \emph{persistent homology}, to describe the topological features of a space at different spatial resolutions. We briefly describe persistent homology in our limited context. Nevertheless, understanding persistent homology can be daunting for those who are unfamiliar with it. For a high-level overview, see~\cite{weinberger2011persistent}, or for detailed background, see~\cite{EdelsbrunnerHarer2008}.

To calculate the persistent homology of a single facial pose, we first calculate the Euclidean distance between all 83 landmarks. We then apply a geometric construction, the Rips complex, $\Rips(r)$, on the point set. In brief, for a given distance, $r$, the Rips complex has all $0$-simplicies, i.e., points, for all values of $r$. A $1$-simplex, i.e., an edge, between two points is formed \textit{iff} $r$ is greater than or equal to their distance. A $2$-simplex, i.e., a triangle, is formed among three points \textit{iff} $r$ is greater than or equal to every pairwise distance between the points.

To extract the persistent homology (see \autoref{fig:ph:rips}), we consider a finite sequence of increasing distances, $0 = r_0 \leq r_1 \leq \cdots \leq r_m=\infty$. A sequence of Rips complexes, known as a Rips filtration, is connected by inclusions, $\Rips(r_0) \to \Rips(r_1) \to \cdots \to \Rips(r_m)$, and the homology of each is calculated, tracking the homomorphisms induced by the inclusions, $\Hgroup(\Rips(r_0)) \to \Hgroup(\Rips(r_1)) \to \cdots \to \Hgroup(\Rips(r_m))$. As the distance increases, topological features, i.e., components and tunnels, appear and disappear. The appearance is known as a \emph{birth} event, $r_{b_i}$, and the disappearance is known as a \emph{death} event, $r_{d_i}$. The birth and death of all features are stored as a multi-set of points in the plane, $(r_{b_i},r_{d_i})$, known as the \textit{persistence diagram}, which is often visualized in the scatterplot display (see \autoref{fig:ph:pd}). From the points, we devise an importance measure, called \textit{persistence}, which helps to differentiate signal from noise. The persistence is simply the difference between the birth and death of a feature, i.e., $r_{d_i}-r_{b_i}$. Furthermore, in visualizations of the persistence diagram, such as \autoref{fig:ph:pd}, distance from the diagonal dotted line represents the persistence of a feature.

In addition to considering all the topology of all landmarks, we provide the user the functionality to consider only related subsets of features. In particular, they have the option of including/excluding jawline, mouth, nose, left/right eyes, and left/right eyebrows in the calculation of the topology.

\subsection{Interpolating Known Geometry}

\begin{figure*}[!th]
    \centering
    \includegraphics[width=0.925\linewidth]{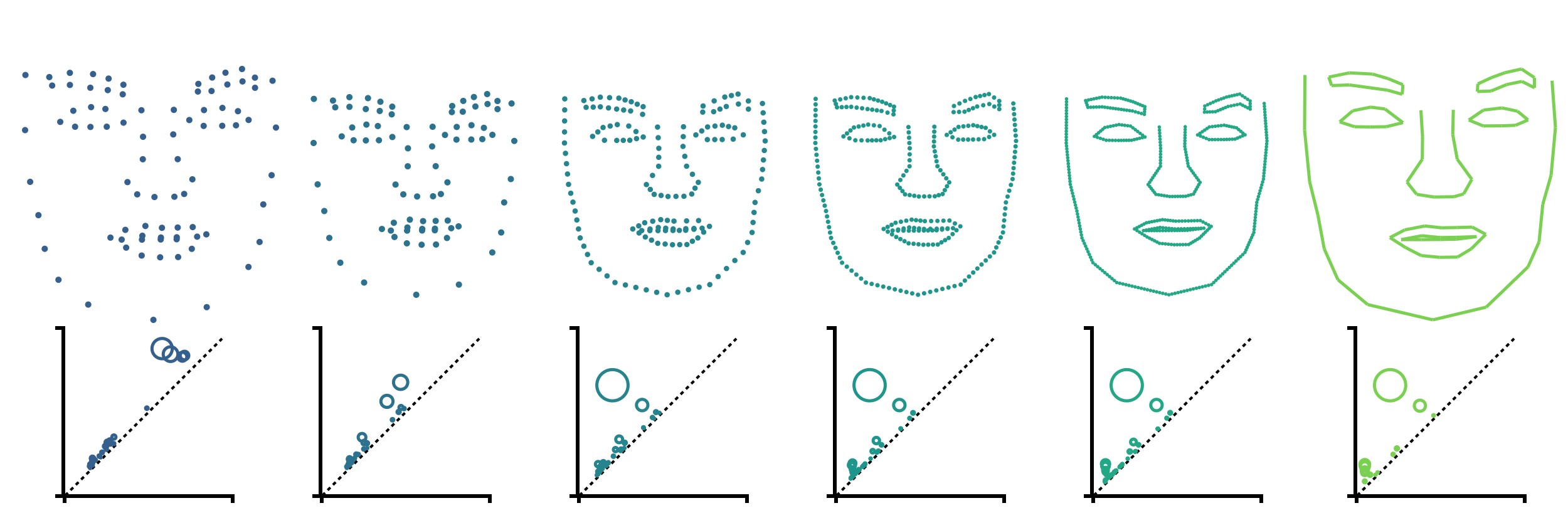}
    
    \vspace{-15pt}
    \hspace{33pt}
    \subfigure[Original Data\label{fig:nonmetric:a}]{\hspace{55pt}}\hfill
    \subfigure[$\epsilon=8$\label{fig:nonmetric:b}]{\hspace{55pt}}\hfill
    \subfigure[$\epsilon=4$\label{fig:nonmetric:c}]{\hspace{55pt}}\hfill
    \subfigure[$\epsilon=2$\label{fig:nonmetric:d}]{\hspace{55pt}}\hfill
    \subfigure[$\epsilon=1$\label{fig:nonmetric:e}]{\hspace{55pt}}\hfill
    \subfigure[Non-metric\label{fig:nonmetric:f}]{\hspace{55pt}}
    \hspace{30pt}
    
    \vspace{-1pt}
    \caption{Illustration of supersampling and non-metric persistent homology shows the data and the persistence diagram of $\Hgroup_1$ features. (a) The original 83 landmarks from a single pose on the female subject F001. (b-e) Supersampling the landmarks with different $\epsilon$ values shows a significant difference in the persistence diagram between the (a) original and (e) supersampled data. (f)~Our non-metric representation of the data requires significantly less data and produces a persistence diagram similar to that of (e) supersampling.}
    \label{fig:nonmetric}
    \vspace{-1pt}
\end{figure*}

Our computation using facial landmarks ignores an important aspect of the data, namely the known connectivity between landmarks. In other words, landmarks of, e.g., the mouth, have known connectivity to their neighboring landmarks. \autoref{fig:nonmetric:f} shows this connectivity. This raises two questions. First, does our failure to consider this connectivity impact the features we extract, and second, how do we efficiently consider the connectivity?

We first consider using interpolation of the connectivity to supersample additional landmarks. For our experiment, we take the known connectivity and interpolate across each edge, such that points are no further than a user-defined $\epsilon$ apart. \autoref{fig:nonmetric:b} through \autoref{fig:nonmetric:e} show four examples with ever-smaller $\epsilon$ values. As expected, as $\epsilon$ gets smaller, the data looks increasingly similar to the known connectivity in \autoref{fig:nonmetric:f}. 

We now consider the impact of the connectivity by comparing the persistence diagrams of $\Hgroup_1$ features in the original data in \autoref{fig:nonmetric:a} to the lowest $\epsilon$ data in \autoref{fig:nonmetric:e}. The persistence diagrams are clearly different (the $\Hgroup_0$ features are also different but more difficult to observe pictorially). The difference is exceedingly important because it means \textit{using the 83 landmark points alone is insufficient to capture the topological structure of the data}.

\begin{figure}[!b]
    \centering
    \vspace{-2pt}
    \subfigure[Calculation Time in Seconds\label{fig:nonmetric_exp:compute}]{\hspace{5pt}\includegraphics[width=0.39\linewidth]{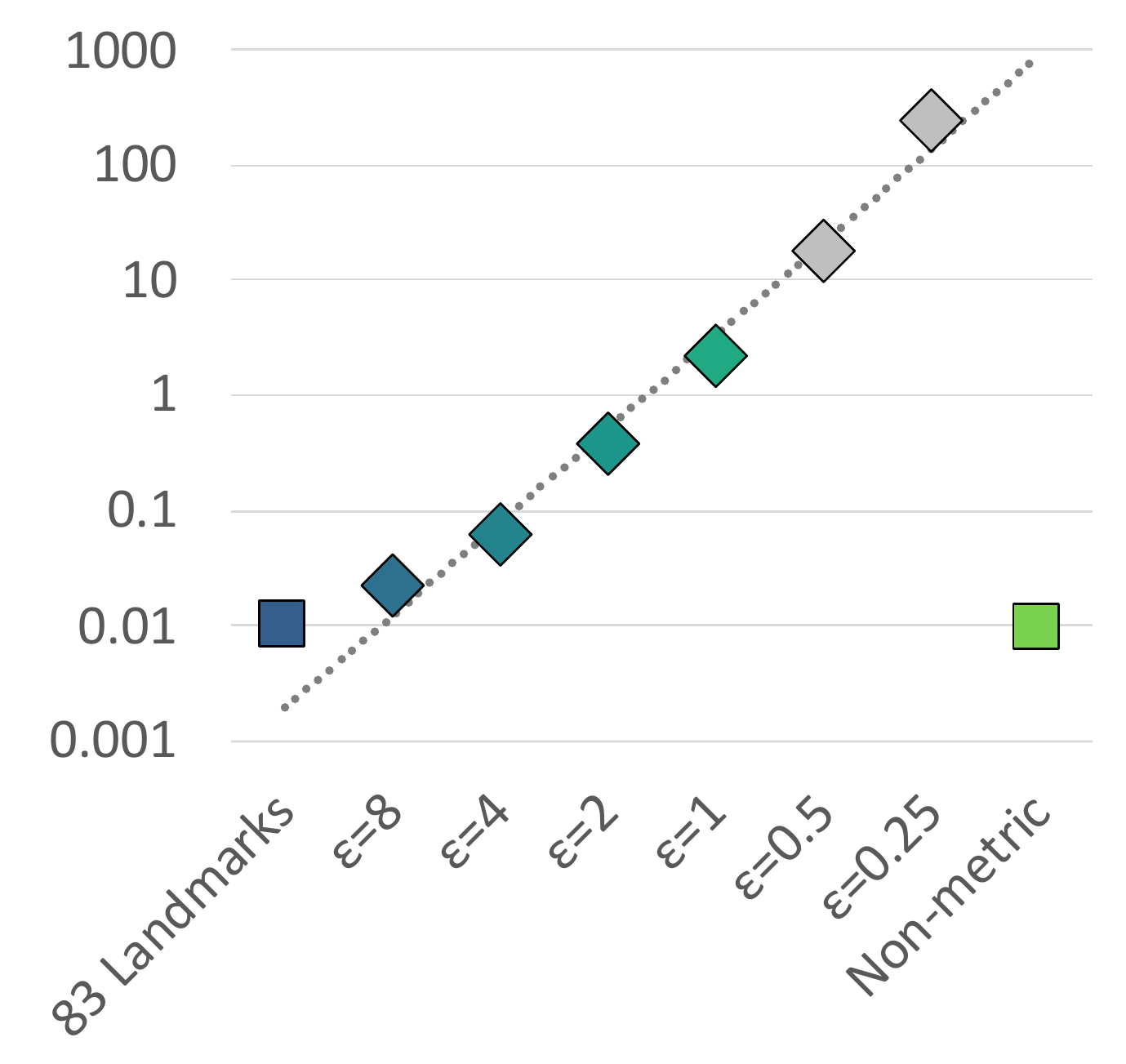}\hspace{5pt}}
    \hspace{10pt}
    \subfigure[Number of Features Generated\label{fig:nonmetric_exp:features}]{\hspace{5pt}\includegraphics[width=0.39\linewidth]{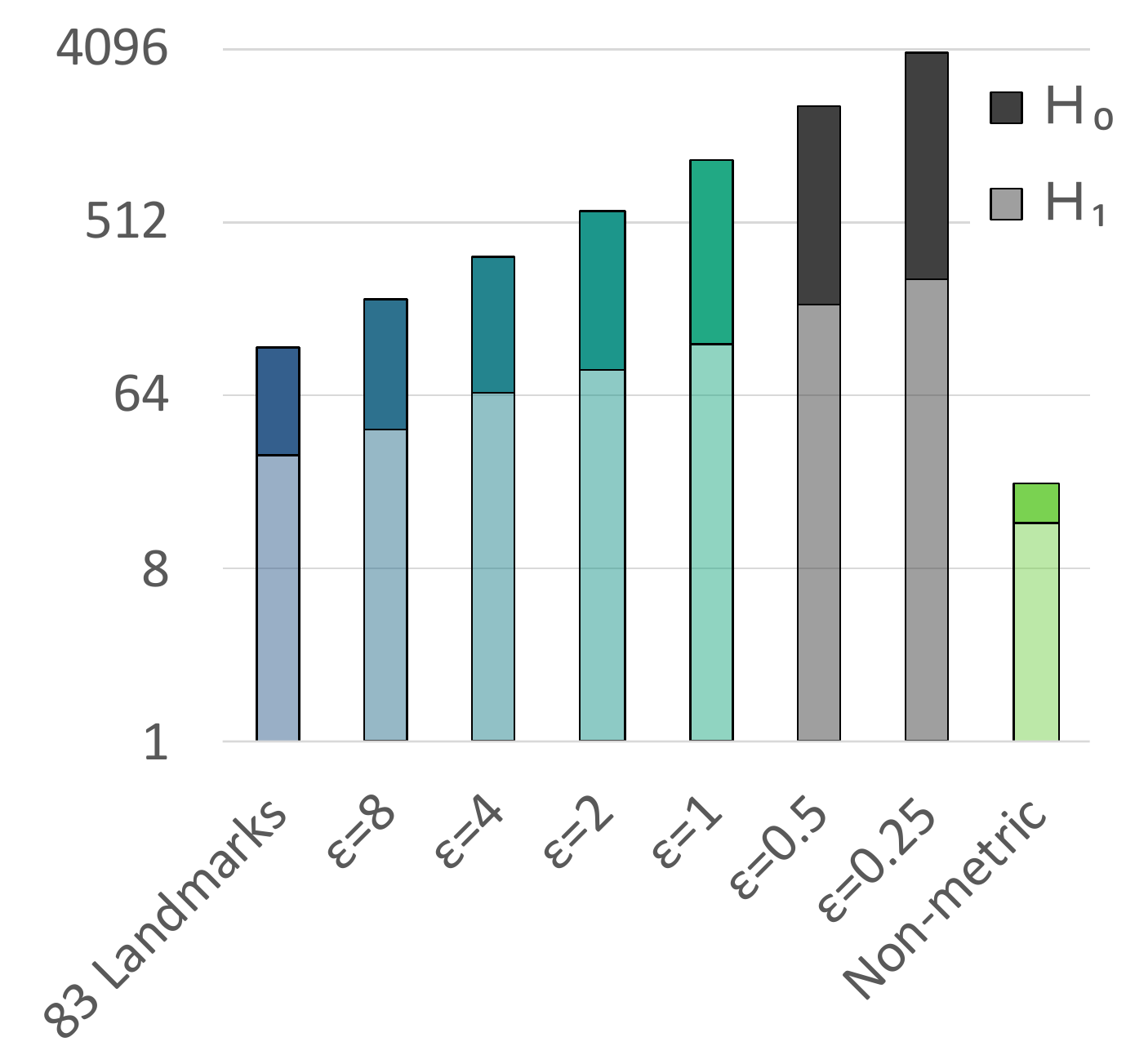}\hspace{5pt}}
    
    \vspace{-1pt}
    \caption{Plots of (a) the compute time and (b) the number of topological features generated for the original 83 landmarks, six levels of supersampling ($\epsilon=8$, $4$, $2$, $1$, $0.5$, and $0.25$), and our non-metric representation. The regression line in (a) only considers the supersampling data points.}
    \label{fig:nonmetric_exp}
\end{figure}

To overcome this limitation, we considered using the supersampled landmarks for calculations. However, there are three interrelated problems to this approach. (1)~The first is the challenge of selecting an appropriate $\epsilon$ value. The smaller the value, the closer the representation is to the geometric structure. For example, \autoref{fig:nonmetric:c} appears sufficient for this example, but it is unclear if this is sufficient for all of the data, leading one to perhaps select an even smaller $\epsilon$. (2)~However, the second challenge is that the smaller the $\epsilon$, the longer the computation time for detecting the topological features. \autoref{fig:nonmetric_exp:compute} shows that as $\epsilon$ is divided in half, the compute time grows exponentially. (3)~The third related challenge is that the smaller $\epsilon$, the greater the number of topological features generated. \autoref{fig:nonmetric_exp:features} shows this extreme growth. To make matters worse, the vast majority of these features are \textit{topological noise} with very low persistence. In other words, they do not contribute to our understanding of the shape of the face.

\subsection{A Non-metric Variant of Persistent Homology}
\label{sec:tda:nonmetric}

We instead use a novel modification to the persistent homology calculation to utilize this connectivity as follows. Instead of considering 83 landmark points, we consider the relationship between 81 landmark edges formed by the known connectivity of the landmark points (see \autoref{fig:nonmetric:f}). We calculate a distance matrix representation of the landmark edges, where the distance is the \textit{shortest Euclidean distance between line segments}. Finally, we run persistent homology calculations on this distance matrix.

One immediate question should be the appropriateness of this configuration for persistent homology calculations, particularly considering that this representation breaks two important axioms of a metric space, namely the identity of indiscernibles and the triangle inequality. Fortunately, persistent homology calculations themselves do not explicitly require a metric space---they have a weaker requirement of inclusion~\cite{EdelsbrunnerHarer2008}. In other words, as long as in the filtration $R(r_i)\subset R(r_{i+1})$, the calculation can proceed. 

The challenge is that the Rips complex does require that the underlying space is metric. We define a new non-metric Rips complex that satisfies the inclusion property, where: 0-simplicies, representing landmark edges, are present for all values of $r$; 1-simplicies appear when $r$ is \textit{strictly greater than} the non-metric distance between a pair of 0-simplicies; and 2-simplices appear when $r$ is \textit{strictly greater than} all of the non-metric distances of the three related 1-simplicies. Fortunately, this definition is similar enough to the standard Rips complex that careful ordering of inclusions (i.e., observing the strictly greater than cases) in the filtration allows us to utilize conventional persistent homology tools on our non-metric distances.

\autoref{fig:nonmetric:f} shows the landmark edges and the persistence diagram of the associated $\Hgroup_1$ features. Our non-metric approach overcomes all three limitations of supersampling. (1)~The result is very similar to the output of the supersampling in \autoref{fig:nonmetric:e} without the need for specifying any $\epsilon$ parameter. (2)~Furthermore, \autoref{fig:nonmetric_exp:compute} shows that the compute time for our non-metric approach is approximately the same as that of the original 83 landmark points. (3)~Finally, \autoref{fig:nonmetric_exp:features} shows the number of topological features output is small (i.e., we avoid outputting extraneous topological noise).

\section{Comparing Facial Pose Topology}

Thus far, we have introduced a method for extracting the topological features from a single facial pose. We now describe how we compare the topology of multiple facial poses. We start by describing the notion of topological distance between persistence diagrams, which serves as a pairwise dissimilarity between them (see \autoref{sec:tda:topoDist}). Next, we discuss how dimension reduction is used on all pairwise dissimilarities to cluster, compare, and summarize changes in topology (see \autoref{sec:tda:dimRed}).

\subsection{Dissimilarity Between Poses}
\label{sec:tda:topoDist}

Once persistence diagrams are calculated, we wish to explore the relationship between them by performing pairwise comparisons of features of the persistence diagrams. This type of pairwise comparison is commonly performed using bottleneck or Wasserstein distance~\cite{EdelsbrunnerHarer2010}. \textit{Intuitively speaking, these measures find the best match between the features of two persistence diagrams and report the topological feature of the largest distortion, in the case of bottleneck distance, or the average topological distortion, in the case of Wasserstein distance.}

Technically speaking, consider two persistence diagrams, $X$ and $Y$, let $\eta$ be a bijection, with all diagonal points, $(x,x)$, added for infinite cardinality~\cite{kerber2017geometry}. The bottleneck distance is $W_{\infty}(X,Y) = \inf_{\eta: X \rightarrow Y} \sup_{x \in X} \left\lVert x-\eta(x) \right\rVert_\infty,$ and the 1-Wasserstein distance, which we use, is $W_1(X,Y) = \inf_{\eta:X \rightarrow Y}  \Sigma_{x\in X} \left\lVert x-\eta(x) \right\rVert_\infty.$ Our implementation computes the bottleneck and 1-Wasserstein distance for $\Hgroup_0$ and $\Hgroup_1$ features separately, and combines the results. In other words, for bottleneck, $\overline{W_{\infty}}(X,Y) = max( W_{\infty}(X_{\Hgroup_0},Y_{\Hgroup_0}), W_{\infty}(X_{\Hgroup_1},Y_{\Hgroup_1}) )$, and for 1-Wasserstein, $\overline{W_1}(X,Y) = W_1(X_{\Hgroup_0},Y_{\Hgroup_0}) + W_1(X_{\Hgroup_1},Y_{\Hgroup_1})$.

\subsection{Summarizing Topological Dissimilarity}
\label{sec:tda:dimRed}

Once the set of all persistence diagrams is calculated, we explore the relationship between them by calculating all pairwise dissimilarities between poses, forming a dissimilarity matrix representing all of the topological variations between facial poses. However, a dissimilarity matrix, such as this, is difficult to explore directly. We investigated several options to represent and evaluate the relationship between different facial poses and emotions. Importantly, each technique preserves a different aspect of the dissimilarity matrix, providing different perspectives on the data.

The first approach we used is 1D relative distance. In this approach, a keyframe or focal pose is selected by the user. All other facial poses are positioned by their relative distance (i.e., pairwise distance) to that keyframe. \textit{Relative distance perfectly preserves the relationship between the keyframe and all other frames. It does not, however, provide information about the relationship between other pairs of frames.}

Next, we consider two dimension reduction techniques, with each using the pairwise dissimilarity matrix directly. We first consider Multidimensional Scaling (MDS)~\cite{kruskal1964multidimensional}, which \textit{tries to preserve pairwise distances between the topology of poses}. Second, we use t-SNE~\cite{maaten2008visualizing} and UMAP~\cite{mcinnes2018umap}, which attempt to \textit{preserve the clustering structure by considering a local neighbor}. Both t-SNE and UMAP contain hyperparameters that can impact the structures visible to the user. We have performed a structured evaluation of various hyperparameters and found that the structures visible in our results are, by-and-large, stable across a wide variety of parameter values (see our supplement for an example). Therefore, we use the default parameters in our evaluation. To measure the dimension reduction quality for all methods, we calculate the goodness-of-fit using the Spearman rank correlation of the Shepard diagram (denoted in the lower right of images as \textit{Rank}). 

Note that \textit{none of these approaches directly consider time}. Nevertheless, the temporal components of the data are presented in the visualization when relevant.

\setstretch{0.98}

\section{Visualization}
\label{sec:vis}

\begin{figure}[!b]
    \centering
    \includegraphics[width=0.975\linewidth]{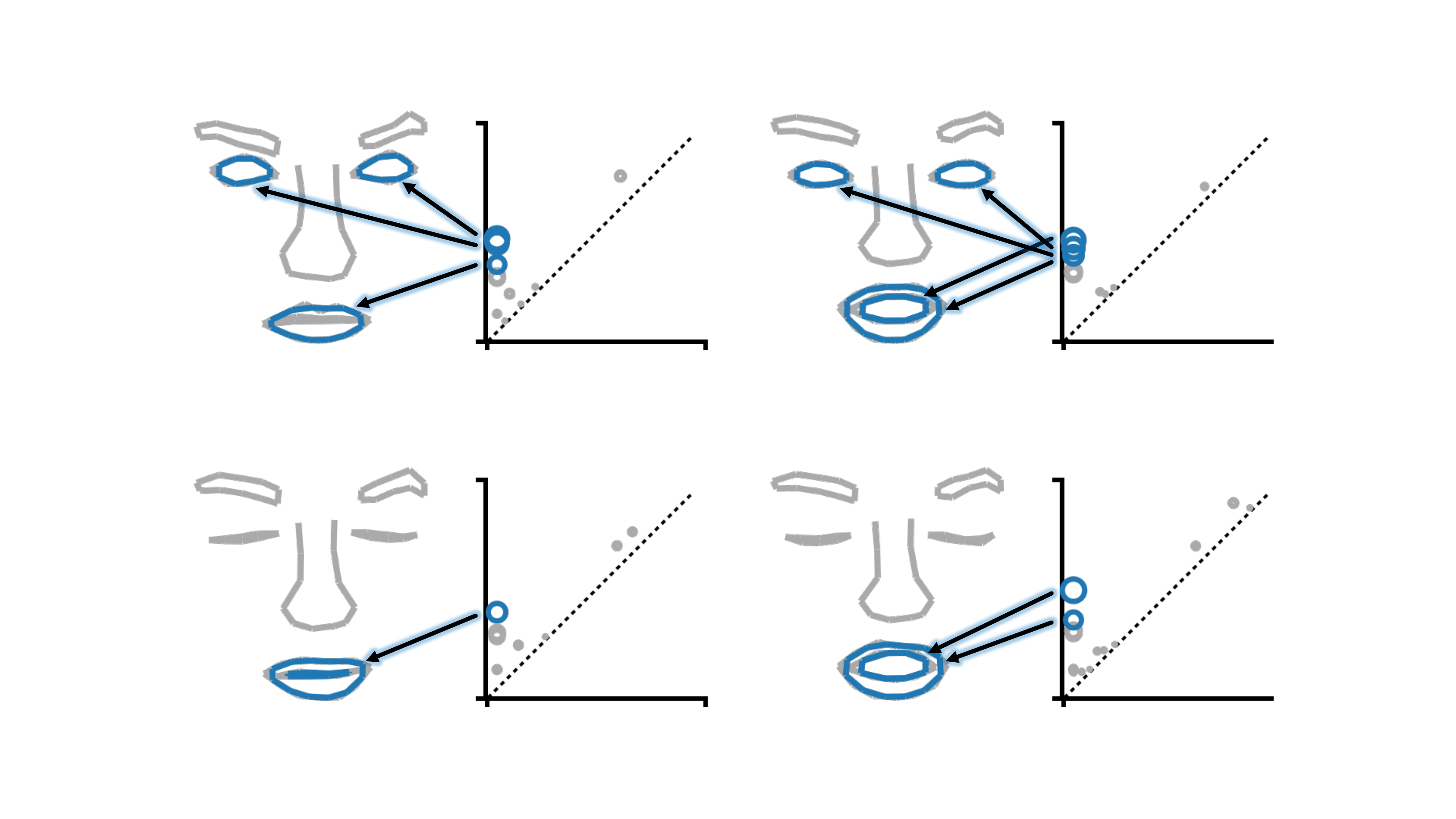}
    \begin{minipage}[t]{1\linewidth}
    \centering
    \vspace{-97pt}
    \subfigure[Eyes Open/Mouth Closed]{\hspace{100pt}}
    \hspace{20pt}
    \subfigure[Eyes Open/Mouth Open]{\hspace{100pt}}
    
    \vspace{58pt}
    \subfigure[Eyes Closed/Mouth Closed]{\hspace{100pt}}
    \hspace{20pt}
    \subfigure[Eyes Closed/Mouth Open]{\hspace{100pt}}
    \end{minipage}
    \caption{Illustration of the explainability of our TDA-based approach using four poses from F001 \surprise. For each, the persistence diagrams are shown (right), along with the highest persistence representative cycles (left). The number and persistence of features explains the difference between poses, while the representative cycles explain which landmarks generated those topological features.}
    \label{fig:representative_ex}
\end{figure}

To examine the topological structure of facial landmark data, we built a visualization (see \autoref{fig:teaser} and \autoref{fig:representative_ex}) with the following design criteria:

\vspace{4pt}
\begin{description}[noitemsep,nolistsep]
    \setlength\itemsep{4pt}
    \item \hspace{-5pt}\labeltext{[D1]}{D:plot} provide multiple ways to evaluate temporal and non-temporal aspects of the data (e.g., animated, static, and non-temporal visualizations);   
    \item \hspace{-5pt}\labeltext{[D2]}{D:dr} provide multiple conditional perspectives (e.g., bottleneck vs.\ Wasserstein, MDS vs.\ t-SNE vs.\ UMAP, etc.) on the topology;
    \item \hspace{-5pt}\labeltext{[D3]}{D:comp} allow comparison of data between two or more emotions;
    \item \hspace{-5pt}\labeltext{[D4]}{D:sub} allow for investigating subsets of landmarks; and
    \item \hspace{-5pt}\labeltext{[D5]}{D:detail} provide direct explanations for the topological differences between facial poses.
\end{description}
\vspace{5pt}

\para{Small Multiples (\autoref{fig:teaser} left)}
Our interface features a small multiples display for comparing different data conditions. The interface features a comparison between two conditions, including comparing two subjects, bottleneck vs.\ Wasserstein distance, metric vs.\ non-metric topology, t-SNE vs.\ MDS, etc.~\ref{D:dr}. The interface further divides each column into comparisons of different subsets of facial features, including full face, eyes+nose, mouth+nose, and eyebrows+nose~\ref{D:sub}. Finally, each row represents one of the six main emotions, \anger, \disgust, \fear, \happiness, \sadness, and \surprise. The user selects the data for the embedding by selecting a column and enabling/disabling specific emotions of interest by selecting rows~\ref{D:comp}. Each small multiple is shown using the visualization modality chosen in the settings at the bottom.

\para{Embedding Graph (\autoref{fig:teaser} top right)}
The primary visualization tool in our approach is the embedding graph, which is either a line chart or scatterplot representation of time-varying topological data for the selected emotions~\ref{D:plot}.
For the line chart, time is plotted horizontally, while the 1D \textit{relative distance} is plotted vertically (see \autoref{fig:au_example} top)~\ref{D:dr}. Each selected emotion is overlaid for time-dependent comparison~\ref{D:comp}. The keyframe is user-selectable, and the visualization updates as the keyframe is modified.
The scatterplot representation uses 2D dimension reduction for both the horizontal and vertical axes. The choice of MDS, t-SNE, or UMAP is provided for the user. The data are either shown as points or connected via a path (see \autoref{fig:expressiveness_tsne:female}) if the user wants a temporal context~\ref{D:plot}. 

The plot is also interactive---selecting a point updates the time-index used in other visualizations, e.g., the 3D landmarks.

\para{3D Landmarks (\autoref{fig:teaser} lower right)}
The 3D landmarks represent the data of the current time-index for the respective emotion~\ref{D:plot}. The faces are placed side-by-side for comparison~\ref{D:comp}\ref{D:detail}. The sliders beneath each can be used to animate or adjust their time-index~\ref{D:plot}.

\para{Persistence Diagrams (\autoref{fig:representative_ex})}
When additional details about a given facial pose are desired, the persistence diagram captured by persistent homology is represented by a scatterplot~\ref{D:detail}. The persistence diagram plots feature birth horizontally and death vertically. In this context, $\Hgroup_0$ features are represented as solid squares, and $\Hgroup_1$ features are represented as rings. The size of each element is proportional to its persistence (i.e., importance). Furthermore, the distance from the dashed diagonal to a feature is also a measure of persistence.

\para{Representative Components and Cycles (\autoref{fig:representative_ex})}
A byproduct of the calculation of persistent homology is a structure known as generators, which are the landmark elements that generated a particular topological feature.
For $\Hgroup_0$, the generators are the 0-simplices representing the joining of two components. For $\Hgroup_1$ features, the data are output in the form of a representative cycle\footnote{For reasons outside the scope of this paper, there is no single generator for a cycle but instead a class of generators. A representative cycle, which is a byproduct of a process called boundary matrix reduction, is output instead~\cite{edelsbrunner2000topological}.}. Each topological feature is associated with a generator that we use to \textit{identify what input data generated that topological feature}, with a general focus on the high persistence features in the data~\ref{D:detail}.

\setstretch{0.97}

\section{Evaluation}
\label{sec:eval}

We evaluate our approach by first performing a detailed evaluation of two individuals---one female (`F001') and one male (`M001') from the BU4DFE expression dataset~\cite{yin126high}. We then evaluate the ability of our approach to differentiate individuals using the entire dataset of 101 subjects. Each of these individuals has approximately 600 facial poses (6 emotions $\times$ $\sim100$ frames per emotion). Since the data provided are large and time-varying, our approach allows conditional observation of the topology of emotions based upon individuals, emotions, selected subset of facial features (full face, eyes+nose, mouth+nose, eyebrows+nose), topological dissimilarity, and dimension reduction technique. Our evaluation looks at how these conditional comparisons can be matched to known phenomena in affective computing. We note that one of the coauthors of this paper, Shaun Canavan, is a researcher in affective computing and provided detailed feedback at every stage of the design.

\subsection{Implementation and Performance}

We implemented our approach using Python for data management, non-metric distance, and dimension reduction calculations, ripser~\cite{bauer2019ripser} for persistent homology calculations, Hera~\cite{kerber2017geometry} for topological distance, and D3.js for the user interface. Persistent homology and topological dissimilarity are pre-calculated for all combinations of landmark subsets. Dimension reduction is performed at run-time, taking at most a few seconds; as this data is calculated, it is also stored in a short-term cache to improve performance. The user interface is interactive. Our source code is available at \url{https://github.com/USFDataVisualization/AffectiveTDA}.

We evaluated the computational performance of the persistent homology and bottleneck and Wasserstein dissimilarity matrix calculations for F001 and M001 in \autoref{tab:performance}. The calculations were performed on a Linux workstation with a 3.40GHz Intel i7-6700 CPU and 48 GB of RAM. In this table, we compare the metric landmark point-based approach and our novel non-metric landmark edge-based approach. Comparing persistent homology calculations, our non-metric approach took approximately twice as long as the metric approach. This is entirely attributable to the extra cost of calculating segment-segment distance (instead of point-point distance). The performance benefit of the non-metric approach comes with the calculation of the dissimilarity matrices, which saw a $10x-15x$ speedup over the metric approach. This is attributable to the reduced number of noise features created by the non-metric approach, as described in \autoref{sec:tda:nonmetric}. Overall, our approach saw a speedup of $\sim7.5x$.

\renewcommand{\arraystretch}{1.3}

\begin{table}[!ht]
    \centering
    \caption{Computation time for metric (M) and non-metric (NM) approaches to extract persistent homology features and calculate the dissimilarity matrix for all frames from each subject, including subsets of facial landmarks (full face, eyes+nose, mouth+nose, eyebrows+nose). The number of landmarks input to each method is similar, 83 for full face metric and 81 for non-metric. In addition, we show the average number of $\Hgroup_0$ and $\Hgroup_1$ features generated per frame.}
    \label{tab:performance}
    \begin{adjustbox}{width=0.975\linewidth}
        \begin{tabular}{c@{\hspace{2pt}}c|c|c||c|c|c||c|}
             & & $|\Hgroup_0|$ & $|\Hgroup_1|$ & Persistent & \multicolumn{2}{c||}{Topological Distance} & \multirow{2}{*}{Total} \\
             &  & (avg) & (avg) & Homology & Bottle. & Wasser. & \\
             \cline{3-8}
            Female & M     & 43.7 & 12.8 & \ \ 84.3 $s$ & 1833.8 $s$ & 1851.7 $s$ & 3769.8 $s$ \\
            \cline{2-8}
            (F001) & NM  & 4.3 & 6.3 & 188.5 $s$ & \ \ 189.8 $s$  & \ \ 133.7 $s$ & \ \ 512.1 $s$ \\
            \hline
            Male & M       & 43.7  & 13.4 & \ \ 91.6 $s$ & 1877.6 $s$  & 2039.2 $s$ & 4008.4 $s$ \\
            \cline{2-8}
            (M001) & NM  & 4.3 & 6.5 & 188.1 $s$ & \ \ 207.5  $s$ & \ \ 136.4 $s$ & \ \ 532.0  $s$ \\
        \end{tabular}
    \end{adjustbox}
\end{table}

\renewcommand{\arraystretch}{1}

\subsection{Relative Distance Topology and Action Units (AUs)}
\label{sec:eval:AUs}

In affective computing, there are various approaches for recognizing expressions, as detailed in \autoref{sec:expRecRW}. One promising approach is the use of action units (AUs)~\cite{FACS}, which are facial muscle movements linked to expression. AUs are represented as an intensity from $[0,5]$, where 0 is inactive, and $>0$ is an active AU, with higher values representing more intense movement. Specific configurations of active AUs have been shown to be useful for recognizing facial expressions~\cite{liu2020saanet, lien1998automated, xu2020exploring, lucey2010automatically, sikander2020novel}.

\begin{table}[!b]
    \caption{Action Units (AUs) and their corresponding facial muscle movements as used in our evaluation.}
    \label{table:AUs}
    \centering 
    \begin{adjustbox}{width=0.88\linewidth}
    \begin{tabular}{|c|c|c|}
    %\hline
     \multicolumn{1}{c|}{\textbf{Action Unit}} & \textbf{Facial Muscles} & \multicolumn{1}{c}{\textbf{Description}} \\
     \thickhline
     AU1 & 	Frontalis, pars medialis & Inner eyebrow raise\\
     \hline
     AU2 & Frontalis, pars lateralis & Outer eyebrow raise\\
     \hline
     \multirow{3}{*}{AU4} & {Depressor Glabellae,} & \multirow{3}{*}{Eyebrow lower}\\
        & {Depressor Supercilli,} &  \\
        & {Currugator} & \\
     \hline
     \multirow{2}{*}{AU5} & {Levator palpebrae} & \multirow{2}{*}{Upper eyelid raise} \\
        & {superioris} &  \\
     \hline
     \multirow{2}{*}{AU6} & {Orbicularis oculi,} & \multirow{2}{*}{Cheek raise}\\
        & {pars orbitalis} & \\
     \hline
     \multirow{2}{*}{AU7} & {Orbicularis oculi,} & \multirow{2}{*}{Eyelid tighten} \\
        &  {pars palpebralis} & \\
     \hline
     \multirow{2}{*}{AU9} & {Levator labii superioris} & \multirow{2}{*}{Nose wrinkle} \\
        & {alaquae nasi} & \\
     \hline
     \multirow{2}{*}{AU10} & {Levator Labii Superioris,} & \multirow{2}{*}{Upper lip raise} \\
        & {Caput infraorbitalis} & \\
     \hline
     AU12 & Zygomatic Major & Lip corner pull \\
     \hline
     AU14 & Buccinator & Dimple \\
     \hline
     AU15 & Depressor anguli oris & Lip corner depress \\
     \hline
     AU17 & Mentalis & Chin raise \\
     \hline
     AU20 & Risorius  & Lip stretch \\
     \hline
     AU23 & Orbicularis oris & Lip tighten \\
     \hline
     \multirow{3}{*}{AU25} & {Depressor Labii,} & \multirow{3}{*}{Lips part} \\
        &  {Relaxation of Mentalis,} & \\
        & {Orbicularis Oris} & \\
     \hline
     \multirow{2}{*}{AU26}  & {Masetter,} & \multirow{2}{*}{Jaw drop} \\
        &  Temporal/Internal Pterygoid & \\
     \hline
     \multirow{3}{*}{AU45} & {Levator Palpebrae,} & \multirow{3}{*}{Blink} \\
        & {Orbicularis Oculi,} & \\
        & { Pars Palpebralis} &\\
     \hline
    \end{tabular}
    \end{adjustbox}
\end{table}

AUs are generally created in one of two ways. Either an expert manually annotates video frames, or a machine learning algorithm extracts them from the data. While the former is a slow and tedious process, the latter is fast but lacks any explainability in measuring the activity of AUs. We automatically detect 17 AUs (see \autoref{table:AUs}) using the publicly available OpenFace toolkit~\cite{baltruvsaitis2016openface}, which is commonly used in affective computing literature~\cite{uddin2020ICPR}. However, coming from a machine learning model, the extracted AUs lack specific explainability.

We now demonstrate how our topology-based system can explain certain AU features detected by OpenFace by comparing the output of each. Our approach is as follows, since all sequences begin with a neutral pose, we consider relative distance with respect to the first frame of the sequence. We hypothesized that we would observe similar signals in the AUs and their associated facial features using our topology-based approach. \autoref{fig:au_example} shows two examples of this relationship. In \autoref{fig:au_example:a}, we compare the activity of the nose and eyes to AU45 (blink) for the F001 \disgust emotion. The relative distance shows three clear spikes at the same frames as AU45 (approximately frames 28, 52, and 75). However, AU45 does not tell the entire story of the activity that the topology is capturing. 

Instead, we hypothesize that the topology is a combination of multiple AUs. \autoref{fig:au_example:b} shows an example comparing the mouth+nose to AU14 (dimple) and AU25 (lips part) for the F001 \fear emotion. In this case, a linear combination of both AUs seems to capture a more complete picture of the activity represented in the topology. A broader analysis of both subjects, multiple emotions, and multiple facial features, as seen in \autoref{tab:auF001} and \autoref{tab:auM001}, revealed these relationships are widely observable. This confirms our hypothesis of a strong similarity between topology features and AUs.

Nevertheless, there is still not a perfect one-to-one relationship between the topology and AUs. The AUs go through further contextual processing than the topology does, e.g., to separate the activity of AU7 (eyelid tightening) from AU45 (blinking) and other related movements. One challenge with the contextual processing in the state-of-the-art in affective computing is the lack of explainability, which our topology-based approach provides (see \autoref{sec:discussion}).

\newcommand{\auEX}[5]{
    \begin{minipage}[t]{0.292\linewidth}
        \begin{minipage}[t]{\linewidth}
            \hfill
            \begin{minipage}[t]{0.96\linewidth}
                \includegraphics[width=\linewidth]{#3}
            \end{minipage}
            \begin{minipage}[t]{0pt}
                \vspace{#1}
                \hspace{#2}
                \fcolorbox{lightgray}{white}{\includegraphics[width=20pt]{#4}}
            \end{minipage}
        \end{minipage}   
        
        \vspace{1pt}
        \hspace{2pt}\includegraphics[height=0.775in, width=0.945\linewidth]{#5}
    \end{minipage}   
}

\newcommand{\auEXX}[5]{
    \begin{minipage}[t]{0.2275\linewidth}
        \begin{minipage}[t]{\linewidth}
            \hfill
            \begin{minipage}[t]{0.975\linewidth}
                \includegraphics[width=\linewidth]{#3}
            \end{minipage}
            \begin{minipage}[t]{0pt}
                \vspace{#1}
                \hspace{#2}
                \fcolorbox{lightgray}{white}{\includegraphics[width=18pt]{#4}}
            \end{minipage}
        \end{minipage}   
        
        \vspace{1pt}
        \hspace{2pt}\includegraphics[height=0.775in, width=0.945\linewidth]{#5}
    \end{minipage}   
}

\newcommand{\auFIG}[5]{
    \begin{minipage}[t]{0.475\linewidth}
        \begin{minipage}[t]{\linewidth}
            \hfill
            \begin{minipage}[t]{0.96\linewidth}
                \includegraphics[width=\linewidth]{#3}
            \end{minipage}
            \begin{minipage}[t]{0pt}
                \vspace{#1}
                \hspace{#2}
                \fcolorbox{lightgray}{white}{\includegraphics[width=18pt]{#4}}
            \end{minipage}
        \end{minipage}   
        \includegraphics[width=1\linewidth]{#5}
    \end{minipage}   
}

\begin{figure}[!th]
    \centering

    \auFIG{-53pt}{-30pt}{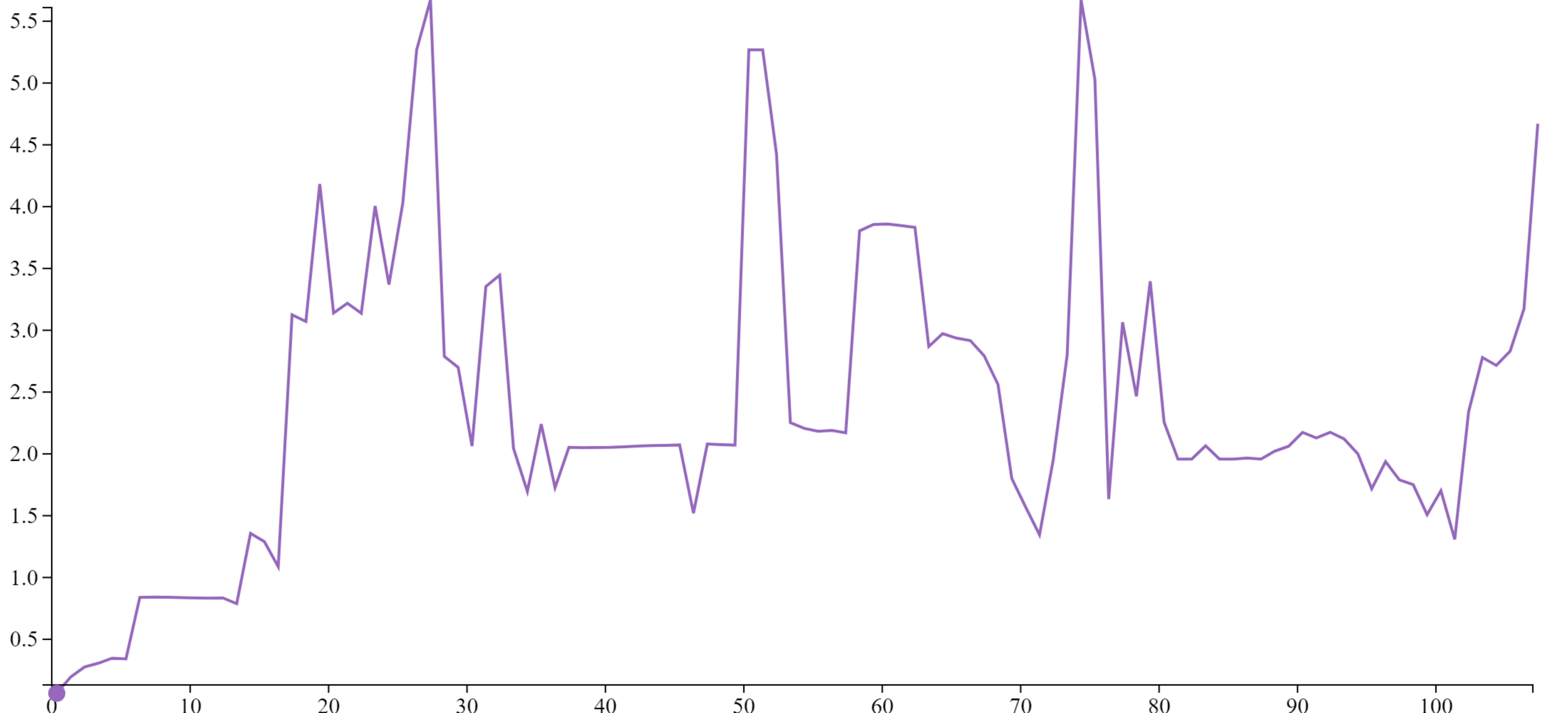}{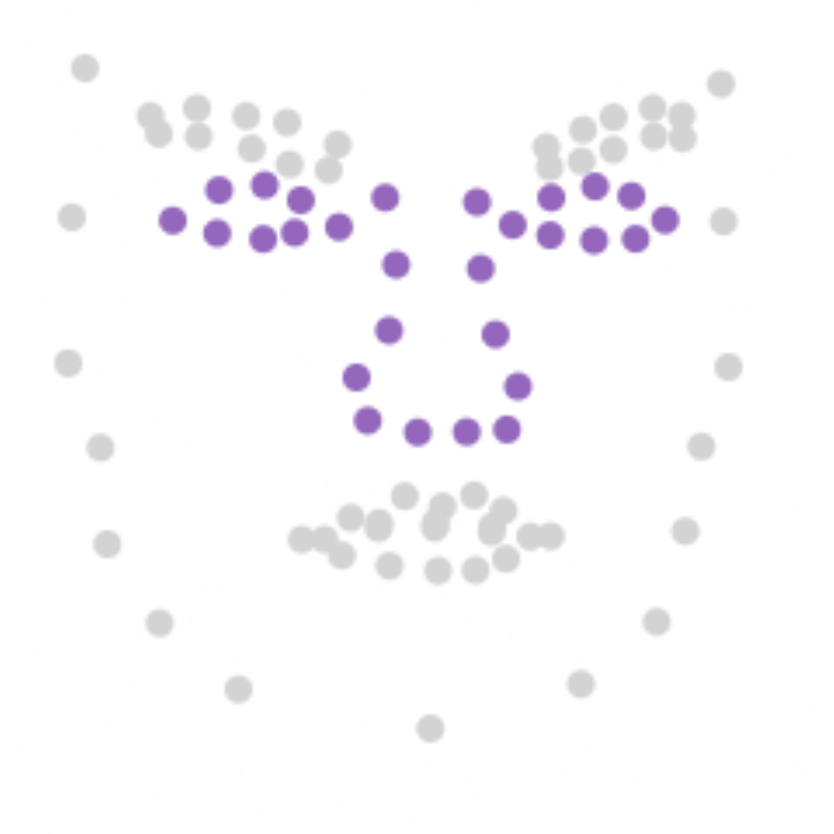}{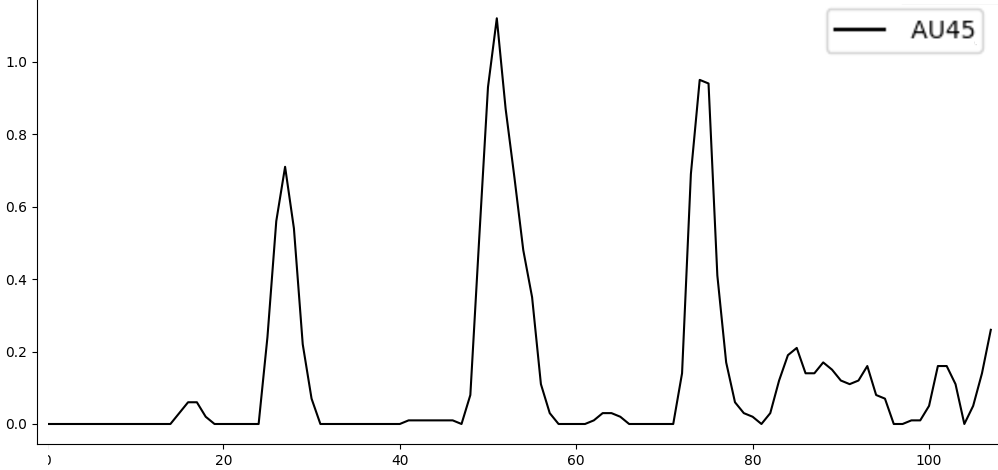}
    \auFIG{-53pt}{-30pt}{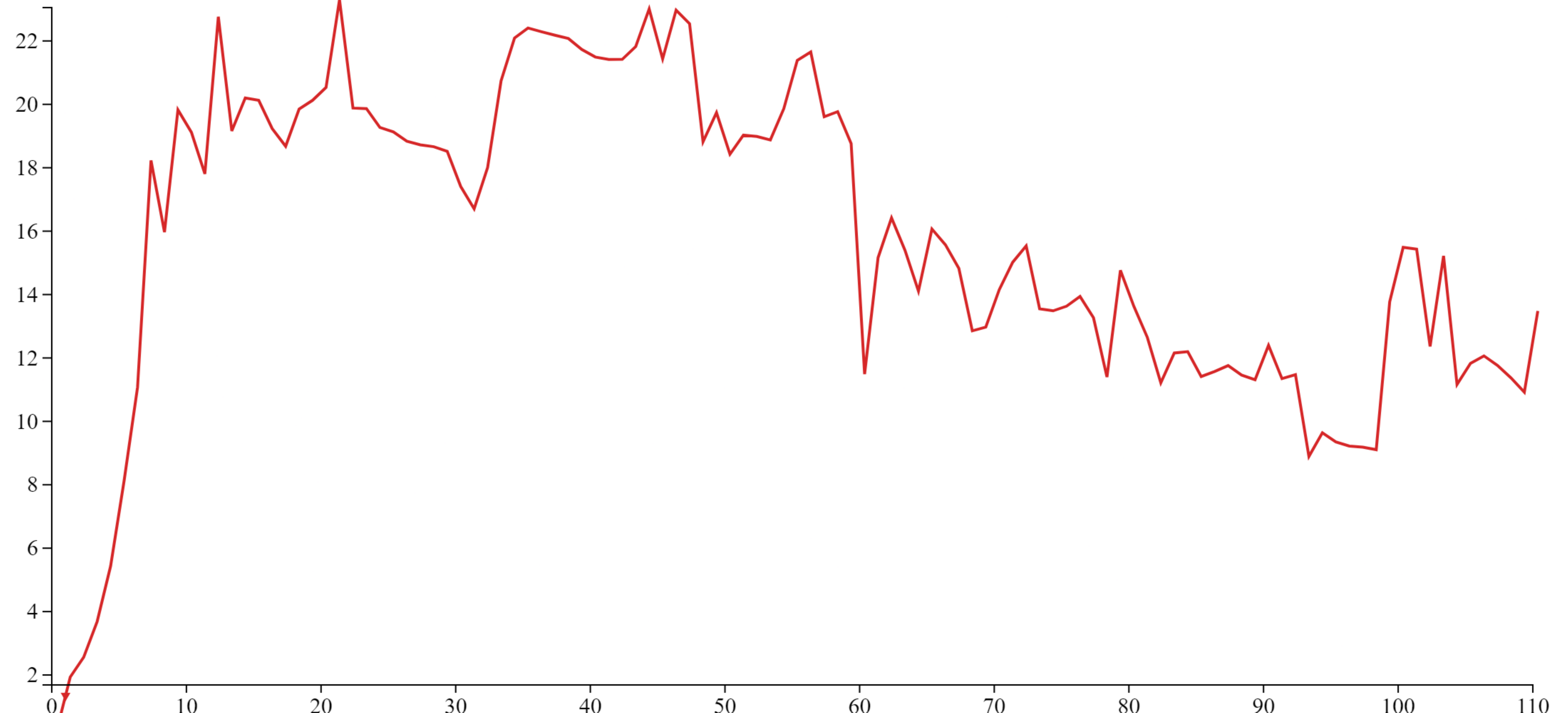}{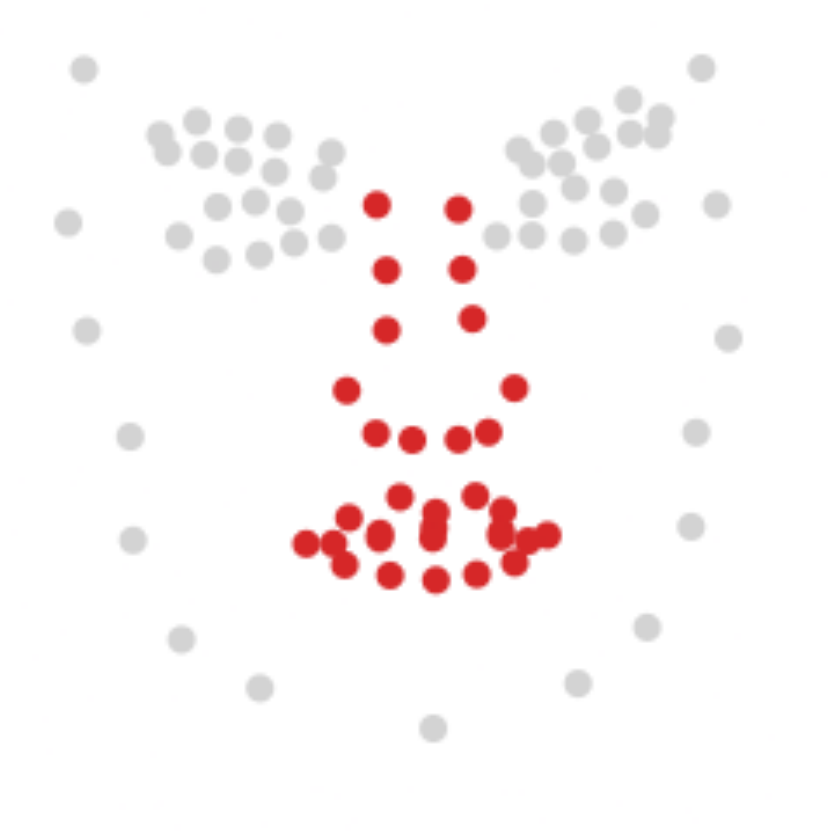}{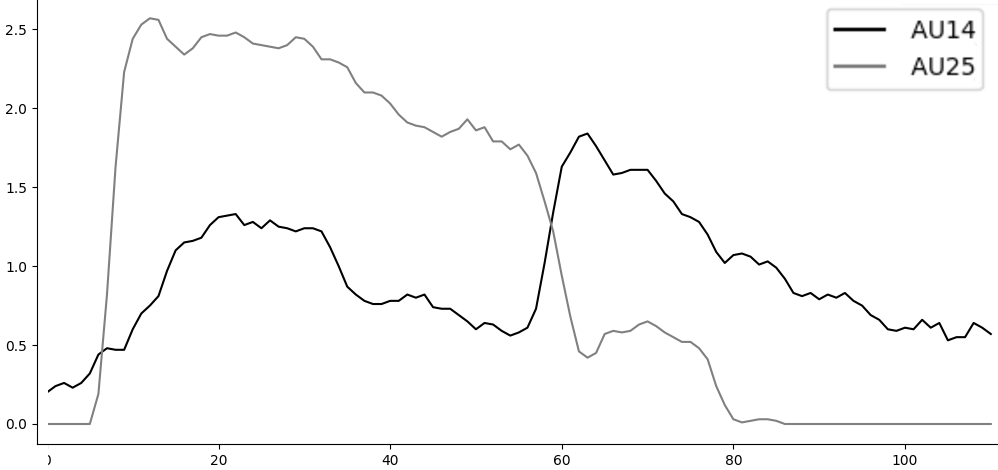}
    
    \vspace{-10pt}
    \hspace{5pt}
    \subfigure[\Disgust Bottleneck Eyes+Nose / AU45\label{fig:au_example:a}]{\hspace{73pt}}
    \hspace{35pt}
    \subfigure[\Fear Wasserstein \newline Mouth+Nose / AU14+25\label{fig:au_example:b}]{\hspace{89pt}}

    \caption{A comparison of relative distance on non-metric topology (top) to Action Units (AUs) (bottom) on F001. The results demonstrate the similarity between the features extracted by the topology and AUs, which are commonly used in affective computing.}
    \label{fig:au_example}
\end{figure}

\begin{figure}[!b]
    \centering
    
    \subfigure[F001 Wasserstein t-SNE\label{fig:expressiveness_tsne:female}]{{\begin{minipage}[t]{0.475\linewidth}
        \includegraphics[trim=515pt 154pt 3pt 56pt, clip, width=\linewidth]{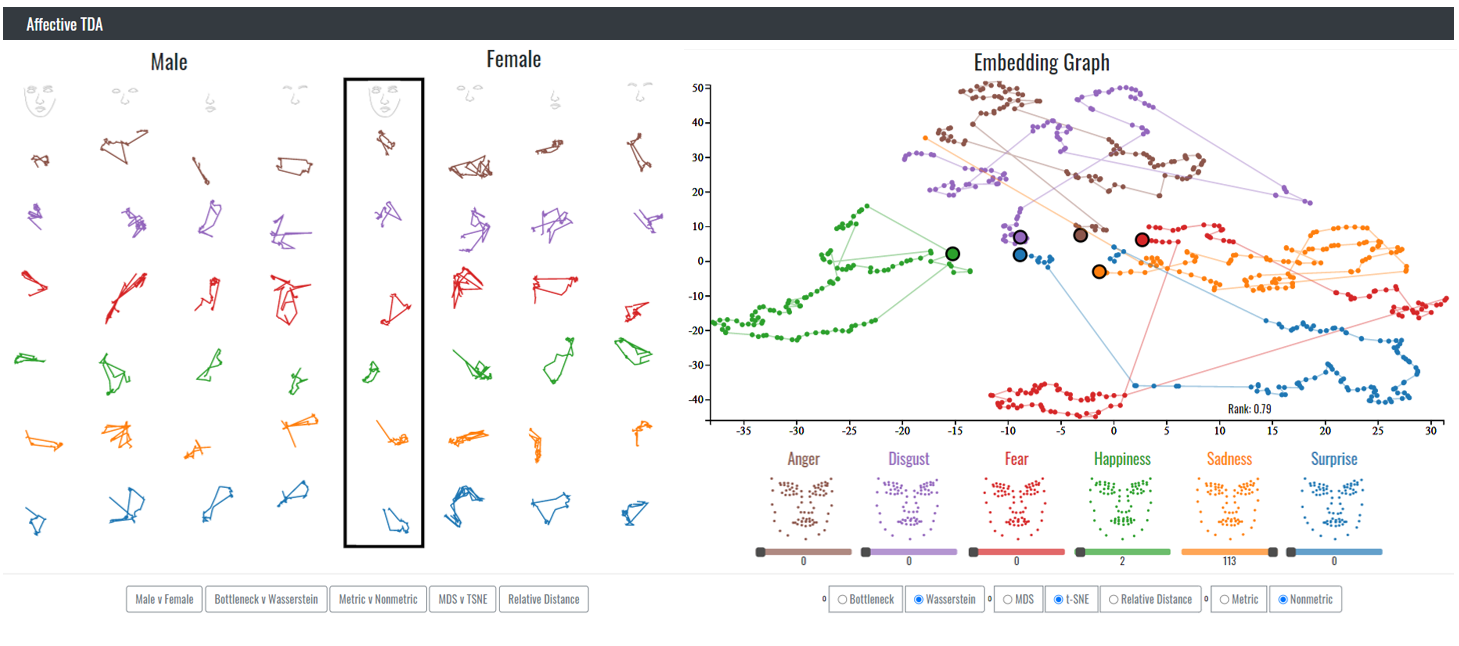}
        \includegraphics[trim=545pt 74pt 33pt 330pt, clip, width=\linewidth]{figures/tsne_expressiveness_female.PNG}
    \end{minipage}
    }}
    \hfill
    \subfigure[M001 Wasserstein t-SNE\label{fig:expressiveness_tsne:male}]{{\begin{minipage}[t]{0.475\linewidth}
        \includegraphics[trim=495pt 147pt 0 51pt, clip, width=\linewidth]{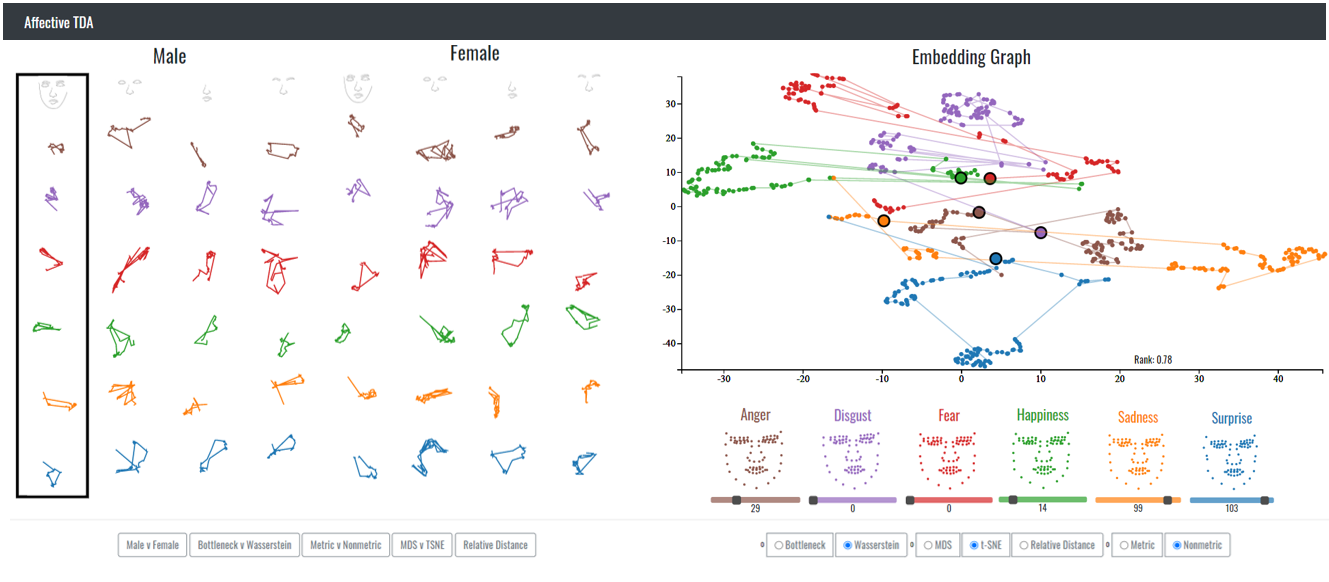}
        \includegraphics[trim=525pt 67pt 30pt 300pt, clip, width=\linewidth]{figures/tsne_expressiveness_male.PNG}
    \end{minipage}
    }}

    \subfigure[M001 Wasserstein MDS\label{fig:expressiveness_mds:male}]{{\includegraphics[width=0.975\linewidth]{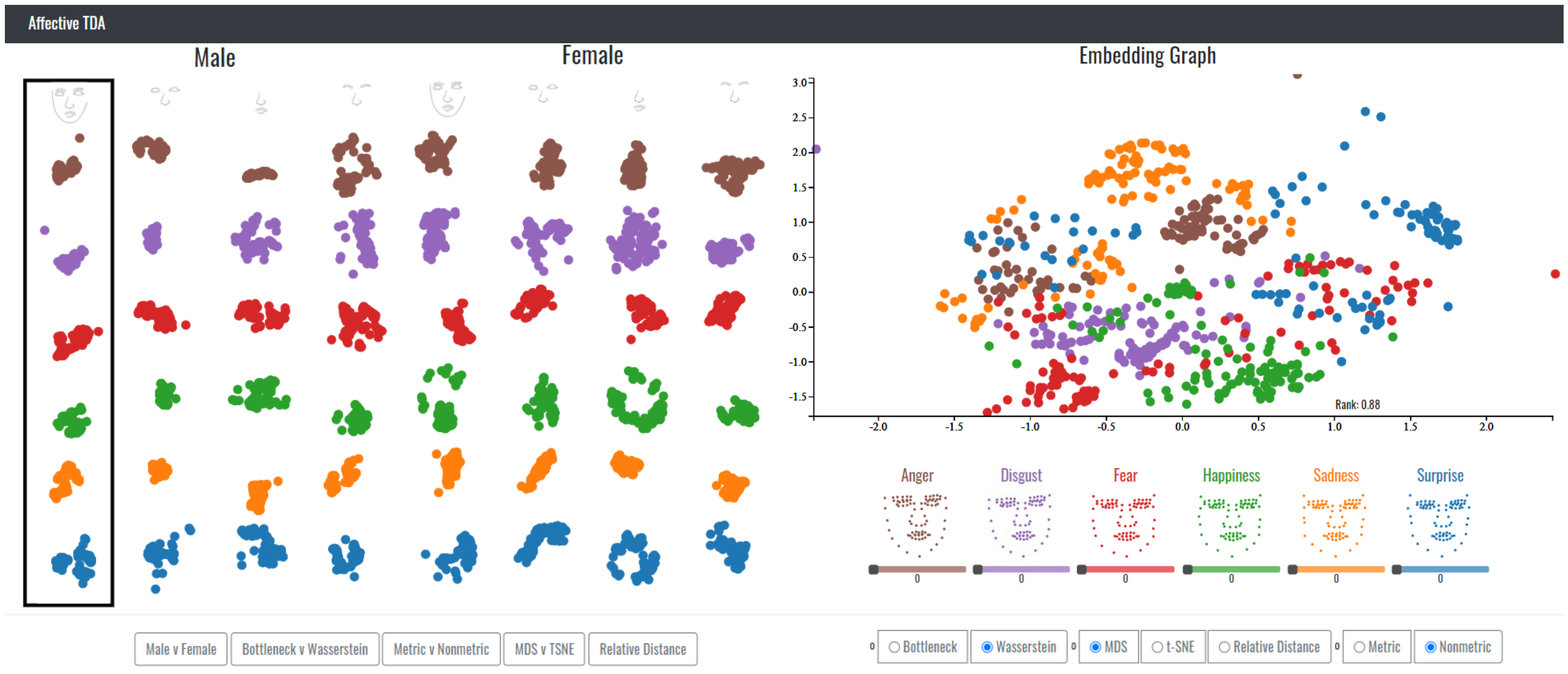}}}

    \caption{Evaluation using (a-b) t-SNE and (c) MDS to determine how effective our non-metric topology-based approach is at differentiating emotions. F001 MDS can be found in \autoref{fig:teaser}.}
    \label{fig:expressiveness}
\end{figure}

\begin{figure*}[!t]
    \centering
    \subfigure[\Anger Bottleneck AU7+45]{
        \auEXX{-55pt}{-30pt}{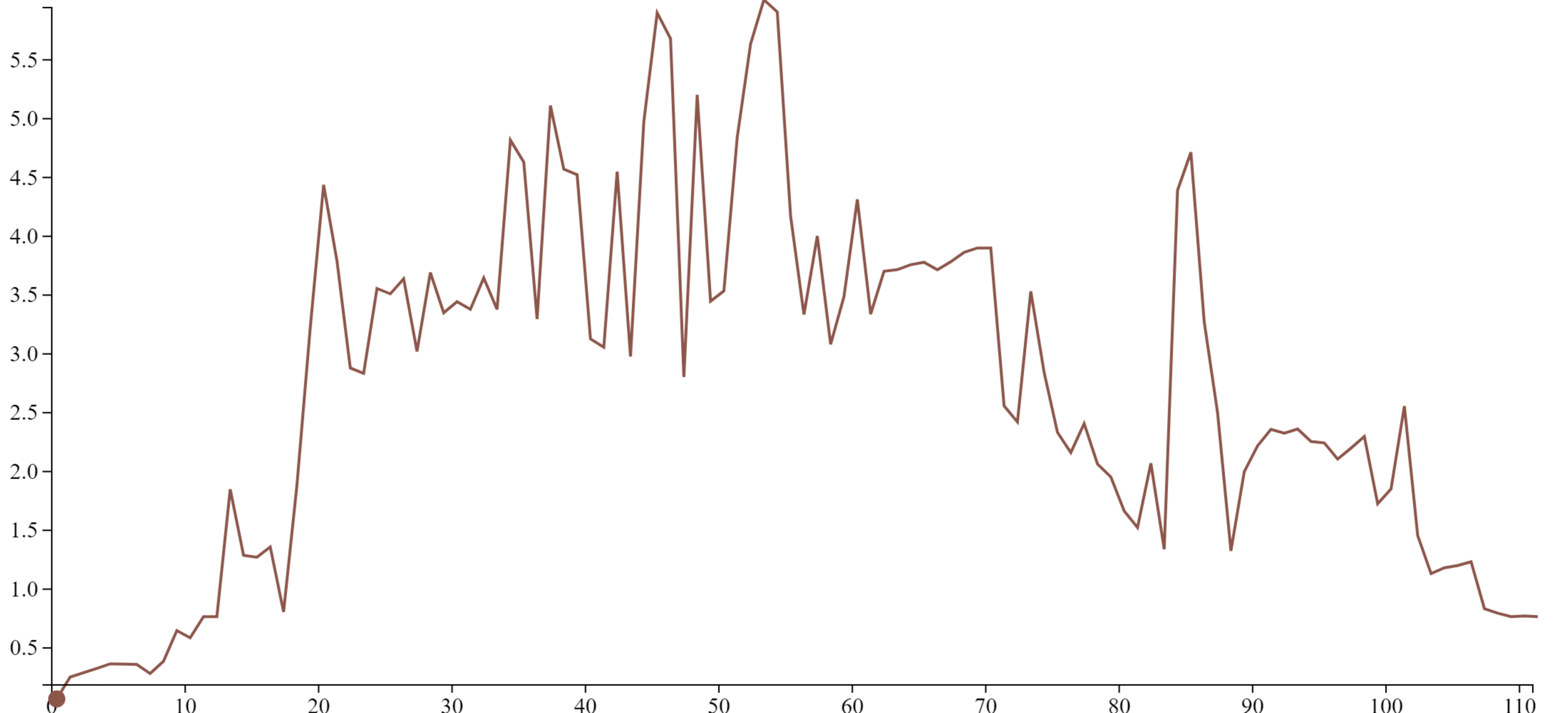}{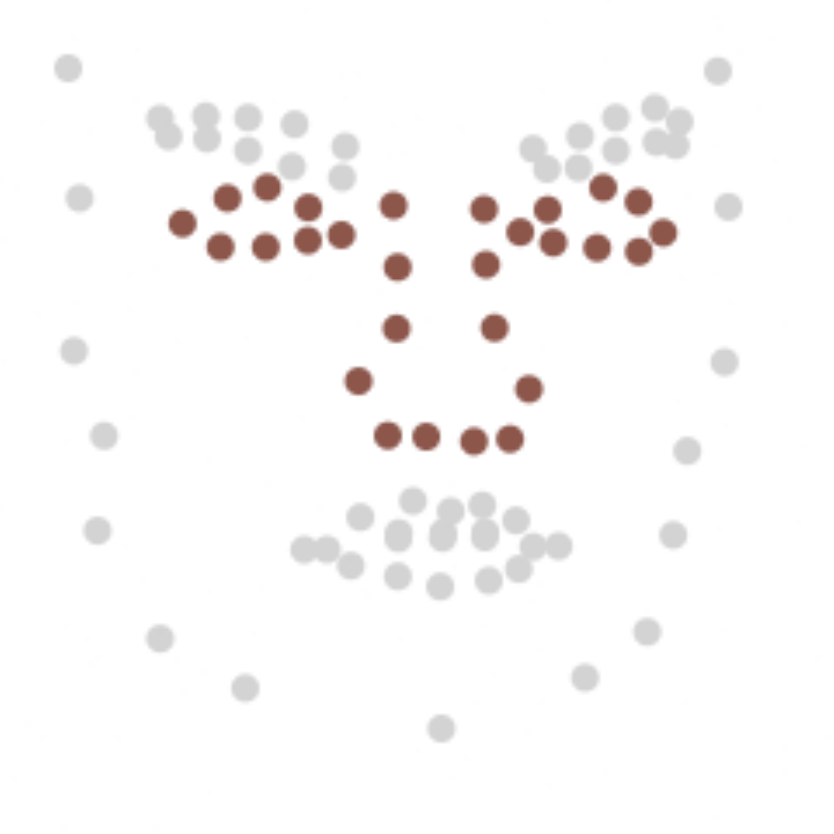}{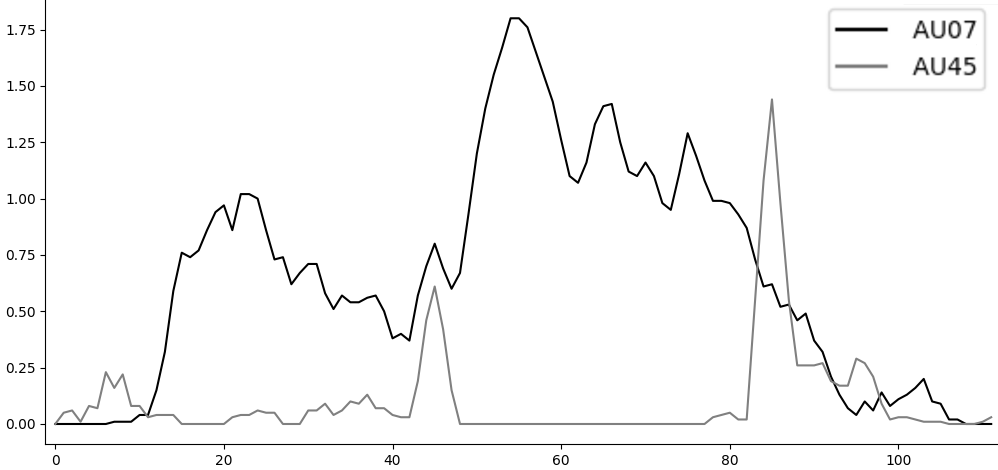}
    }
    \subfigure[\Anger Wasserstein AU4+7]{
        \auEXX{-55pt}{-30pt}{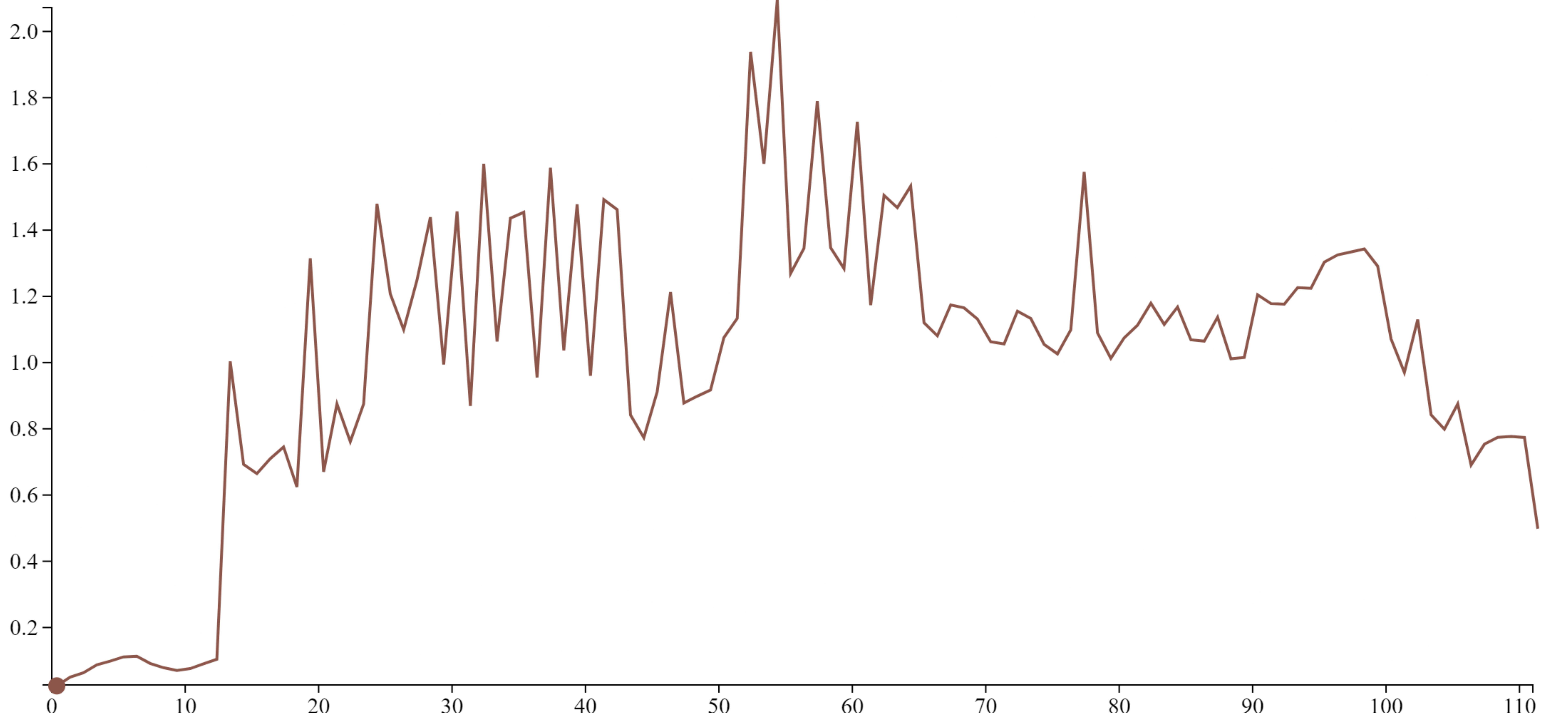}{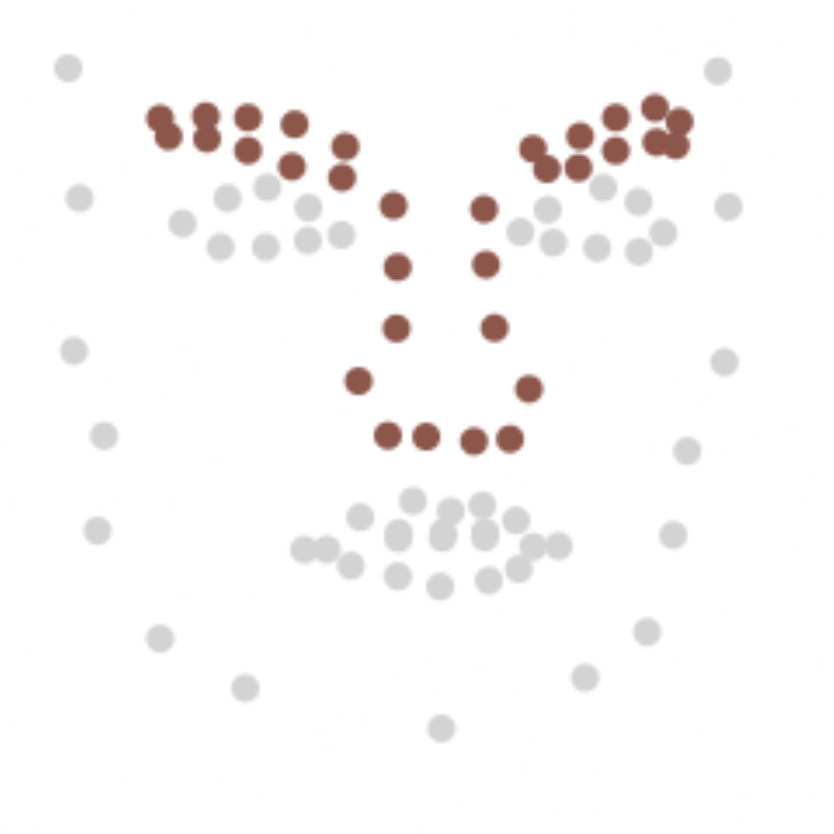}{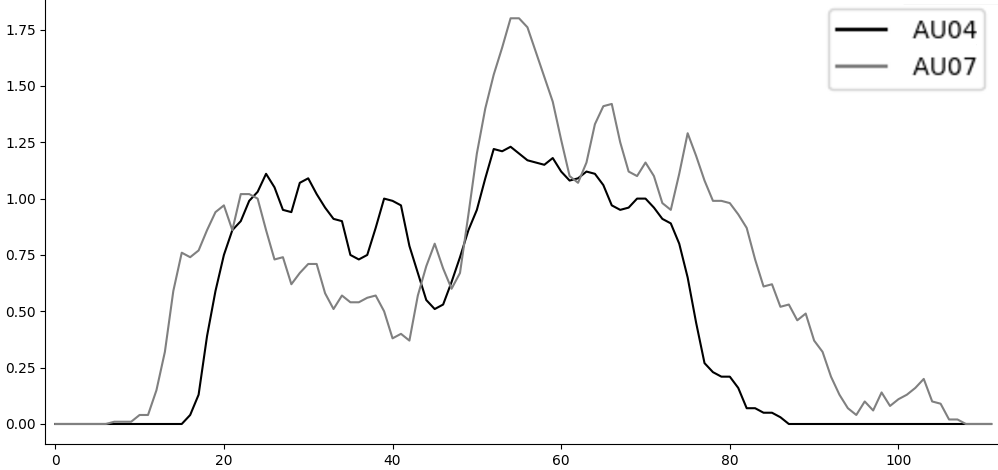}
    }    
    \subfigure[\Fear Wasserstein AU7+45]{
        \auEXX{-55pt}{-30pt}{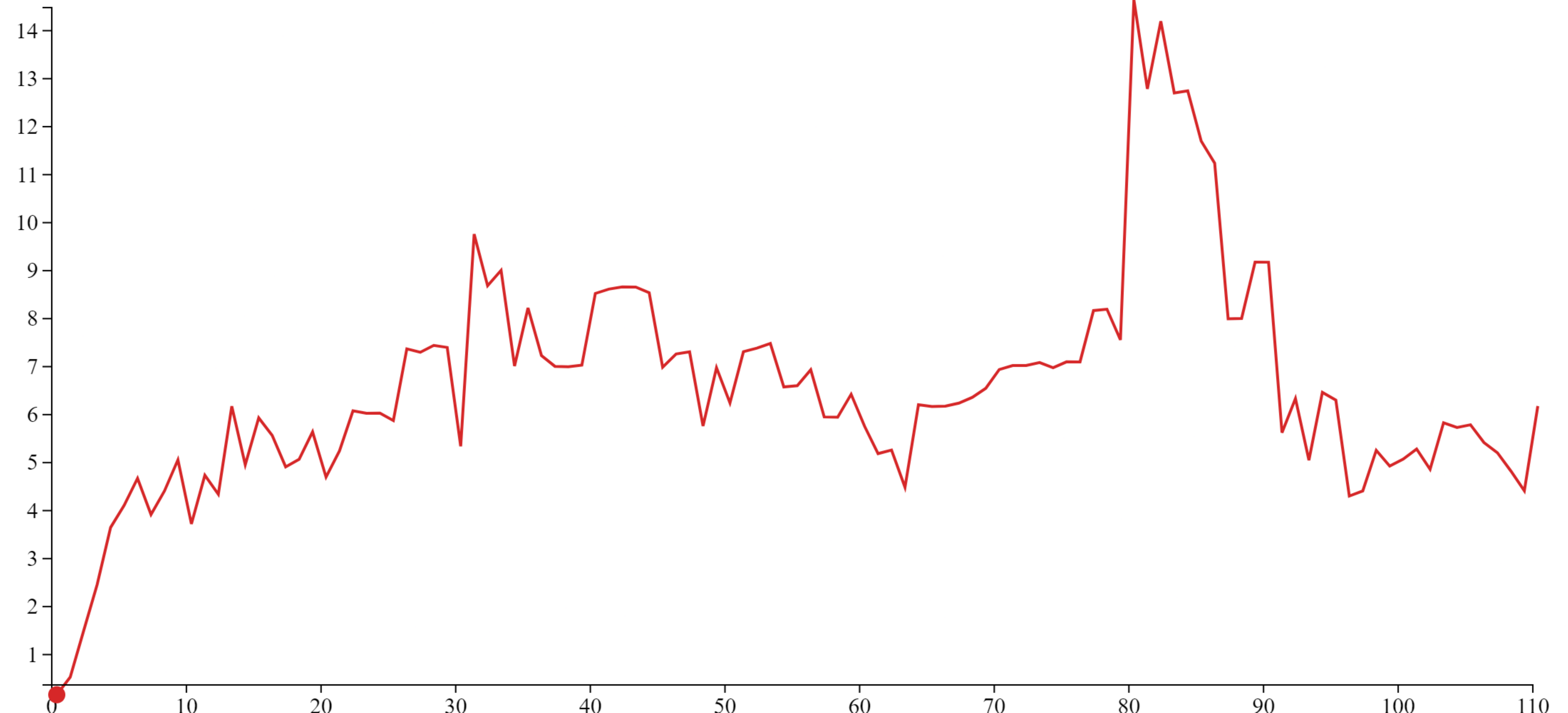}{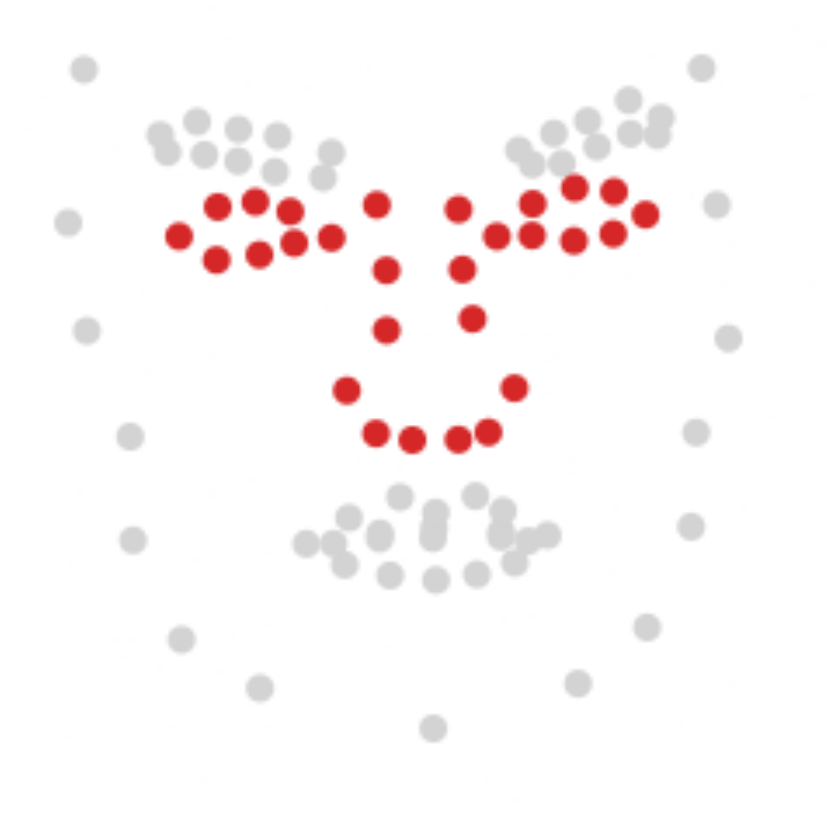}{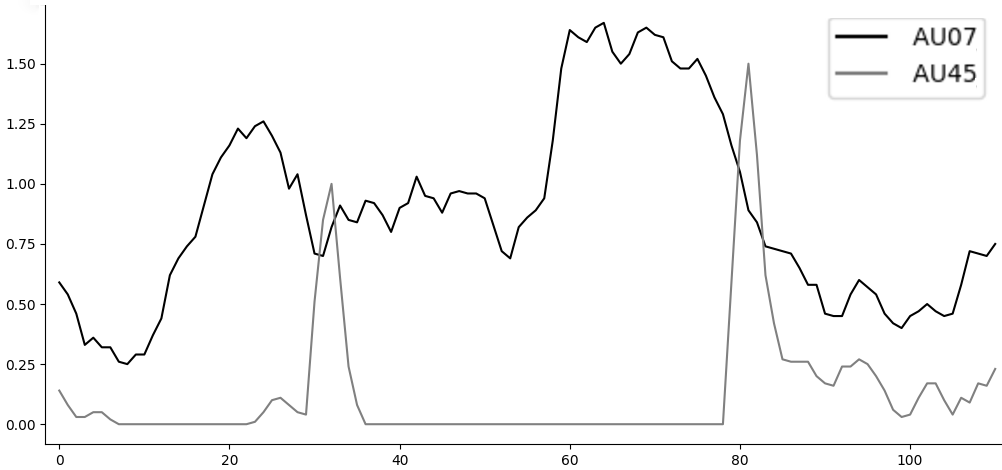}
    }
    \subfigure[\Fear Bottleneck AU1+2+7]{
        \auEXX{-55pt}{-30pt}{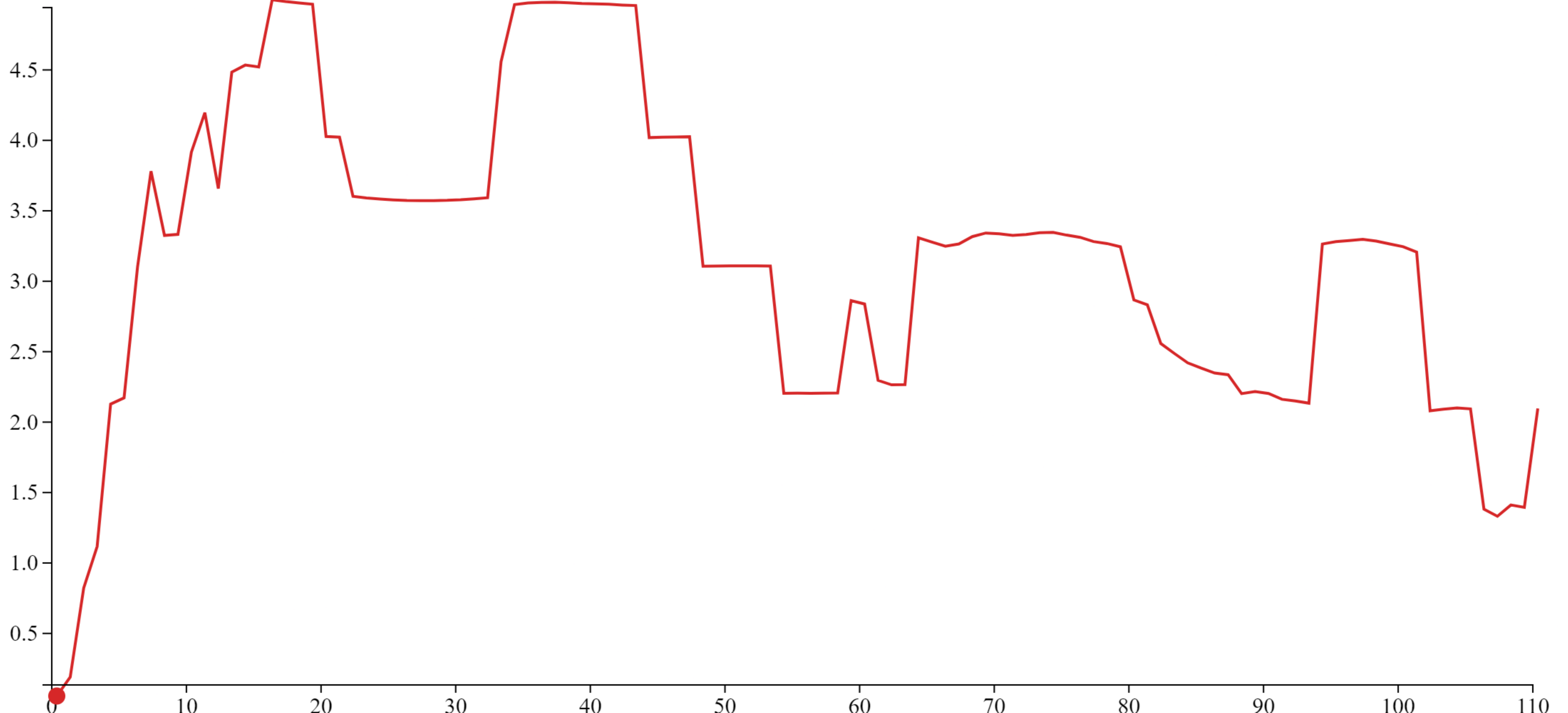}{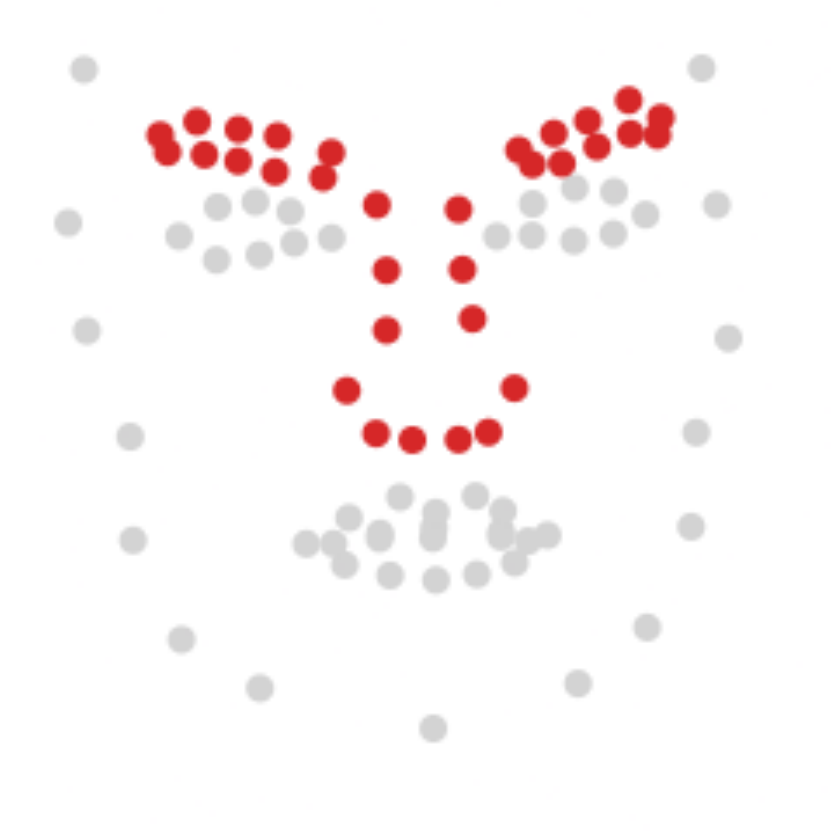}{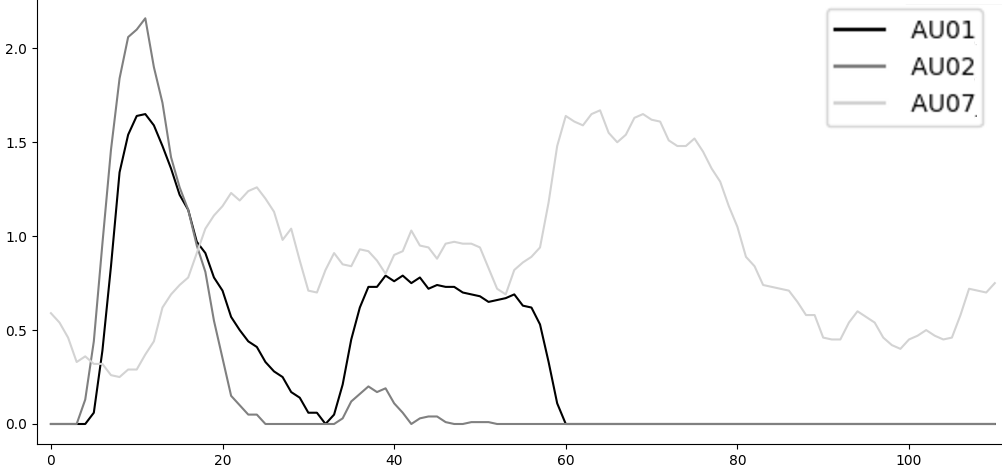}
    }

    \subfigure[\Disgust Bottleneck AU25]{
        \auEXX{-55pt}{-30pt}{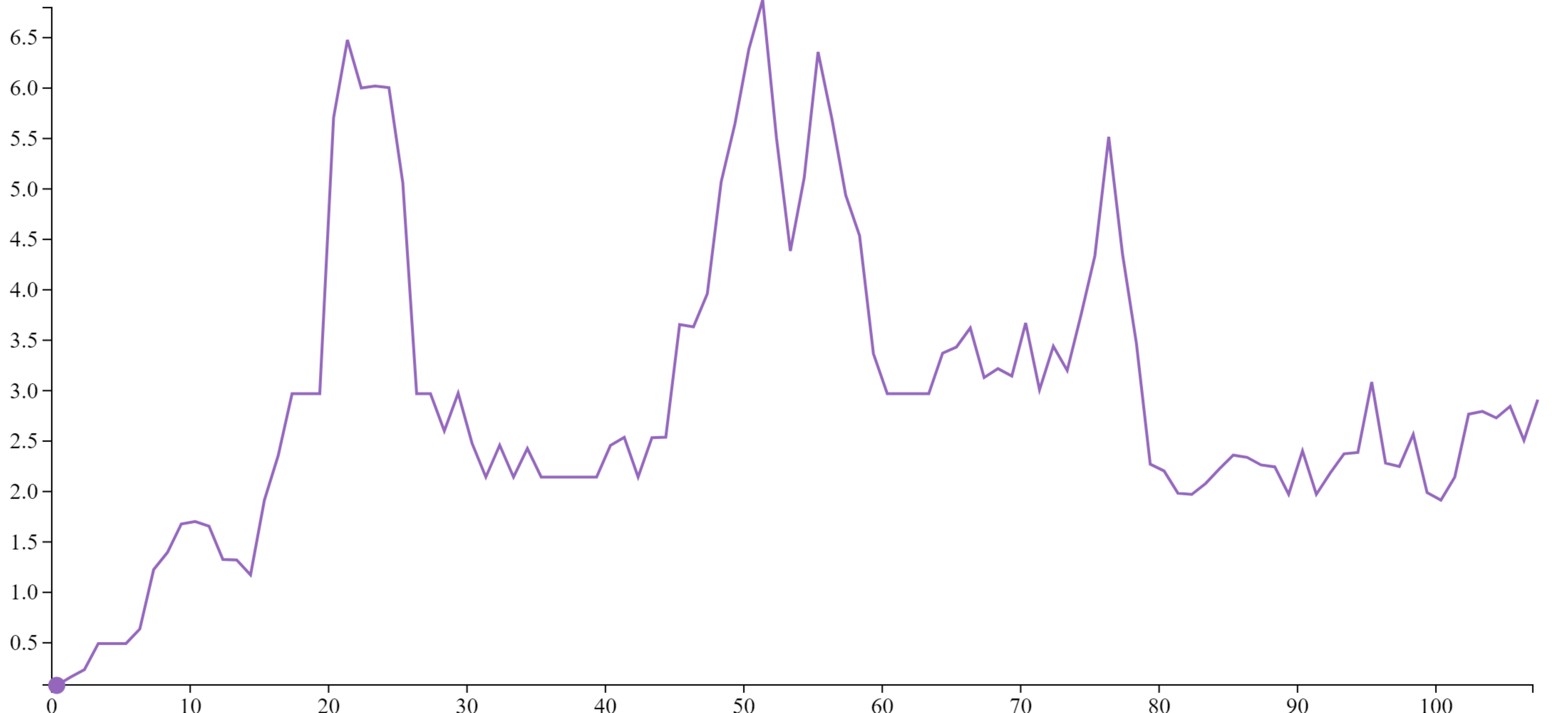}{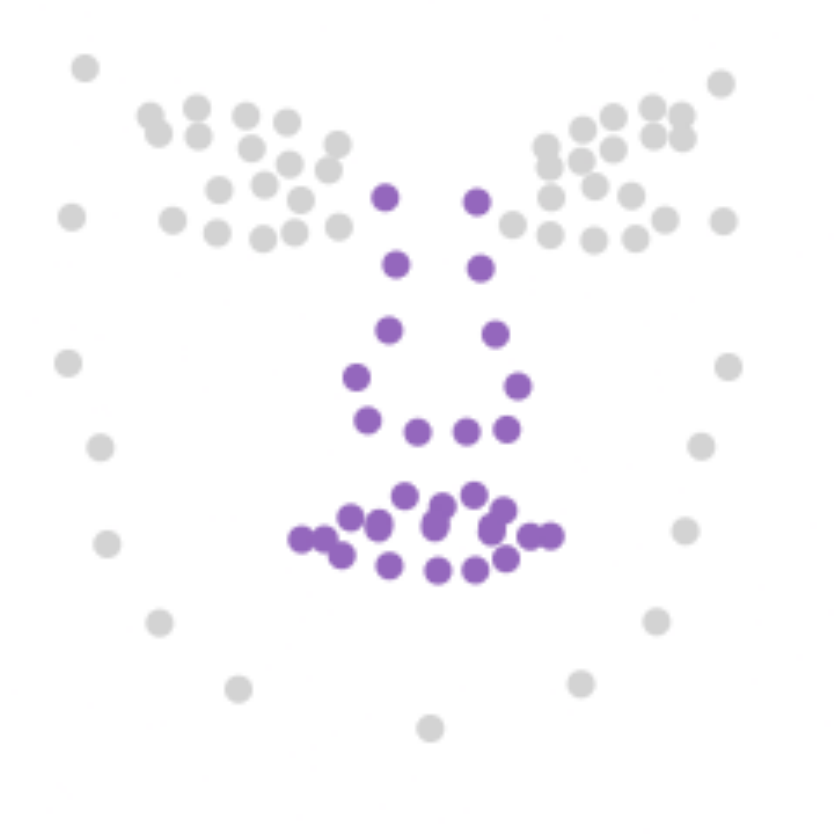}{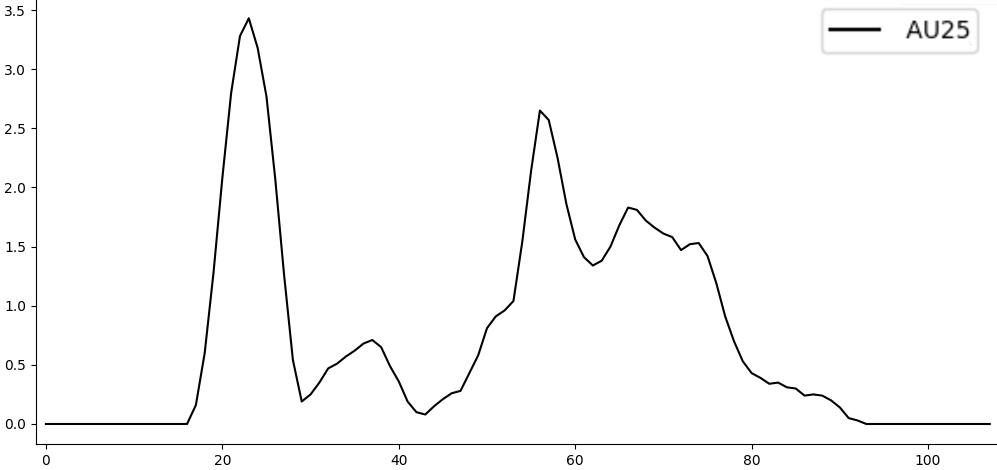}
    }
    \subfigure[\Disgust Bottleneck AU4+7]{
        \auEXX{-55pt}{-30pt}{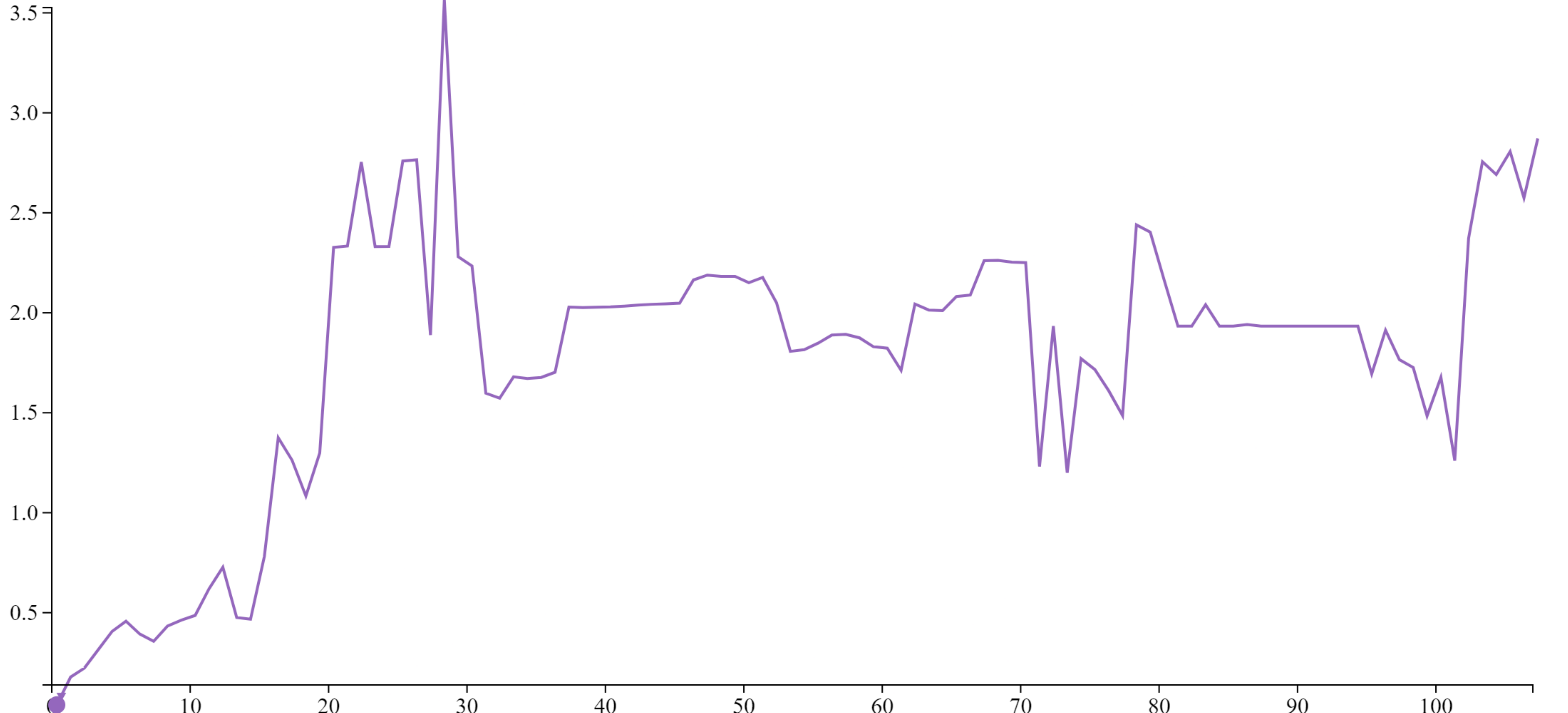}{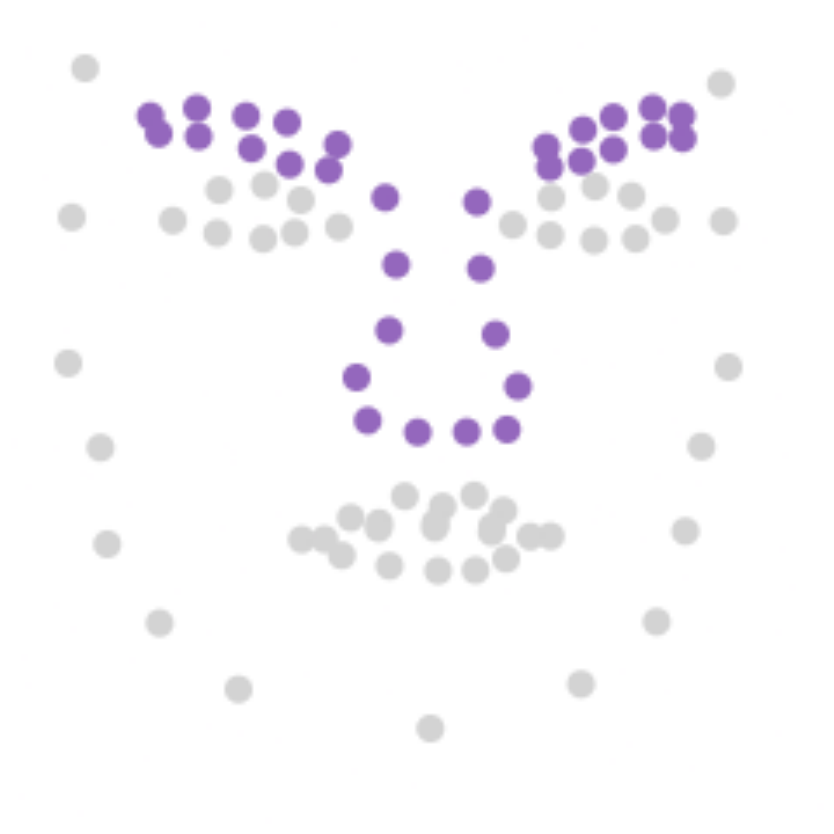}{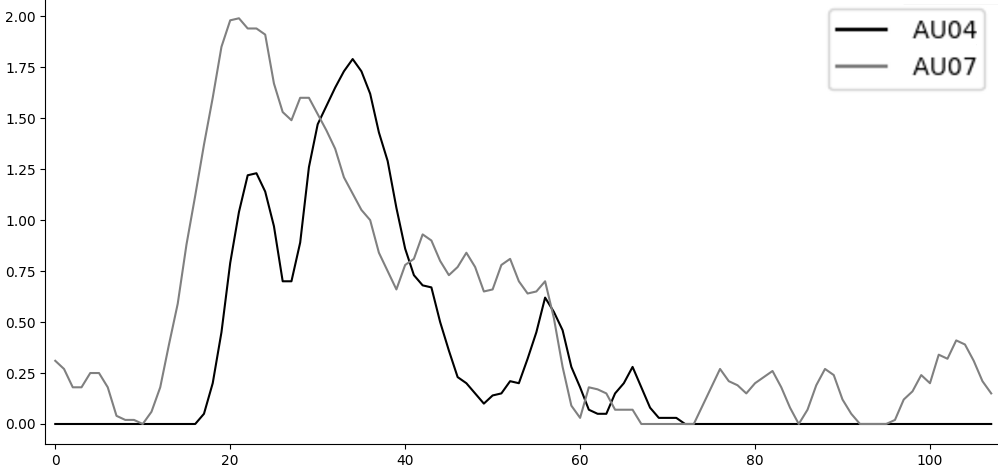}
    }    
    \subfigure[\Happiness Wass.~AU6+7+45]{
        \auEXX{-55pt}{-30pt}{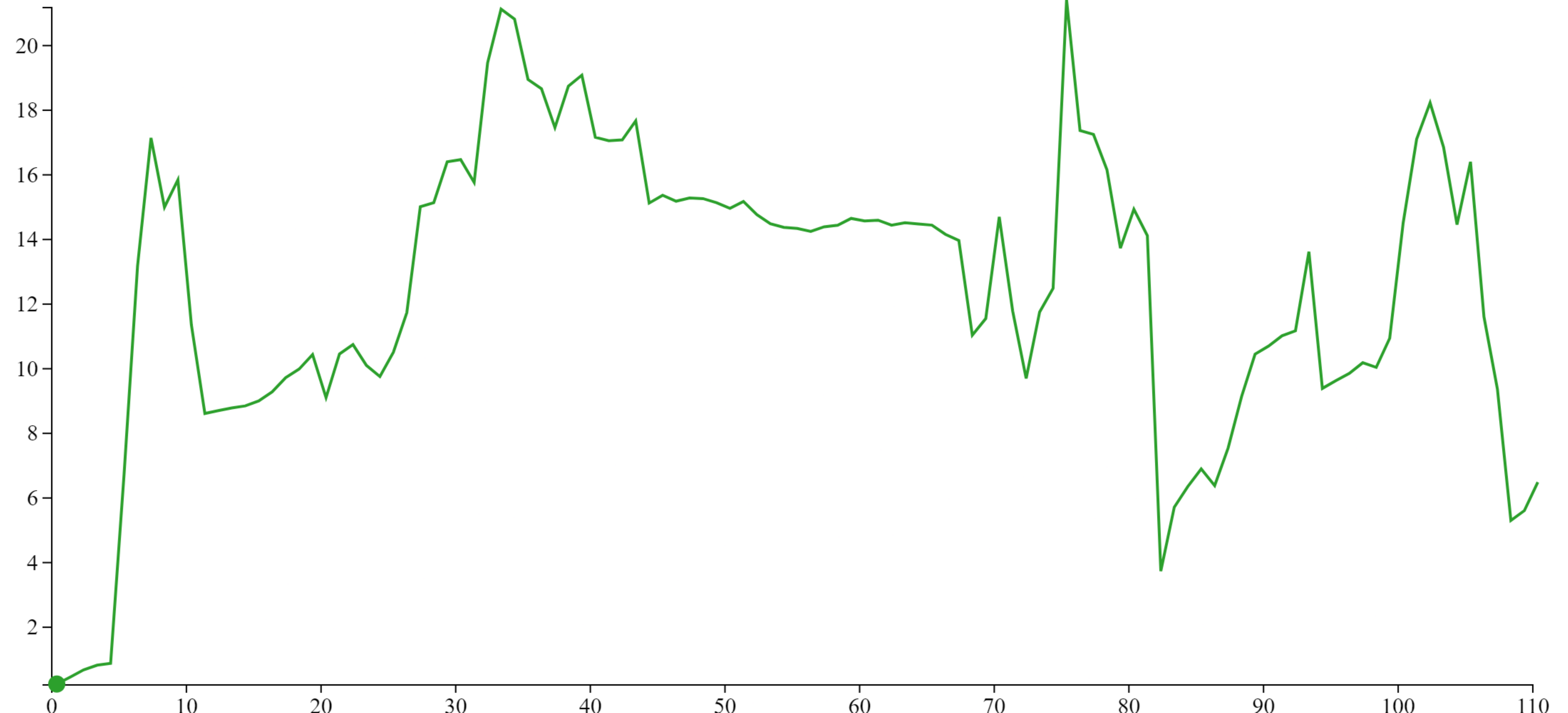}{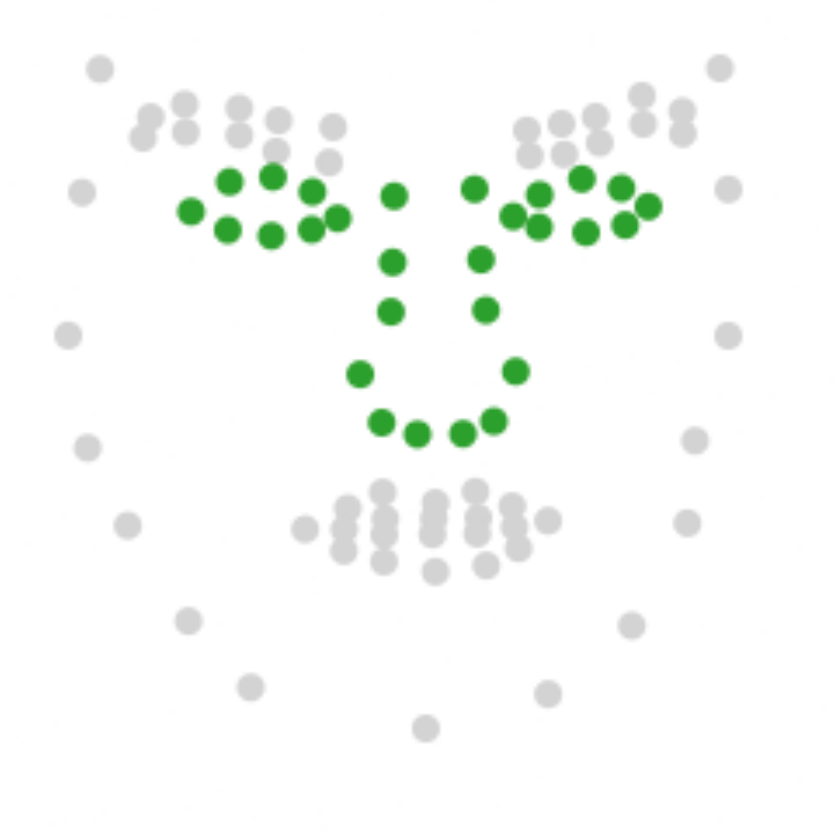}{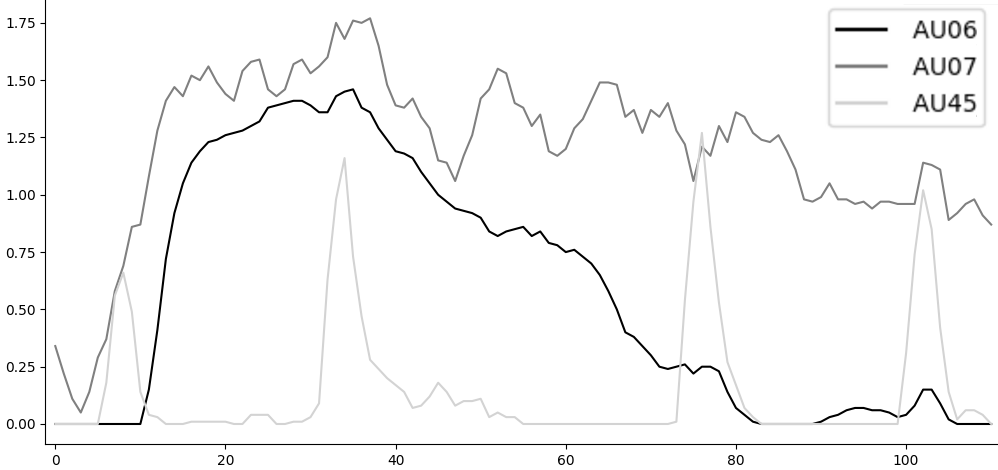}
    }
    \subfigure[\Happiness Wass.~AU12+20+25]{
        \auEXX{-55pt}{-30pt}{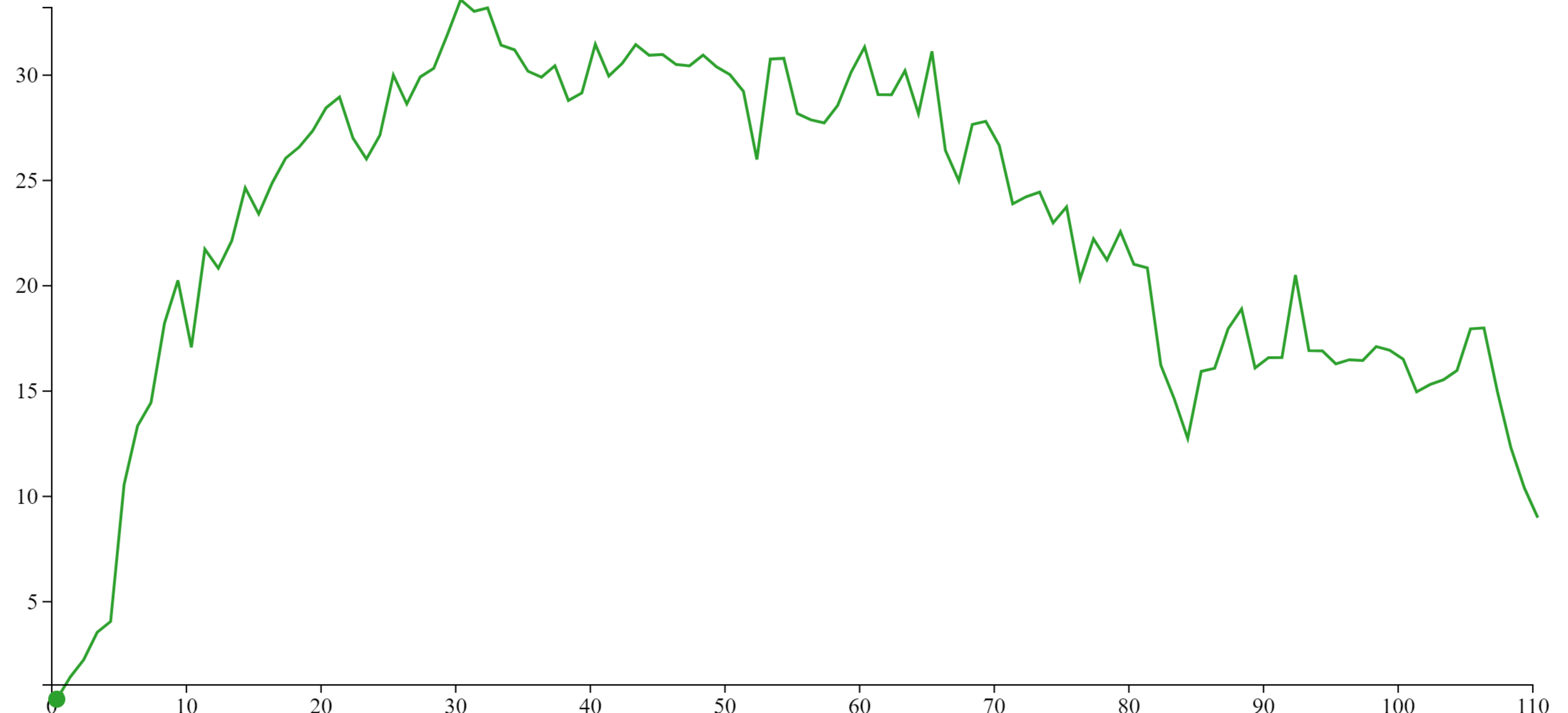}{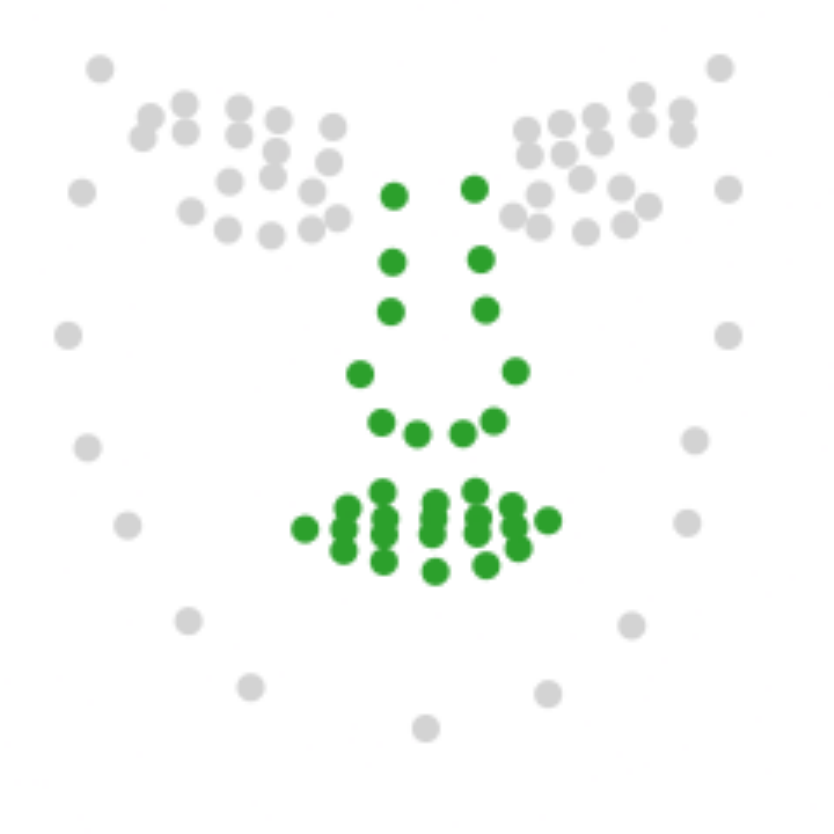}{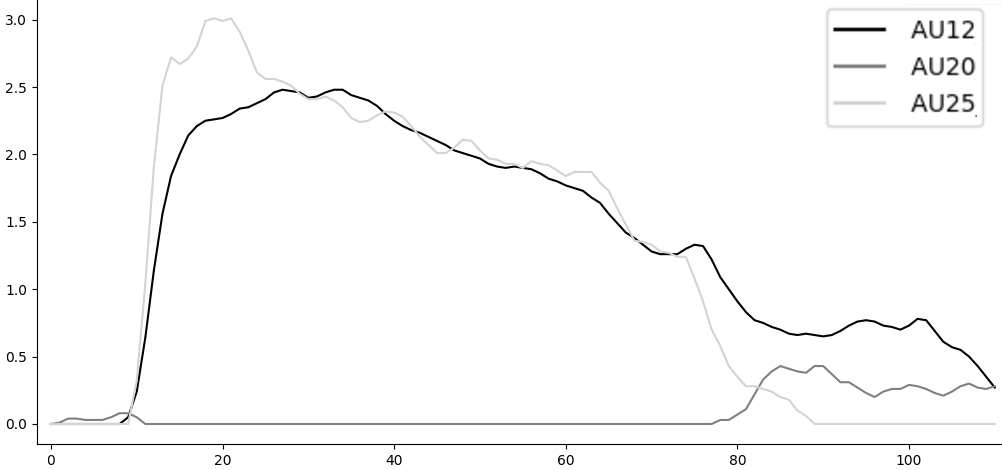}
    }

    \subfigure[\Sadness Bottleneck AU14+17]{
        \auEXX{-55pt}{-30pt}{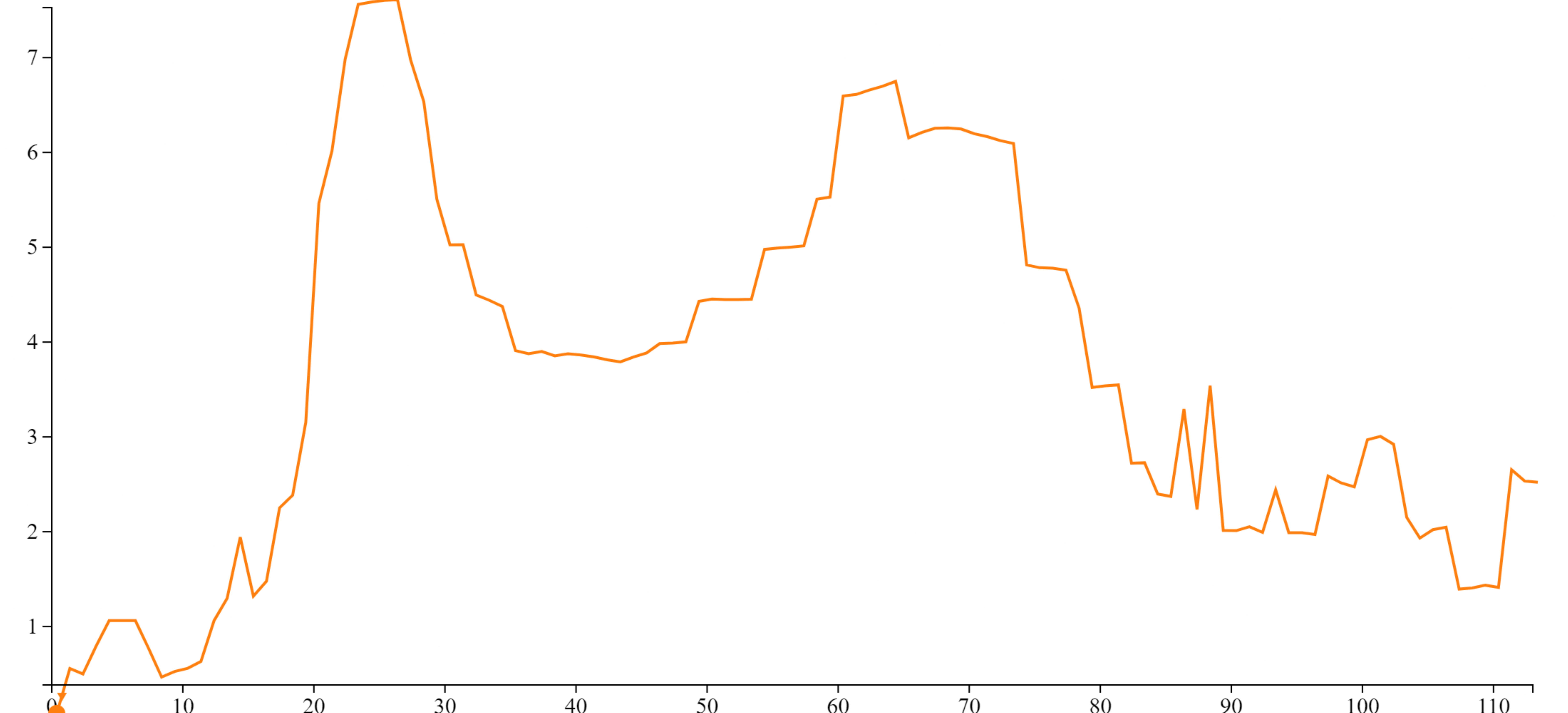}{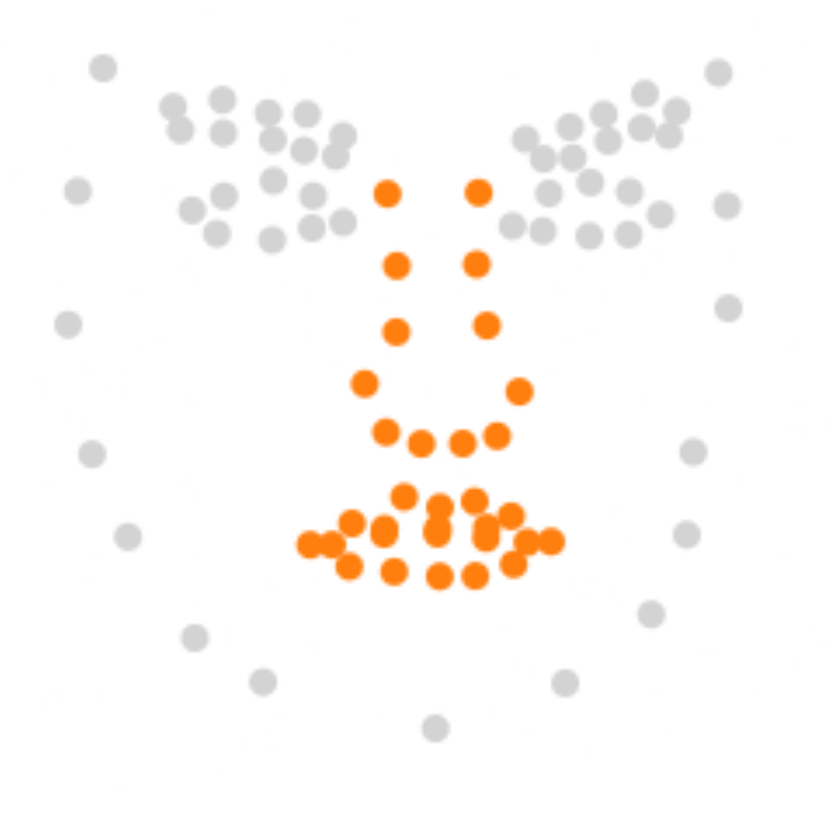}{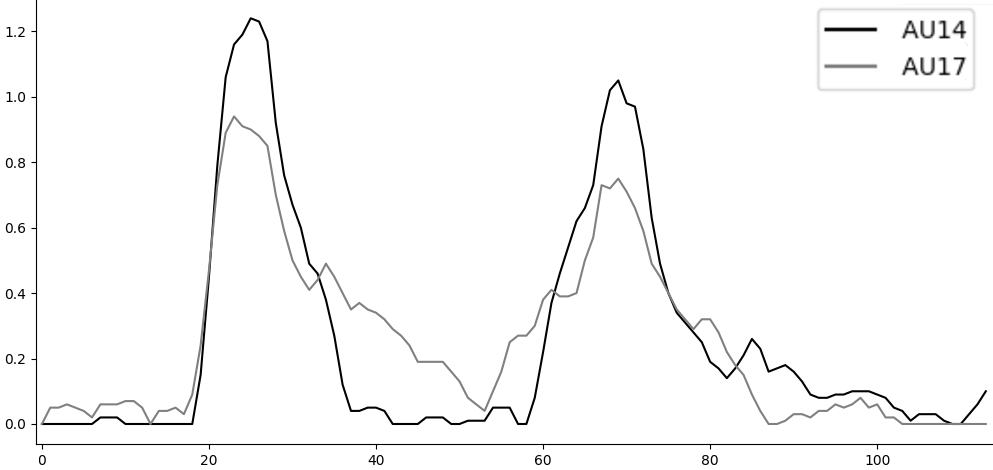}
    }
    \subfigure[\Surprise Wasserstein AU7+45]{
        \auEXX{-55pt}{-30pt}{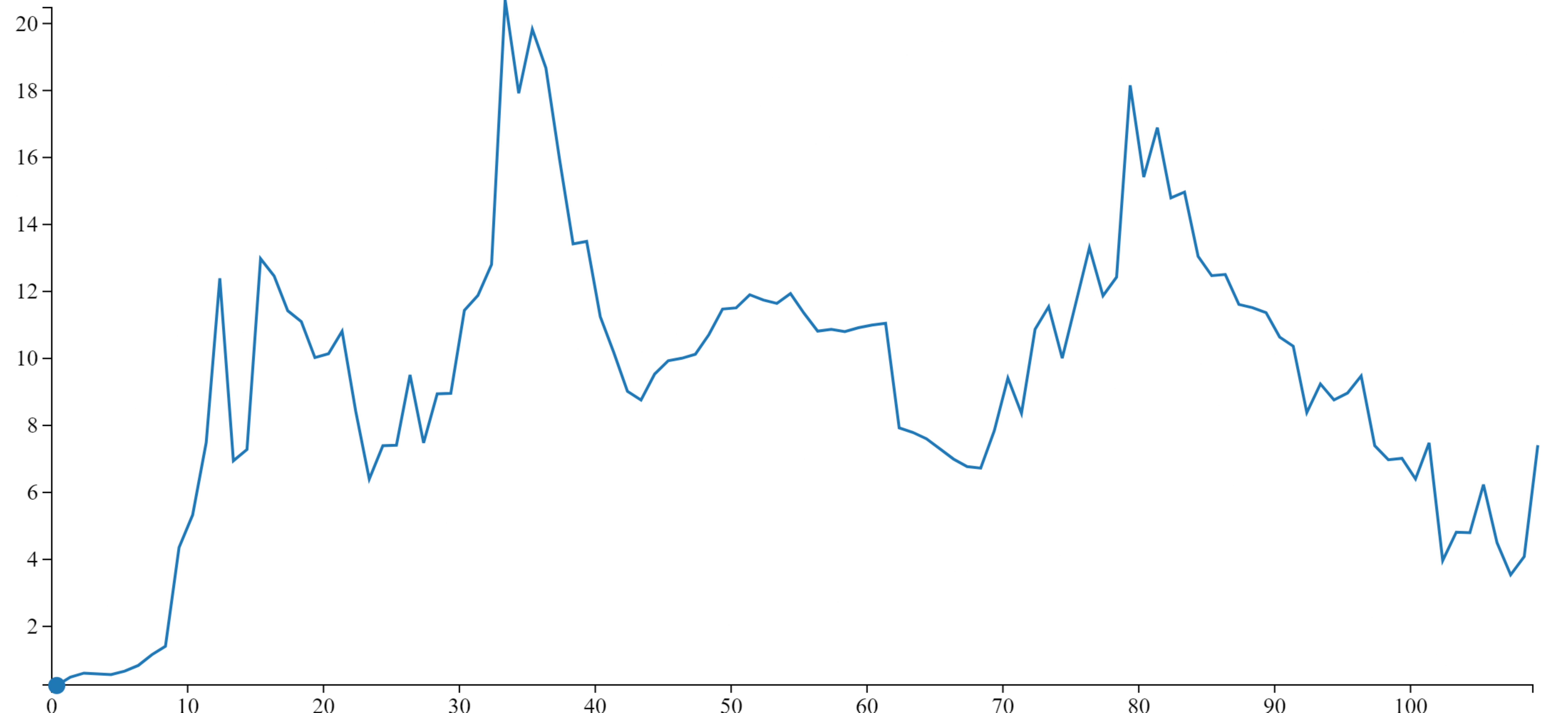}{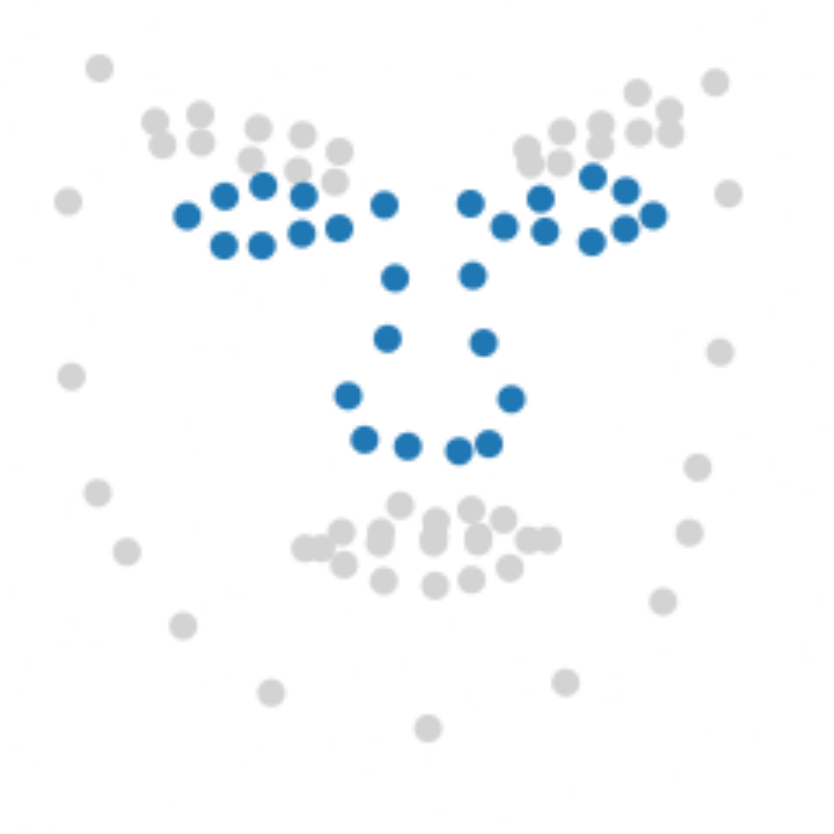}{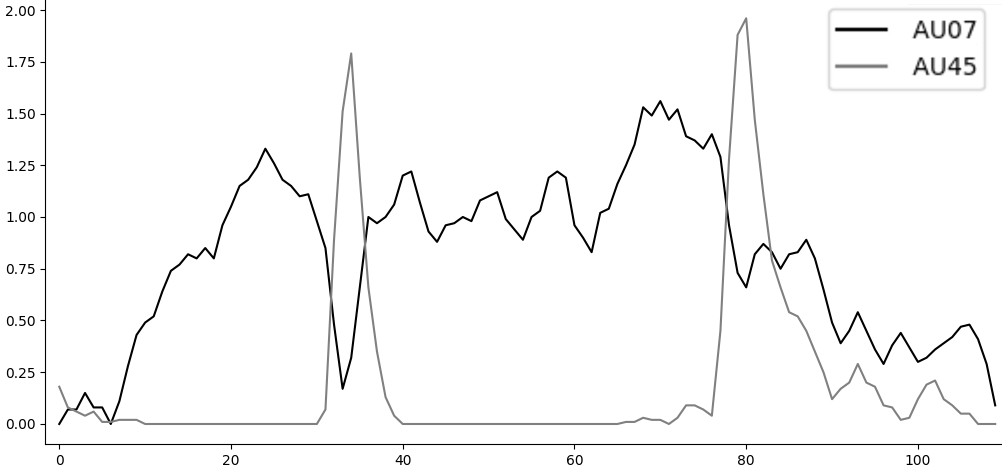}
    }
    \subfigure[\Surprise Wasserstein AU14+25]{
        \auEXX{-55pt}{-30pt}{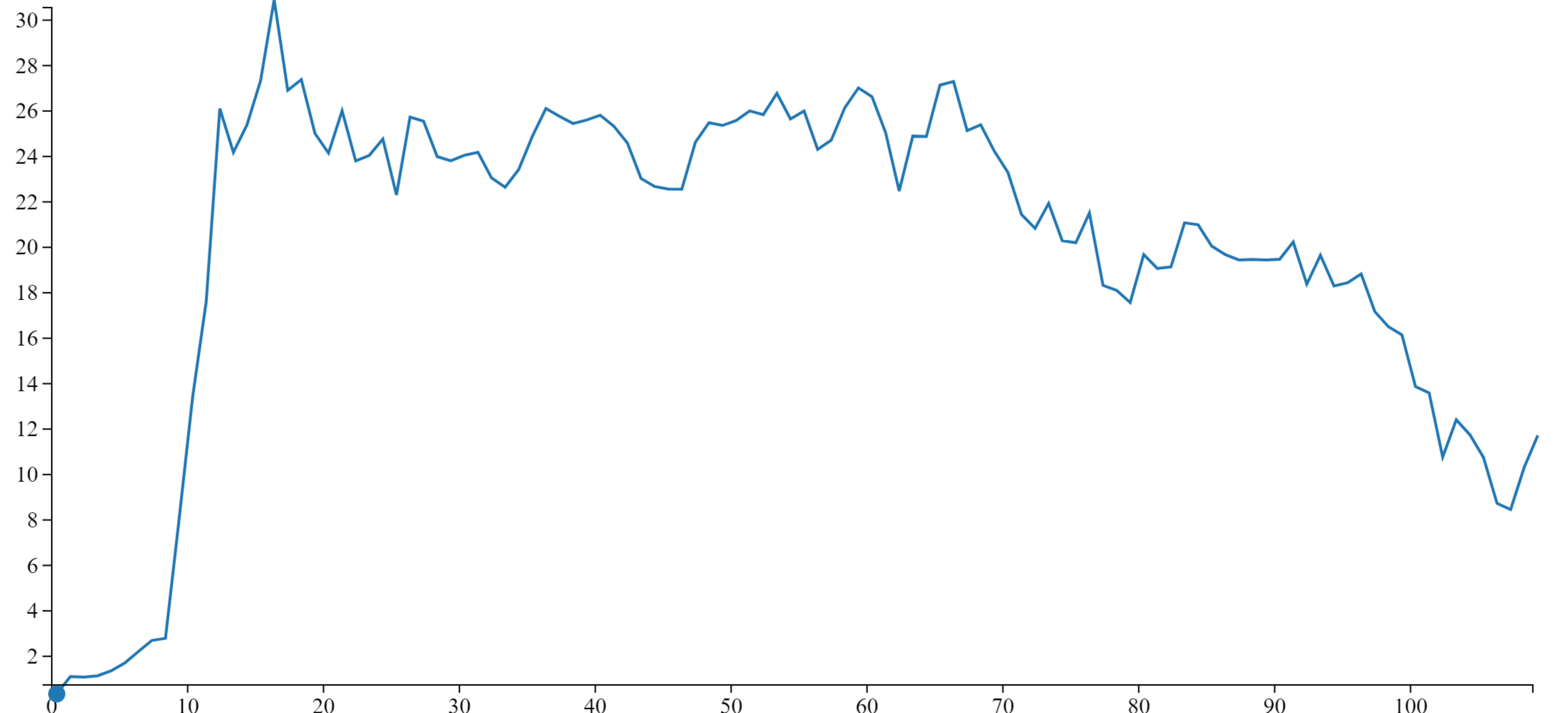}{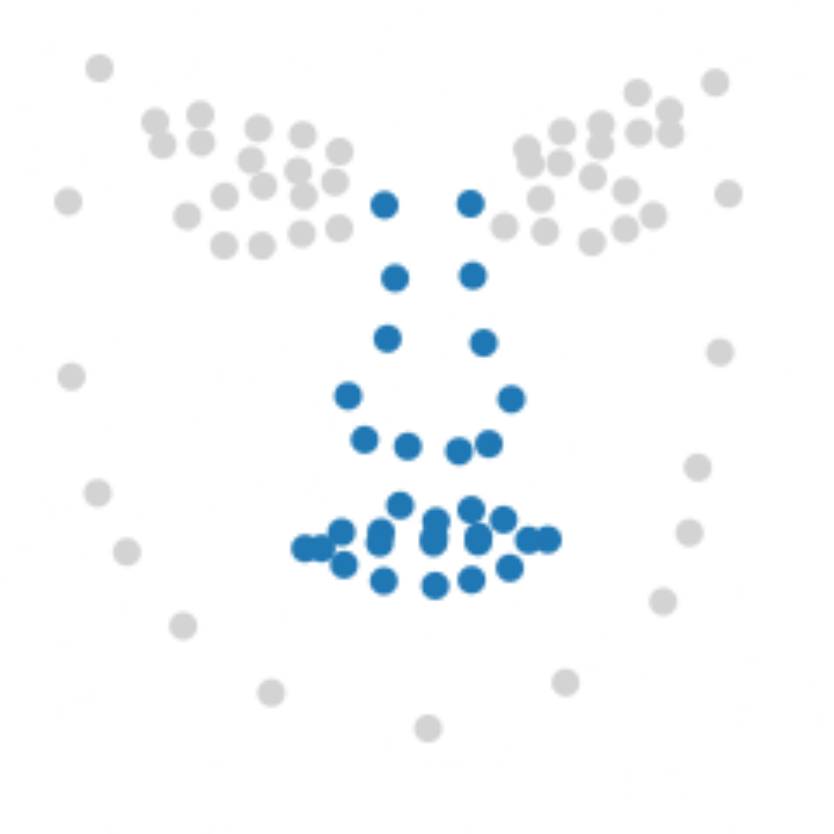}{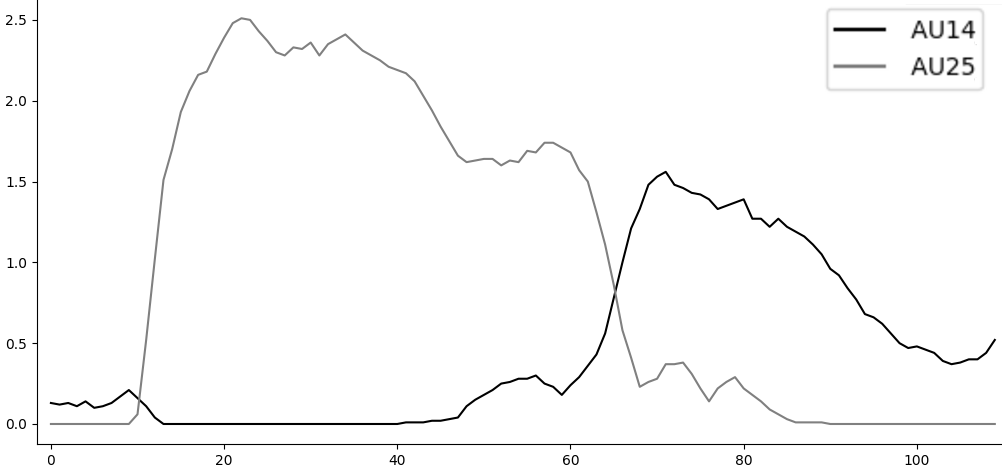}
    }
    \subfigure[\Surprise Bottleneck AU1+2+7]{
        \auEXX{-55pt}{-30pt}{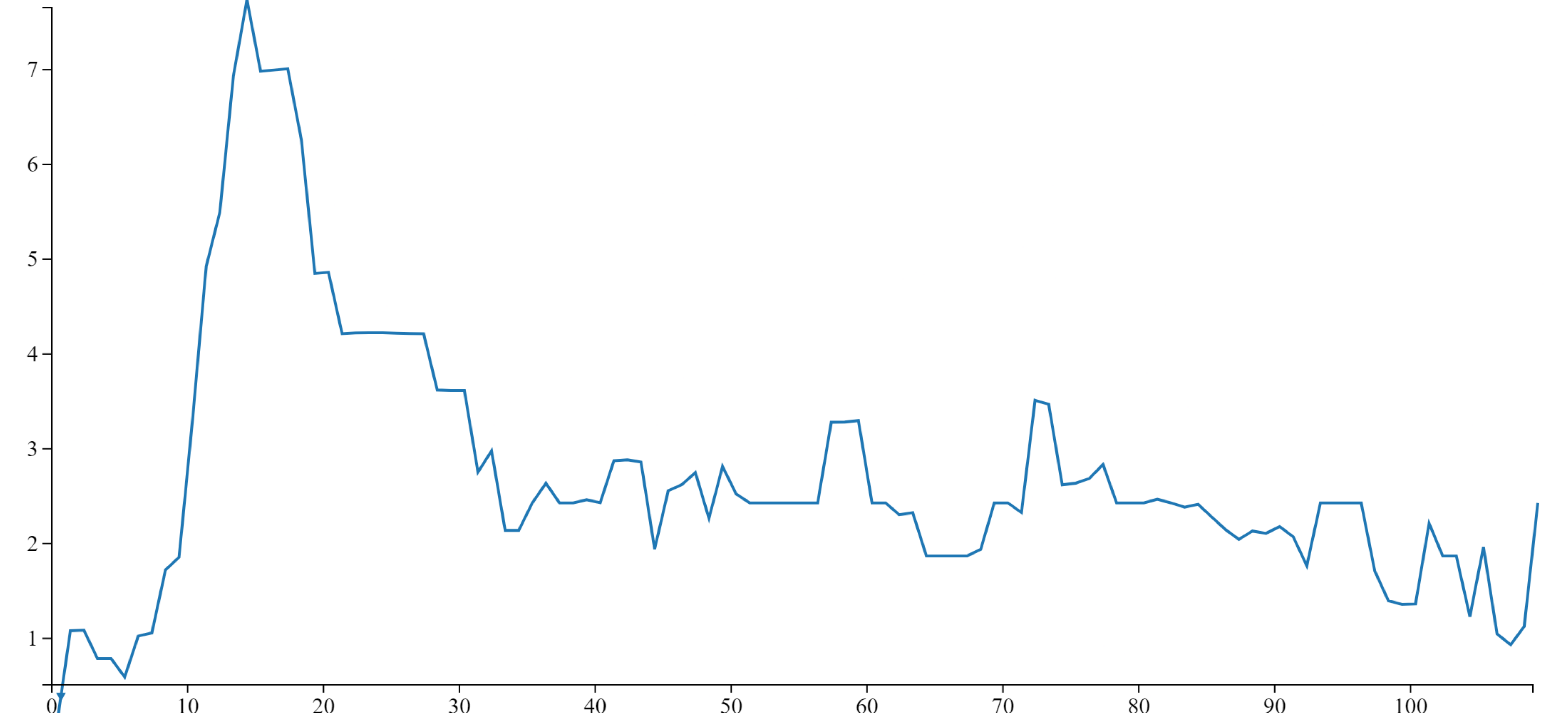}{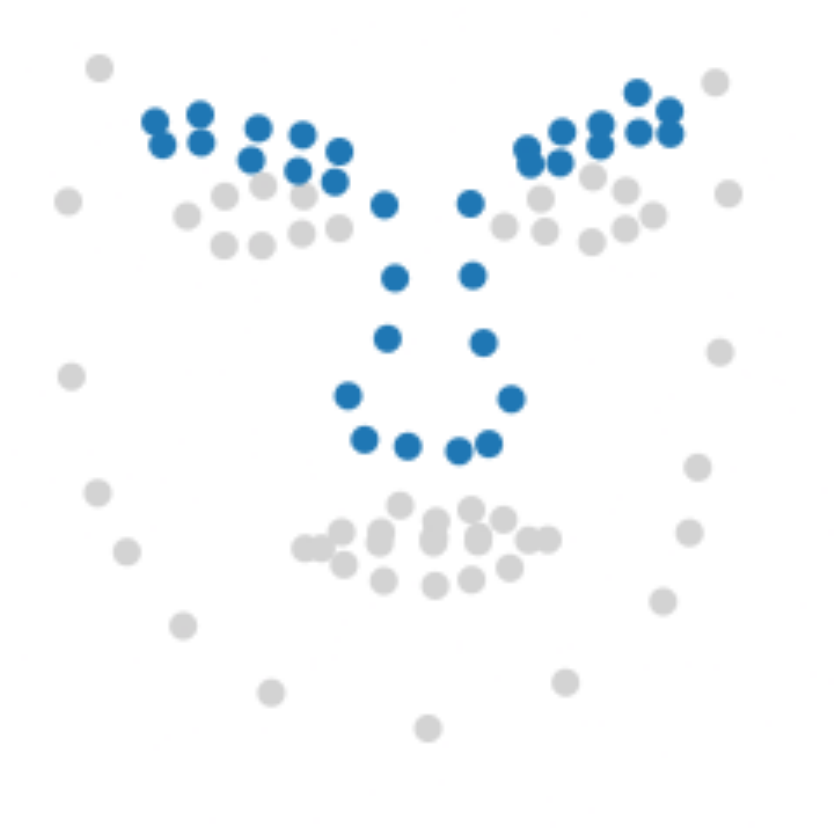}{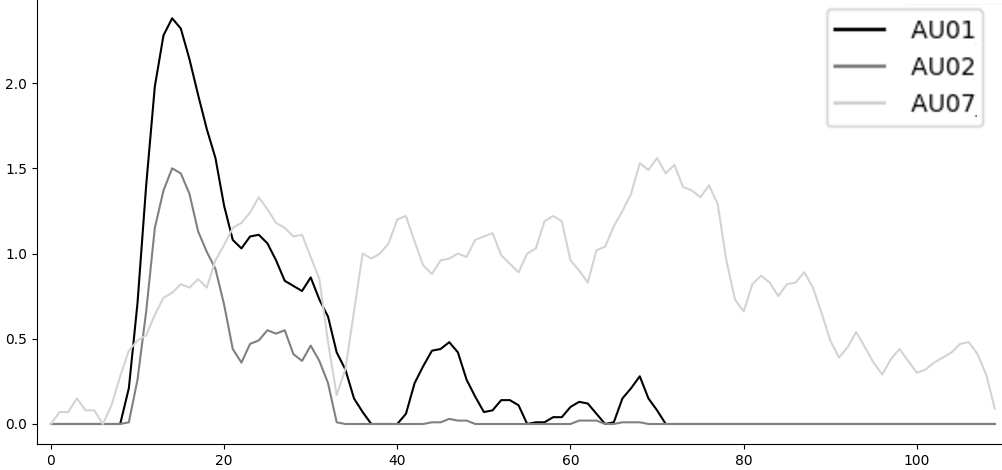}
    }
    
    \vspace{-5pt}    
    \caption{Comparison of F001 non-metric topology (top) and AUs (bottom) shows a similarity between eyes+nose (column 1), mouth+nose (column 2), and eyebrows+nose (column 3) and AUs associated with those facial regions.}
  \label{tab:auF001}
\end{figure*}

\begin{figure*}[!t]
    \centering

    \subfigure[\Disgust Wasserstein AU7]{
        \auEX{-30pt}{-35pt}{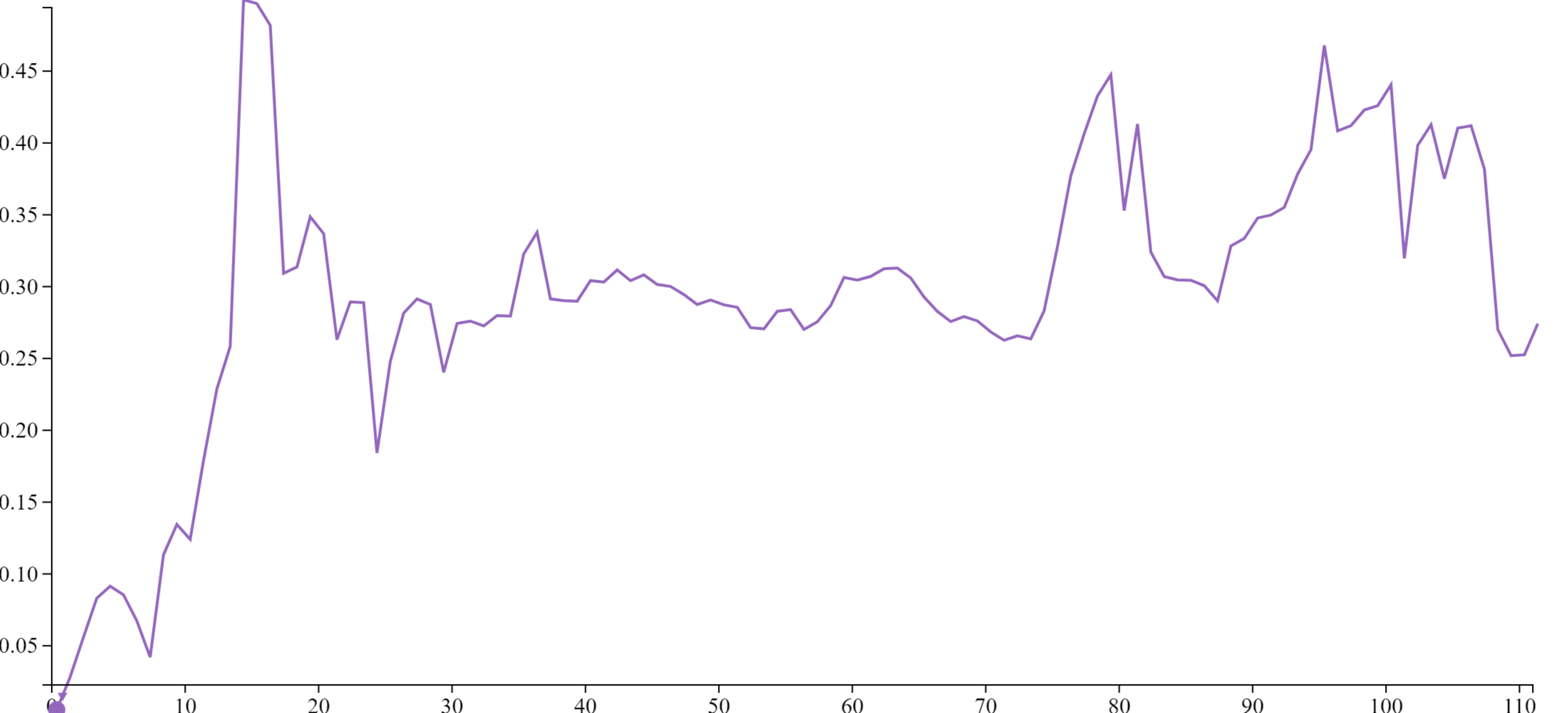}{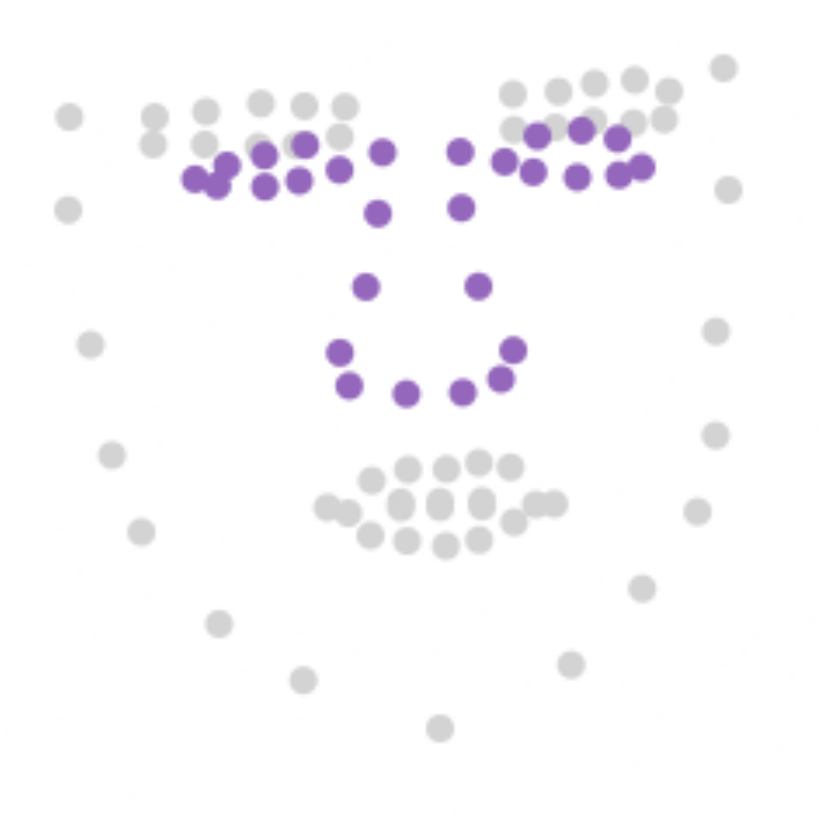}{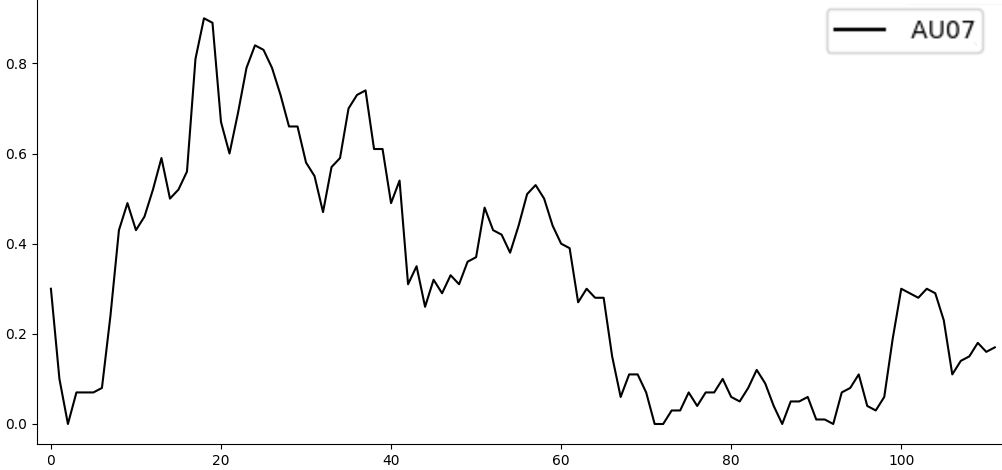}
    }    
    \subfigure[\Fear Wasserstein AU10+14+25]{
        \auEX{-65pt}{-35pt}{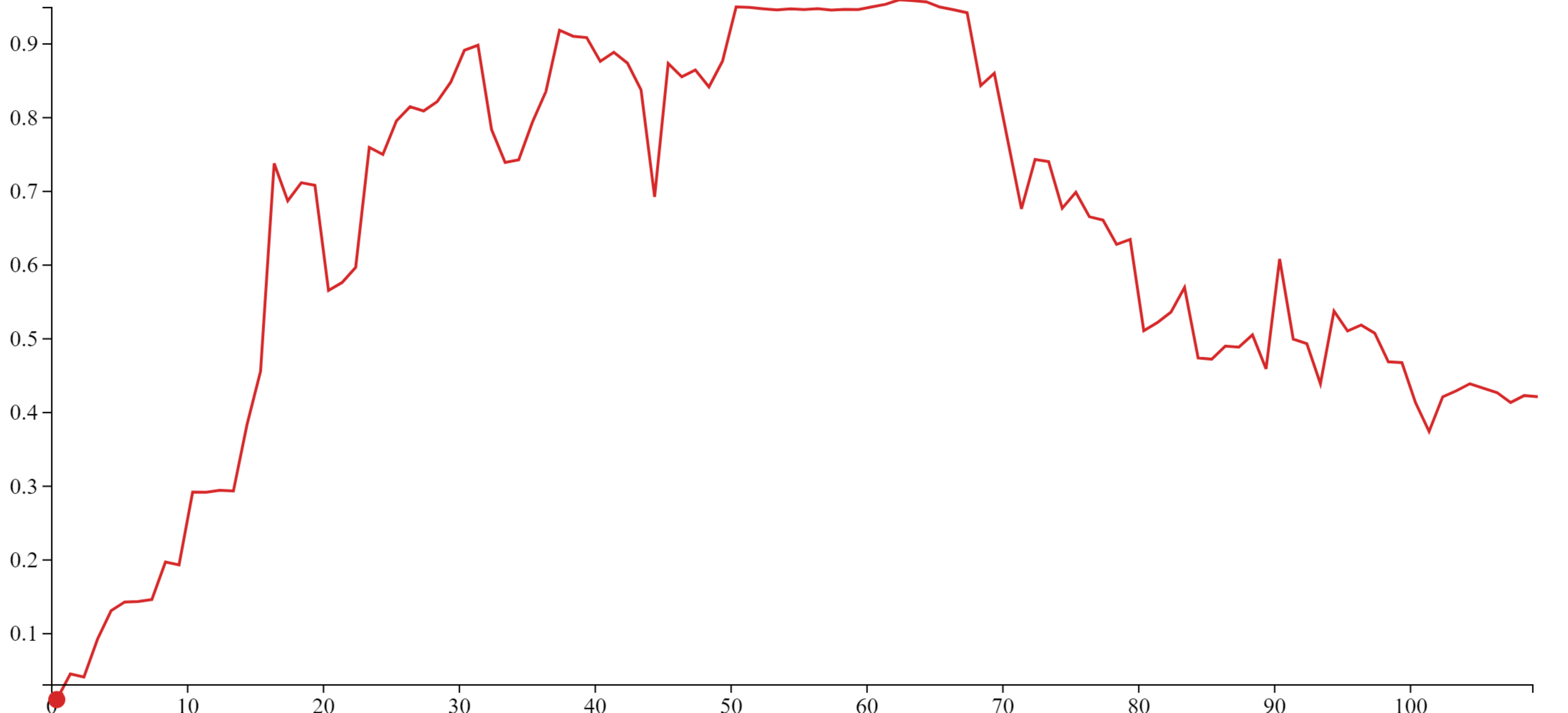}{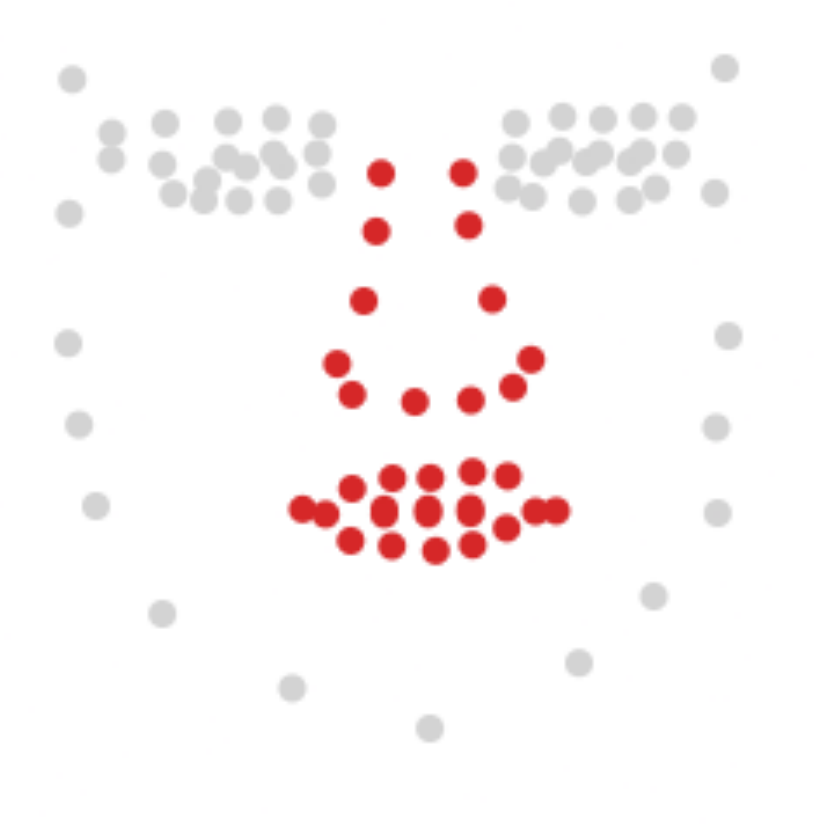}{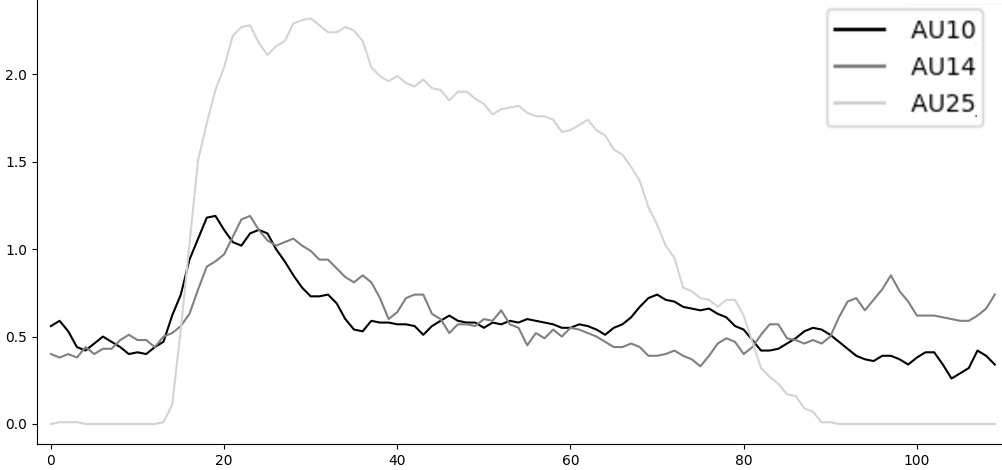}
    }
    \subfigure[\Disgust Wasserstein AU4+7]{
        \auEX{-30pt}{-35pt}{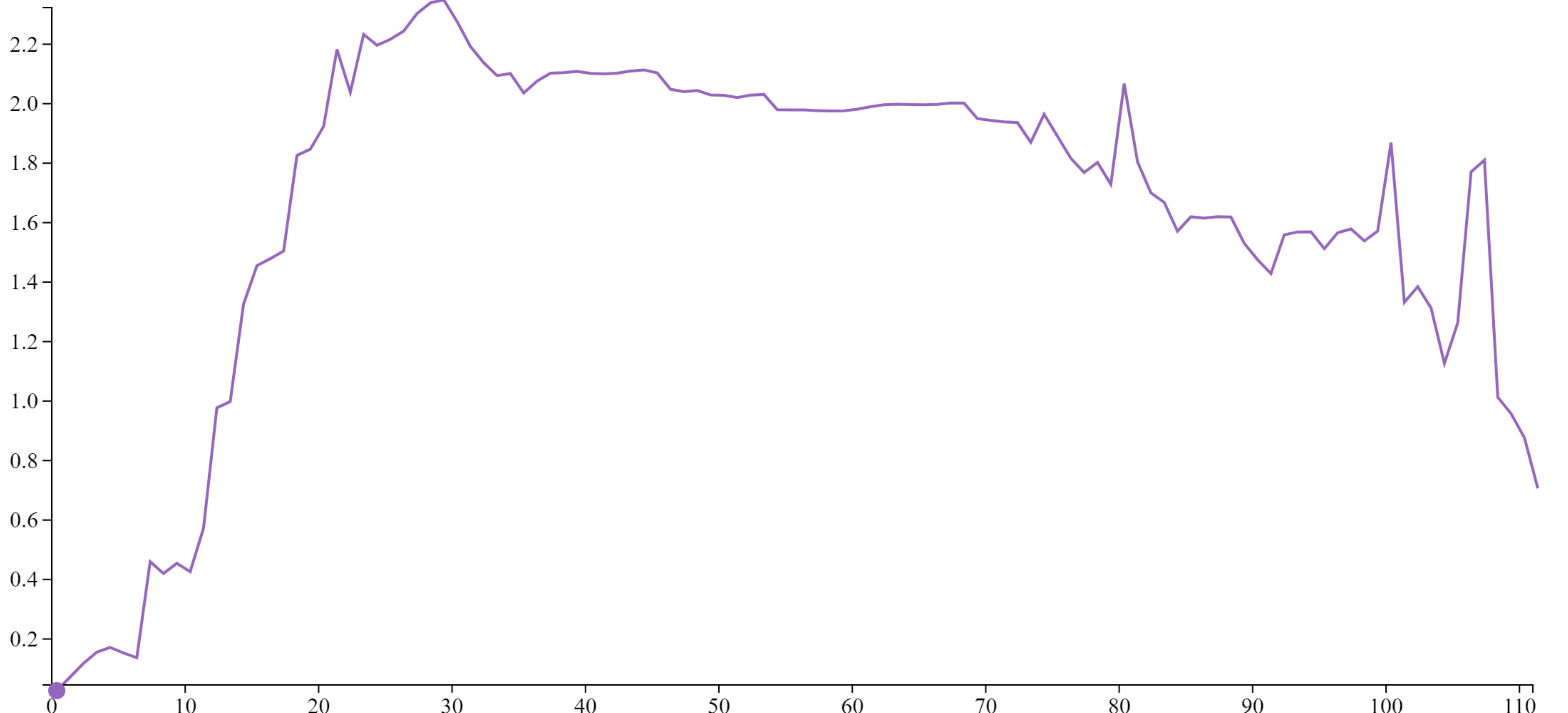}{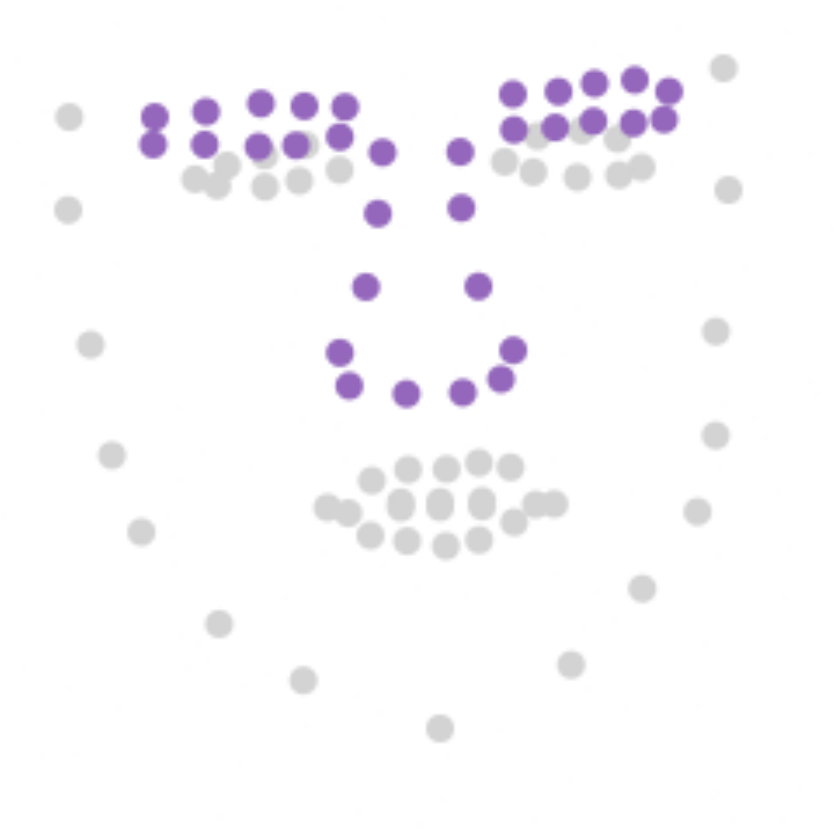}{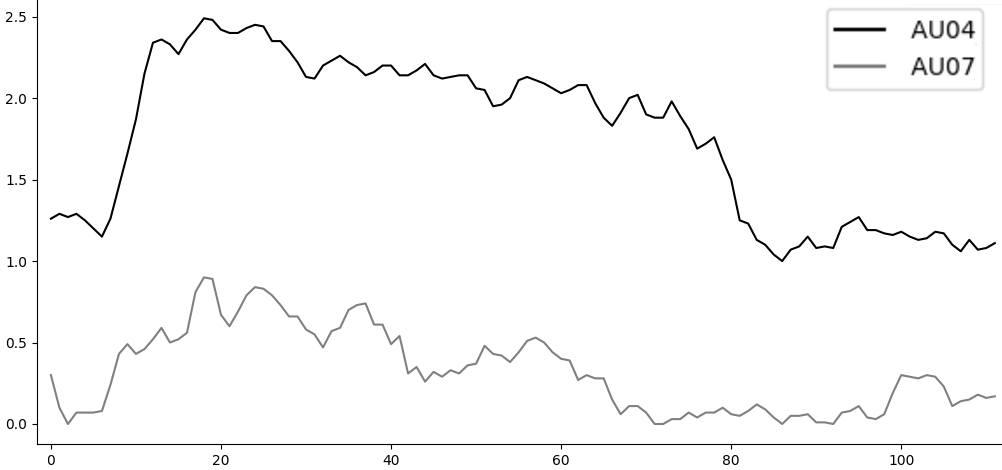}
    }

    \subfigure[\Surprise Wasserstein AU45]{
        \auEX{-65pt}{-50pt}{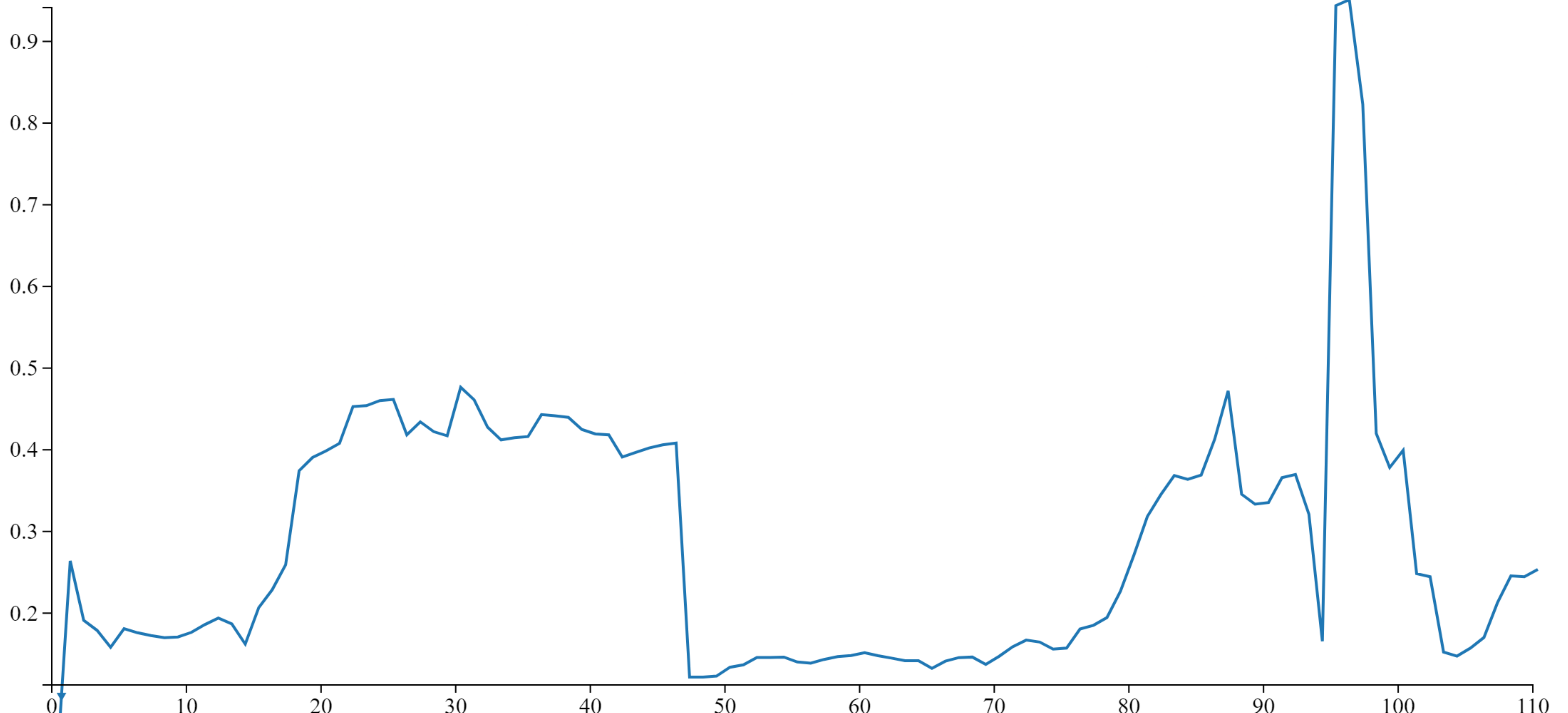}{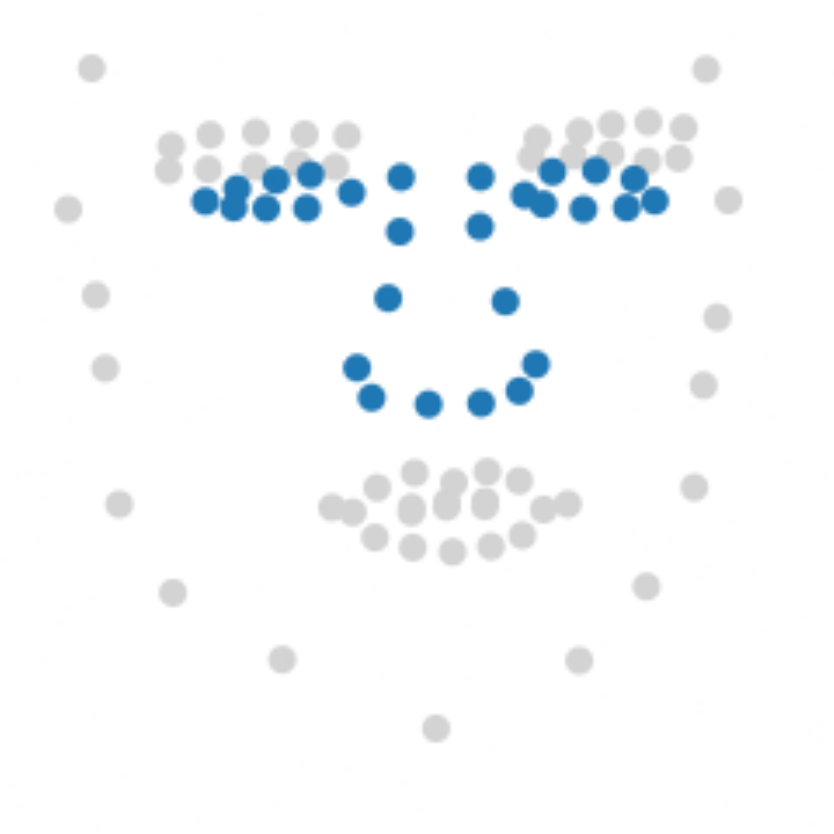}{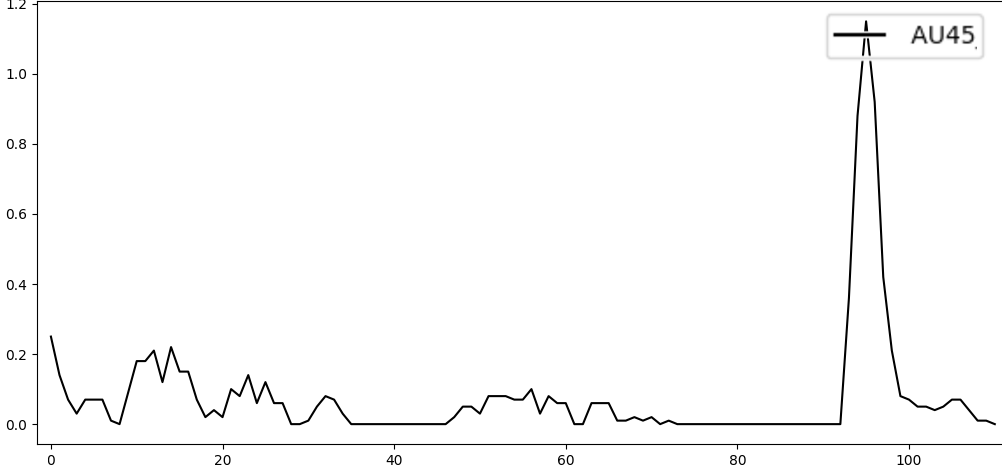}
    }
    \subfigure[\Surprise Wasserstein AU10]{
        \auEX{-65pt}{-35pt}{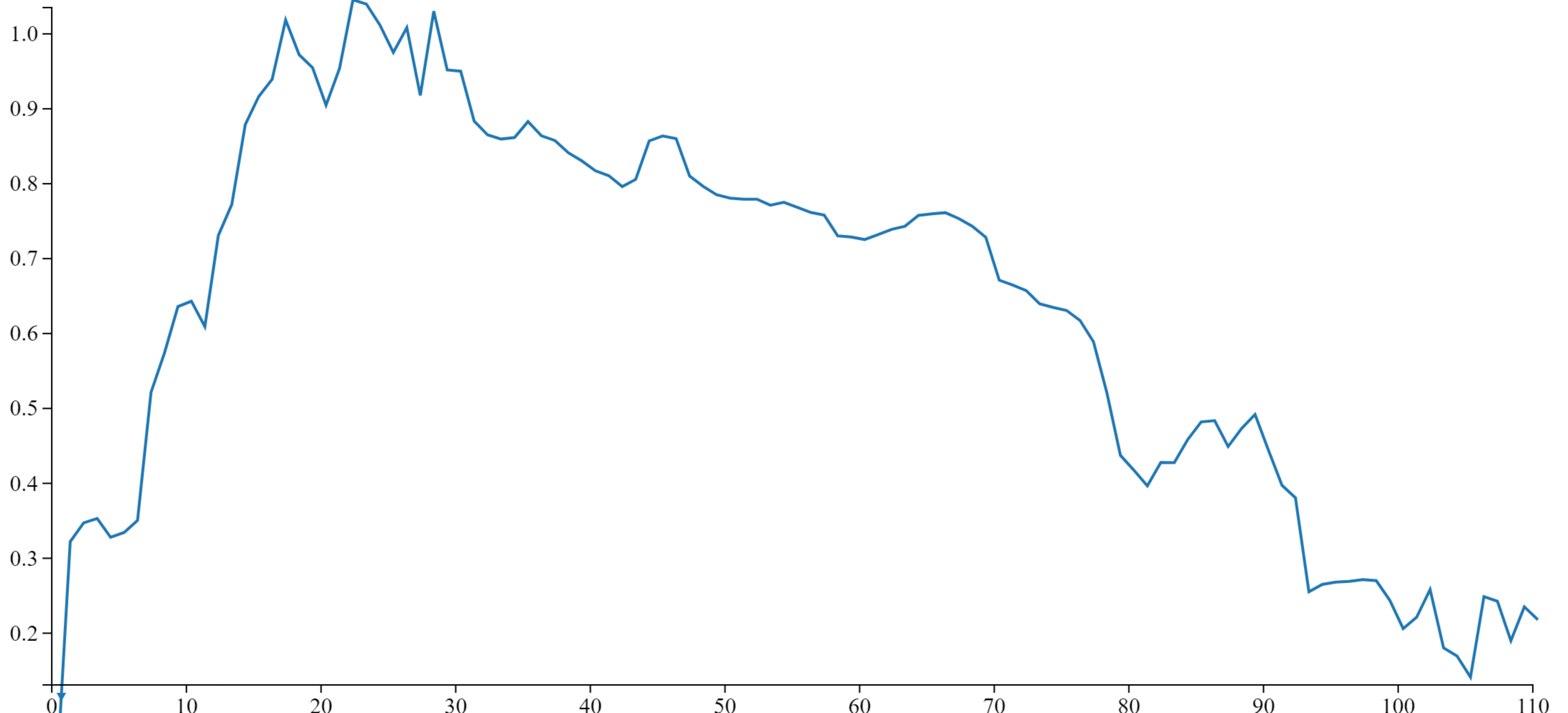}{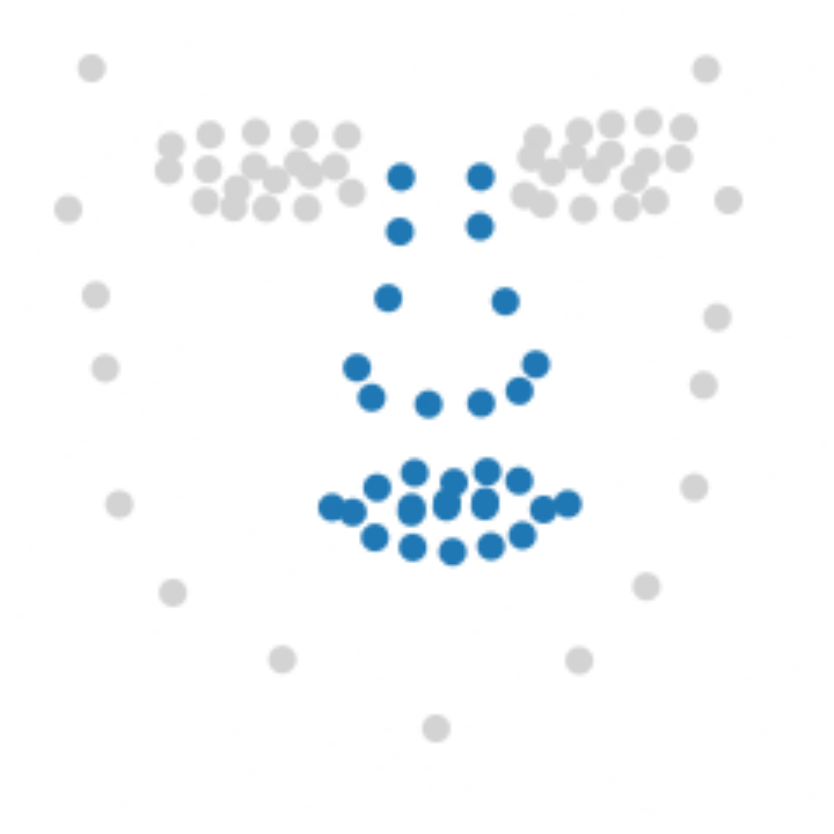}{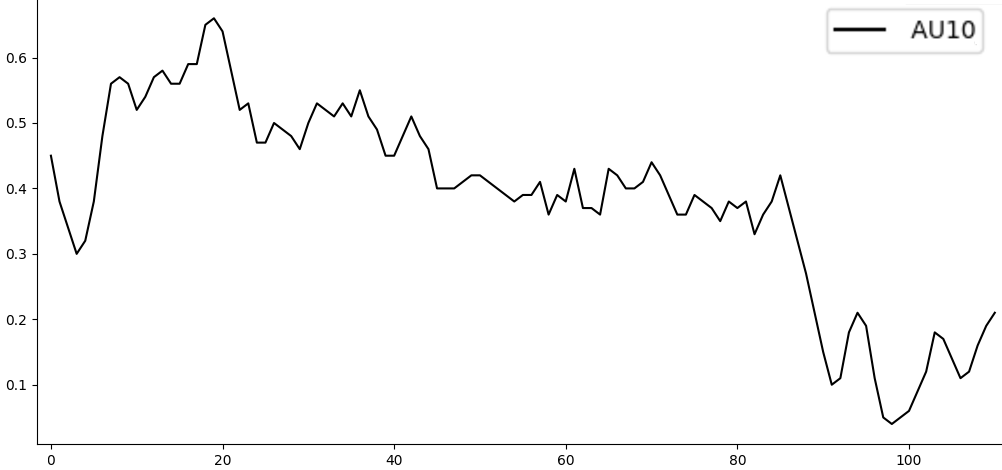}
    }
    \subfigure[\Sadness Wasserstein AU4]{
        \auEX{-30pt}{-45pt}{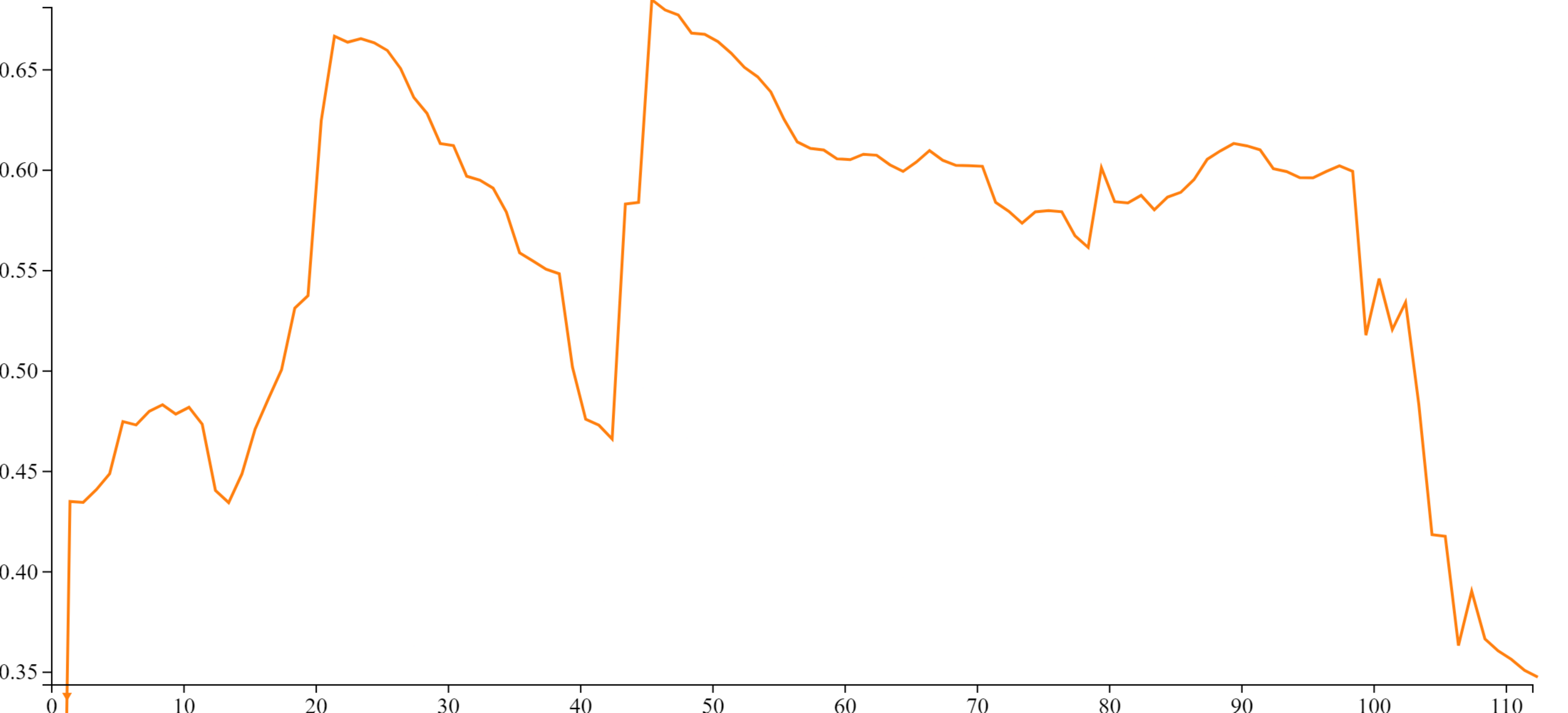}{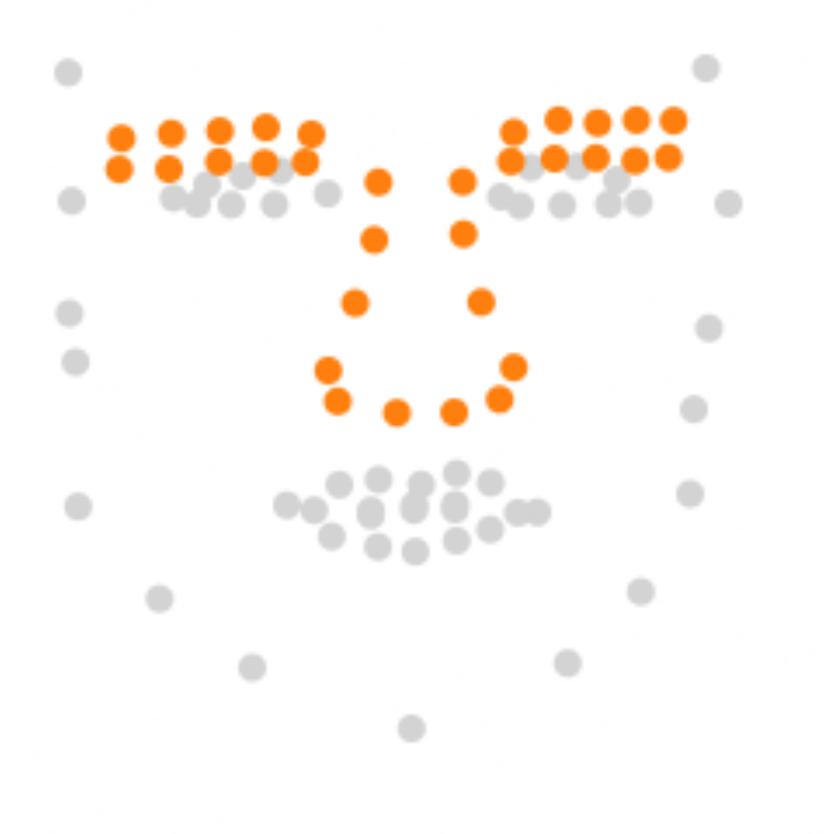}{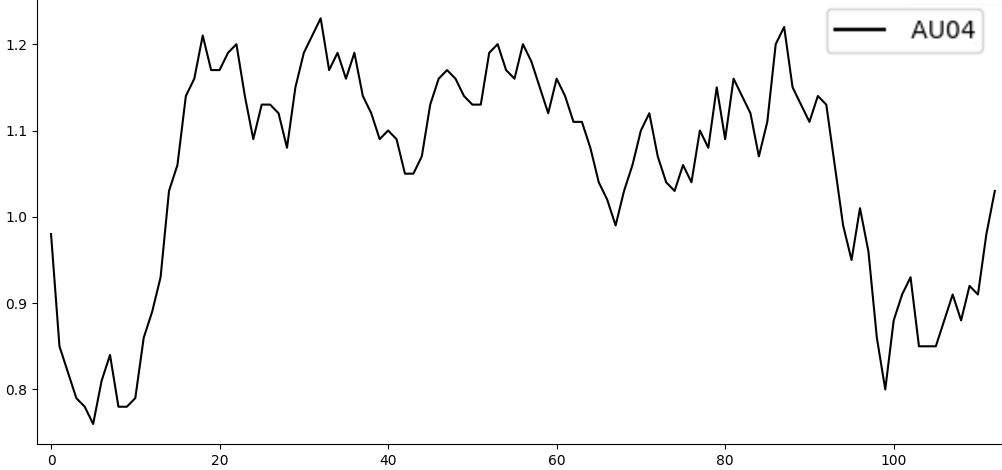}
    }
    
    %\vspace{-3pt}
    \caption{Comparison of M001 non-metric topology (top) and AUs (bottom) shows a similarity between eyes+nose (column 1), mouth+nose (column 2), and eyebrows+nose (column 3) and AUs associated with those facial regions.}
  \label{tab:auM001}
\end{figure*}

\subsection{Comparing and Differentiating Expressions}
\label{sec:eval:exp}

Next, we consider whether the topological features are sufficient for differentiating the six different emotions present in the data. To perform this evaluation, we look at the full face topology for the female subject using t-SNE (\autoref{fig:expressiveness_tsne:female}, Shepard fitness: 0.79) and MDS (\autoref{fig:teaser}, Shepard fitness: 0.88), and male subject using t-SNE (\autoref{fig:expressiveness_tsne:male}, Shepard fitness: 0.78) and MDS (\autoref{fig:expressiveness_mds:male}, Shepard fitness: 0.88).

We begin by examining the female and male subjects using t-SNE, as seen in \autoref{fig:expressiveness_tsne:female} and \autoref{fig:expressiveness_tsne:male}, respectively. We make three important observations about the resulting images. (1)~First, for both subjects, the emotional states tend to form separate clusters, indicating that they are indeed differentiable. This is particularly important if the topology were to be used for predicting unknown emotional states. (2)~Second, most of the emotions begin and end towards the centers of the plots. This colocation is caused by the neutral facial pose that subjects were asked to begin and end with each sequence. (3)~The final observation is that the facial poses form temporally coherent `strings.' This observation is particularly poignant, considering that \textit{nowhere in calculating the topological dissimilarity does it utilize temporal information}. 

We next consider the MDS projections for the female subject and male subject in \autoref{fig:teaser} and \autoref{fig:expressiveness_mds:male}, respectively. With the female subject, we observe that \happiness, \surprise, \fear, and \sadness largely cluster into separate regions of the plot with limited overlap or mixing. The \anger and \disgust emotions, on the other hand, overlap significantly, which corresponds with the recent literature in the affective computing community that considers them to be similar expressions~\cite{molho2017disgust}. Another interesting finding is that the emotional states of the male are less differentiated than those of the female. Interestingly, it is commonly accepted in affective computing that, in general, the expressiveness of females is more differentiable than that of males~\cite{deng2016gender}. Given that we are only observing two subjects, no broad gender-based conclusions can be made in our case. Nevertheless, \textit{this} female's expressions are more differentiated than \textit{this} male's expressions.

\subsection{Comparing and Differentiating Individuals}
\label{sec:eval:ind}

Finally, we consider how topological features allow the differentiation of each of the 101 subjects in the BU4DFE dataset. 

\begin{figure*}[!t]
    \centering

    \rotatebox{90}{\hspace{11pt}10 subjects}
    \subfigure[\Anger (SF: 0.48)\label{fig:clustering:angry_10}]{\fbox{\includegraphics[trim= 92pt 81pt 73pt 89pt, clip, width=0.15\linewidth]{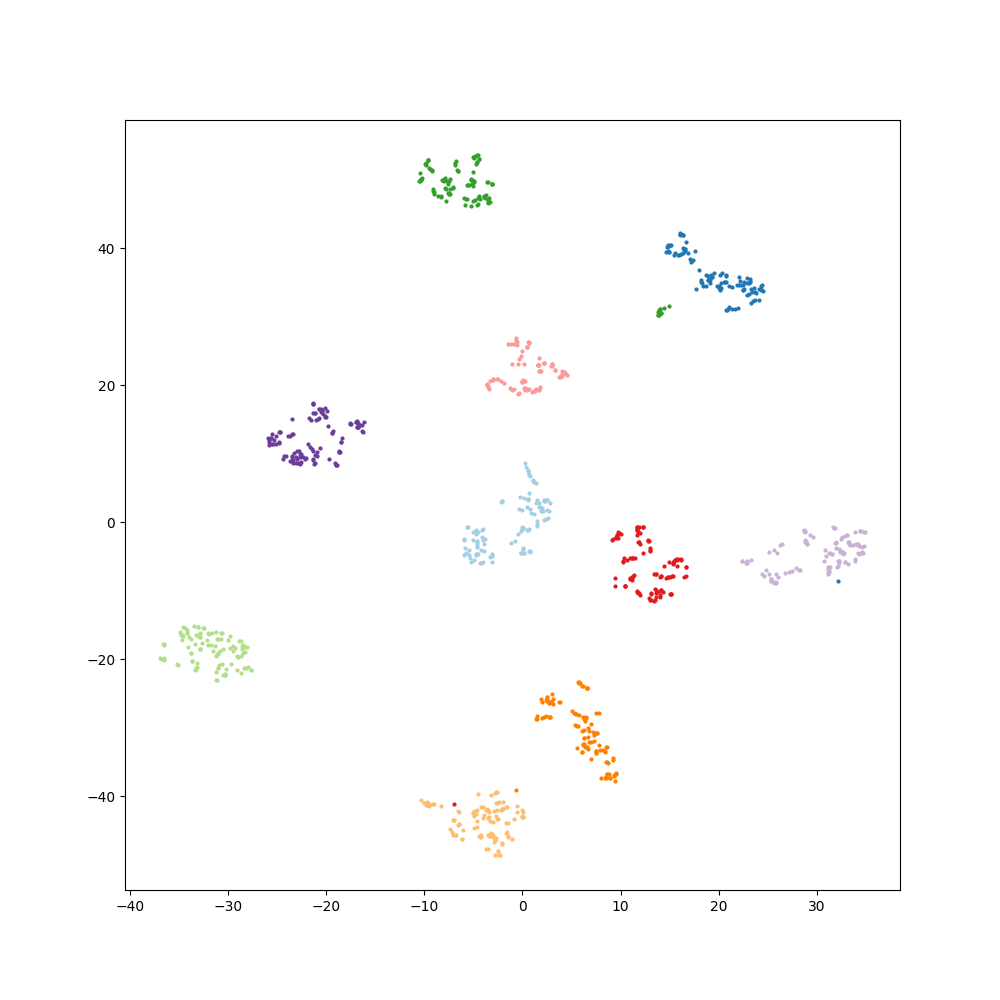}}}
    \hfill
    \subfigure[\Disgust (SF: 0.53)\label{fig:clustering:disgust_10}]{\fbox{\includegraphics[trim= 92pt 81pt 73pt 89pt, clip, width=0.15\linewidth]{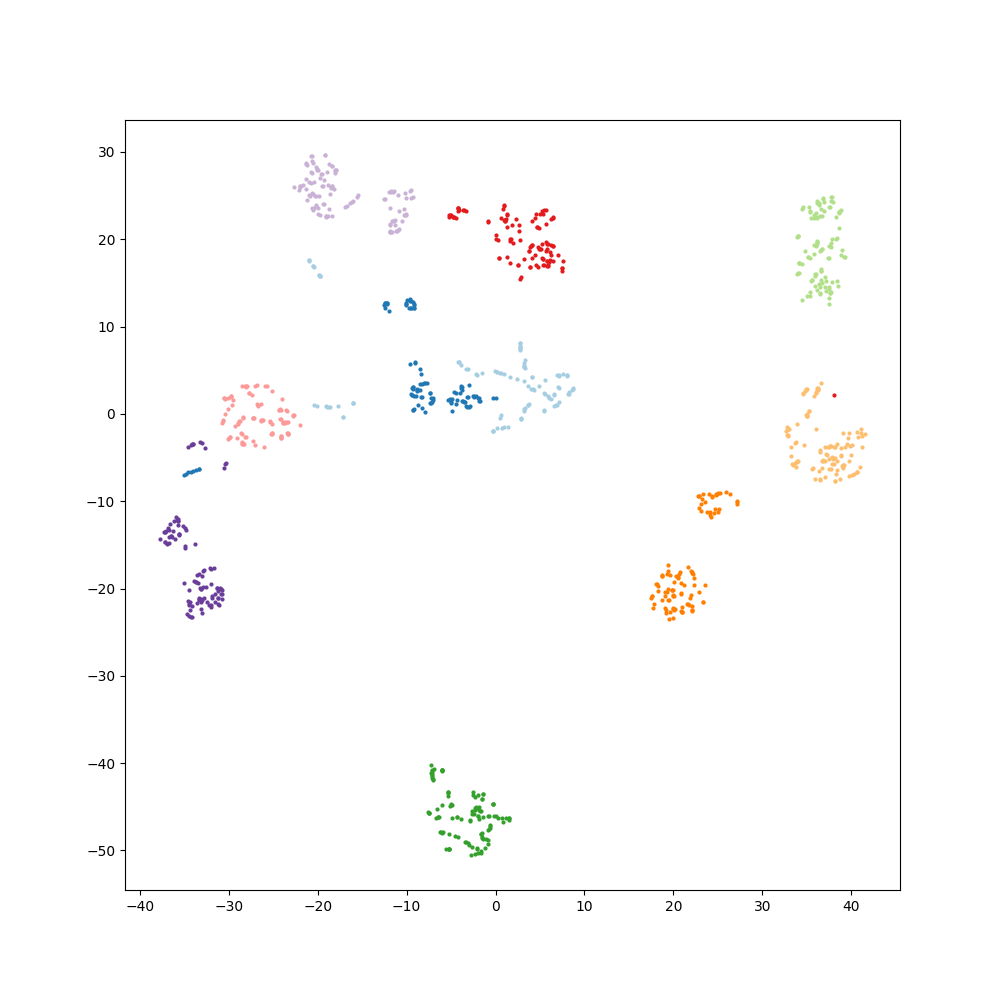}}}
    \hfill
    \subfigure[\Fear (SF: 0.49)\label{fig:clustering:fear_10}]{\fbox{\includegraphics[trim= 92pt 81pt 73pt 89pt, clip, width=0.15\linewidth]{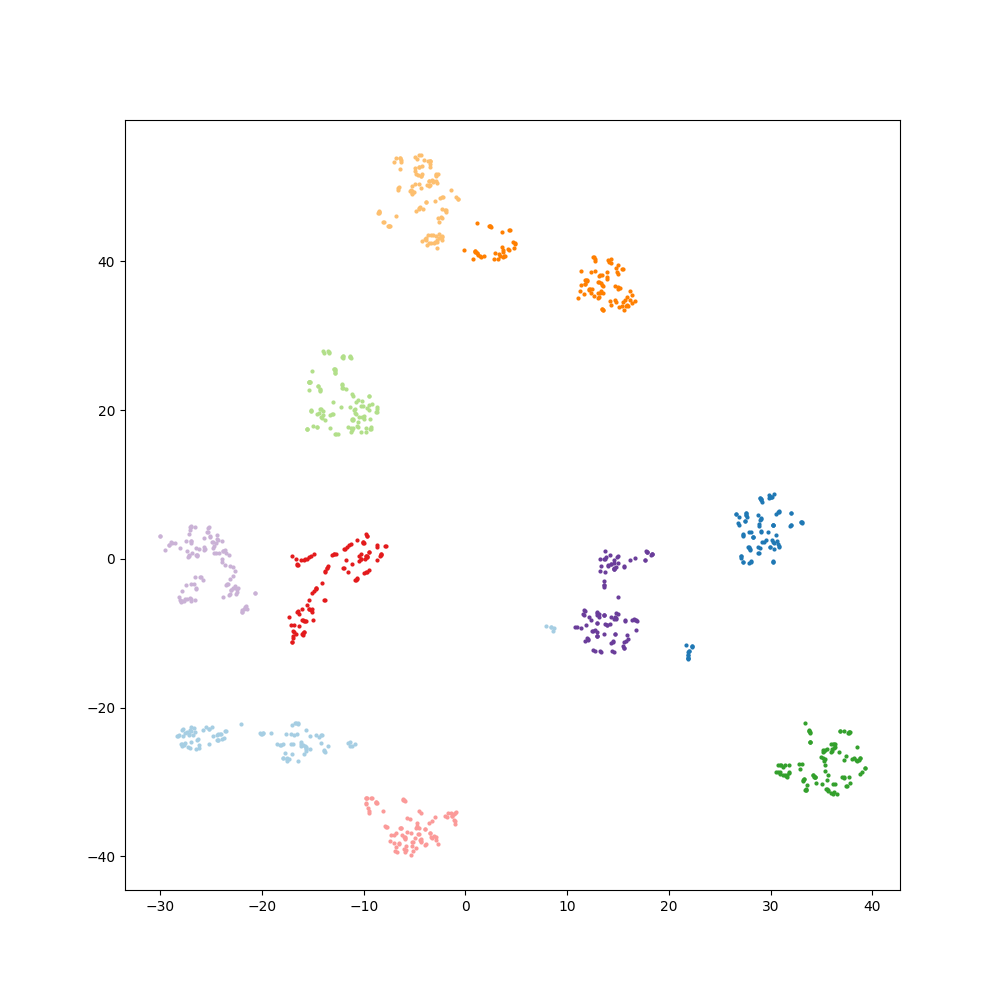}}}
    \hfill
    \subfigure[\Happiness (SF: 0.62)\label{fig:clustering:happy_10}]{\fbox{\includegraphics[trim= 92pt 81pt 73pt 89pt, clip, width=0.15\linewidth]{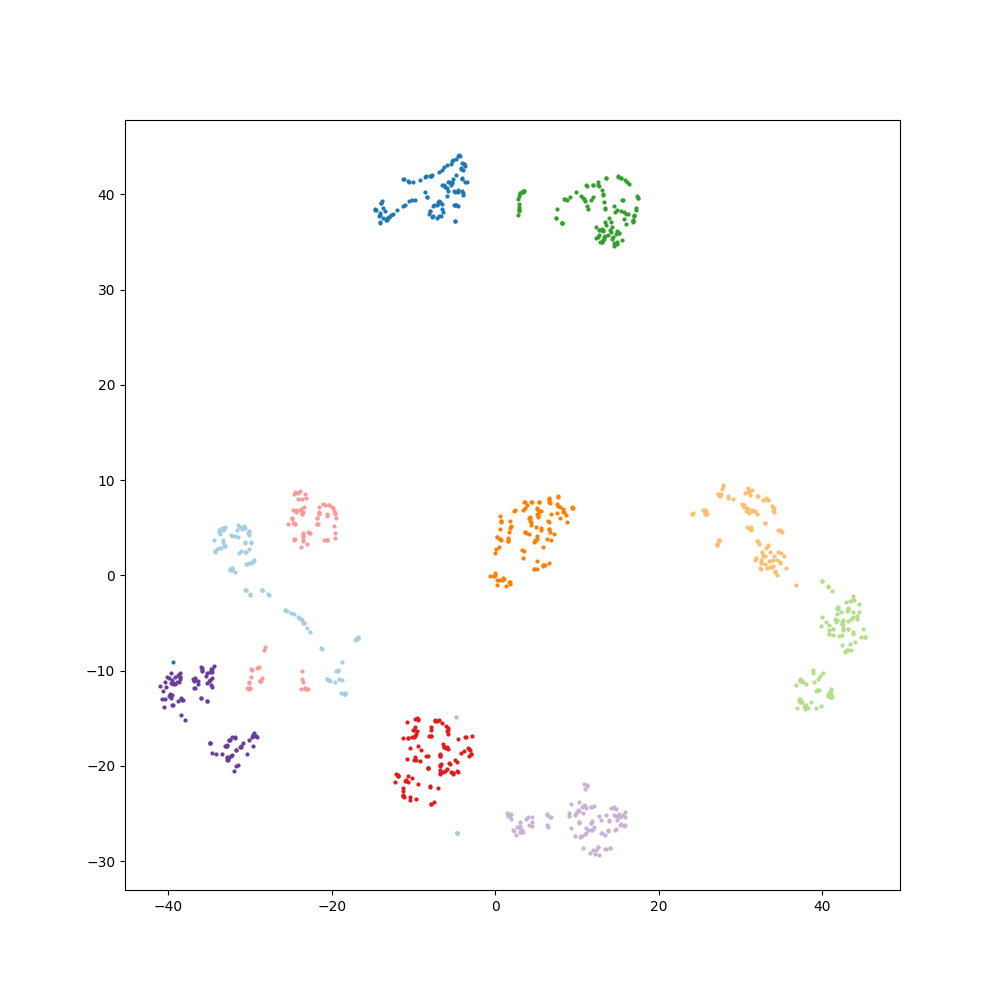}}}
    \hfill
    \subfigure[\Sadness (SF: 0.57)\label{fig:clustering:sad_10}]{\fbox{\includegraphics[trim= 92pt 81pt 73pt 89pt, clip, width=0.15\linewidth]{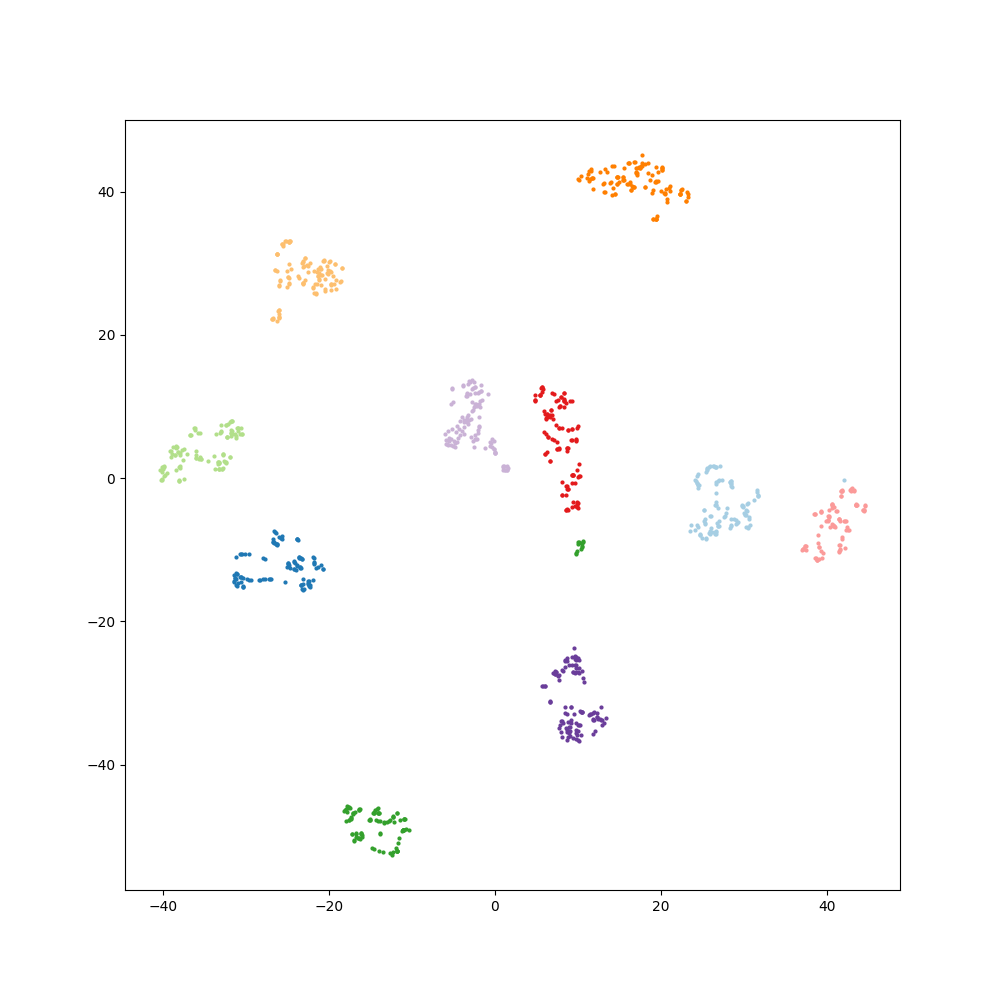}}}
    \hfill
    \subfigure[\Surprise (SF: 0.38)\label{fig:clustering:surprise_10}]{\fbox{\includegraphics[trim= 92pt 81pt 73pt 89pt, clip, width=0.15\linewidth]{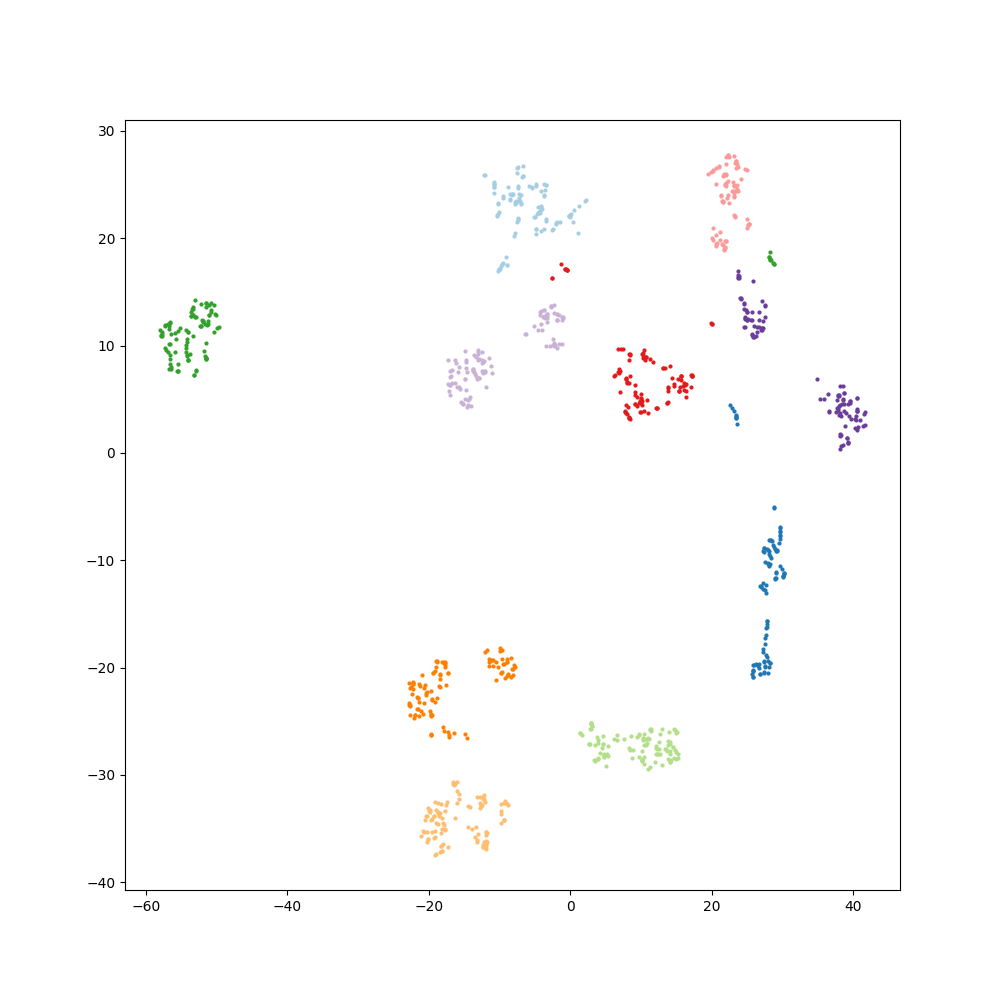}}}

    %\autoref{fig:clustering:angry_full} Shepard fitness: 0.65, \autoref{fig:clustering:disgust_full} Shepard fitness: 0.62, \autoref{fig:clustering:fear_full} Shepard fitness: 0.59, \autoref{fig:clustering:happy_full} Shepard fitness: 0.56, \autoref{fig:clustering:sad_full} Shepard fitness: 0.55, \autoref{fig:clustering:surprise_full}) Shepard fitness: 0.61
    
    \rotatebox{90}{\hspace{11pt}101 subjects}
    \subfigure[\Anger (SF: 0.65)\label{fig:clustering:angry_full}]{\fbox{\includegraphics[trim= 92pt 81pt 73pt 89pt, clip, width=0.15\linewidth]{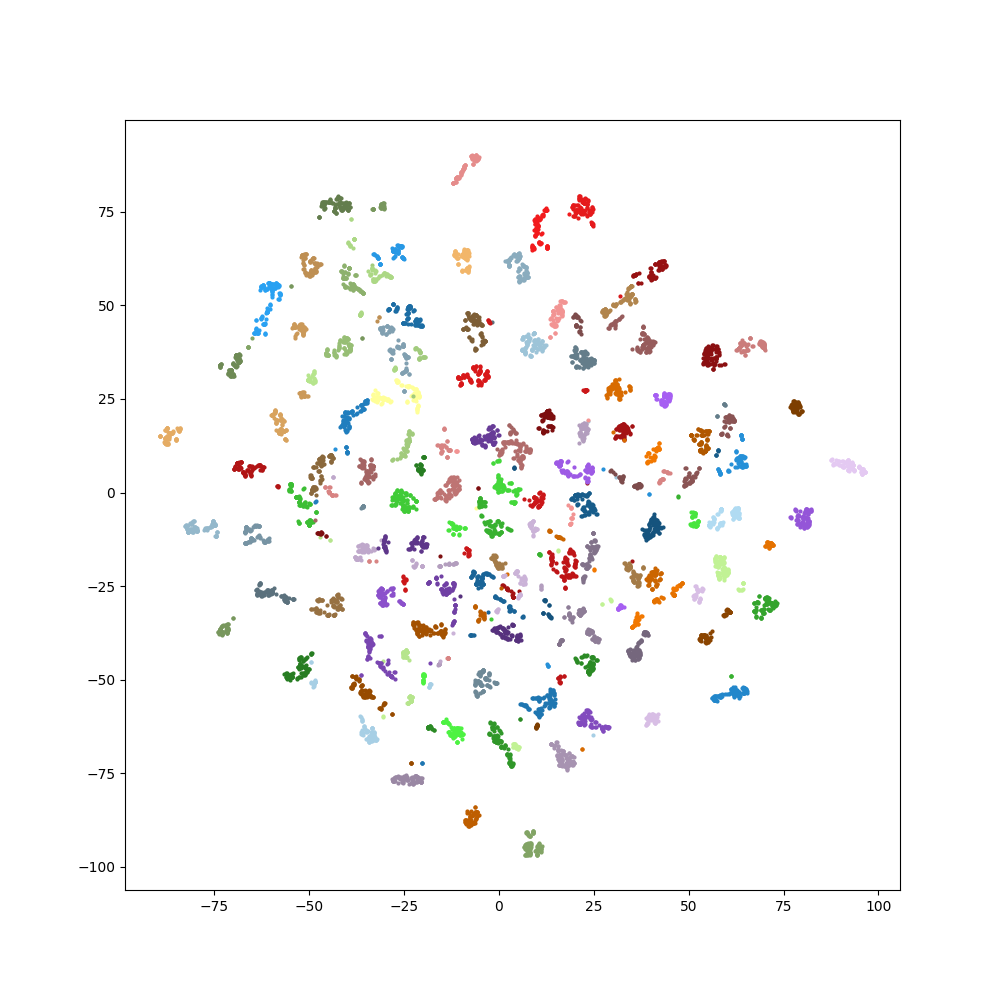}}}
    \hfill
    \subfigure[\Disgust (SF: 0.62)\label{fig:clustering:disgust_full}]{\fbox{\includegraphics[trim= 92pt 81pt 73pt 89pt, clip, width=0.15\linewidth]{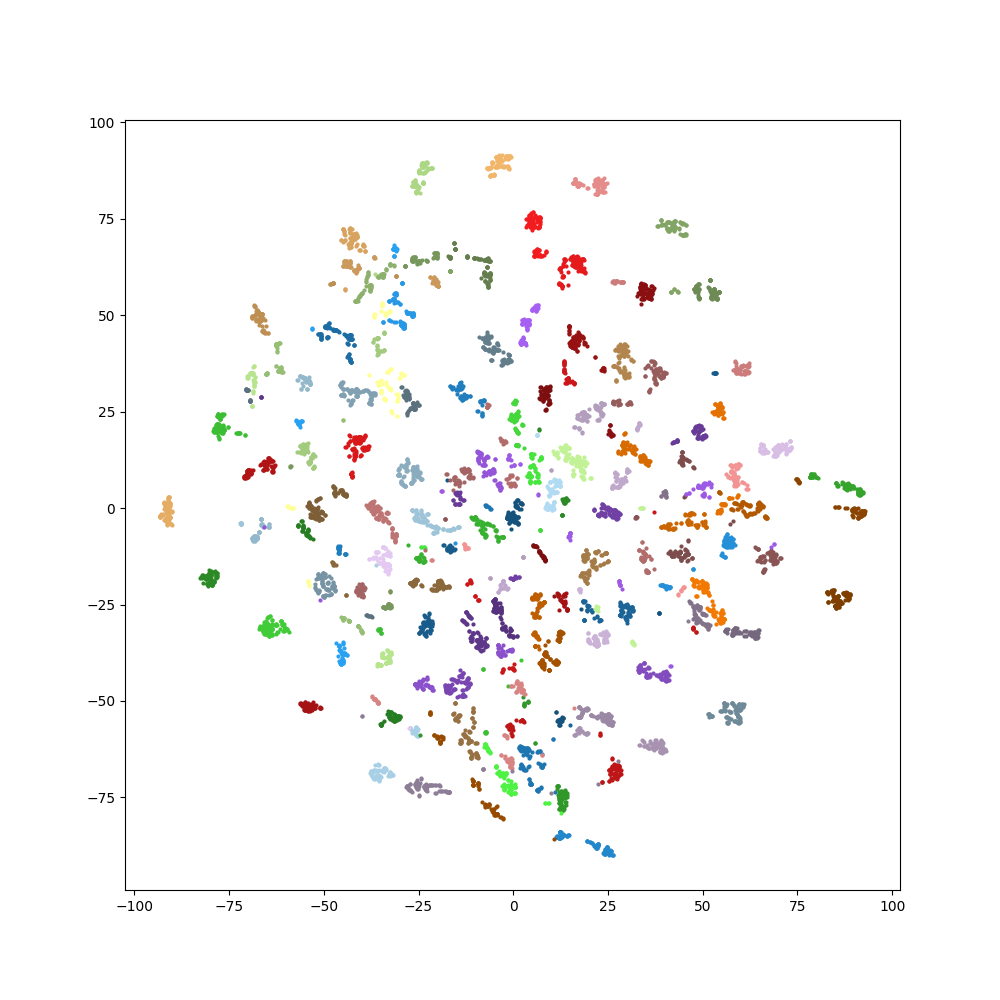}}}
    \hfill
    \subfigure[\Fear (SF: 0.59)\label{fig:clustering:fear_full}]{\fbox{\includegraphics[trim= 92pt 81pt 73pt 89pt, clip, width=0.15\linewidth]{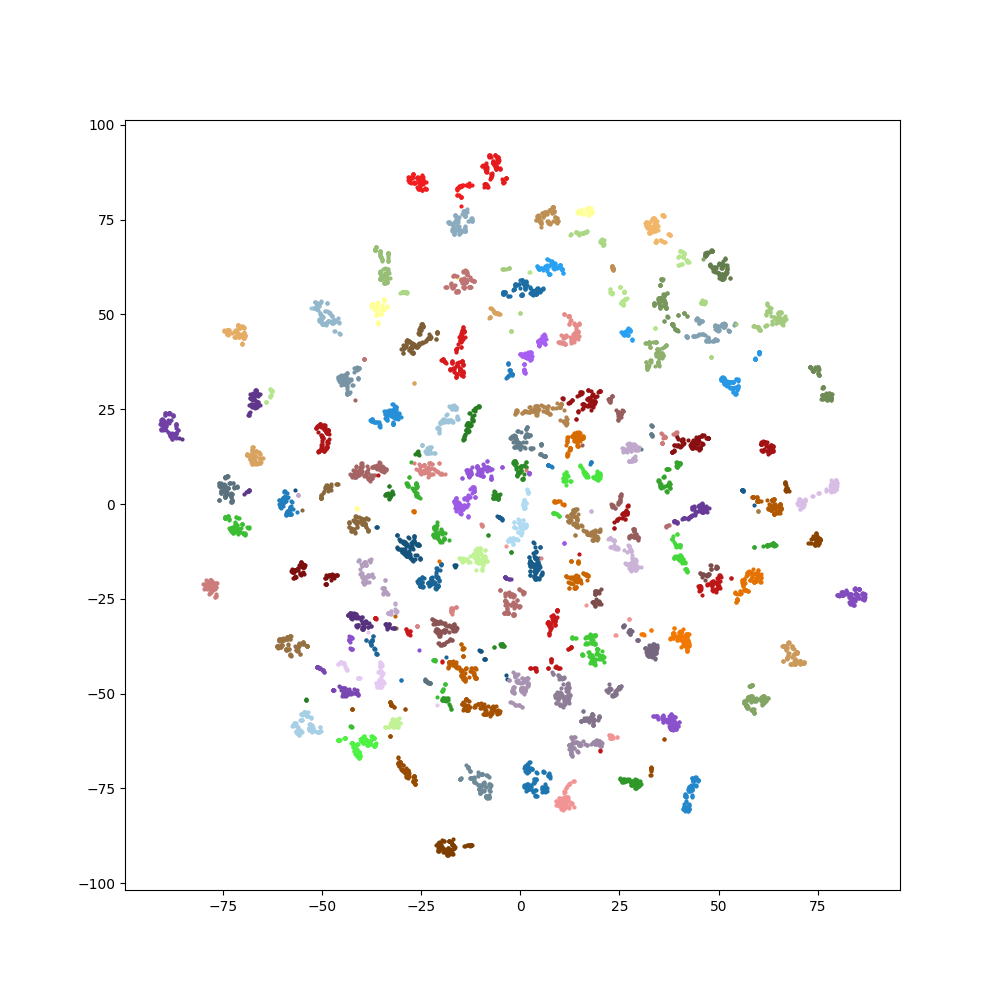}}}
    \hfill
    \subfigure[\Happiness (SF: 0.56)\label{fig:clustering:happy_full}]{\fbox{\includegraphics[trim= 92pt 81pt 73pt 89pt, clip, width=0.15\linewidth]{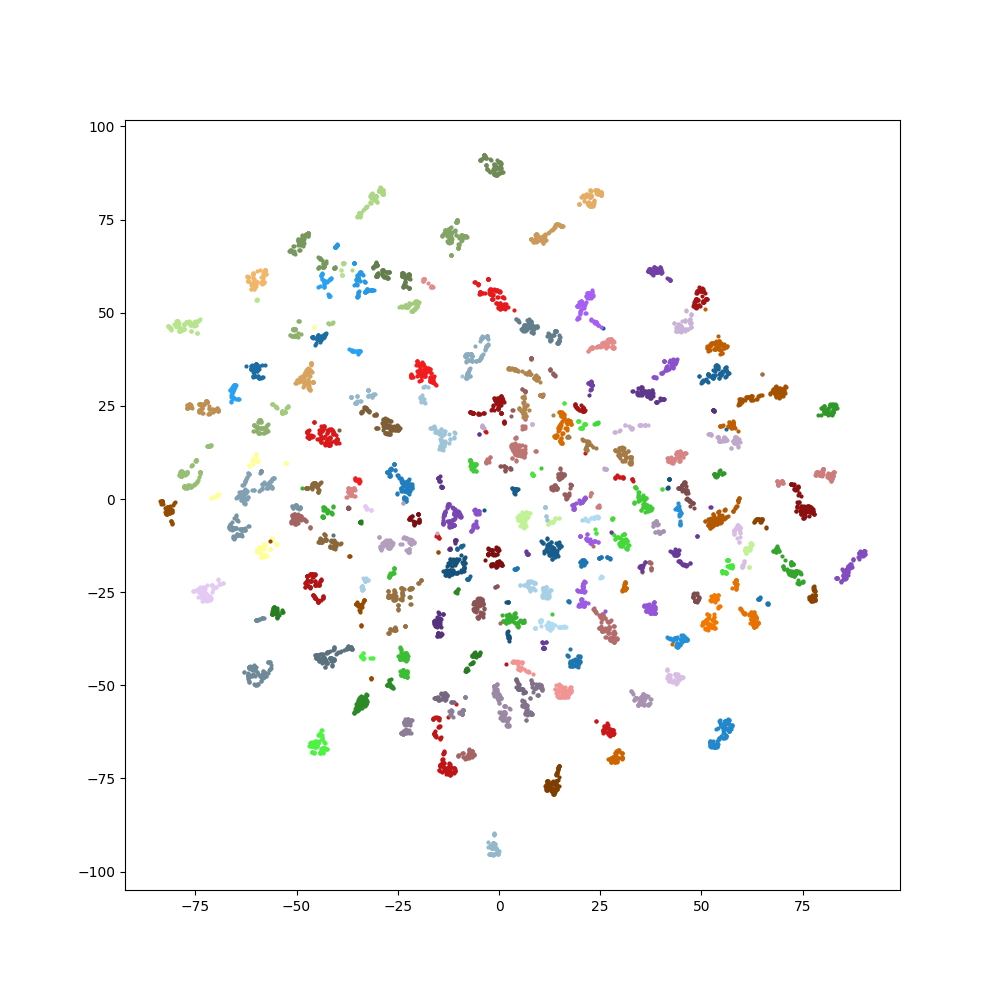}}}
    \hfill
    \subfigure[\Sadness (SF: 0.55)\label{fig:clustering:sad_full}]{\fbox{\includegraphics[trim= 92pt 81pt 73pt 89pt, clip, width=0.15\linewidth]{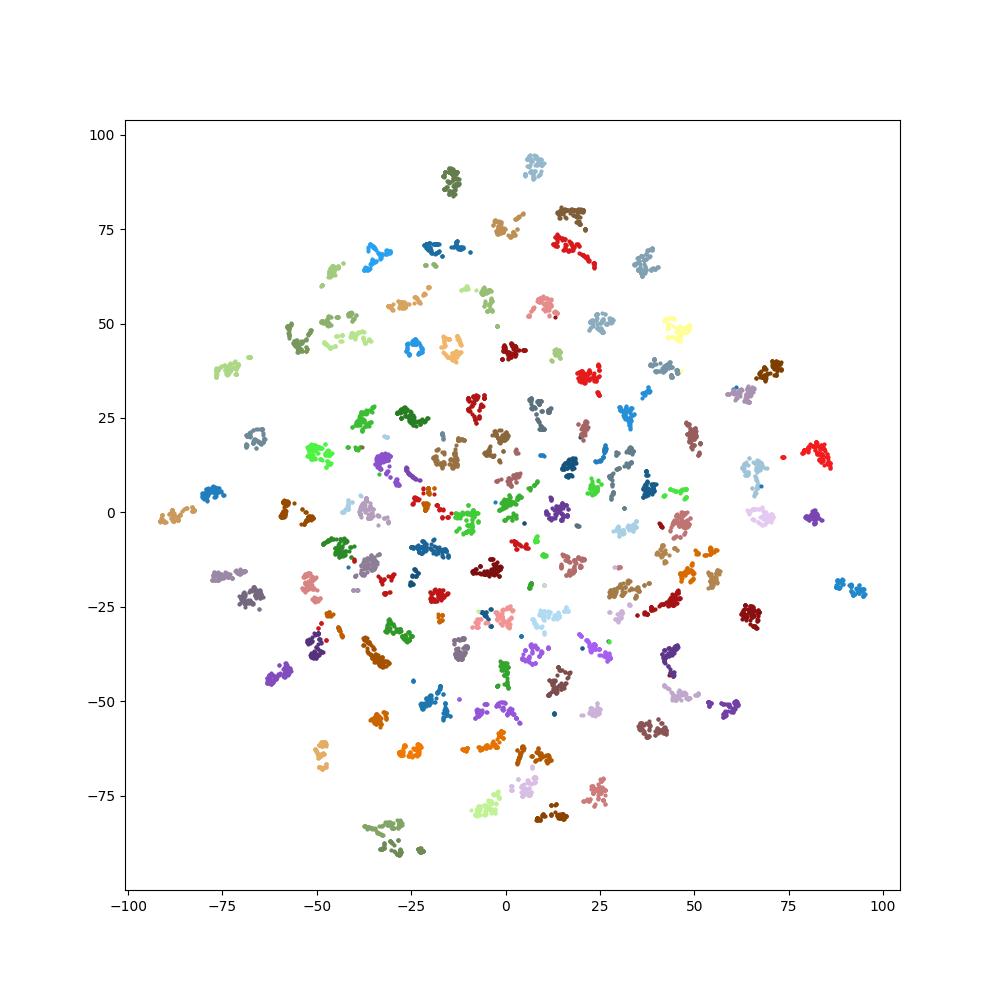}}}
    \hfill
    \subfigure[\Surprise (SF: 0.61)\label{fig:clustering:surprise_full}]{\fbox{\includegraphics[trim= 92pt 81pt 73pt 89pt, clip, width=0.15\linewidth]{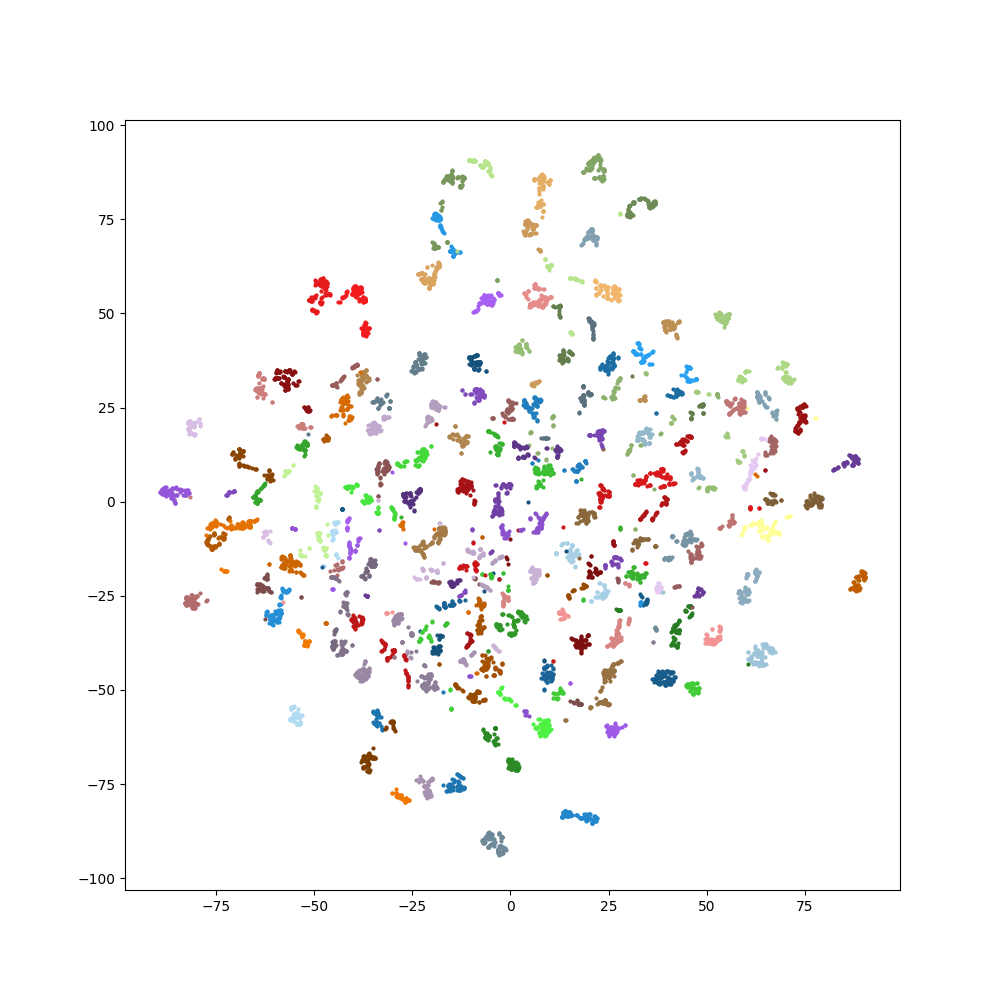}}}

    \caption{Clustering of 10 subjects (F001-F010) on the top and all 101 subjects on the bottom. For 10 subjects, each subject is colored differently. For 101 subject tests, 12 colors were mapped to 101 by creating roughly 10 shades per color. Each plot includes the associated Shepard fitness (SF).}
    \label{fig:clustering}
\end{figure*}

To perform this evaluation, we look at the full face topology of a subset of 10 subjects (F001-F010) using t-SNE (\autoref{fig:clustering:angry_10}-\ref{fig:clustering:surprise_10}). We notice that, for all six emotions, all ten subjects form relatively independent clusters; this is particularly true of the \anger (\autoref{fig:clustering:angry_10}) and \sadness (\autoref{fig:clustering:sad_10}) emotions, while some minor overlap occurs for a few of the individuals in the other emotions. 

We performed a similar evaluation for all 101 subjects (58 female and 43 male) (\autoref{fig:clustering:angry_full}-\ref{fig:clustering:surprise_full}). We can see that the clustering behavior seen with only 10 subjects scales to all the subjects of the dataset. To test the robustness of this t-SNE result to variations in hyperparameters, we ran the tests with four different perplexities (30, 40, 50, and 100) and found that the clusters remained roughly constant throughout (see supplemental material). The clustering phenomenon present in the t-SNE dimension reduction images was also present when we used UMAP (see supplemental materials).

\setstretch{0.97}

\section{Discussion}
\label{sec:discussion}

\subsection{Contribution to Affective Computing}

 Due to the challenging nature of detecting AUs and determining emotion from expression, we hypothesize that our TDA-based approach can be used to provide new insights into these challenges. As it has been shown that temporal AU information can make recognizing emotions easier, our approach evaluates temporal facial expressions (i.e., AUs), which allows us to visualize a new representation of this data. As shown in \autoref{sec:eval}, this representation shows the similarity between the topological signals and the AU signals over time, which provides the following insight, as validated by the coauthor on this paper, who is a researcher in affective computing.

\para{Validation That AUs Are Correctly Detected} A limitation of current machine learning AU detection models is their accuracy, as little improvement has been made compared to previous models~\cite{hinduja2020impact}. This is mainly due to the models detecting AUs that are not active, as well as not detecting AUs that are active. Our TDA-based approach can validate that the detected AUs are correctly capturing the muscle movement of the face. More specifically, the proposed approach will ensure that the AUs that have been detected are correct. As shown in \autoref{fig:au_example:a}, AU45 has high-intensity values three times during the sequence. This correctly corresponds to the three blinks that occur in the data, which are also captured in the topological signal. If the blinks were not captured in the topological signal, then the spikes in the AU intensity signal could be attributed to mislabelling or noise. This could facilitate more intelligent active learning~\cite{ahmed2018wild} that would improve the machine learning detection models.

\para{Relationship Between Which AUs Occur Together} AU co-occurrences~\cite{song2015exploiting} and patterns~\cite{hinduja2020impact} can have a significant impact on the accuracy of machine learning models. Our approach can also give insight into multiple AUs that occur together, including specific AUs that occur when an emotion is expressed. This is detailed in \autoref{fig:au_example:b}, where AU14 and AU25 are active at different intensities at different times during the sequence. The topological signal shows a combination of the two AU signals corresponding to the most intense segments of the active AUs. AU14 is a dimple, and AU25 is active when the lips part, which are common muscle movements that could occur during a wide smile. When a smile occurs, AU6 and AU12 are commonly found together, according to the Facial Action Coding
System (FACS)~\cite{FACS}. However, according to Barrett et al.~\cite{barrett2019emotional}, expressions vary across cultures and situations meaning, AU6 and AU12 may not be active in all smiles. The proposed approach can provide insight into this phenomenon, allowing investigations of the relationships of new AUs, over time, for different expressions.

\para{Detecting Facial Expressions} There are many successful approaches to detecting facial expressions in affective computing~\cite{li2020deep,yang2018facial,zeng2018facial,li2018occlusion,minaee2021deep}. Considering this, the purpose of the proposed approach is \textit{not} to detect facial expressions but to provide greater analysis and explainability of the data. The insight provided by our visualization will allow new insight not previously seen in affective computing, as we can analyze the movement of the face using the proposed TDA-based approach, which directly corresponds to AU movements. This will result in new, more accurate ways of building facial expression detection models.

\para{Explainability of Machine Learning Models} Machine learning has given us many advancements in fields as diverse as medicine~\cite{willemink2020preparing}, security~\cite{albiero2020analysis}, and education~\cite{dhall2020emotiw}. However, one of the main limitations is the lack of explainability~\cite{dovsilovic2018explainable}. Considering this, one of the key advantages of TDA over machine learning is the explainability of the features identified in the process. We demonstrate this using an example of four facial poses from the female \surprise data, as shown in \autoref{fig:representative_ex}. These examples focus on the opening and closing of the eyes and mouth, which is commonly associated with a surprised expression. Given a machine learning model that successfully detected the AUs associated with this expression, with the long list of possible muscle movements (e.g., AU1, AU2, AU5, AU25, and AU26), it is difficult to understand \textit{why} such a model detected them. This is especially true given the black-box nature of neural networks~\cite{oh2019towards}. In \autoref{fig:representative_ex}, we can directly see the features that change the data. In the persistence diagrams, the number and persistence of the most important features are clearly different. Furthermore, when evaluating the representative cycles, we can further associate the landmark geometry of each high persistence feature in the data. This explainability will facilitate more accurate emotion recognition systems. This is due to the insight that the explainability will give affective computing researchers to better understand and tune their models, resulting in more accurate models.

\subsection{Topology Doesn't Capture Everything}

\para{Limits of Topology} Some of the advantages of topology over geometry are also its biggest weaknesses. There are certain shapes of the data that may not be captured by topology alone. For example, smiles and frowns \textit{may} have the same topological shape. That said, the relationship between the smile, nose, and jawline may be sufficient enough to disambiguate between smiles and frowns. Furthermore, smiles and frowns will also be associated with other changes in facial features, e.g., changes in eye or eyebrow shape. At the very least, our evaluation showed that smile emotions, e.g., \happiness, and frown emotions, e.g., \sadness, were differentiable in \autoref{sec:eval:exp}.

\para{Differences Between TDA and Machine Learning}
Topology does not capture all features of AUs. The AU extraction may utilize other data, nonlinearities, knowledge of physiological relationships of AUs, etc., to determine the extent of the activation of the AUs. That said, it is also important to understand that \textit{the AU information is not ground truth}. It is the output of a machine learning technique, and it may, in fact, not be showing genuine AU activation. On the other hand, all of the topological features we observed are in the data, and for any matching feature, TDA provides evidence as to why the machine learning algorithm classified AU activation as it did.

\subsection{Challenges and Limitations of Automatic AU Detection}

The use of AUs for expression recognition is a promising approach. However, there are significant challenges in the detection of them~\cite{ertugrul2020crossing}. Many works that have developed approaches for automatically detecting AUs using machine learning have focused on learning single AUs. However, it has been shown that patterns of AUs can have a significant impact on detection~\cite{hinduja2020impact}. This leads to a bigger limitation of current machine learning approaches to AU detection, namely the data. State-of-the-art machine learning models require a large amount of good data to be accurate. Current models are trained on data that has biases in ethnicity~\cite{mavadati2013disfa}, as well as significant imbalances in the distribution of AUs~\cite{zhang2014bp4d, zhang2016multimodal}. Along with these data biases and imbalances, the ground truth data is often manually annotated, which is subjective and can lead to errors~\cite{o2020promoting}. This results in machine learning models that are not learning how to represent an AU but learning the distribution of data~\cite{hinduja2020impact}. In machine learning, many times, the solution to the problem is to collect more data. In the case of AUs, it has the additional challenge that multiple AUs occur simultaneously~\cite{song2015exploiting}, resulting in an unresolvable imbalance of data for the AUs that occur more often (e.g., AU6 or AU12)~\cite{hinduja2020impact}. These challenges lead to machine learning models that often fail to recognize AUs that are active, as well as recognize AUs that are not active~\cite{zhang2020region}. 

Along with challenges in detecting AUs, there is a larger discussion of how humans learn and express emotions~\cite{hoemann2020developing}. Broadly, this discussion can be categorized into two hypotheses. The first hypothesis states that emotions can be recognized from facial expressions (AUs)~\cite{ekman1987universals}. This hypothesis is the basis for the Facial Action Coding System (FACS) AUs~\cite{FACS} and is a significant motivation for the field of affective computing. The second hypothesis contradicts that and states that expressions vary across cultures, situations, and people in the same situation. Because of this, emotions cannot be recognized from facial expressions alone~\cite{barrett2019emotional}. Instead, other factors such as context, physiology, age, and gender should be considered. While these two hypotheses contradict one another, recent work has shown validity in both hypotheses. More specifically, while it is difficult to determine emotion from AUs given a single facial image when temporal AU information and context are considered, the task becomes easier~\cite{hindujaPatterns2020}. Along with this, it has also been shown that the fusion of physiological signals, such as heart rate and respiration, along with AUs can be used to accurately recognize pain in subjects~\cite{hinduja2020multimodal}. Although it is a challenging problem to automatically detect AUs using machine learning, these works show that AUs are still a promising approach to solve the challenging problem of emotion from expression.

\section{Conclusions}

In this paper, we have demonstrated that TDA can be used to discern and better understand patterns that exist between emotions. Paired with machine learning approaches to affective computing, TDA provides a means to evaluate particular aspects of the data to discern what parts of the face may be causing the machine learning algorithm to recognize the data as a particular emotion or explain the shortcomings or misdiagnoses that the machine learning algorithm provides, e.g., if the algorithm recognizes a \happiness emotion as \anger, TDA may help to discern what led to this misdiagnosis. The next phase of this work is to begin evaluating these affective computing hypotheses, particularly those discussed in \autoref{sec:discussion}, using our TDA-based analysis and interface.

\acknowledgments{
We thank the anonymous reviewers for their feedback. This project was supported in part by the National Science Foundation (IIS-1845204).}

\setstretch{1}

\bibliographystyle{abbrv-doi}

\bibliography{main}

\begin{thebibliography}{10}

\bibitem{ahmed2018wild}
M.~U. Ahmed, K.~J. Woo, K.~Y. Hyeon, M.~R. Bashar, and P.~K. Rhee.
\newblock Wild facial expression recognition based on incremental active
  learning.
\newblock {\em Cognitive Systems Research}, 52:212--222, 2018.

\bibitem{albiero2020analysis}
V.~Albiero, K.~KS, K.~Vangara, K.~Zhang, M.~C. King, and K.~W. Bowyer.
\newblock Analysis of gender inequality in face recognition accuracy.
\newblock In {\em Proceedings of the IEEE Winter Conference on Applications of
  Computer Vision Workshops}, pp. 81--89, 2020.

\bibitem{aouada20203d}
D.~Aouada, K.~Cherenkova, G.~Gusev, B.~Ottersten, et~al.
\newblock {3D Deformation Signature for Dynamic Face Recognition}.
\newblock In {\em IEEE International Conference on Acoustics, Speech and Signal
  Processing (ICASSP)}, pp. 2138--2142, 2020.

\bibitem{baltruvsaitis2016openface}
T.~Baltru{\v{s}}aitis et~al.
\newblock Openface: an open source facial behavior analysis toolkit.
\newblock In {\em IEEE Winter Conference on Applications of Computer Vision
  (WACV)}, pp. 1--10, 2016.

\bibitem{barrett2019emotional}
L.~F. Barrett, R.~Adolphs, S.~Marsella, A.~M. Martinez, and S.~D. Pollak.
\newblock Emotional expressions reconsidered: Challenges to inferring emotion
  from human facial movements.
\newblock {\em Psychological Science in the Public Interest}, 20(1):1--68,
  2019.

\bibitem{bauer2019ripser}
U.~Bauer.
\newblock {Ripser: Efficient Computation of Vietoris-Rips Persistence
  Barcodes}.
\newblock {\em arXiv preprint arXiv:1908.02518}, 2019.

\bibitem{Bradley:1994}
M.~Bradley and P.~Lang.
\newblock {Measuring Emotion: The Self-Assessment Manikin and the Semantic
  Differential}.
\newblock {\em Journal of Behavior Therapy and Experimental Psychiatry}, 1994.

\bibitem{canavan2015landmark}
S.~Canavan, P.~Liu, X.~Zhang, and L.~Yin.
\newblock {Landmark Localization on 3D/4D Range Data Using a Shape Index-Based
  Statistical Shape Model with Global and Local Constraints}.
\newblock {\em Computer Vision and Image Understanding}, 139:136--148, 2015.

\bibitem{Cernea:2013}
D.~Cernea, C.~Weber, A.~Ebert, and A.~Kerren.
\newblock {Emotion Scents - A Method of Representing User Emotions on GUI
  Widgets}.
\newblock {\em SPIE Digital Library}, 2013.

\bibitem{cernea2015emotion}
D.~Cernea, C.~Weber, A.~Ebert, and A.~Kerren.
\newblock Emotion-prints: Interaction-driven emotion visualization on
  multi-touch interfaces.
\newblock In {\em Visualization and Data Analysis 2015}, vol. 9397, p. 93970A.
  International Society for Optics and Photonics, 2015.

\bibitem{cernea2014group}
D.~Cernea, C.~Weber, A.~Kerren, and A.~Ebert.
\newblock Group affective tone awareness and regulation through virtual agents.
\newblock In {\em Proceeding of IVA 2014 Workshop on Affective Agents, Boston,
  MA, USA, 27-29 August}, pp. 9--16, 2014.

\bibitem{chinaev2018mobileface}
N.~Chinaev, A.~Chigorin, and I.~Laptev.
\newblock {Mobileface: 3D Face Reconstruction with Efficient CNN Regression}.
\newblock In {\em European Conference on Computer Vision (ECCV)}, pp. 0--0,
  2018.

\bibitem{cohen2007stability}
D.~Cohen-Steiner, H.~Edelsbrunner, and J.~Harer.
\newblock Stability of persistence diagrams.
\newblock {\em Discrete \& computational geometry}, 37(1):103--120, 2007.

\bibitem{cootes2001active}
T.~Cootes, G.~Edwards, and C.~Taylor.
\newblock {Active Appearance Models}.
\newblock {\em IEEE Transactions on Pattern Analysis and Machine Intelligence},
  23(6):681--685, 2001.

\bibitem{deng2016gender}
Y.~Deng, L.~Chang, M.~Yang, M.~Huo, and R.~Zhou.
\newblock Gender differences in emotional response: Inconsistency between
  experience and expressivity.
\newblock {\em PloS one}, 11(6):e0158666, 2016.

\bibitem{dhall2020emotiw}
A.~Dhall, G.~Sharma, R.~Goecke, and T.~Gedeon.
\newblock Emotiw 2020: Driver gaze, group emotion, student engagement and
  physiological signal based challenges.
\newblock In {\em International Conference on Multimodal Interaction}, pp.
  784--789, 2020.

\bibitem{di3d}
{\em Di3D Inc.}
\newblock \url{http://www.di3d.com}.

\bibitem{dovsilovic2018explainable}
F.~K. Do{\v{s}}ilovi{\'c}, M.~Br{\v{c}}i{\'c}, and N.~Hlupi{\'c}.
\newblock Explainable artificial intelligence: A survey.
\newblock In {\em International Convention on Information and Communication
  Technology, Electronics and Microelectronics (MIPRO)}, pp. 0210--0215, 2018.

\bibitem{EdelsbrunnerHarer2008}
H.~Edelsbrunner and J.~Harer.
\newblock {Persistent Homology - A Survey}.
\newblock {\em Contemporary Mathematics}, 453:257--282, 2008.

\bibitem{EdelsbrunnerHarer2010}
H.~Edelsbrunner and J.~Harer.
\newblock {\em {Computational Topology: An Introduction}}.
\newblock American Mathematical Society, 2010.

\bibitem{edelsbrunner2000topological}
H.~Edelsbrunner, D.~Letscher, and A.~Zomorodian.
\newblock Topological persistence and simplification.
\newblock In {\em Proceedings 41st Annual Symposium on Foundations of Computer
  Science}, pp. 454--463, 2000.

\bibitem{ekman1999basic}
P.~Ekman.
\newblock {Basic Emotions}.
\newblock {\em Handbook of Cognition and Emotion}, 98(45-60):16, 1999.

\bibitem{ekman1969repertoire}
P.~Ekman and W.~Friesen.
\newblock {The Repertoire of Nonverbal Behavior: Categories, Origins, Usage,
  and Coding}.
\newblock {\em Semiotica}, 1(1):49--98, 1969.

\bibitem{ekman1987universals}
P.~Ekman, W.~Friesen, M.~O'Sullivan, A.~Chan, I.~Diacoyanni-Tarlatzis,
  K.~Heider, R.~Krause, W.~A. LeCompte, T.~Pitcairn, P.~Ricci-Bitti, et~al.
\newblock {Universals and Cultural Differences in the Judgments of Facial
  Expressions of Emotion}.
\newblock {\em Journal of Personality and Social Psychology}, 53(4):712, 1987.

\bibitem{FACS}
R.~Ekman.
\newblock {\em What the face reveals: Basic and applied studies of spontaneous
  expression using the Facial Action Coding System (FACS)}.
\newblock Oxford University Press, USA, 1997.

\bibitem{ertugrul2020crossing}
I.~O. Ertugrul, J.~F. Cohn, L.~A. Jeni, Z.~Zhang, L.~Yin, and Q.~Ji.
\newblock Crossing domains for au coding: Perspectives, approaches, and
  measures.
\newblock {\em IEEE Transactions on Biometrics, Behavior, and Identity
  Science}, 2(2):158--171, 2020.

\bibitem{fabiano2019deformable}
D.~Fabiano and S.~Canavan.
\newblock {Deformable Synthesis Model for Emotion Recognition}.
\newblock In {\em IEEE International Conference on Automatic Face and Gesture
  Recognition}, pp. 1--5, 2019.

\bibitem{fabiano2020impact}
D.~Fabiano, M.~Jaishanker, and S.~Canavan.
\newblock {Impact of Multiple Modalities on Emotion Recognition: Investigation
  into 3d Facial Landmarks, Action Units, and Physiological Data}.
\newblock {\em arXiv preprint arXiv:2005.08341}, 2020.

\bibitem{fan2020facial}
Y.~Fan, V.~Li, and J.~C. Lam.
\newblock {Facial Expression Recognition with Deeply-Supervised Attention
  Network}.
\newblock {\em IEEE Transactions on Affective Computing}, 2020.

\bibitem{lawrence20143d}
S.~L. Fernandes and G.~J. Bala.
\newblock {3D and 4D Face Recognition: A Comprehensive Review}.
\newblock {\em Recent Patents on Engineering}, 8(2), 2014.

\bibitem{fleureau2012physiological}
J.~Fleureau, P.~Guillotel, and Q.~Huynh-Thu.
\newblock {Physiological-Based Affect Event Detector for Entertainment Video
  Applications}.
\newblock {\em IEEE Transactions on Affective Computing}, 3(3):379--385, 2012.

\bibitem{gao20193d}
W.~Gao, X.~Zhao, Z.~Gao, J.~Zou, P.~Dou, and I.~Kakadiaris.
\newblock {3D Face Reconstruction From Volumes of Videos Using a MapReduce
  Framework}.
\newblock {\em IEEE Access}, 7:165559--165570, 2019.

\bibitem{grasshof2020multilinear}
S.~Grasshof, H.~Ackermann, S.~Brandt, and J.~Ostermann.
\newblock {Multilinear Modelling of Faces and Expressions}.
\newblock {\em IEEE Transactions on Pattern Analysis and Machine Intelligence},
  2020.

\bibitem{hajij2020fast}
M.~Hajij, E.~Munch, and P.~Rosen.
\newblock Fast and scalable complex network descriptor using pagerank and
  persistent homology.
\newblock In {\em 2020 International Conference on Intelligent Data Science
  Technologies and Applications (IDSTA)}, pp. 110--114. IEEE, 2020.

\bibitem{hajij2018visual}
M.~Hajij, B.~Wang, C.~Scheidegger, and P.~Rosen.
\newblock {Visual Detection of Structural Changes in Time-Varying Graphs Using
  Persistent Homology}.
\newblock In {\em IEEE Pacific Visualization}, pp. 125--134, 2018.

\bibitem{hariri20173d}
W.~Hariri, H.~Tabia, N.~Farah, A.~Benouareth, and D.~Declercq.
\newblock {3D Facial Expression Recognition Using Kernel Methods on Riemannian
  Manifold}.
\newblock {\em Engineering Applications of Artificial Intelligence}, 64:25--32,
  2017.

\bibitem{hernandez2016facial}
A.~Hernandez-Matamoros, A.~Bonarini, E.~Escamilla-Hernandez,
  M.~Nakano-Miyatake, and H.~Perez-Meana.
\newblock {Facial Expression Recognition with Automatic Segmentation of Face
  Regions Using a Fuzzy-Based Classification Approach}.
\newblock {\em Knowledge-Based Systems}, 110:1--14, 2016.

\bibitem{hindujaPatterns2020}
S.~Hinduja, S.~Aathreya, S.~Canavan, J.~Cohn, and L.~Yin.
\newblock Recognizing context using facial expression dynamics from action unit
  patterns.
\newblock {\em IEEE Transactions on Affective Computing (Under Review)}, 2020.

\bibitem{hinduja2020impact}
S.~Hinduja, S.~Canavan, and S.~Aathreya.
\newblock Impact of action unit occurrence patterns on detection.
\newblock {\em arXiv preprint arXiv:2010.07982}, 2020.

\bibitem{hinduja2020multimodal}
S.~Hinduja, S.~Canavan, and G.~Kaur.
\newblock Multimodal fusion of physiological signals and facial action units
  for pain recognition.
\newblock In {\em IEEE International Conference on Automatic Face and Gesture
  Recognition}, pp. 387--391, 2020.

\bibitem{hoemann2020developing}
K.~Hoemann, R.~Wu, V.~LoBue, L.~M. Oakes, F.~Xu, and L.~F. Barrett.
\newblock Developing an understanding of emotion categories: Lessons from
  objects.
\newblock {\em Trends in Cognitive Sciences}, 24(1):39--51, 2020.

\bibitem{jannat2020subject}
S.~Jannat, D.~Fabiano, S.~Canavan, and T.~Neal.
\newblock {Subject Identification across Large Expression Variations Using 3D
  Facial Landmarks}.
\newblock {\em arXiv preprint arXiv:2005.08339}, 2020.

\bibitem{kalam2019facial}
A.~Kalam, M.~Haque, M.~Jashem, M.~Hasan, M.~Ibrahim, and T.~Jabid.
\newblock {Facial Expression Recognition Using Local Composition Pattern}.
\newblock In {\em International Conference on Computer and Communications
  Management}, pp. 63--67, 2019.

\bibitem{kerber2017geometry}
M.~Kerber, D.~Morozov, and A.~Nigmetov.
\newblock {Geometry Helps to Compare Persistence Diagrams}.
\newblock {\em Journal of Experimental Algorithmics}, 22:1--4, 2017.

\bibitem{kim2020contverb}
H.~Kim, J.~Ben-Othman, L.~Mokdad, and K.~Lim.
\newblock {CONTVERB: Continuous Virtual Emotion Recognition Using Replaceable
  Barriers for Intelligent Emotion-based IoT Services and Applications}.
\newblock {\em IEEE Network}, 2020.

\bibitem{koenderink1992surface}
J.~Koenderink and A.~V. Doorn.
\newblock {Surface Shape and Curvature Scales}.
\newblock {\em Image and Vision Computing}, 10(8):557--564, 1992.

\bibitem{Kovacevik:2020}
N.~Kovacevic, R.~Wampfler, B.~Solenthaler, M.~Gross, and T.~G{\"u}nther.
\newblock {Glyph-Based Visualization of Affective States}.
\newblock {\em EuroVis: Short Papers}, 2020.

\bibitem{kruskal1964multidimensional}
J.~Kruskal.
\newblock {Multidimensional Scaling by Optimizing Goodness of Fit to a
  Non-Metric Hypothesis}.
\newblock {\em Psychometrika}, 29(1):1--27, 1964.

\bibitem{li2020deep}
S.~Li and W.~Deng.
\newblock {Deep Facial Expression Recognition: A Survey}.
\newblock {\em IEEE Transactions on Affective Computing}, 2020.

\bibitem{li2018occlusion}
Y.~Li, J.~Zeng, S.~Shan, and X.~Chen.
\newblock Occlusion aware facial expression recognition using cnn with
  attention mechanism.
\newblock {\em IEEE Transactions on Image Processing}, 28(5):2439--2450, 2018.

\bibitem{lien1998automated}
J.~J. Lien, T.~Kanade, J.~F. Cohn, and C.-C. Li.
\newblock Automated facial expression recognition based on facs action units.
\newblock In {\em IEEE International Conference on Automatic Face and Gesture
  Recognition}, pp. 390--395, 1998.

\bibitem{liu2020saanet}
D.~Liu, X.~Ouyang, S.~Xu, P.~Zhou, K.~He, and S.~Wen.
\newblock Saanet: Siamese action-units attention network for improving dynamic
  facial expression recognition.
\newblock {\em Neurocomputing}, 413:145--157, 2020.

\bibitem{liu20193d}
F.~Liu, L.~Tran, and X.~Liu.
\newblock {3D Face Modeling from Diverse Raw Scan Data}.
\newblock In {\em IEEE International Conference on Computer Vision}, pp.
  9408--9418, 2019.

\bibitem{lucey2010automatically}
P.~Lucey, J.~F. Cohn, I.~Matthews, S.~Lucey, S.~Sridharan, J.~Howlett, and
  K.~M. Prkachin.
\newblock Automatically detecting pain in video through facial action units.
\newblock {\em IEEE Transactions on Systems, Man, and Cybernetics, Part B
  (Cybernetics)}, 41(3):664--674, 2010.

\bibitem{mavadati2013disfa}
S.~M. Mavadati, M.~H. Mahoor, K.~Bartlett, P.~Trinh, and J.~F. Cohn.
\newblock Disfa: A spontaneous facial action intensity database.
\newblock {\em IEEE Transactions on Affective Computing}, 4(2):151--160, 2013.

\bibitem{mcduff2012affectaura}
D.~McDuff, A.~Karlson, A.~Kapoor, A.~Roseway, and M.~Czerwinski.
\newblock Affectaura: an intelligent system for emotional memory.
\newblock In {\em Proceedings of the SIGCHI Conference on Human Factors in
  Computing Systems}, pp. 849--858, 2012.

\bibitem{mcinnes2018umap}
L.~McInnes, J.~Healy, and J.~Melville.
\newblock Umap: Uniform manifold approximation and projection for dimension
  reduction.
\newblock {\em arXiv preprint arXiv:1802.03426}, 2018.

\bibitem{minaee2021deep}
S.~Minaee, M.~Minaei, and A.~Abdolrashidi.
\newblock Deep-emotion: Facial expression recognition using attentional
  convolutional network.
\newblock {\em Sensors}, 21(9):3046, 2021.

\bibitem{molho2017disgust}
C.~Molho, J.~M. Tybur, E.~G{\"u}ler, D.~Balliet, and W.~Hofmann.
\newblock Disgust and anger relate to different aggressive responses to moral
  violations.
\newblock {\em Psychological Science}, 28(5):609--619, 2017.

\bibitem{o2020promoting}
K.~O'Connor, A.~Sarker, J.~Perrone, and G.~G. Hernandez.
\newblock Promoting reproducible research for characterizing nonmedical use of
  medications through data annotation: Description of a twitter corpus and
  guidelines.
\newblock {\em Journal of Medical Internet Research}, 22(2):e15861, 2020.

\bibitem{oh2019towards}
S.~J. Oh, B.~Schiele, and M.~Fritz.
\newblock Towards reverse-engineering black-box neural networks.
\newblock In {\em Explainable AI: Interpreting, Explaining and Visualizing Deep
  Learning}, pp. 121--144. Springer, 2019.

\bibitem{otberdout2019automatic}
N.~Otberdout, A.~Kacem, M.~Daoudi, L.~Ballihi, and S.~Berretti.
\newblock Automatic analysis of facial expressions based on deep covariance
  trajectories.
\newblock {\em IEEE transactions on neural networks and learning systems},
  31(10):3892--3905, 2019.

\bibitem{patil2017emotion}
G.~Patil and P.~Suja.
\newblock {Emotion Recognition from 3D Videos Using Optical Flow Method}.
\newblock In {\em IEEE International Conference On Smart Technologies For Smart
  Nation (SmartTechCon)}, pp. 825--829, 2017.

\bibitem{pham2016robust}
H.~Pham, V.~Pavlovic, J.~Cai, and T.~jen Cham.
\newblock {Robust Real-Time Performance-Driven 3D Face Tracking}.
\newblock In {\em IEEE International Conference on Pattern Recognition (ICPR)},
  pp. 1851--1856, 2016.

\bibitem{picard2000affective}
R.~Picard.
\newblock {\em {Affective Computing}}.
\newblock MIT press, 2000.

\bibitem{qin2020heartbees}
C.~Y. Qin, M.~Constantinides, L.~M. Aiello, and D.~Quercia.
\newblock Heartbees: Visualizing crowd affects.
\newblock {\em arXiv preprint arXiv:2010.07209}, 2020.

\bibitem{rieck2017clique}
B.~Rieck, U.~Fugacci, J.~Lukasczyk, and H.~Leitte.
\newblock {Clique Community Persistence: A Topological Visual Analysis Approach
  for Complex Networks}.
\newblock {\em IEEE Transactions on Visualization and Computer Graphics},
  24(1):822--831, 2017.

\bibitem{rieck2012multivariate}
B.~Rieck, H.~Mara, and H.~Leitte.
\newblock {Multivariate Data Analysis Using Persistence-Based Filtering and
  Topological Signatures}.
\newblock {\em IEEE Transactions on Visualization and Computer Graphics},
  18(12):2382--2391, 2012.

\bibitem{sikander2020novel}
G.~Sikander and S.~Anwar.
\newblock A novel machine vision-based 3d facial action unit identification for
  fatigue detection.
\newblock {\em IEEE Transactions on Intelligent Transportation Systems}, 2020.

\bibitem{song2015exploiting}
Y.~Song, D.~McDuff, D.~Vasisht, and A.~Kapoor.
\newblock Exploiting sparsity and co-occurrence structure for action unit
  recognition.
\newblock In {\em IEEE International Conference on Automatic Face and Gesture
  Recognition}, vol.~1, pp. 1--8, 2015.

\bibitem{suh2019persistent}
A.~Suh, M.~Hajij, B.~Wang, C.~Scheidegger, and P.~Rosen.
\newblock {Persistent Homology Guided Force-Directed Graph Layouts}.
\newblock {\em IEEE Transactions on Visualization and Computer Graphics},
  26(1):697--707, 2019.

\bibitem{sun20083d}
Y.~Sun and L.~Yin.
\newblock {3D Spatio-Temporal Face Recognition Using Dynamic Range Model
  Sequences}.
\newblock In {\em IEEE Computer Society Conference on Computer Vision and
  Pattern Recognition Workshops}, pp. 1--7, 2008.

\bibitem{tierny2017topology}
J.~Tierny, G.~Favelier, J.~Levine, C.~Gueunet, and M.~Michaux.
\newblock {The Topology Toolkit}.
\newblock {\em IEEE Transactions on Visualization and Computer Graphics},
  24(1):832--842, 2017.

\bibitem{tornincasa20193d}
S.~Tornincasa, E.~Vezzetti, S.~Moos, M.~G. Violante, F.~Marcolin, N.~Dagnes,
  L.~Ulrich, and G.~F. Tregnaghi.
\newblock {3D Facial Action Units and Expression Recognition Using a Crisp
  Logic}.
\newblock {\em Computer Aided Design and Applications}, 16:256--268, 2019.

\bibitem{uddin2020ICPR}
M.~T. Uddin and S.~Canavan.
\newblock Quantified facial temporal-expressiveness dynamics for affect
  analysis.
\newblock {\em International Conference on Pattern Recognition}, 2020.

\bibitem{maaten2008visualizing}
L.~van~der Maaten and G.~Hinton.
\newblock {Visualizing Data Using t-SNE}.
\newblock {\em Journal of Machine Learning Research}, 9:2579--2605, 2008.

\bibitem{vapnik2013nature}
V.~Vapnik.
\newblock {\em {The Nature of Statistical Learning Theory}}.
\newblock Springer science \& business media, 2013.

\bibitem{wang2011branching}
B.~Wang, B.~Summa, V.~Pascucci, and M.~Vejdemo-Johansson.
\newblock {Branching and Circular Features in High Dimensional Data}.
\newblock {\em IEEE Transactions on Visualization and Computer Graphics},
  17(12):1902--1911, 2011.

\bibitem{weinberger2011persistent}
S.~Weinberger.
\newblock What is... persistent homology?
\newblock {\em Notices of the AMS}, 58(1):36--39, 2011.

\bibitem{willemink2020preparing}
M.~J. Willemink, W.~A. Koszek, C.~Hardell, J.~Wu, D.~Fleischmann, H.~Harvey,
  L.~R. Folio, R.~M. Summers, D.~L. Rubin, and M.~P. Lungren.
\newblock Preparing medical imaging data for machine learning.
\newblock {\em Radiology}, 295(1):4--15, 2020.

\bibitem{xu2020exploring}
X.~Xu and V.~R. de~Sa.
\newblock Exploring multidimensional measurements for pain evaluation using
  facial action units.
\newblock In {\em IEEE International Conference on Automatic Face and Gesture
  Recognition}, pp. 559--565, 2020.

\bibitem{xue2015automatic}
M.~Xue, A.~Mian, W.~Liu, and L.~Li.
\newblock {Automatic 4D Facial Expression Recognition Using DCT Features}.
\newblock In {\em IEEE Winter Conference on Applications of Computer Vision},
  pp. 199--206, 2015.

\bibitem{yang2018facial}
H.~Yang, U.~Ciftci, and L.~Yin.
\newblock Facial expression recognition by de-expression residue learning.
\newblock In {\em Proceedings of the IEEE conference on computer vision and
  pattern recognition}, pp. 2168--2177, 2018.

\bibitem{yin126high}
L.~Yin, X.~Chen, Y.~Sun, T.~Worm, and M.~Reale.
\newblock {A High-Resolution 3D Dynamic Facial Expression Database}.
\newblock In {\em IEEE International Conference on Automatic Face and Gesture
  Recognition}, vol. 126, p.~6, 2008.

\bibitem{zamzmi2016approach}
G.~Zamzmi, C.-Y. Pai, D.~Goldgof, R.~Kasturi, T.~Ashmeade, and Y.~Sun.
\newblock {An Approach for Automated Multimodal Analysis of Infants' Pain}.
\newblock In {\em IEEE International Conference on Pattern Recognition (ICPR)},
  2016.

\bibitem{zeng2020emotioncues}
H.~Zeng, X.~Shu, Y.~Wang, Y.~Wang, L.~Zhang, T.-C. Pong, and H.~Qu.
\newblock Emotioncues: Emotion-oriented visual summarization of classroom
  videos.
\newblock {\em IEEE transactions on visualization and computer graphics}, 2020.

\bibitem{zeng2018facial}
N.~Zeng, H.~Zhang, B.~Song, W.~Liu, Y.~Li, and A.~M. Dobaie.
\newblock Facial expression recognition via learning deep sparse autoencoders.
\newblock {\em Neurocomputing}, 273:643--649, 2018.

\bibitem{zhang2014bp4d}
X.~Zhang, L.~Yin, J.~F. Cohn, S.~Canavan, M.~Reale, A.~Horowitz, P.~Liu, and
  J.~M. Girard.
\newblock Bp4d-spontaneous: a high-resolution spontaneous 3d dynamic facial
  expression database.
\newblock {\em Image and Vision Computing}, 32(10):692--706, 2014.

\bibitem{zhang2016multimodal}
Z.~Zhang, J.~M. Girard, Y.~Wu, X.~Zhang, P.~Liu, U.~Ciftci, S.~Canavan,
  M.~Reale, A.~Horowitz, H.~Yang, et~al.
\newblock Multimodal spontaneous emotion corpus for human behavior analysis.
\newblock In {\em IEEE Conference on Computer Vision and Pattern Recognition
  (CVPR)}, pp. 3438--3446, 2016.

\bibitem{zhang2020region}
Z.~Zhang, T.~Wang, and L.~Yin.
\newblock Region of interest based graph convolution: A heatmap regression
  approach for action unit detection.
\newblock In {\em ACM International Conference on Multimedia}, pp. 2890--2898,
  2020.

\bibitem{zhen2016muscular}
Q.~Zhen, D.~Huang, Y.~Wang, and L.~Chen.
\newblock {Muscular Movement Model-Based Automatic 3D/4D Facial Expression
  Recognition}.
\newblock {\em IEEE Transactions on Multimedia}, 18(7):1438--1450, 2016.

\end{thebibliography}

% \firstsection{Appendix}

% \maketitle

% \input{sec.appendix}

\newpage

\section*{Additional Examples of Relative Distance Topology}

\begin{figure}[!h]
    \centering

    \subfigure[Bottleneck\hspace{30pt}]{\includegraphics[trim=0 159pt 0 20pt, clip, width=0.95\linewidth]{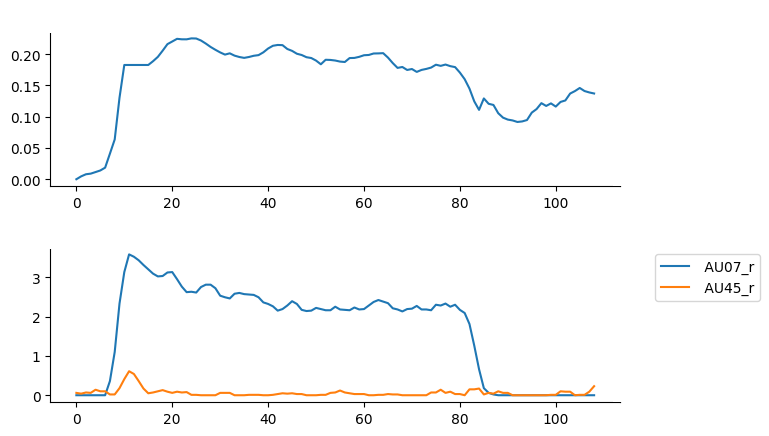}}
    
    \vspace{-5pt}
    \subfigure[Wasserstein\hspace{30pt}]{\includegraphics[trim=0 159pt 0 20pt, clip, width=0.95\linewidth]{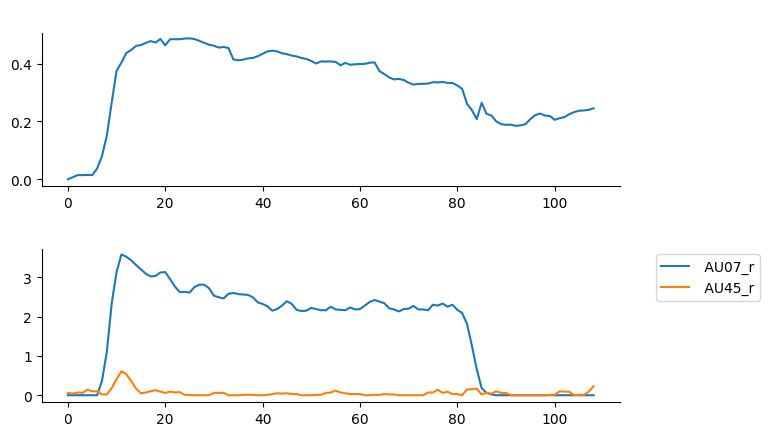}}
    
    \vspace{-5pt}
    \subfigure[Action Units\hspace{30pt}]{\includegraphics[trim=0 4pt 0 175pt, clip, width=0.95\linewidth]{figures/au_examples/F002_wasserstein_leftEye_rightEye_nose_Angry.png}}

    \caption{F002 \Anger using eyes, eyebrows, nose, and mouth}
    \label{fig:F002:anger}
\end{figure}

\begin{figure}[!h]
    \centering

    \subfigure[Bottleneck\hspace{30pt}]{\includegraphics[trim=0 159pt 0 20pt, clip,width=0.95\linewidth]{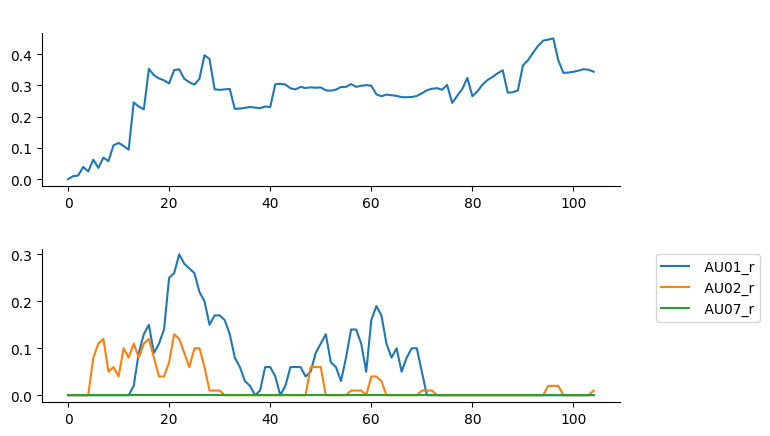}}
    
    \vspace{-5pt}
    \subfigure[Wasserstein\hspace{30pt}]{\includegraphics[trim=0 159pt 0 20pt, clip, width=0.95\linewidth]{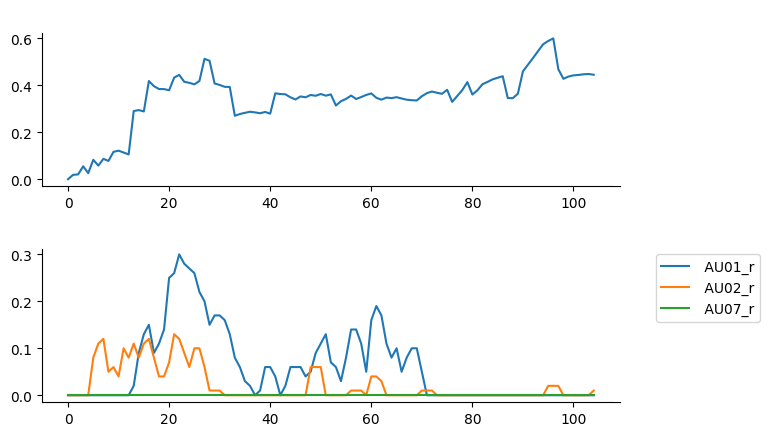}}
    
    \vspace{-5pt}
    \subfigure[Action Units\hspace{30pt}]{\includegraphics[trim=0 4pt 0 175pt, clip, width=0.95\linewidth]{figures/au_examples/F002_wasserstein_leftEyebrow_rightEyebrow_nose_Fear.png}}

    \caption{F002 \Fear using eyes, eyebrows, nose, and mouth}
    \label{fig:F002:fear}
\end{figure}

\newpage
\section*{\hspace{-16pt}and Action Units (AUs)}

\begin{figure}[!h]
    \centering

    \subfigure[Bottleneck\hspace{30pt}]{\includegraphics[trim=0 159pt 0 20pt, clip,width=0.95\linewidth]{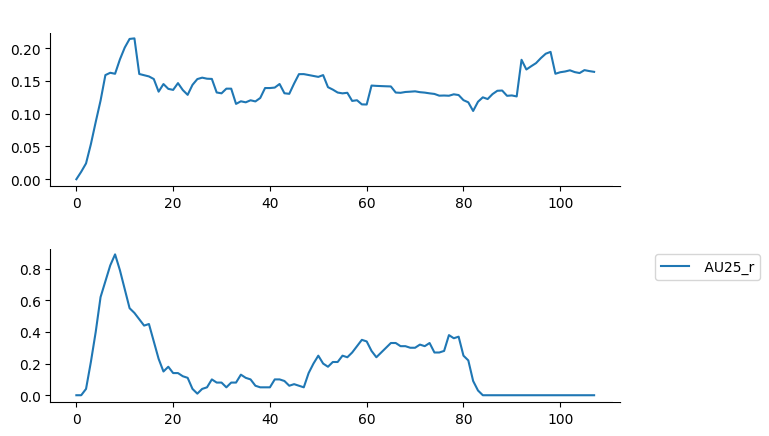}}
    
    \vspace{-5pt}
    \subfigure[Wasserstein\hspace{30pt}]{\includegraphics[trim=0 159pt 0 20pt, clip, width=0.95\linewidth]{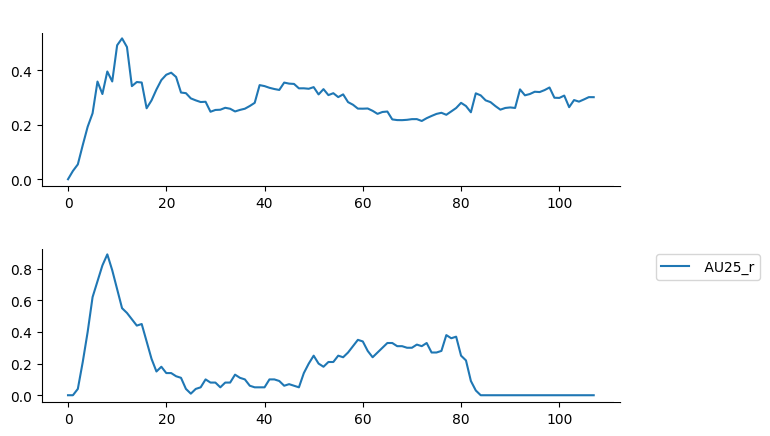}}
    
    \vspace{-5pt}
    \subfigure[Action Units\hspace{30pt}]{\includegraphics[trim=0 4pt 0 175pt, clip, width=0.95\linewidth]{figures/au_examples/F002_wasserstein_nose_mouth_Disgust.png}}
    
    \caption{F002 \Disgust using eyes, eyebrows, nose, and mouth}
    \label{fig:F002:disgust}
\end{figure}

\begin{figure}[!h]
    \centering

    \subfigure[Bottleneck\hspace{30pt}]{\includegraphics[trim=0 159pt 0 20pt, clip, width=0.95\linewidth]{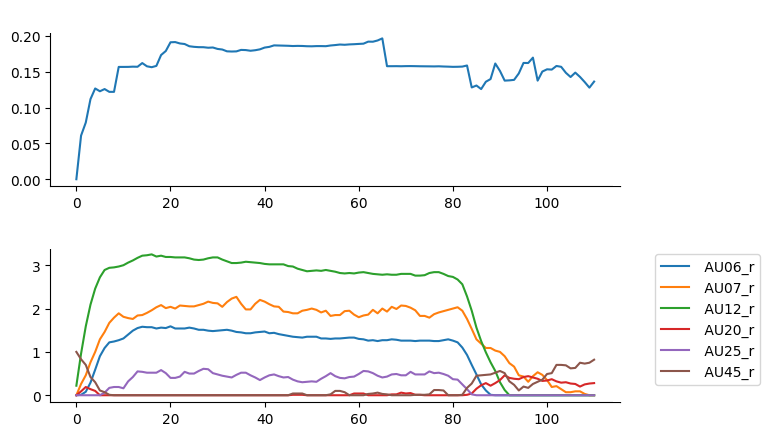}}
    
    \vspace{-5pt}
    \subfigure[Wasserstein\hspace{30pt}]{\includegraphics[trim=0 159pt 0 20pt, clip, width=0.95\linewidth]{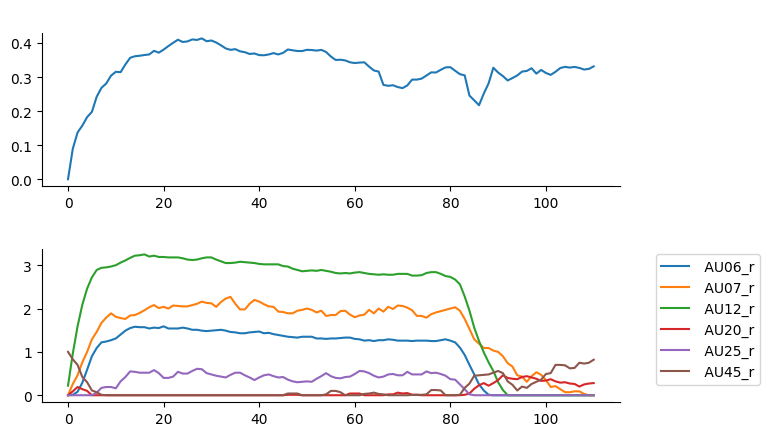}}
    
    \vspace{-5pt}
    \subfigure[Action Units\hspace{30pt}]{\includegraphics[trim=0 4pt 0 175pt, clip,width=0.95\linewidth]{figures/au_examples/F002_wasserstein_leftEye_rightEye_leftEyebrow_rightEyebrow_nose_mouth_Happy.png}}
    
    \caption{F002 \Happiness using eyes, eyebrows, nose, and mouth}
    \label{fig:F002:happiness}
\end{figure}

\newpage

\begin{figure}[!h]
    \centering

    \subfigure[Bottleneck\hspace{30pt}]{\includegraphics[trim=0 159pt 0 20pt, clip, width=0.95\linewidth]{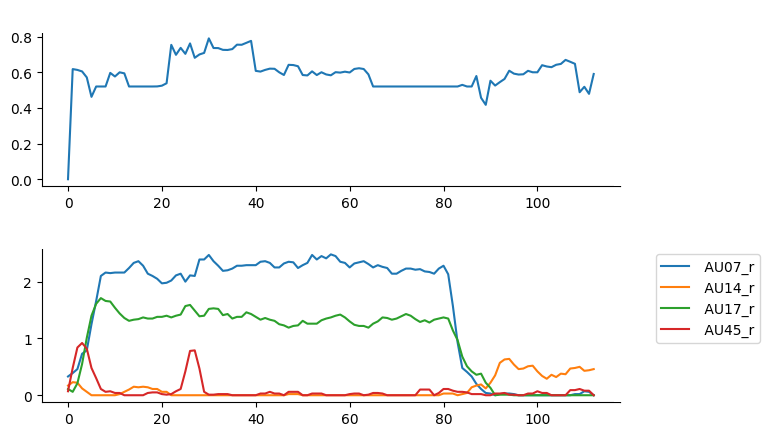}}
    \subfigure[Wasserstein\hspace{30pt}]{\includegraphics[trim=0 159pt 0 20pt, clip, width=0.95\linewidth]{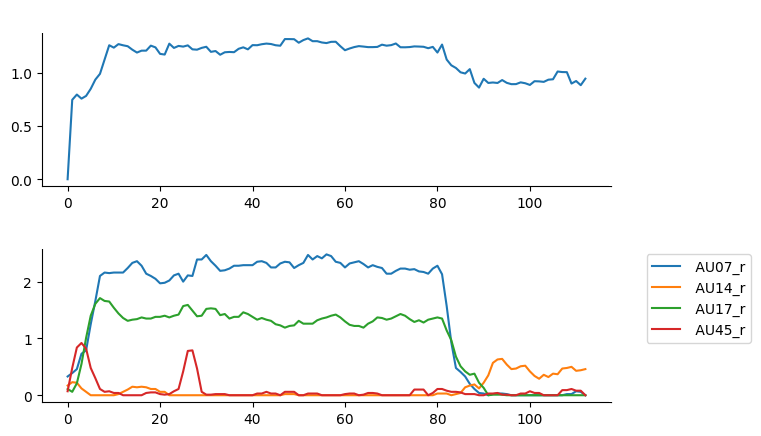}}
    \subfigure[Action Units\hspace{30pt}]{\includegraphics[trim=0 4pt 0 175pt, clip, width=0.95\linewidth]{figures/au_examples/F002_wasserstein_leftEye_rightEye_leftEyebrow_rightEyebrow_nose_mouth_jawline_Sad.png}}
        
    \caption{F002 \Sadness using eyes, eyebrows, nose, and mouth}
    \label{fig:F002:sadness}
\end{figure}

\begin{figure}[!ht]
    
    \subfigure[Bottleneck\hspace{30pt}]{\includegraphics[trim=0 159pt 0 20pt, clip, width=0.95\linewidth]{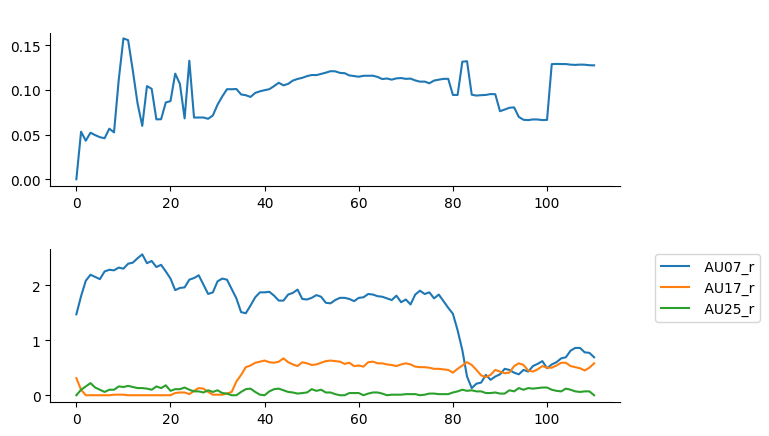}}
    \subfigure[Wasserstein\hspace{30pt}]{\includegraphics[trim=0 159pt 0 20pt, clip, width=0.95\linewidth]{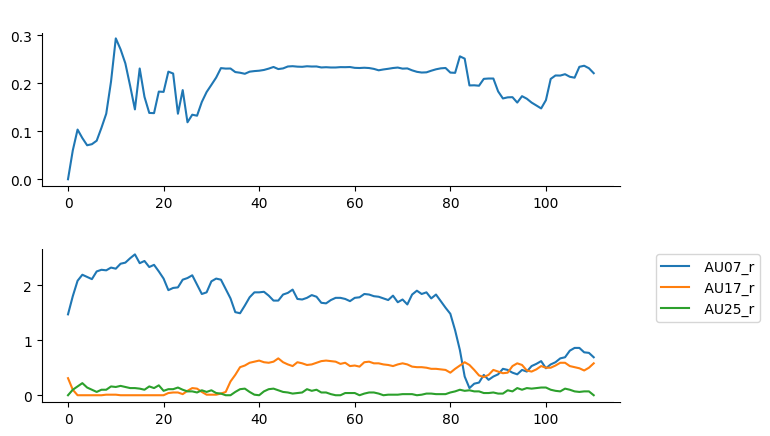}}
    \subfigure[Action Units\hspace{30pt}]{\includegraphics[trim=0 4pt 0 175pt, clip, width=0.95\linewidth]{figures/au_examples/M002_wasserstein_nose_mouth_Angry.png}}
    
    \caption{M002 \Anger using eyes, eyebrows, nose, and mouth}
    \label{fig:M002:anger}
\end{figure}

\newpage

\begin{figure}[!h]
    \centering

    \subfigure[Bottleneck\hspace{30pt}]{\includegraphics[trim=0 159pt 0 20pt, clip, width=0.95\linewidth]{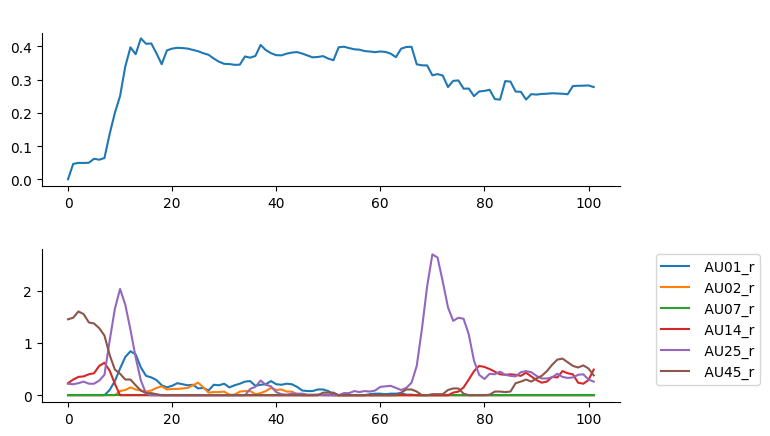}}
    \subfigure[Wasserstein\hspace{30pt}]{\includegraphics[trim=0 159pt 0 20pt, clip, width=0.95\linewidth]{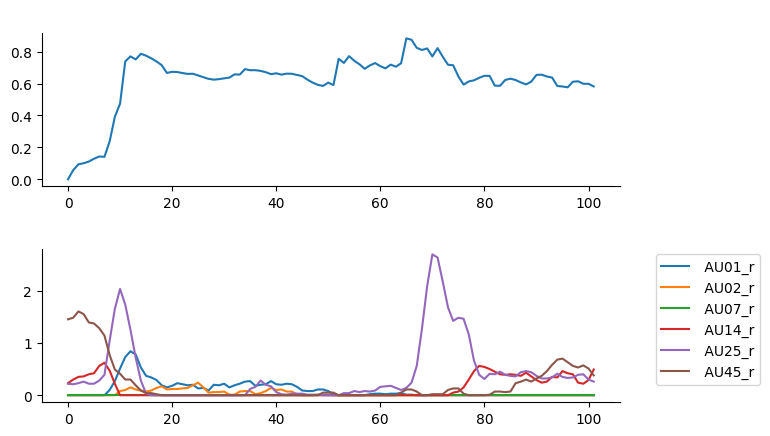}}
    \subfigure[Action Units\hspace{30pt}]{\includegraphics[trim=0 4pt 0 175pt, clip, width=0.95\linewidth]{figures/au_examples/F002_wasserstein_leftEye_rightEye_leftEyebrow_rightEyebrow_nose_mouth_Surprise.png}}

    \caption{F002 \Surprise using eyes, eyebrows, nose, and mouth}
    \label{fig:F002:surprise}
\end{figure}

\begin{figure}[!ht]

    \subfigure[Bottleneck\hspace{30pt}]{\includegraphics[trim=0 159pt 0 20pt, clip, width=0.95\linewidth]{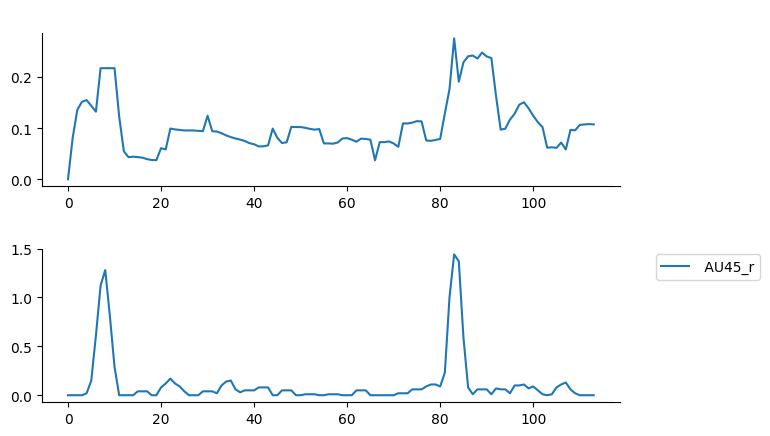}}
    \subfigure[Wasserstein\hspace{30pt}]{\includegraphics[trim=0 159pt 0 20pt, clip, width=0.95\linewidth]{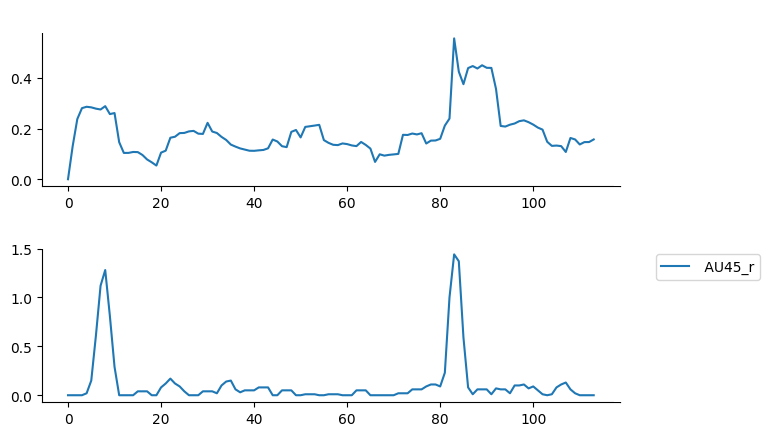}}
    \subfigure[Action Units\hspace{30pt}]{\includegraphics[trim=0 4pt 0 175pt, clip, width=0.95\linewidth]{figures/au_examples/M002_wasserstein_leftEye_rightEye_nose_Disgust.png}}

    \caption{M002 \Disgust using eyes, eyebrows, nose, and mouth}
    \label{fig:M002:disgust}
\end{figure}

\newpage

\begin{figure}[!h]
    \centering

    \subfigure[Bottleneck\hspace{30pt}]{\includegraphics[trim=0 159pt 0 20pt, clip, width=0.95\linewidth]{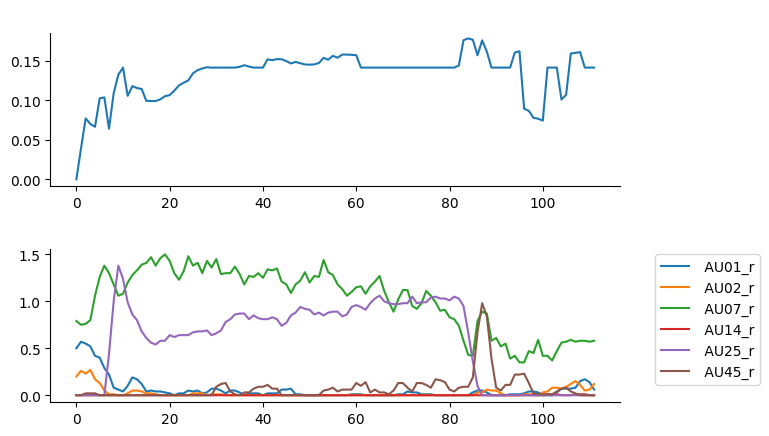}}
    \subfigure[Wasserstein\hspace{30pt}]{\includegraphics[trim=0 159pt 0 20pt, clip, width=0.95\linewidth]{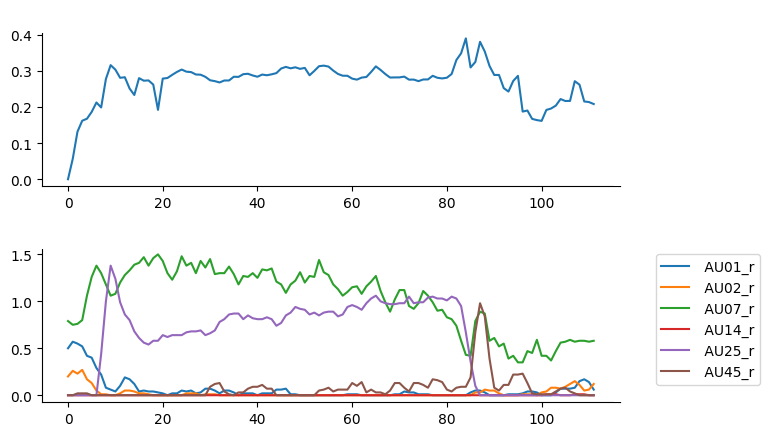}}
    \subfigure[Action Units\hspace{30pt}]{\includegraphics[trim=0 4pt 0 175pt, clip, width=0.95\linewidth]{figures/au_examples/M002_wasserstein_leftEye_rightEye_leftEyebrow_rightEyebrow_nose_mouth_Fear.png}}

    \caption{M002 \Fear using eyes, eyebrows, nose, and mouth}
    \label{fig:M002:fear}
\end{figure}

\begin{figure}[!ht]

    \subfigure[Bottleneck\hspace{30pt}]{\includegraphics[trim=0 159pt 0 20pt, clip, width=0.95\linewidth]{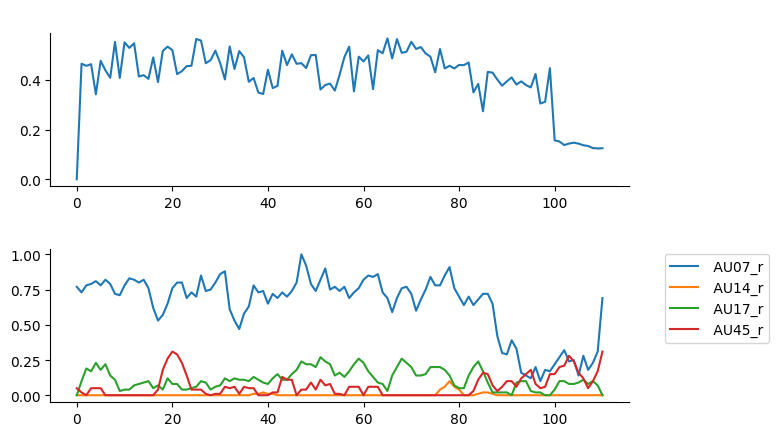}}
    \subfigure[Wasserstein\hspace{30pt}]{\includegraphics[trim=0 159pt 0 20pt, clip, width=0.95\linewidth]{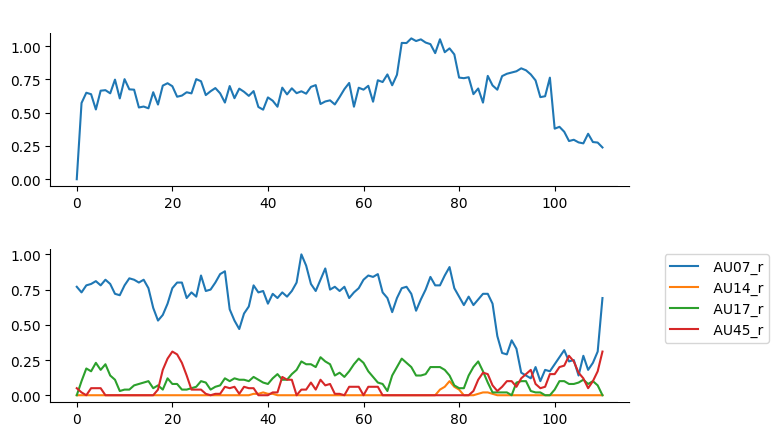}}
    \subfigure[Action Units\hspace{30pt}]{\includegraphics[trim=0 4pt 0 175pt, clip, width=0.95\linewidth]{figures/au_examples/M002_wasserstein_leftEye_rightEye_leftEyebrow_rightEyebrow_nose_mouth_jawline_Sad.png}}

    \caption{M002 \Sadness using eyes, eyebrows, nose, and mouth}
    \label{fig:M002:sadness}
\end{figure}

\newpage

\begin{figure}[!ht]

    \subfigure[Bottleneck\hspace{30pt}]{\includegraphics[trim=0 159pt 0 20pt, clip, width=0.95\linewidth]{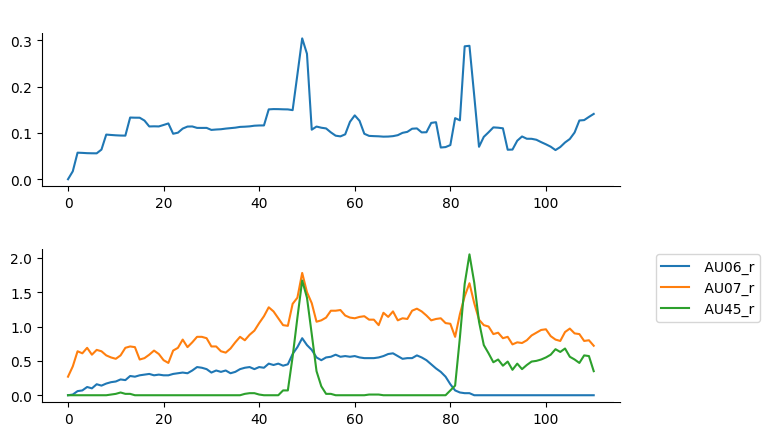}}
    \subfigure[Wasserstein\hspace{30pt}]{\includegraphics[trim=0 159pt 0 20pt, clip, width=0.95\linewidth]{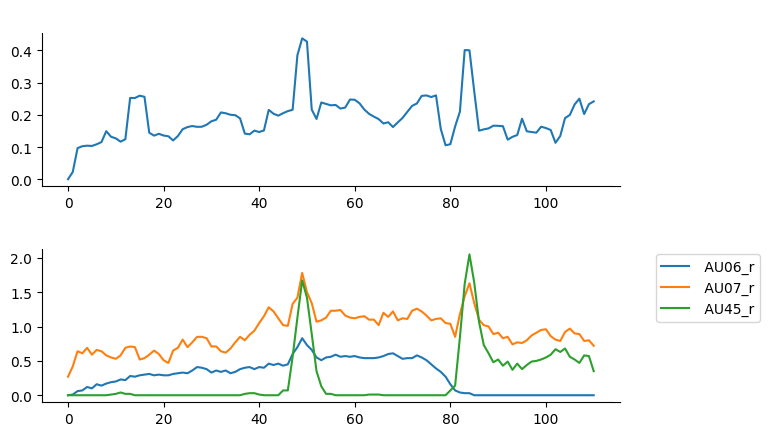}}
    \subfigure[Action Units\hspace{30pt}]{\includegraphics[trim=0 4pt 0 175pt, clip, width=0.95\linewidth]{figures/au_examples/M002_wasserstein_leftEye_rightEye_nose_Happy.png}}
    
    \caption{M002 \Happiness using eyes, eyebrows, nose, and mouth}
    \label{fig:M002:happiness}
\end{figure}

\begin{figure}[!ht]
    
    \subfigure[Bottleneck\hspace{30pt}]{\includegraphics[trim=0 159pt 0 20pt, clip, width=0.95\linewidth]{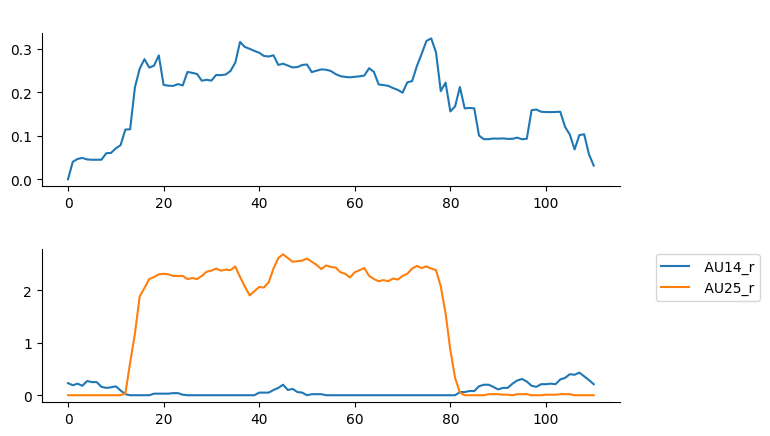}}
    \subfigure[Wasserstein\hspace{30pt}]{\includegraphics[trim=0 159pt 0 20pt, clip, width=0.95\linewidth]{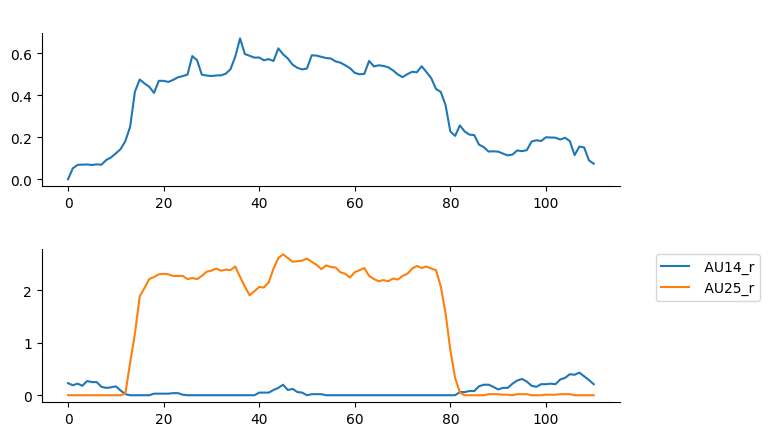}}
    \subfigure[Action Units\hspace{30pt}]{\includegraphics[trim=0 4pt 0 175pt, clip, width=0.95\linewidth]{figures/au_examples/M002_wasserstein_nose_mouth_Surprise.png}}
    
    \caption{M002 \Surprise using eyes, eyebrows, nose, and mouth}
    \label{fig:M002:surprise}
\end{figure}

\newpage

\section*{Additional Examples Comparing and Differentiating}
% add in all the extra figures somewhere here

% will be moved to a separate doc later

%%%%%%%%%%%%%%%%%%%%%%%%%%%%%%%%%%%%%%%%%%%%%%%%%%%%%%%%%%%%%%%%%%%%%%%%%%%%%%%%%%%%%%%%%%%%%%%%%
%%%%%%%%%%%%%%%%%%%%%%%%%%%%%%%%%%%%%%%%%%%%% TSNE %%%%%%%%%%%%%%%%%%%%%%%%%%%%%%%%%%%%%%%%%%%%%%
%%%%%%%%%%%%%%%%%%%%%%%%%%%%%%%%%%%%%%%%%%%%%%%%%%%%%%%%%%%%%%%%%%%%%%%%%%%%%%%%%%%%%%%%%%%%%%%%%

% \Anger

\begin{figure}[!h]
    \centering
    \subfigure[\Anger perplexity: 30 (full) \label{fig:clustering:angry_full_perp_30}]{{\hspace{4pt}\includegraphics[trim= 92pt 81pt 73pt 89pt, clip, height=1.40in, width=0.45\linewidth]{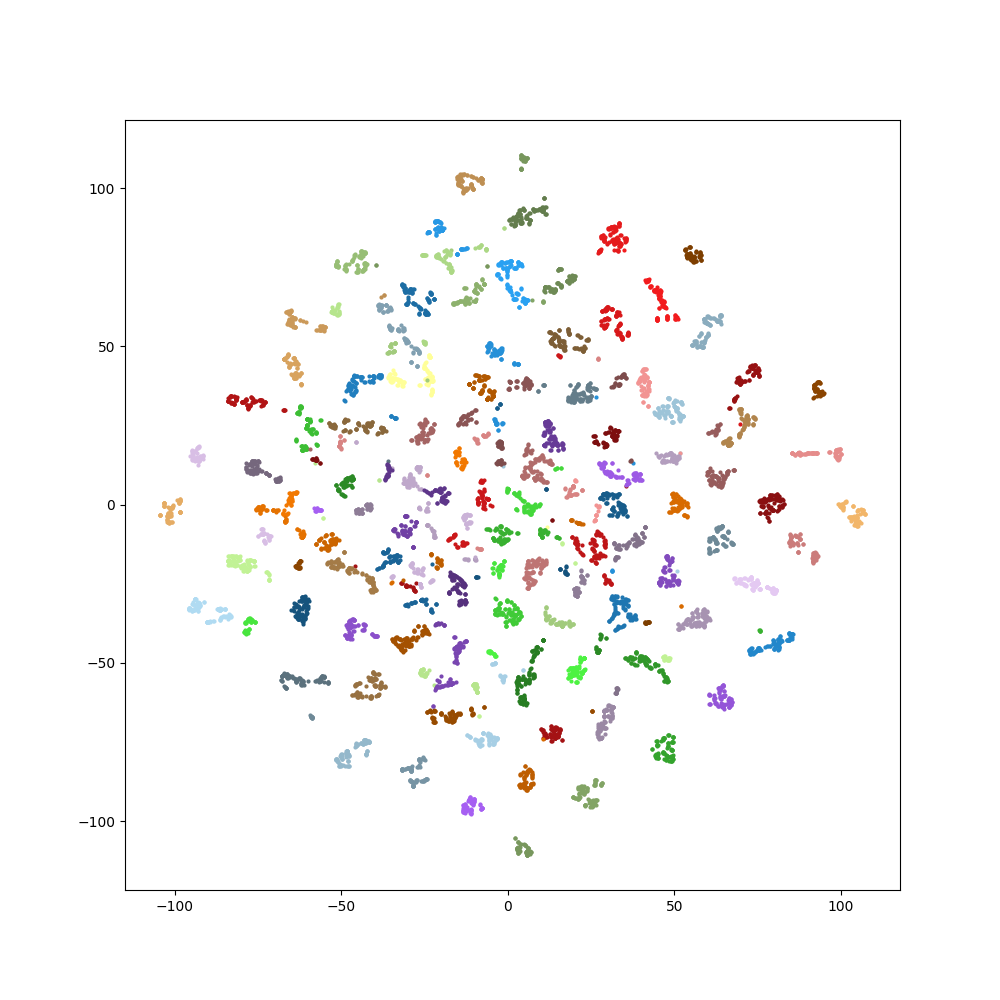}\hspace{4pt}}}
    \hfill
    \subfigure[\Anger perplexity: 30 (10 subj.) \label{fig:clustering:angry_10_perp_30}]{{\hspace{4pt}\includegraphics[trim= 92pt 81pt 73pt 89pt, clip, height=1.40in, width=0.45\linewidth]{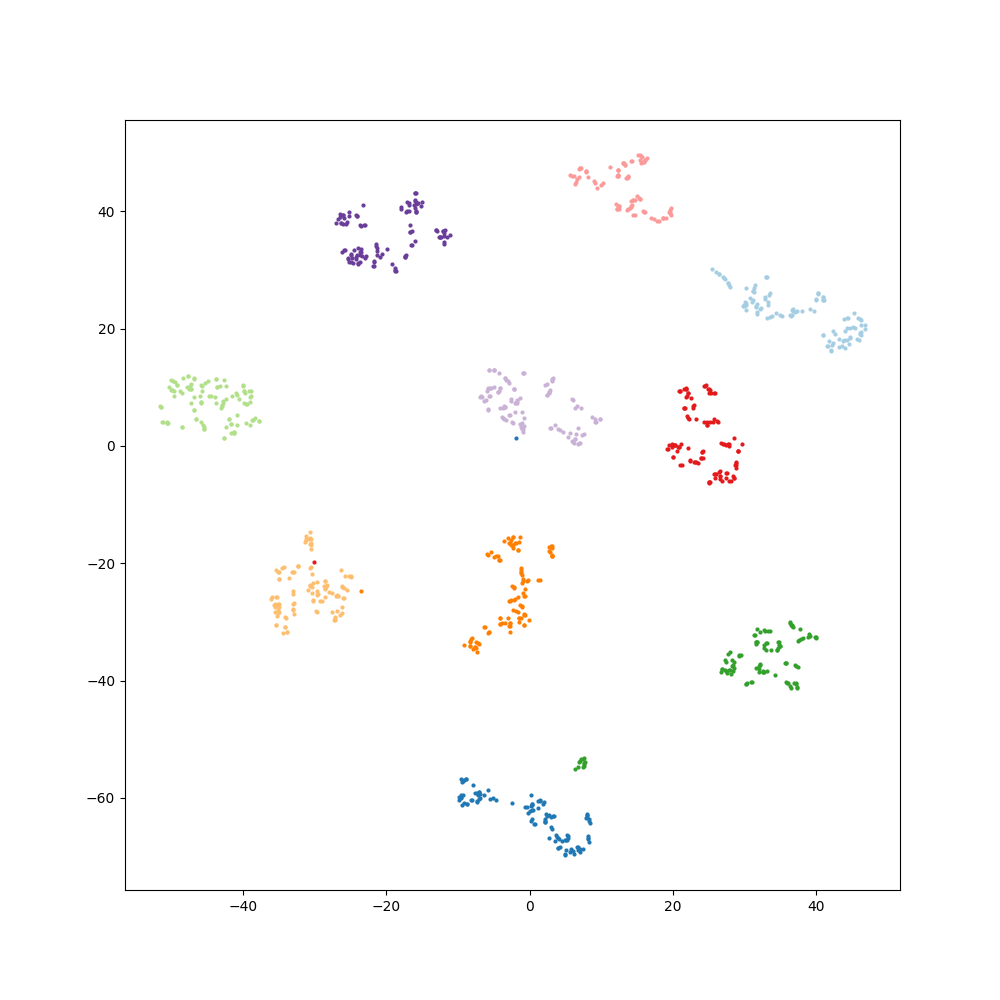}\hspace{4pt}}}
    
    \subfigure[\Anger perplexity: 40 (full) \label{fig:clustering:angry_full_perp_40}]{{\hspace{4pt}\includegraphics[trim= 92pt 81pt 73pt 89pt, clip, height=1.40in, width=0.45\linewidth]{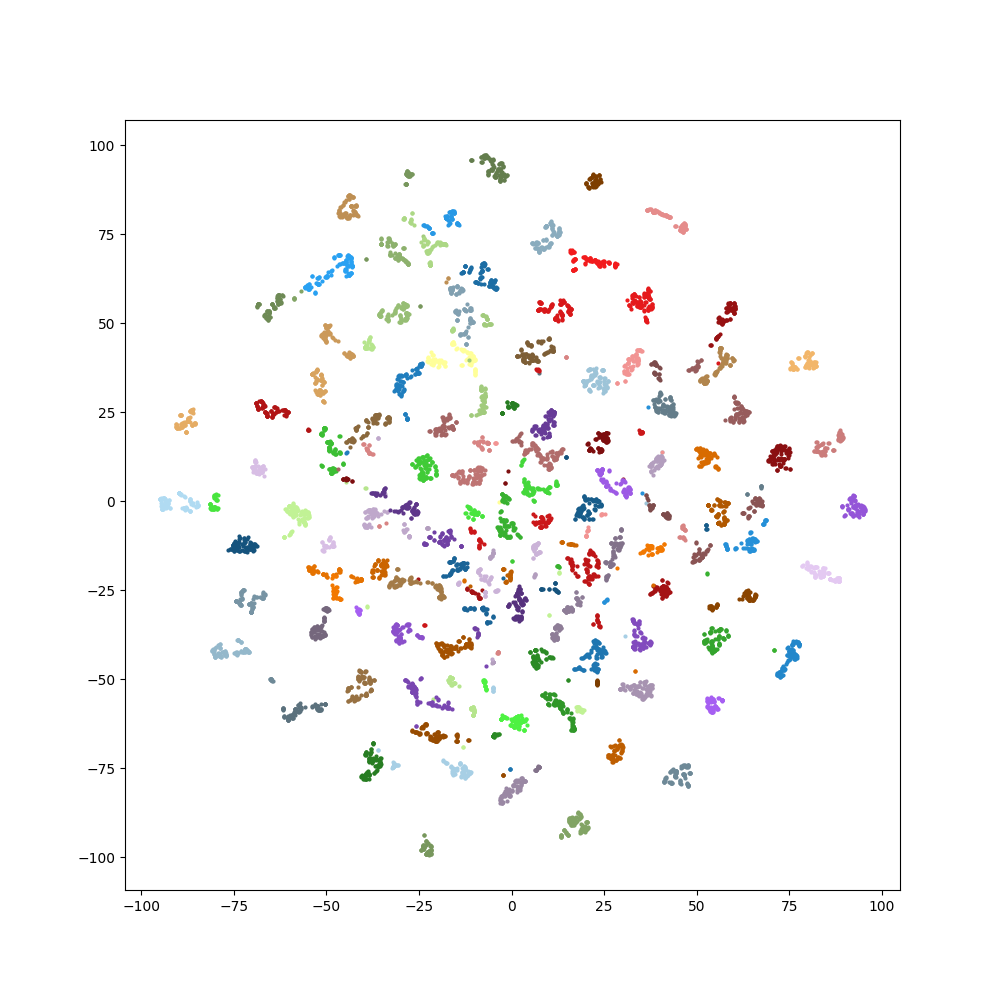}\hspace{4pt}}}
    \hfill
    \subfigure[\Anger (10 subj.)  Perplexity=40 \label{fig:clustering:angry_10_perp_40}]{{\hspace{4pt}\includegraphics[trim= 92pt 81pt 73pt 89pt, clip, height=1.40in, width=0.45\linewidth]{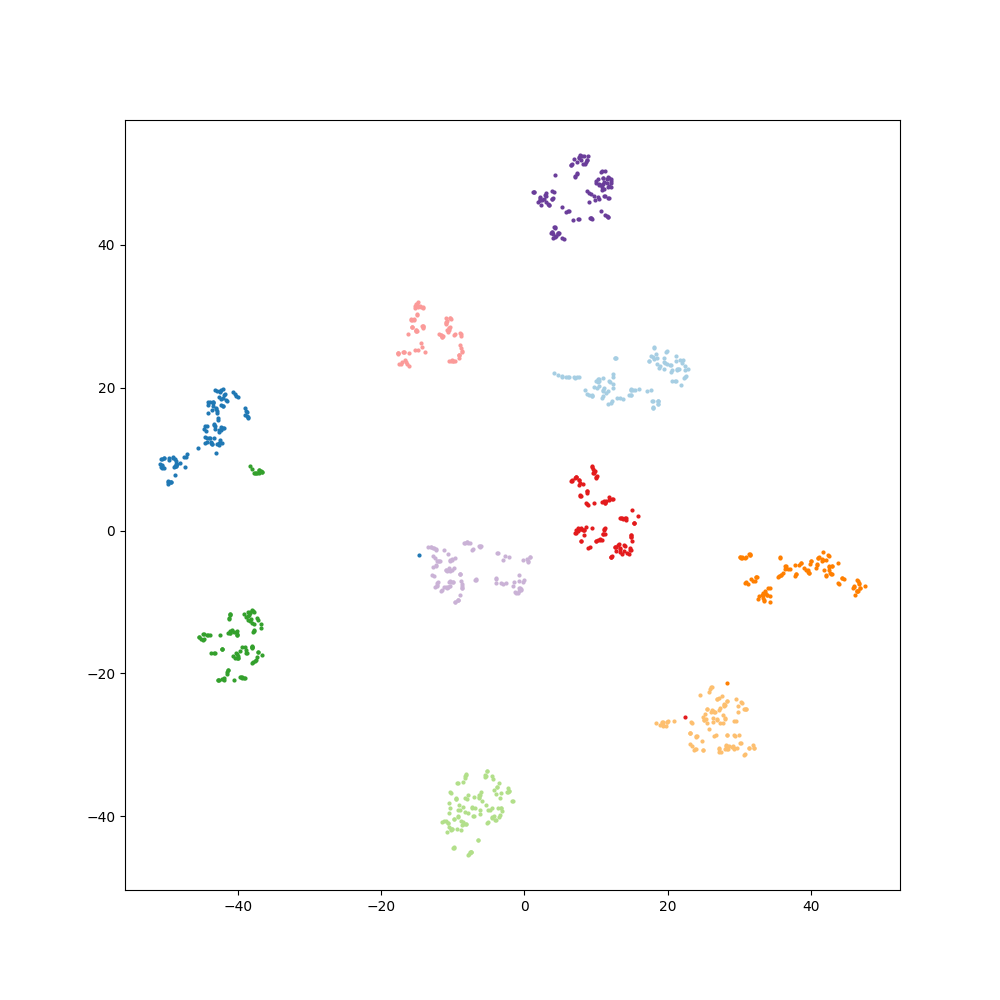}\hspace{4pt}}}
    
    \subfigure[\Anger perplexity: 50 (full) \label{fig:clustering:angry_full_perp_50}]{{\hspace{4pt}\includegraphics[trim= 92pt 81pt 73pt 89pt, clip, height=1.40in, width=0.45\linewidth]{figures/clustering/tsne/Angry_wasserstein_perplexity-50.png}\hspace{4pt}}}
    \hfill
    \subfigure[\Anger perplexity: 50 (10 subj.) \label{fig:clustering:angry_10_perp_50}]{{\hspace{4pt}\includegraphics[trim= 92pt 81pt 73pt 89pt, clip, height=1.40in, width=0.45\linewidth]{figures/clustering/tsne/Angry_wasserstein_perplexity-50_10subjects.png}\hspace{4pt}}}
    
    \subfigure[\Anger perplexity: 100 (full) \label{fig:clustering:angry_full_perp_100}]{{\hspace{4pt}\includegraphics[trim= 92pt 81pt 73pt 89pt, clip, height=1.40in, width=0.45\linewidth]{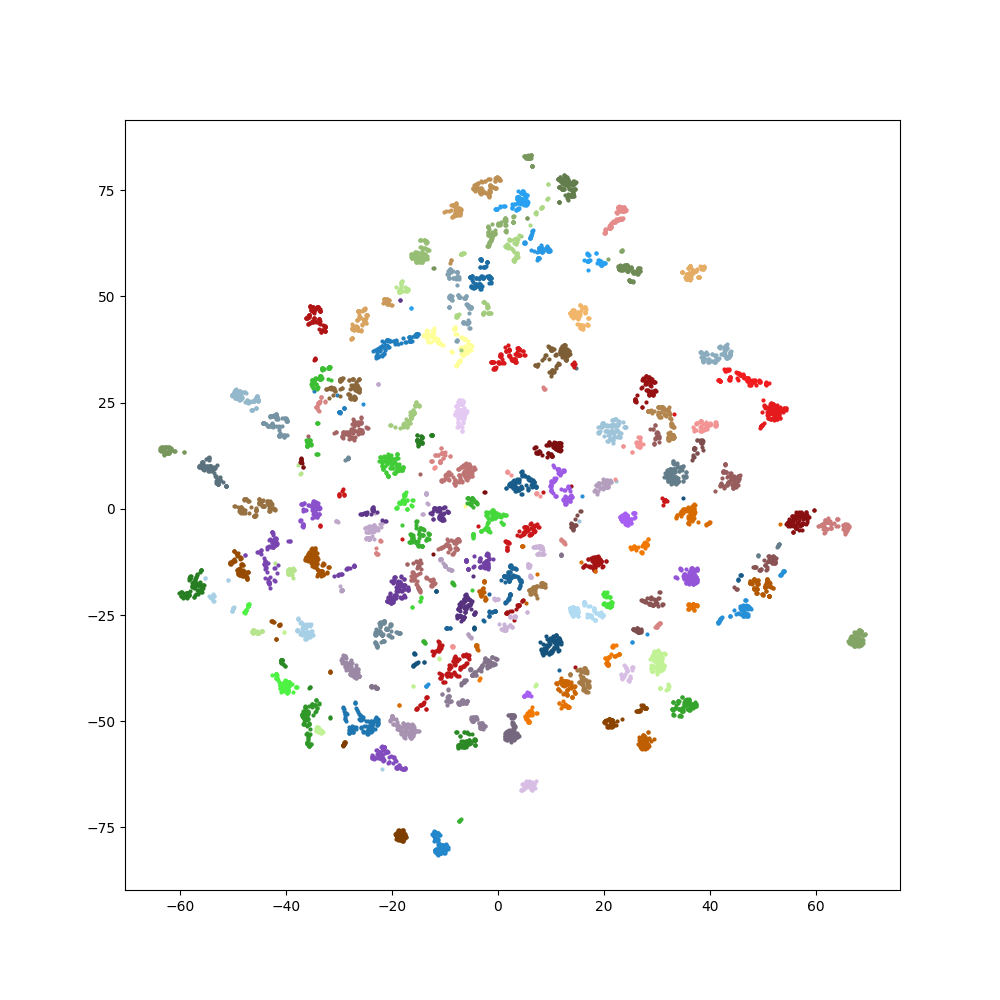}\hspace{4pt}}}
    \hfill
    \subfigure[\Anger perplexity: 100 (10 subj.) \label{fig:clustering:angry_10_perp_100}]{{\hspace{4pt}\includegraphics[trim= 92pt 81pt 73pt 89pt, clip, height=1.40in, width=0.45\linewidth]{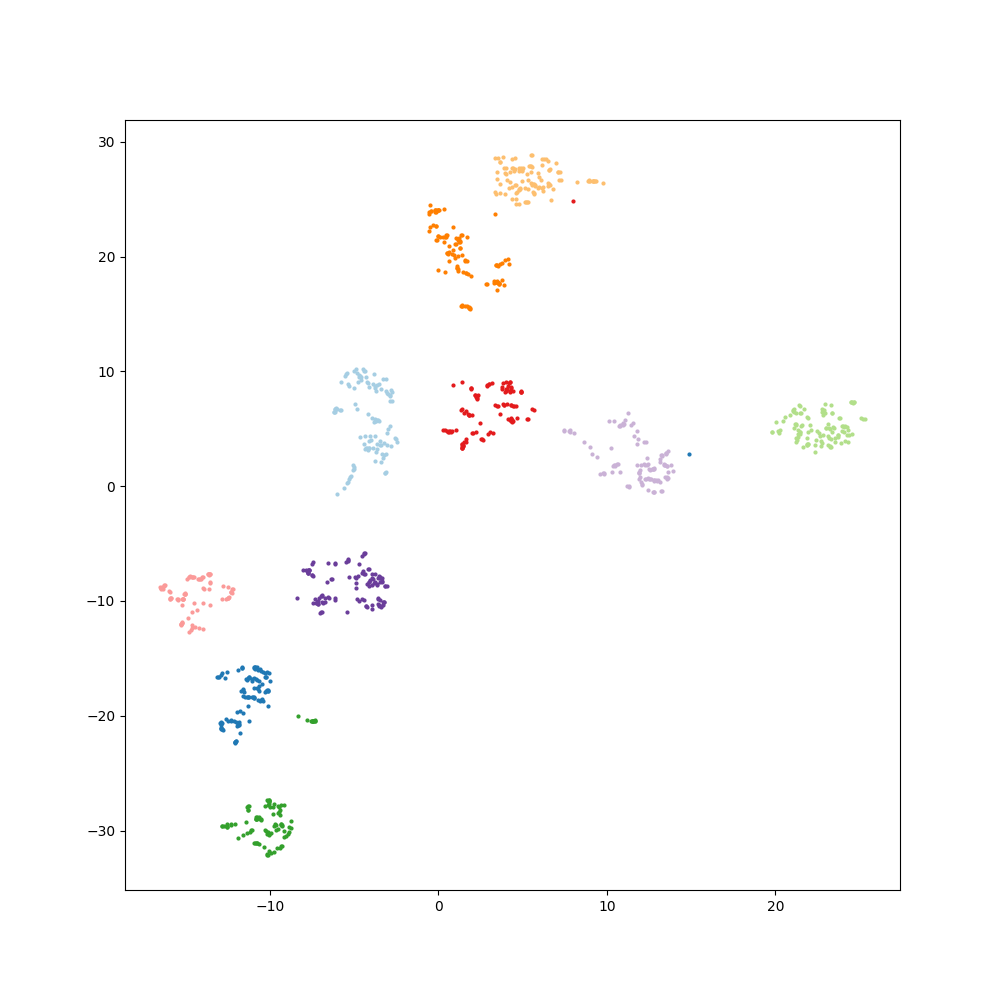}\hspace{4pt}}}
    
    \caption{t-SNE clustering of individual topological data for \Anger emotion at different perplexities.}
    \label{fig:clustering:angry}
\end{figure}

\newpage
\section*{\hspace{-17pt}Individuals}
% \Disgust

\begin{figure}[!h]
    \centering
    \subfigure[\Disgust perplexity: 30 (full) \label{fig:clustering:disgust_full_perp_30}]{{\hspace{4pt}\includegraphics[trim= 92pt 81pt 73pt 89pt, clip, height=1.40in, width=0.45\linewidth]{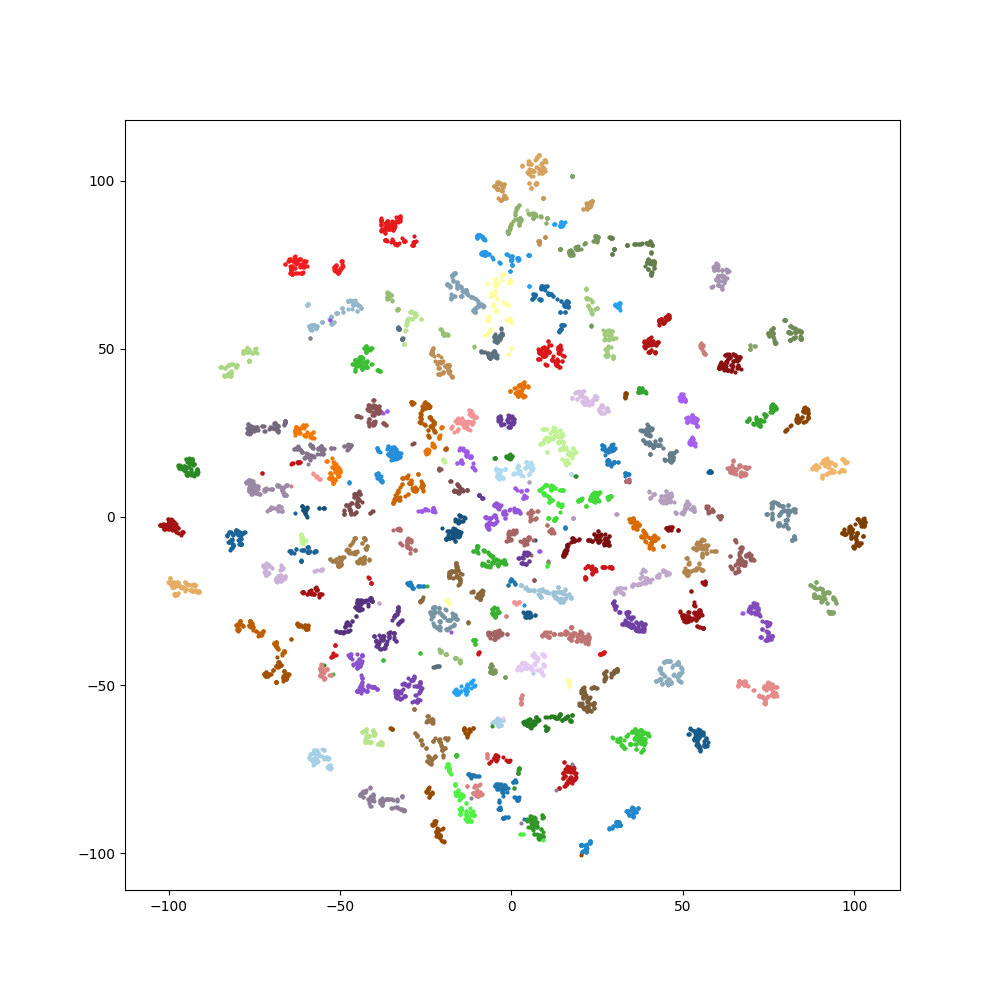}\hspace{4pt}}}
    \hfill
    \subfigure[\Disgust perplexity: 30 (10 subj.) \label{fig:clustering:disgust_10_perp_30}]{{\hspace{4pt}\includegraphics[trim= 92pt 81pt 73pt 89pt, clip, height=1.40in, width=0.45\linewidth]{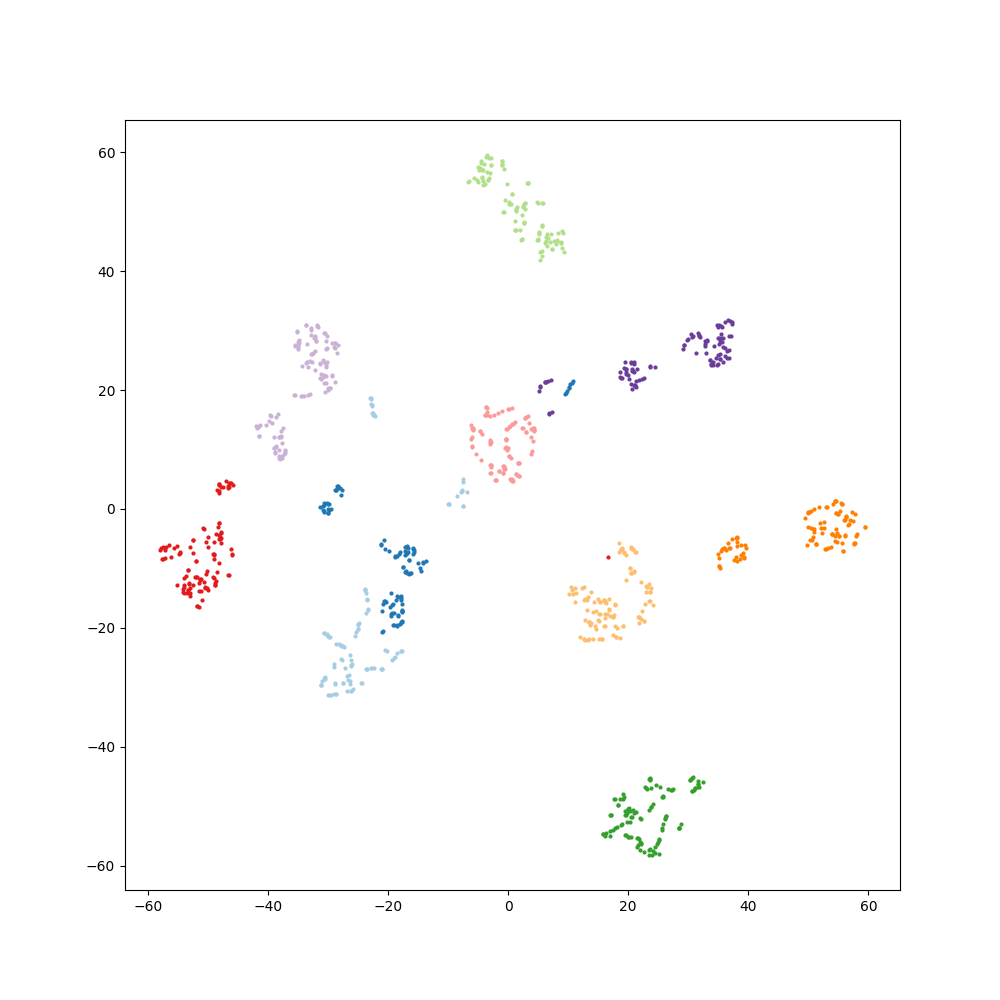}\hspace{4pt}}}
    
    \subfigure[\Disgust perplexity: 40 (full) \label{fig:clustering:disgust_full_perp_40}]{{\hspace{4pt}\includegraphics[trim= 92pt 81pt 73pt 89pt, clip, height=1.40in, width=0.45\linewidth]{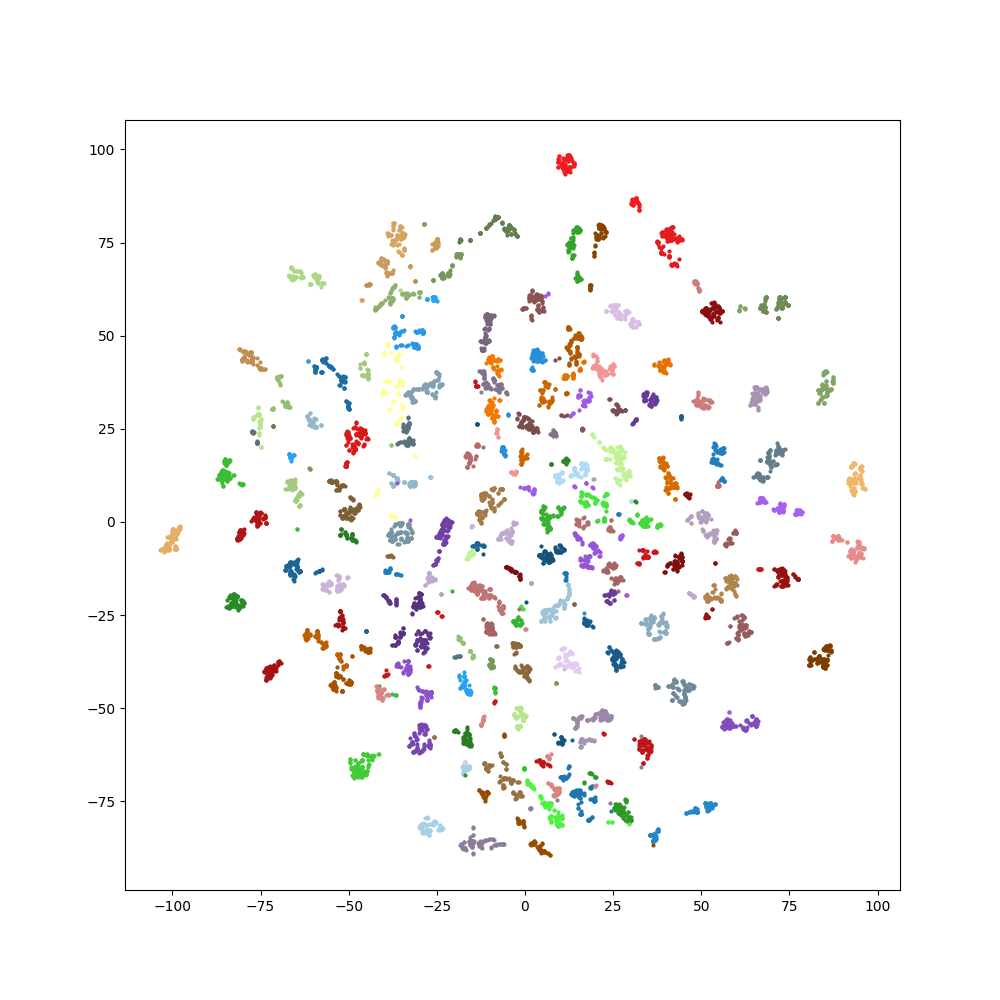}\hspace{4pt}}}
    \hfill
    \subfigure[\Disgust perplexity: 40 (10 subj.) \label{fig:clustering:disgust_10_perp_40}]{{\hspace{4pt}\includegraphics[trim= 92pt 81pt 73pt 89pt, clip, height=1.40in, width=0.45\linewidth]{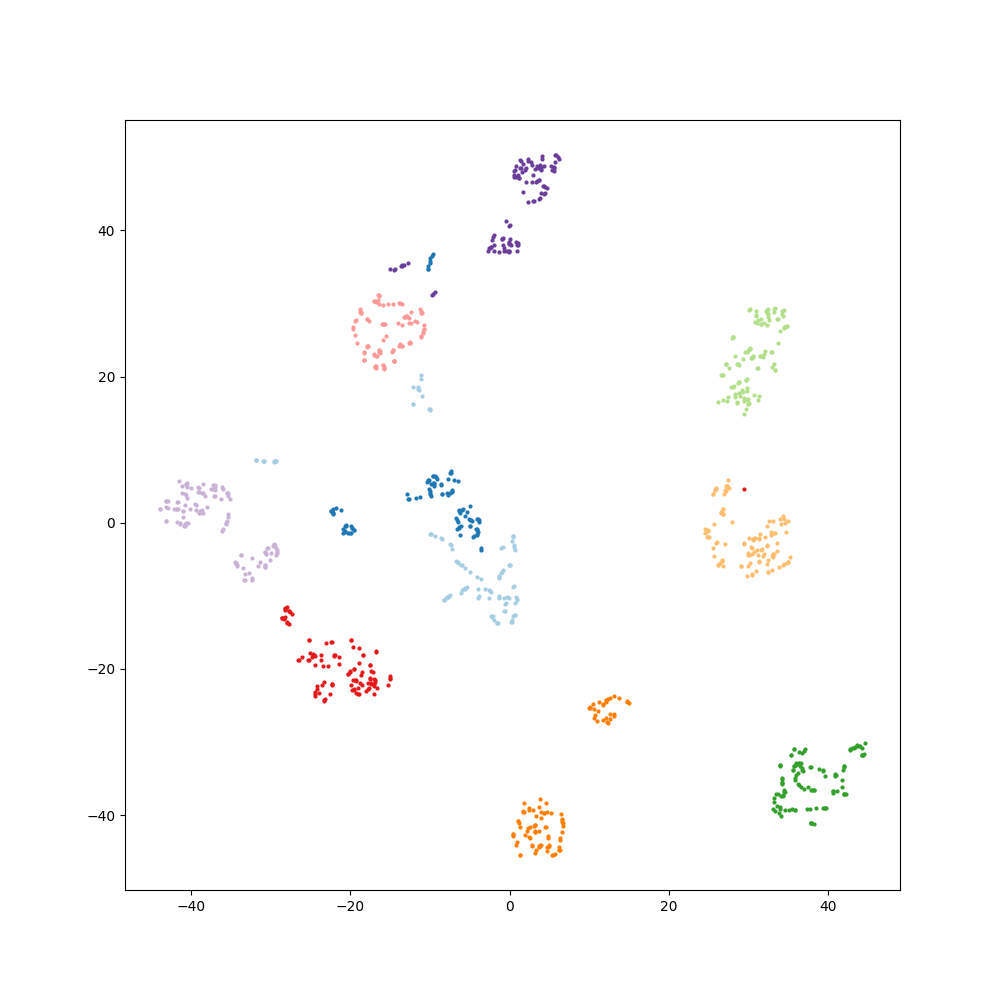}\hspace{4pt}}}
    
    \subfigure[\Disgust perplexity: 50 (full) \label{fig:clustering:disgust_full_perp_50}]{{\hspace{4pt}\includegraphics[trim= 92pt 81pt 73pt 89pt, clip, height=1.40in, width=0.45\linewidth]{figures/clustering/tsne/Disgust_wasserstein_perplexity-50.png}\hspace{4pt}}}
    \hfill
    \subfigure[\Disgust perplexity: 50 (10 subj.) \label{fig:clustering:disgust_10_perp_50}]{{\hspace{4pt}\includegraphics[trim= 92pt 81pt 73pt 89pt, clip, height=1.40in, width=0.45\linewidth]{figures/clustering/tsne/Disgust_wasserstein_perplexity-50_10subjects.png}\hspace{4pt}}}
    
    \subfigure[\Disgust perplexity: 100 (full) \label{fig:clustering:disgust_full_perp_100}]{{\hspace{4pt}\includegraphics[trim= 92pt 81pt 73pt 89pt, clip, height=1.40in, width=0.45\linewidth]{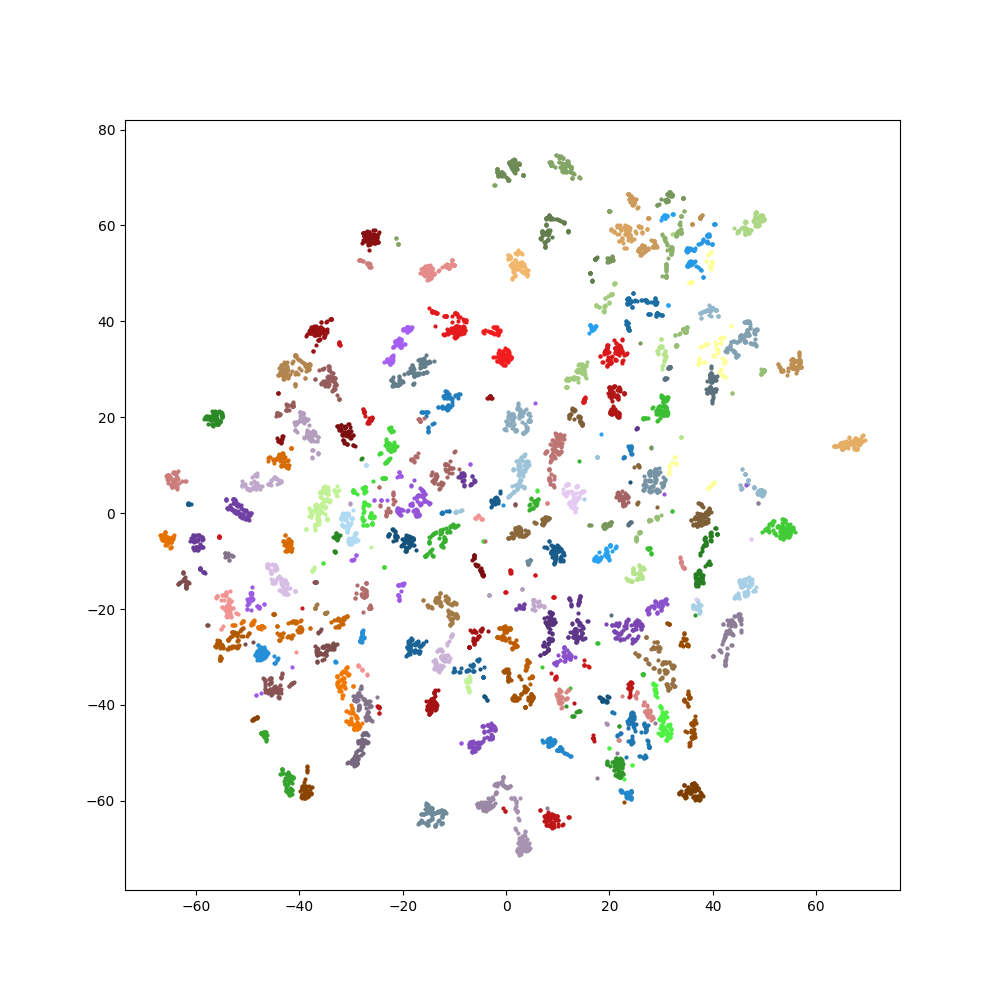}\hspace{4pt}}}
    \hfill
    \subfigure[\Disgust perplexity: 100 (10 subj.) \label{fig:clustering:disgust_10_perp_100}]{{\hspace{4pt}\includegraphics[trim= 92pt 81pt 73pt 89pt, clip, height=1.40in, width=0.45\linewidth]{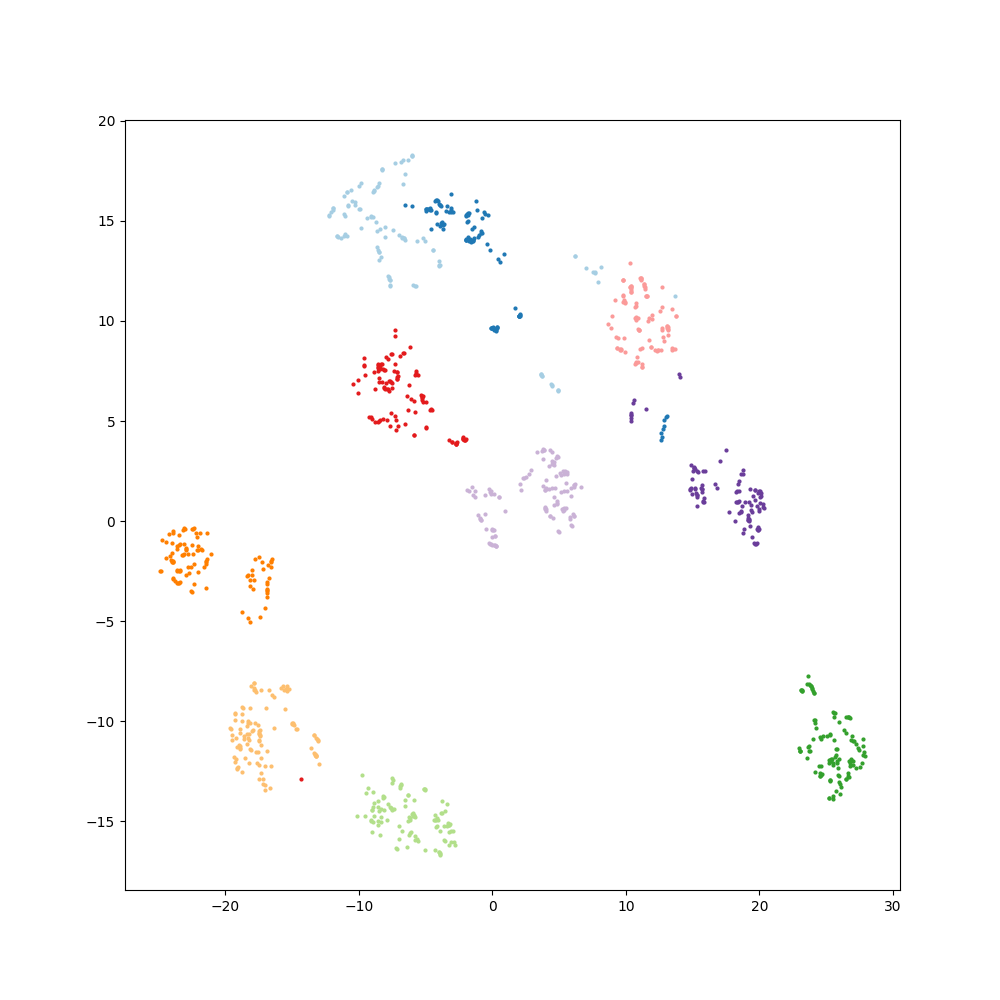}\hspace{4pt}}}
    
    \caption{t-SNE clustering of individual topological data for \Disgust emotion at different perplexities.}
    \label{fig:clustering:disgust}
\end{figure}

% \Fear

\begin{figure}[!b]
    \centering
    \subfigure[\Fear perplexity: 30 (full) \label{fig:clustering:fear_full_perp_30}]{{\hspace{4pt}\includegraphics[trim= 92pt 81pt 73pt 89pt, clip, height=1.40in, width=0.45\linewidth]{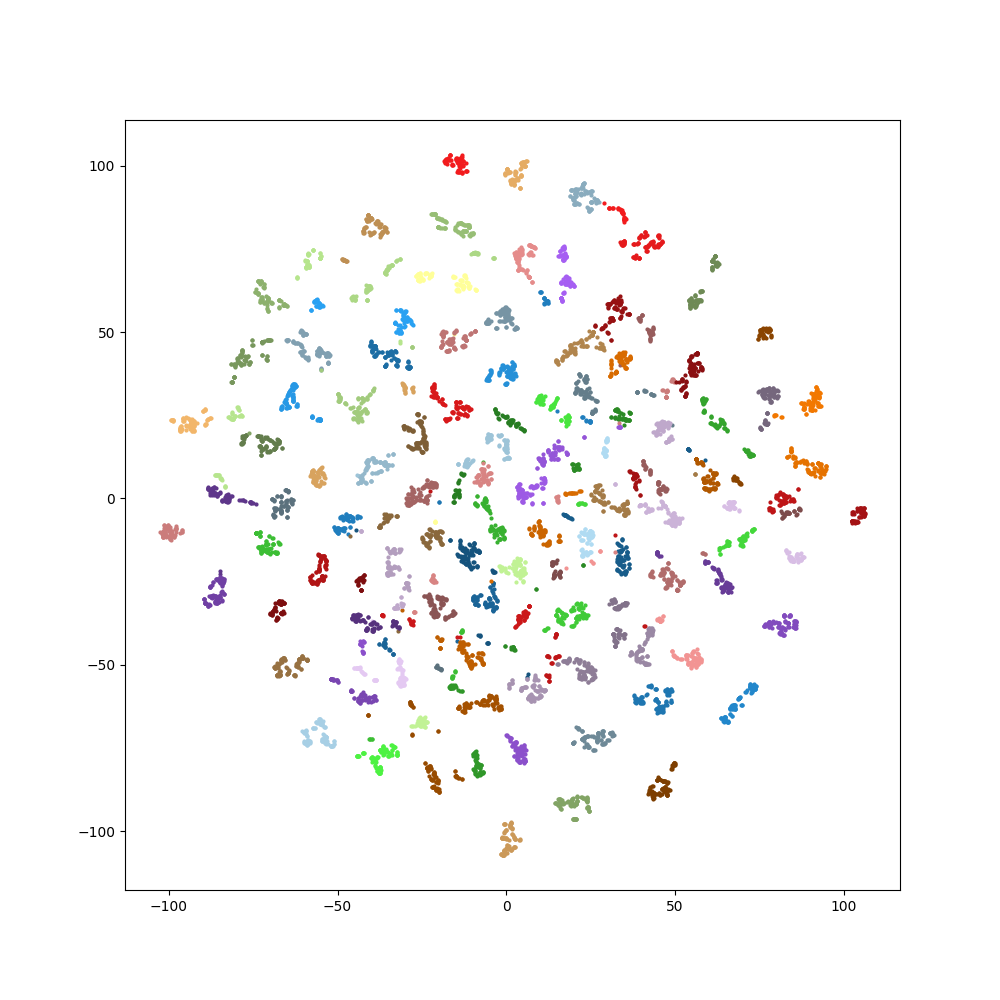}\hspace{4pt}}}
    \hfill
    \subfigure[\Fear perplexity: 30 (10 subj.) \label{fig:clustering:fear_10_perp_30}]{{\hspace{4pt}\includegraphics[trim= 92pt 81pt 73pt 89pt, clip, height=1.40in, width=0.45\linewidth]{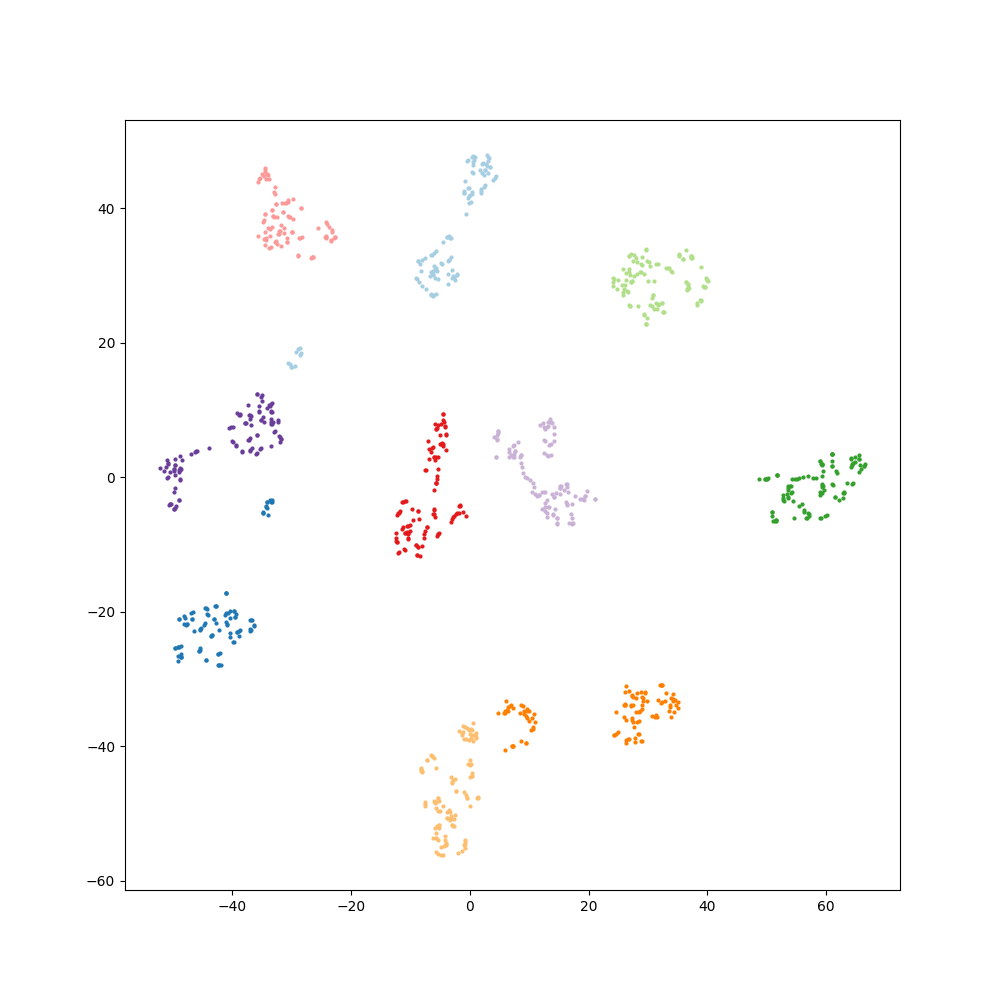}\hspace{4pt}}}
    
    \subfigure[\Fear perplexity: 40 (full) \label{fig:clustering:fear_full_perp_40}]{{\hspace{4pt}\includegraphics[trim= 92pt 81pt 73pt 89pt, clip, height=1.40in, width=0.45\linewidth]{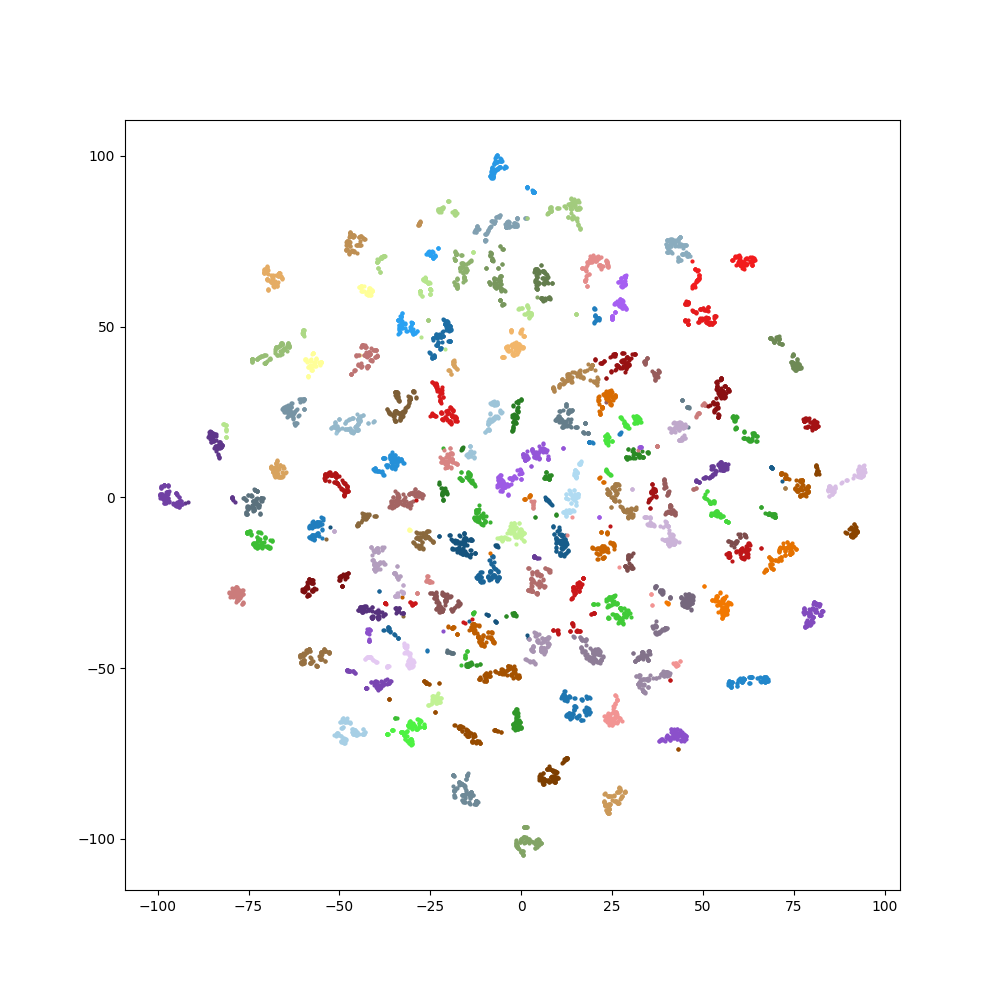}\hspace{4pt}}}
    \hfill
    \subfigure[\Fear perplexity: 40 (10 subj.) \label{fig:clustering:fear_10_perp_40}]{{\hspace{4pt}\includegraphics[trim= 92pt 81pt 73pt 89pt, clip, height=1.40in, width=0.45\linewidth]{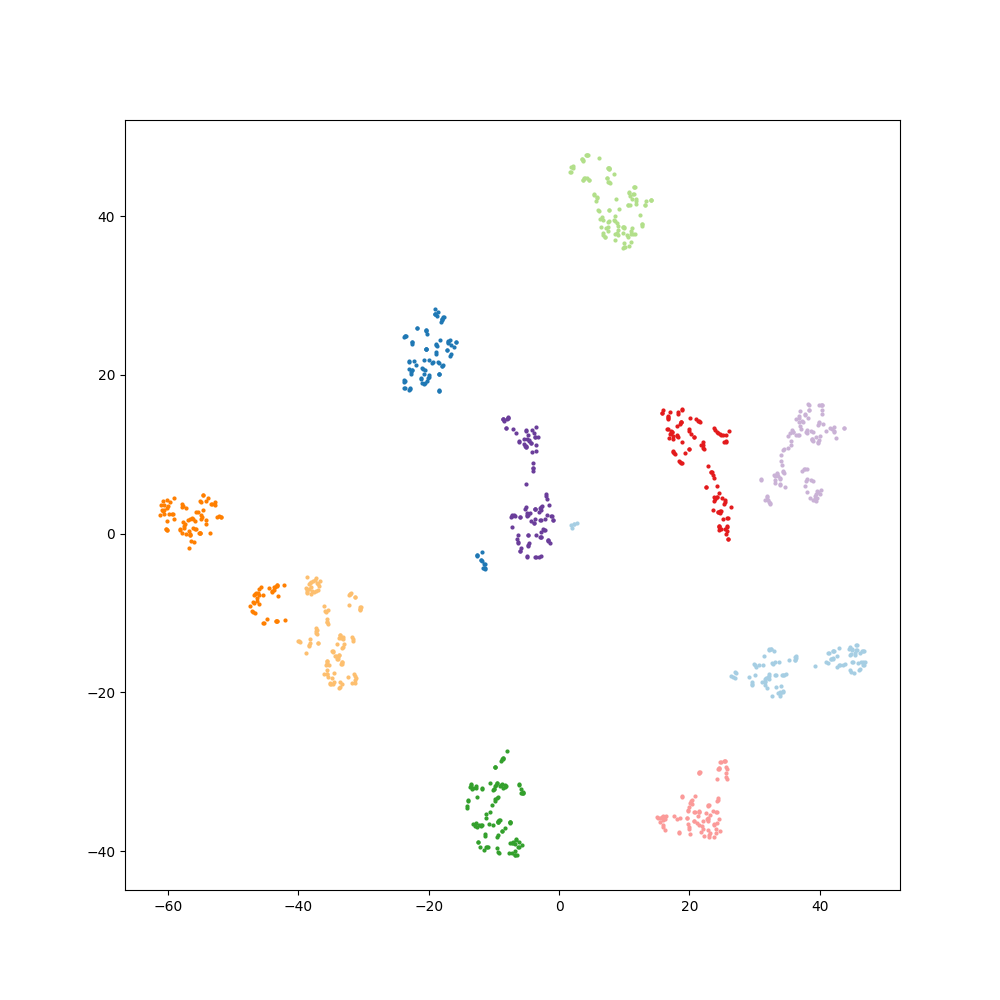}\hspace{4pt}}}
    
    \subfigure[\Fear perplexity: 50 (full) \label{fig:clustering:fear_full_perp_50}]{{\hspace{4pt}\includegraphics[trim= 92pt 81pt 73pt 89pt, clip, height=1.40in, width=0.45\linewidth]{figures/clustering/tsne/Fear_wasserstein_perplexity-50.png}\hspace{4pt}}}
    \hfill
    \subfigure[\Fear perplexity: 50 (10 subj.) \label{fig:clustering:fear_10_perp_50}]{{\hspace{4pt}\includegraphics[trim= 92pt 81pt 73pt 89pt, clip, height=1.40in, width=0.45\linewidth]{figures/clustering/tsne/Fear_wasserstein_perplexity-50_10subjects.png}\hspace{4pt}}}
    
    \subfigure[\Fear perplexity: 100 (full) \label{fig:clustering:fear_full_perp_100}]{{\hspace{4pt}\includegraphics[trim= 92pt 81pt 73pt 89pt, clip, height=1.40in, width=0.45\linewidth]{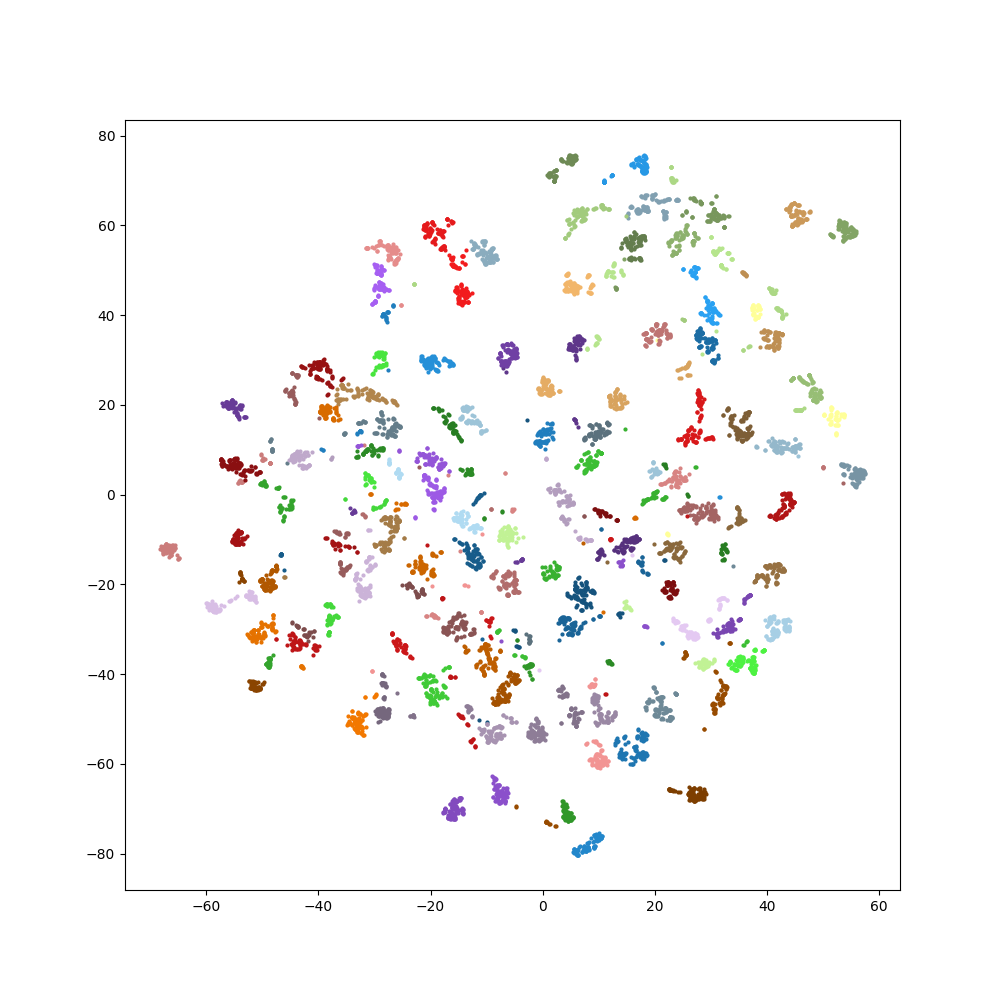}\hspace{4pt}}}
    \hfill
    \subfigure[\Fear perplexity: 100 (10 subj.) \label{fig:clustering:fear_10_perp_100}]{{\hspace{4pt}\includegraphics[trim= 92pt 81pt 73pt 89pt, clip, height=1.40in, width=0.45\linewidth]{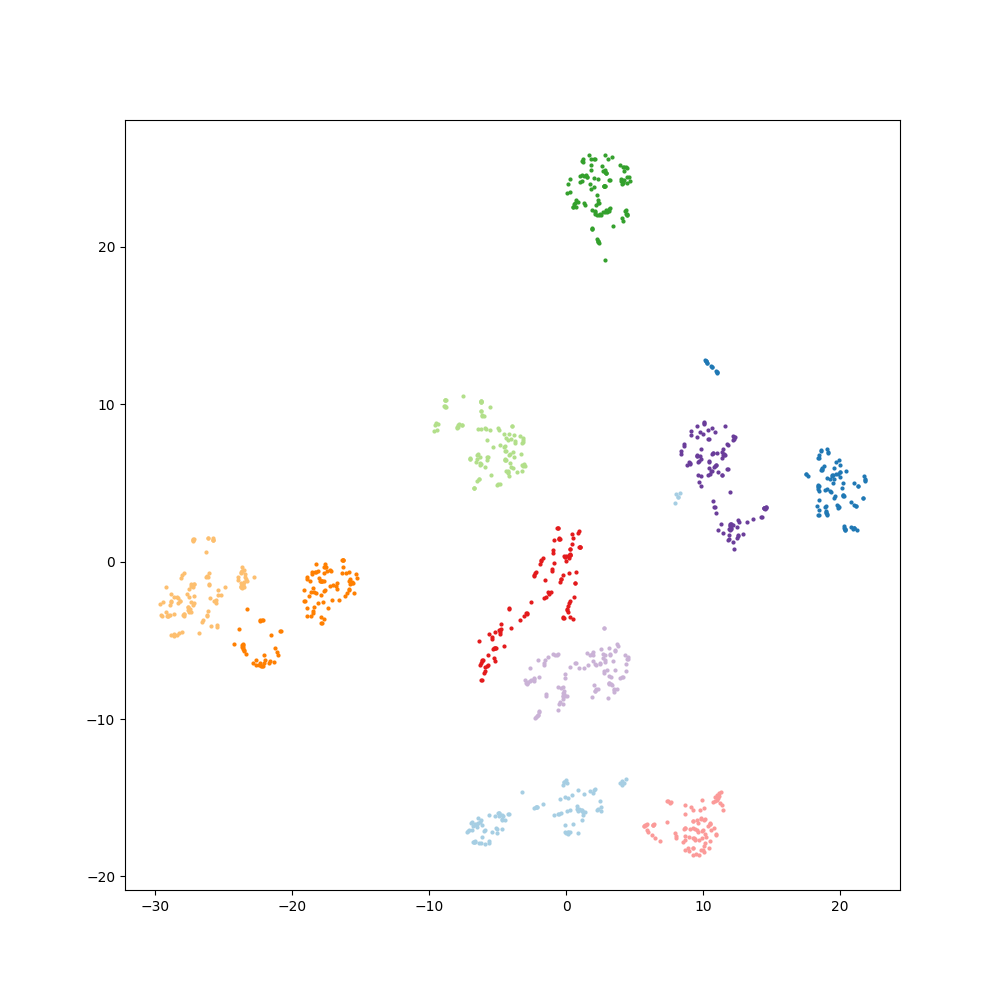}\hspace{4pt}}}
    
    \caption{t-SNE clustering of individual topological data for \Fear emotion at different perplexities.}
    \label{fig:clustering:fear}
\end{figure}

% Happy

\begin{figure}[!b]
    \centering
    \subfigure[\Happiness perp.: 30 (full) \label{fig:clustering:happy_full_perp_30}]{{\hspace{4pt}\includegraphics[trim= 92pt 81pt 73pt 89pt, clip, height=1.40in, width=0.45\linewidth]{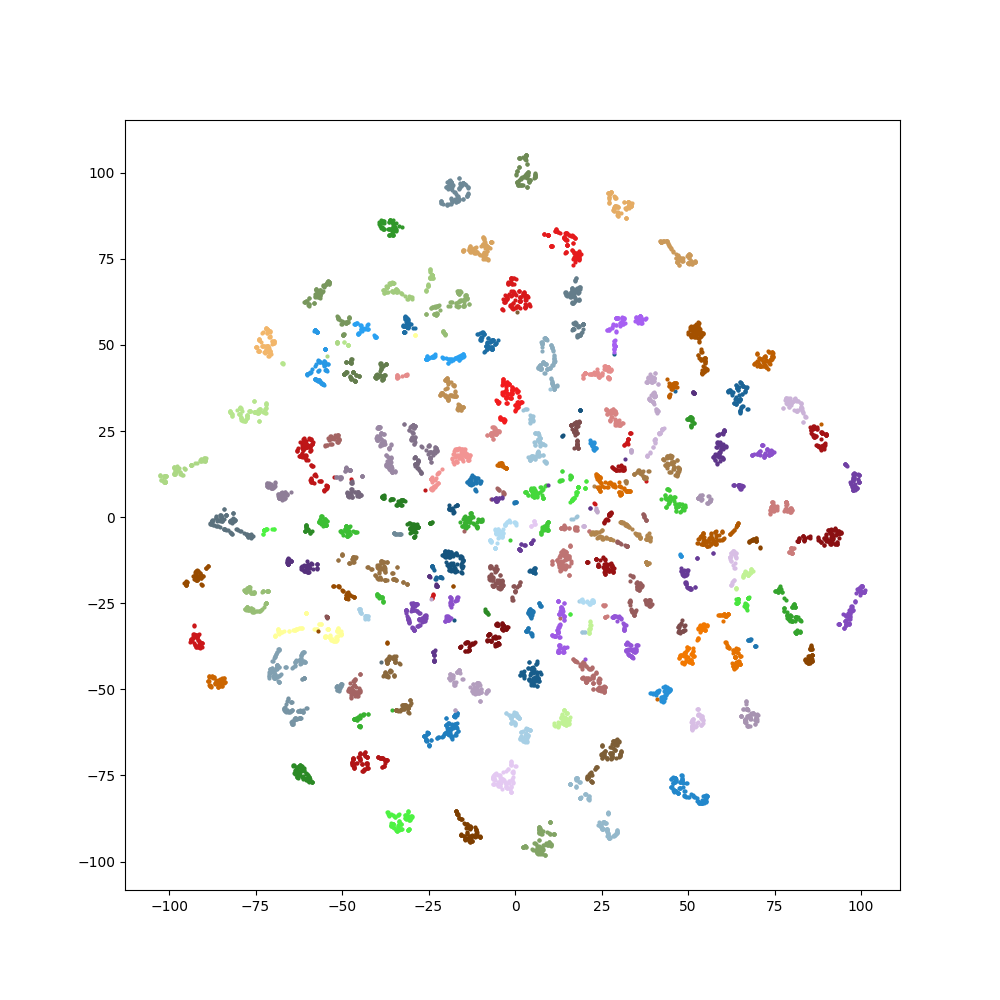}\hspace{4pt}}}
    \hfill
    \subfigure[\Happiness perp.: 30 (10 subj.) \label{fig:clustering:happy_10_perp_30}]{{\hspace{4pt}\includegraphics[trim= 92pt 81pt 73pt 89pt, clip, height=1.40in, width=0.45\linewidth]{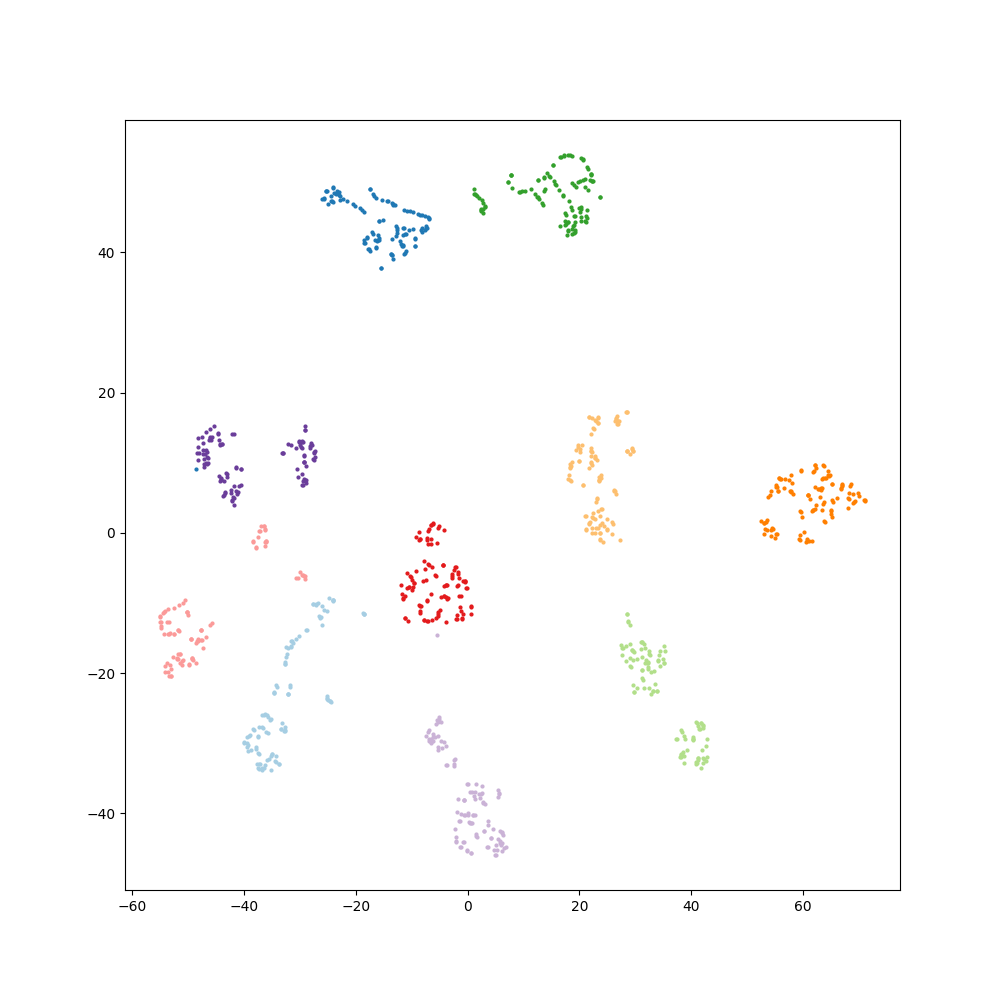}\hspace{4pt}}}
    
    \subfigure[\Happiness perp.: 40 (full) \label{fig:clustering:happy_full_perp_40}]{{\hspace{4pt}\includegraphics[trim= 92pt 81pt 73pt 89pt, clip, height=1.40in, width=0.45\linewidth]{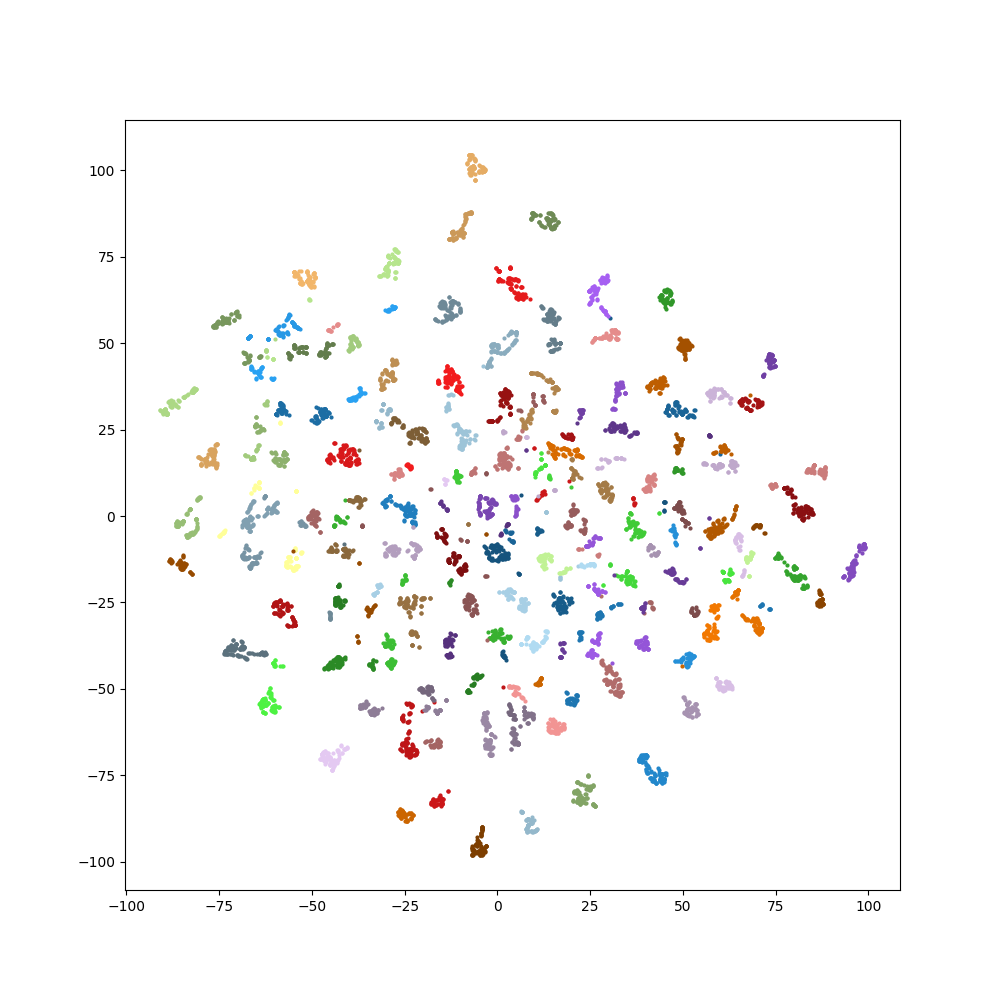}\hspace{4pt}}}
    \subfigure[\Happiness perp.: 40 (10 subj.) \label{fig:clustering:happy_10_perp_40}]{{\hspace{4pt}\includegraphics[trim= 92pt 81pt 73pt 89pt, clip, height=1.40in, width=0.45\linewidth]{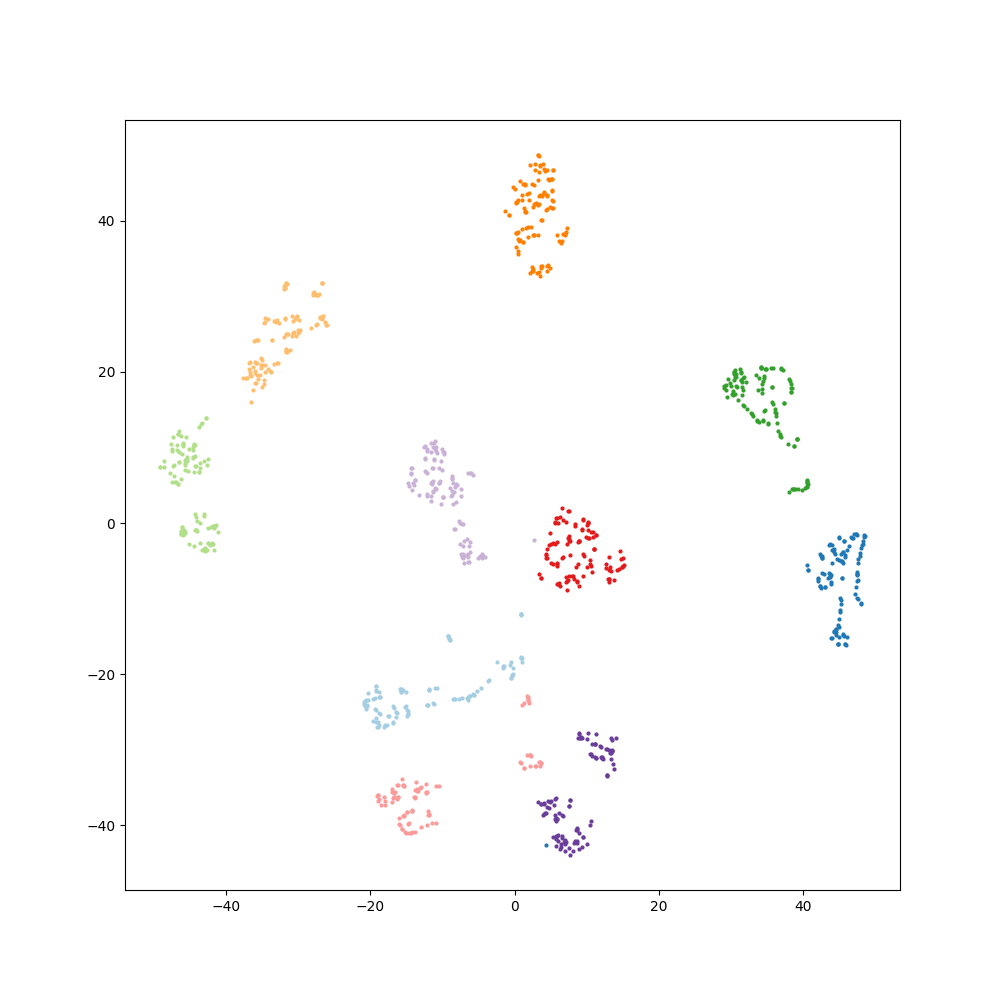}\hspace{4pt}}}
    
    \subfigure[\Happiness perp.: 50 (full) \label{fig:clustering:happy_full_perp_50}]{{\hspace{4pt}\includegraphics[trim= 92pt 81pt 73pt 89pt, clip, height=1.40in, width=0.45\linewidth]{figures/clustering/tsne/Happy_wasserstein_perplexity-50.png}\hspace{4pt}}}
    \subfigure[\Happiness perp.: 50 (10 subj.) \label{fig:clustering:happy_10_perp_50}]{{\hspace{4pt}\includegraphics[trim= 92pt 81pt 73pt 89pt, clip, height=1.40in, width=0.45\linewidth]{figures/clustering/tsne/Happy_wasserstein_perplexity-50_10subjects.png}\hspace{4pt}}}
    
    \subfigure[\Happiness perp.: 100 (full) \label{fig:clustering:happy_full_perp_100}]{{\hspace{4pt}\includegraphics[trim= 92pt 81pt 73pt 89pt, clip, height=1.40in, width=0.45\linewidth]{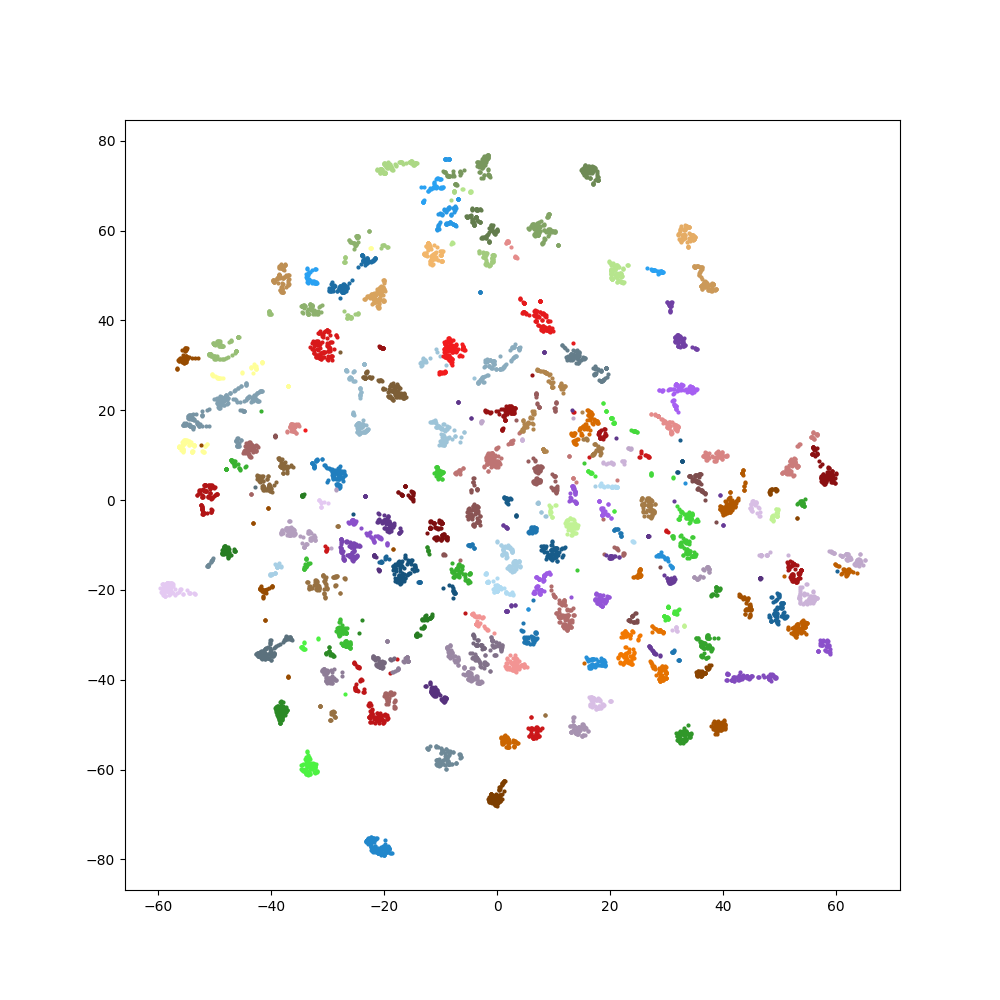}\hspace{4pt}}}
    \subfigure[\Happiness perp.: 100 (10 subj.) \label{fig:clustering:happy_10_perp_100}]{{\hspace{4pt}\includegraphics[trim= 92pt 81pt 73pt 89pt, clip, height=1.40in, width=0.45\linewidth]{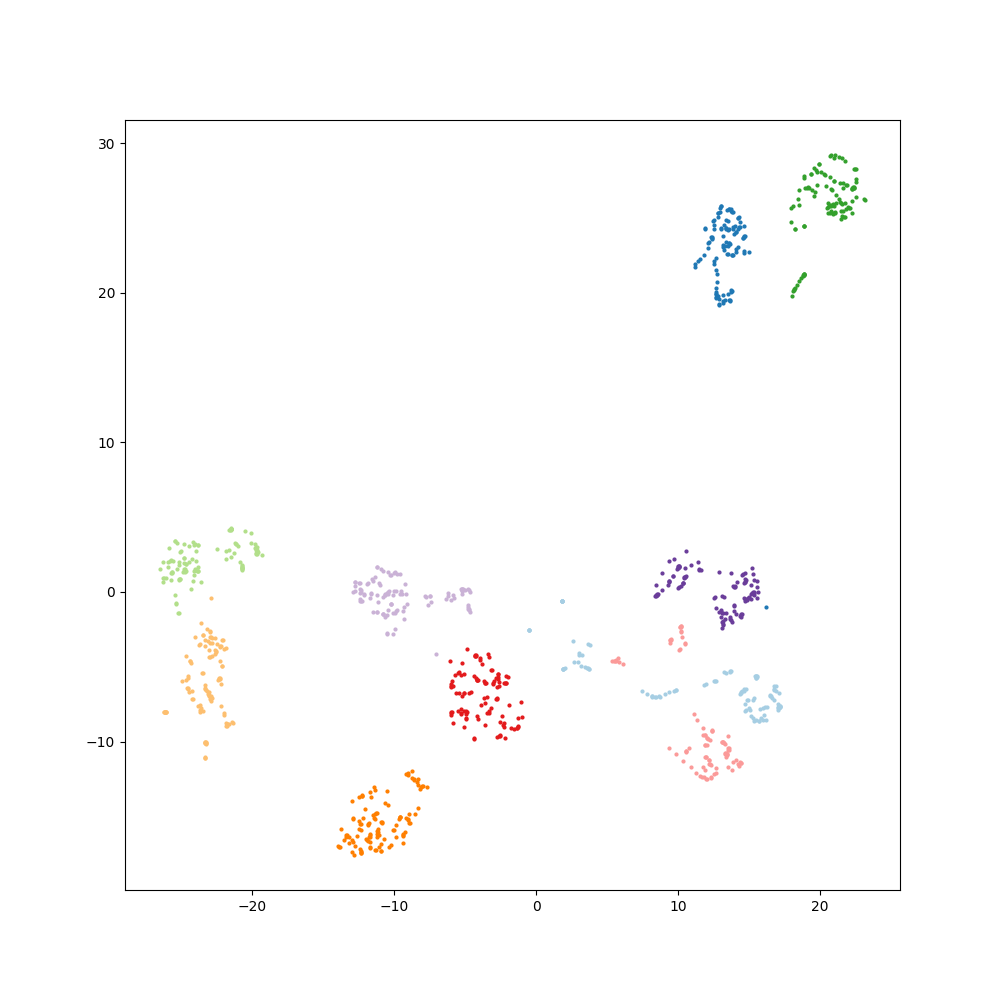}\hspace{4pt}}}
    
    \caption{t-SNE clustering of individual topological data for \Happiness emotion at different perplexities.}
    \label{fig:clustering:happy}
\end{figure}

% \Sadness

\begin{figure}[!b]
    \centering
    \subfigure[\Sadness perplexity: 30 (full) \label{fig:clustering:sad_full_perp_30}]{{\hspace{4pt}\includegraphics[trim= 92pt 81pt 73pt 89pt, clip, height=1.40in, width=0.45\linewidth]{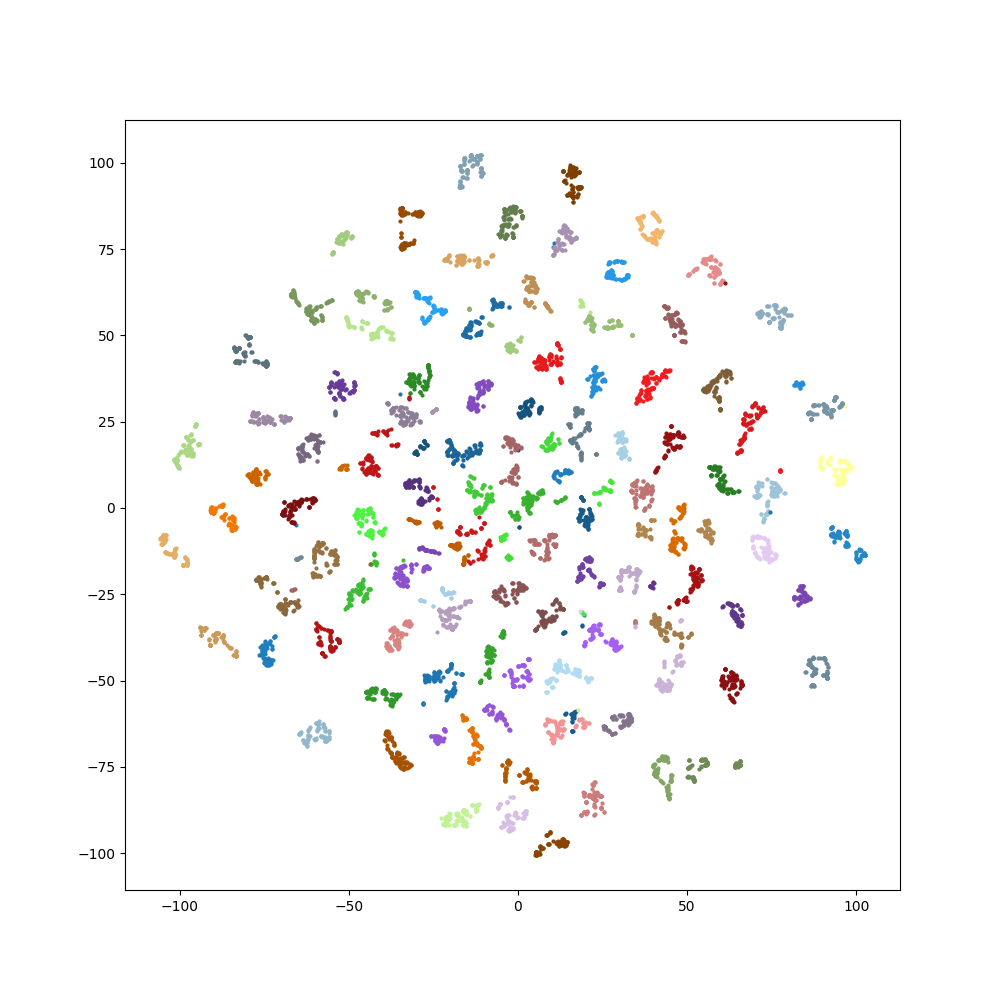}\hspace{4pt}}}
    \subfigure[\Sadness perplexity: 30 (10 subj.) \label{fig:clustering:sad_10_perp_30}]{{\hspace{4pt}\includegraphics[trim= 92pt 81pt 73pt 89pt, clip, height=1.40in, width=0.45\linewidth]{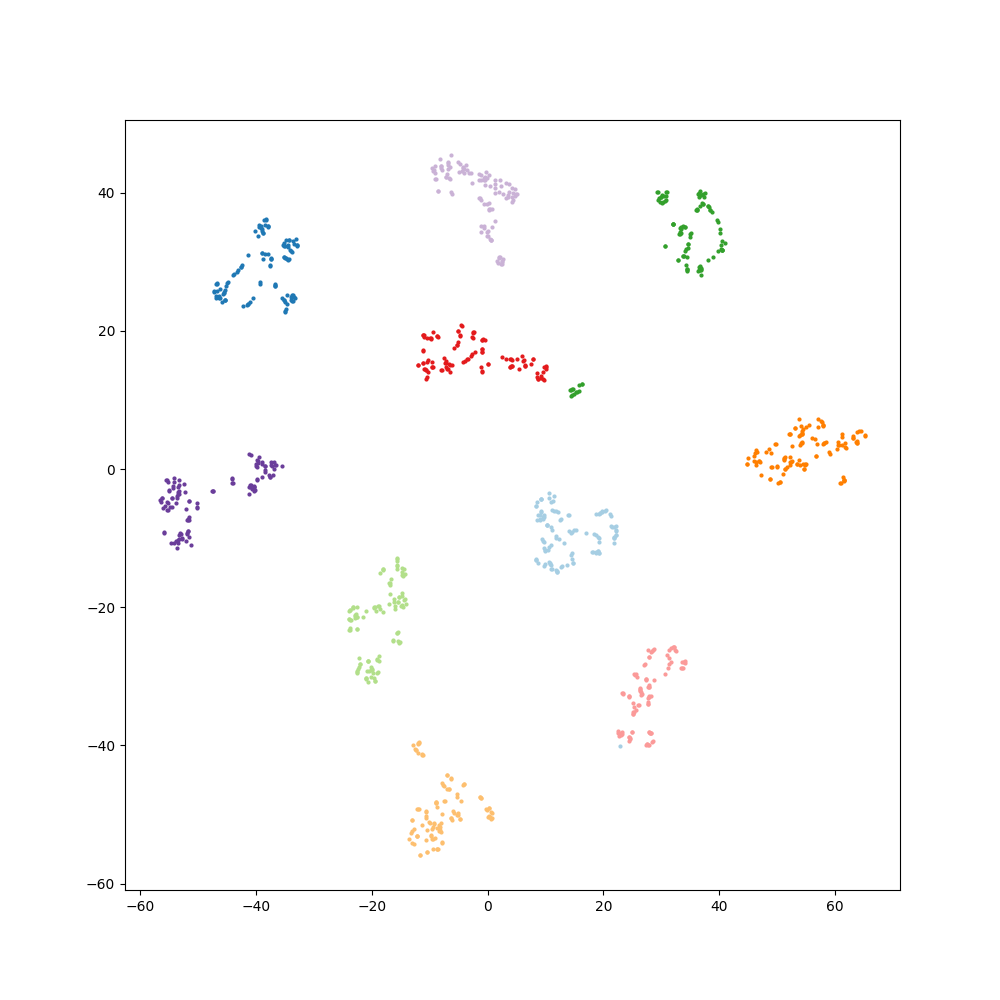}\hspace{4pt}}}
    
    \subfigure[\Sadness perplexity: 40 (full) \label{fig:clustering:sad_full_perp_40}]{{\hspace{4pt}\includegraphics[trim= 92pt 81pt 73pt 89pt, clip, height=1.40in, width=0.45\linewidth]{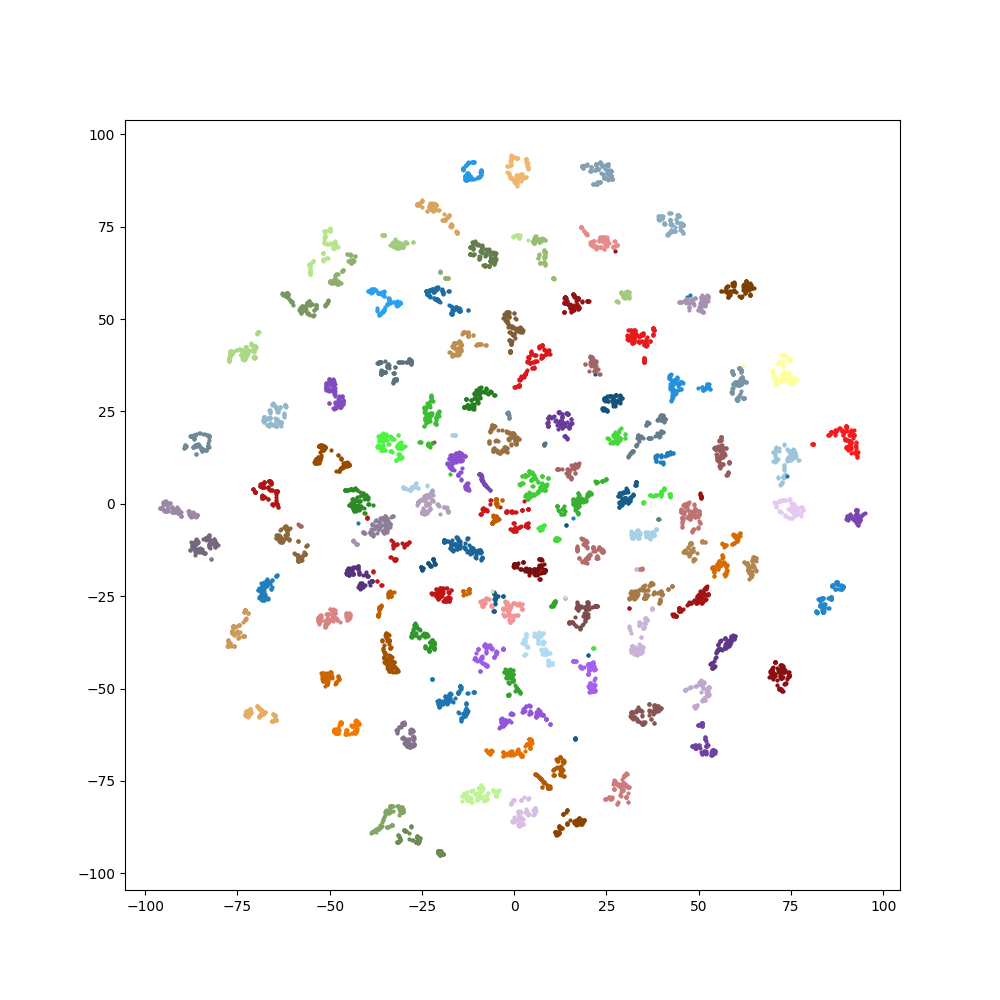}\hspace{4pt}}}
    \subfigure[\Sadness perplexity: 40 (10 subj.) \label{fig:clustering:sad_10_perp_40}]{{\hspace{4pt}\includegraphics[trim= 92pt 81pt 73pt 89pt, clip, height=1.40in, width=0.45\linewidth]{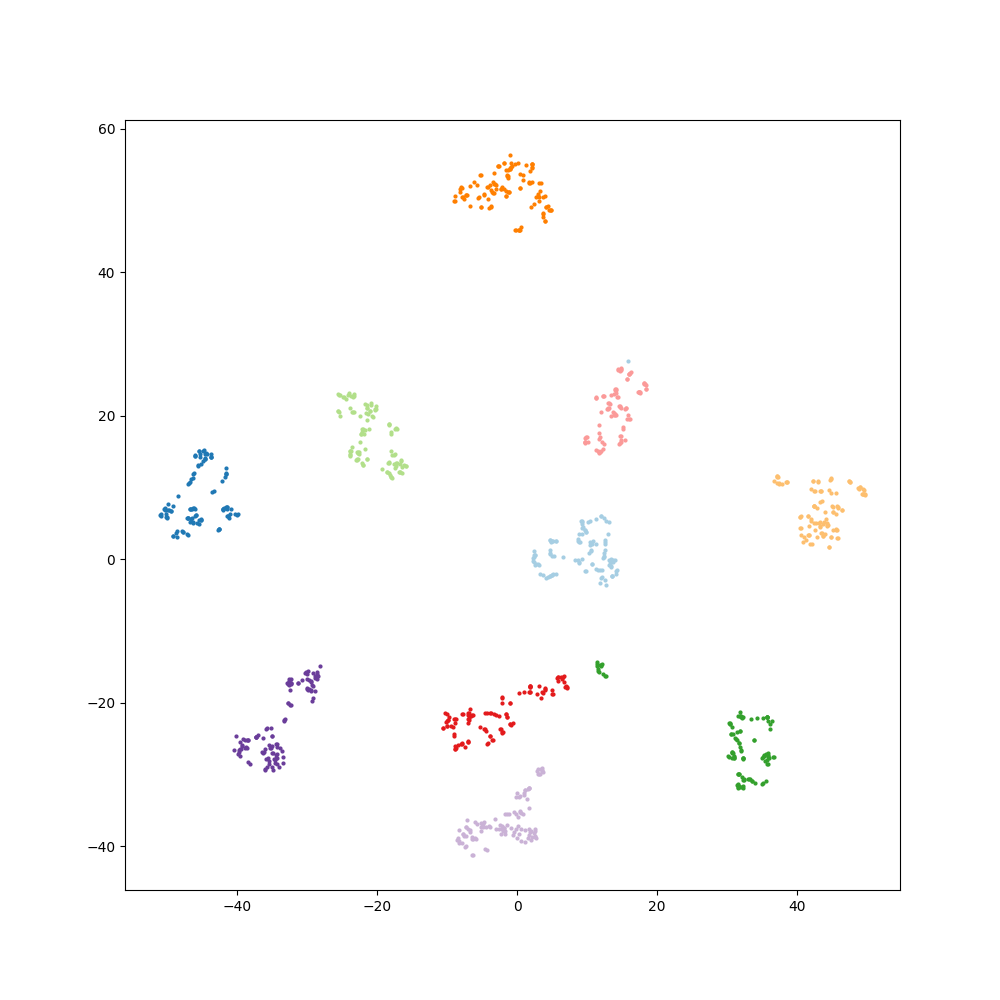}\hspace{4pt}}}
    
    \subfigure[\Sadness perplexity: 50 (full) \label{fig:clustering:sad_full_perp_50}]{{\hspace{4pt}\includegraphics[trim= 92pt 81pt 73pt 89pt, clip, height=1.40in, width=0.45\linewidth]{figures/clustering/tsne/Sad_wasserstein_perplexity-50.png}\hspace{4pt}}}
    \subfigure[\Sadness perplexity: 50 (10 subj.) \label{fig:clustering:sad_10_perp_50}]{{\hspace{4pt}\includegraphics[trim= 92pt 81pt 73pt 89pt, clip, height=1.40in, width=0.45\linewidth]{figures/clustering/tsne/Sad_wasserstein_perplexity-50_10subjects.png}\hspace{4pt}}}
    
    \subfigure[\Sadness perplexity: 100 (full) \label{fig:clustering:sad_full_perp_100}]{{\hspace{4pt}\includegraphics[trim= 92pt 81pt 73pt 89pt, clip, height=1.40in, width=0.45\linewidth]{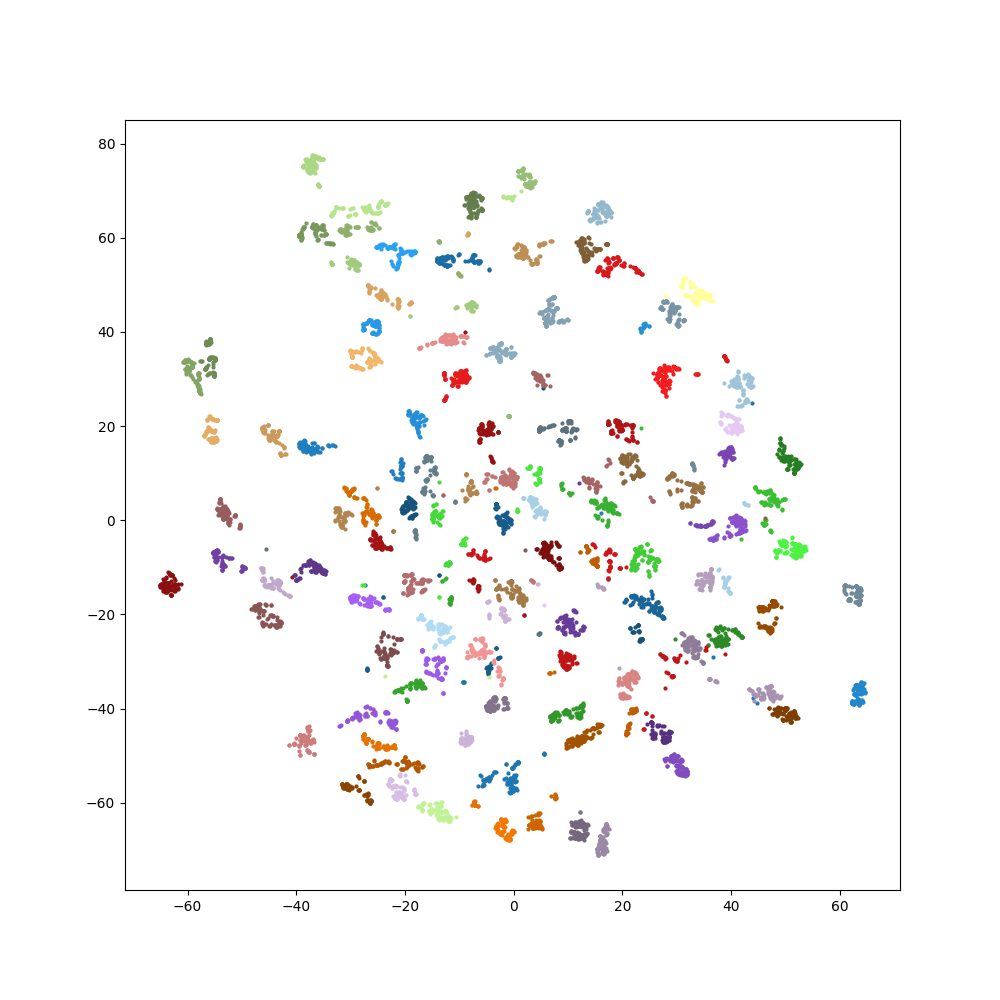}\hspace{4pt}}}
    \subfigure[\Sadness perplexity: 100 (10 subj.) \label{fig:clustering:sad_10_perp_100}]{{\hspace{4pt}\includegraphics[trim= 92pt 81pt 73pt 89pt, clip, height=1.40in, width=0.45\linewidth]{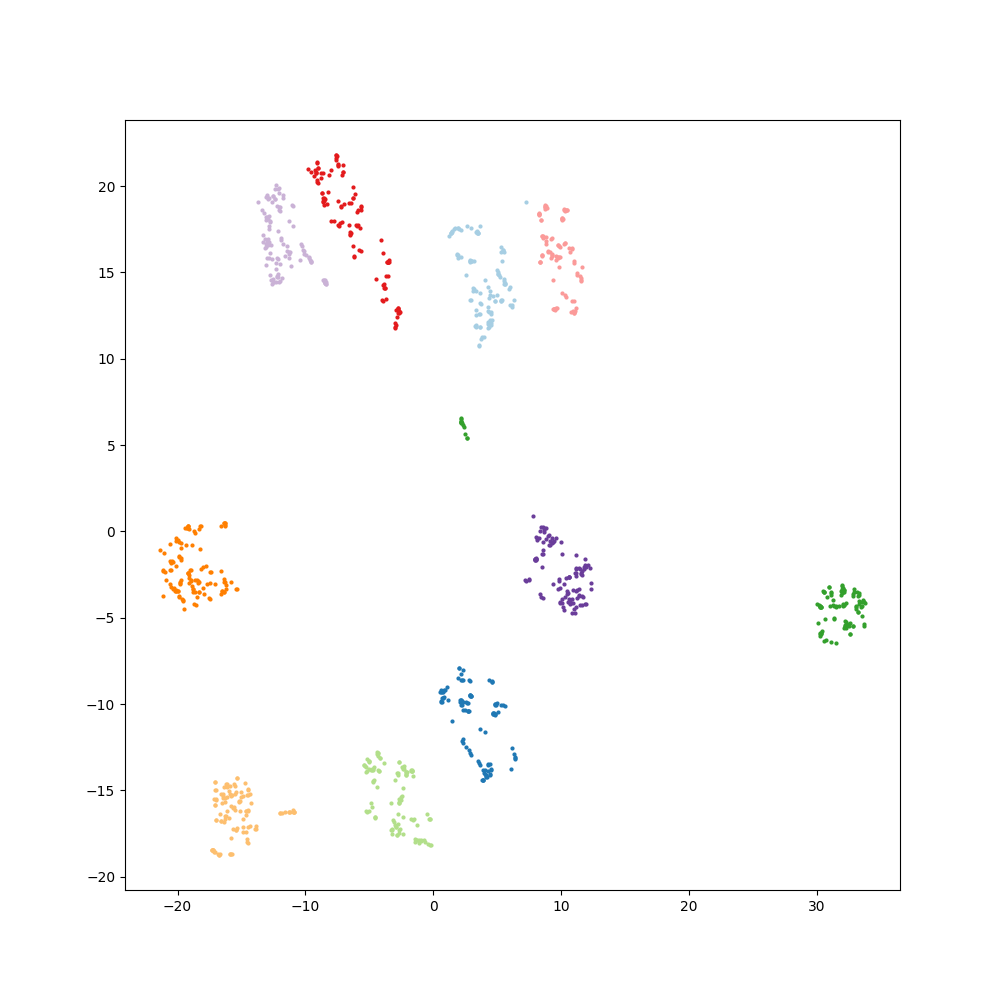}\hspace{4pt}}}
    
    \caption{t-SNE clustering of individual topological data for \Sadness emotion at different perplexities.}
    \label{fig:clustering:sad}
\end{figure}

% \Surprise

\begin{figure}[!b]
    \centering
    \subfigure[\Surprise perplexity: 30 (full) \label{fig:clustering:surprise_full_perp_30}]{{\hspace{4pt}\includegraphics[trim= 92pt 81pt 73pt 89pt, clip, height=1.40in, width=0.45\linewidth]{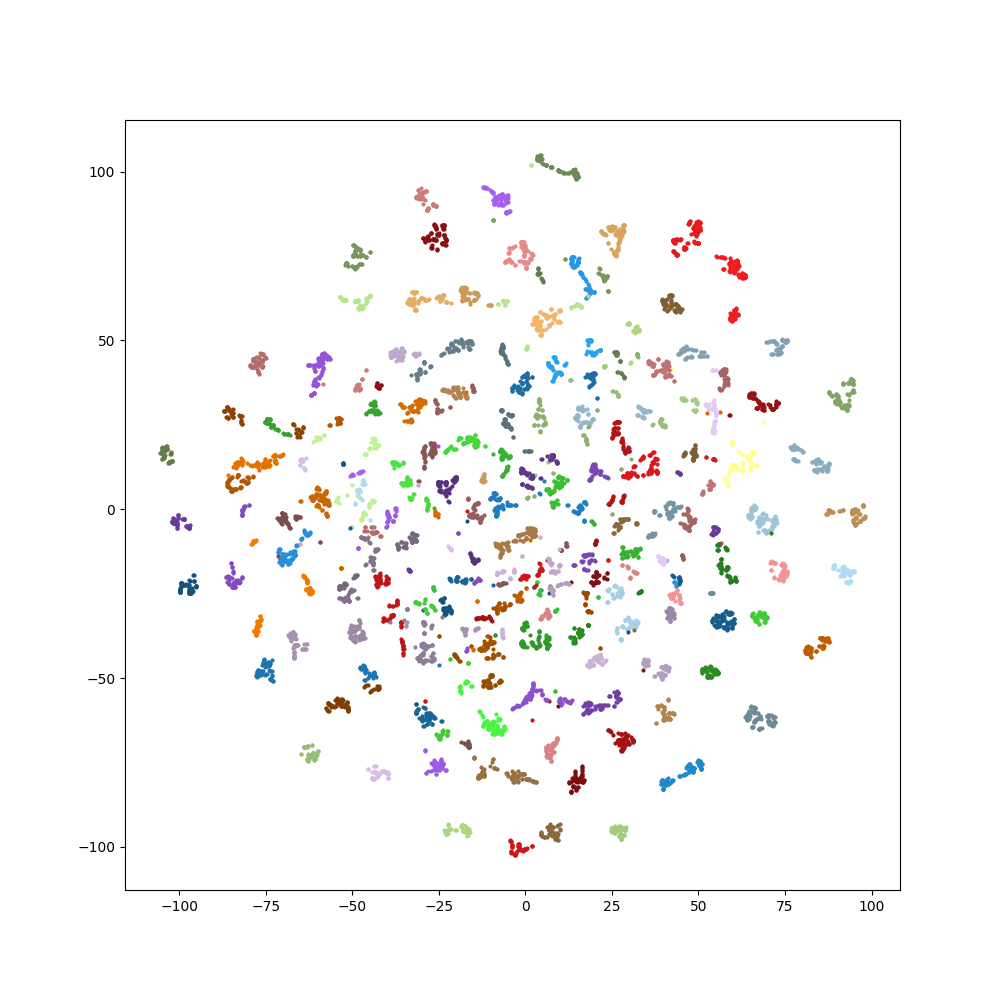}\hspace{4pt}}}
    \subfigure[\Surprise perplexity: 30 (10 subj.) \label{fig:clustering:surprise_10_perp_30}]{{\hspace{4pt}\includegraphics[trim= 92pt 81pt 73pt 89pt, clip, height=1.40in, width=0.45\linewidth]{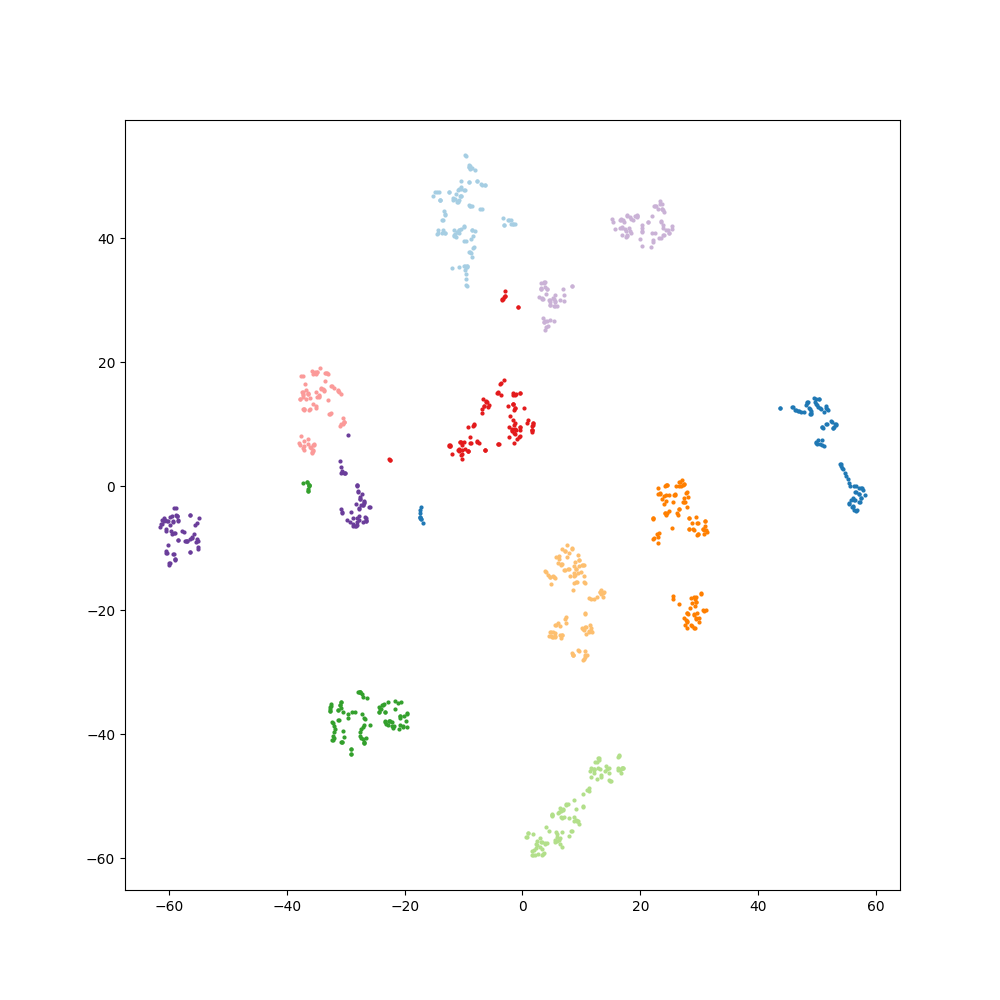}\hspace{4pt}}}
    
    \subfigure[\Surprise perplexity: 40 (full) \label{fig:clustering:surprise_full_perp_40}]{{\hspace{4pt}\includegraphics[trim= 92pt 81pt 73pt 89pt, clip, height=1.40in, width=0.45\linewidth]{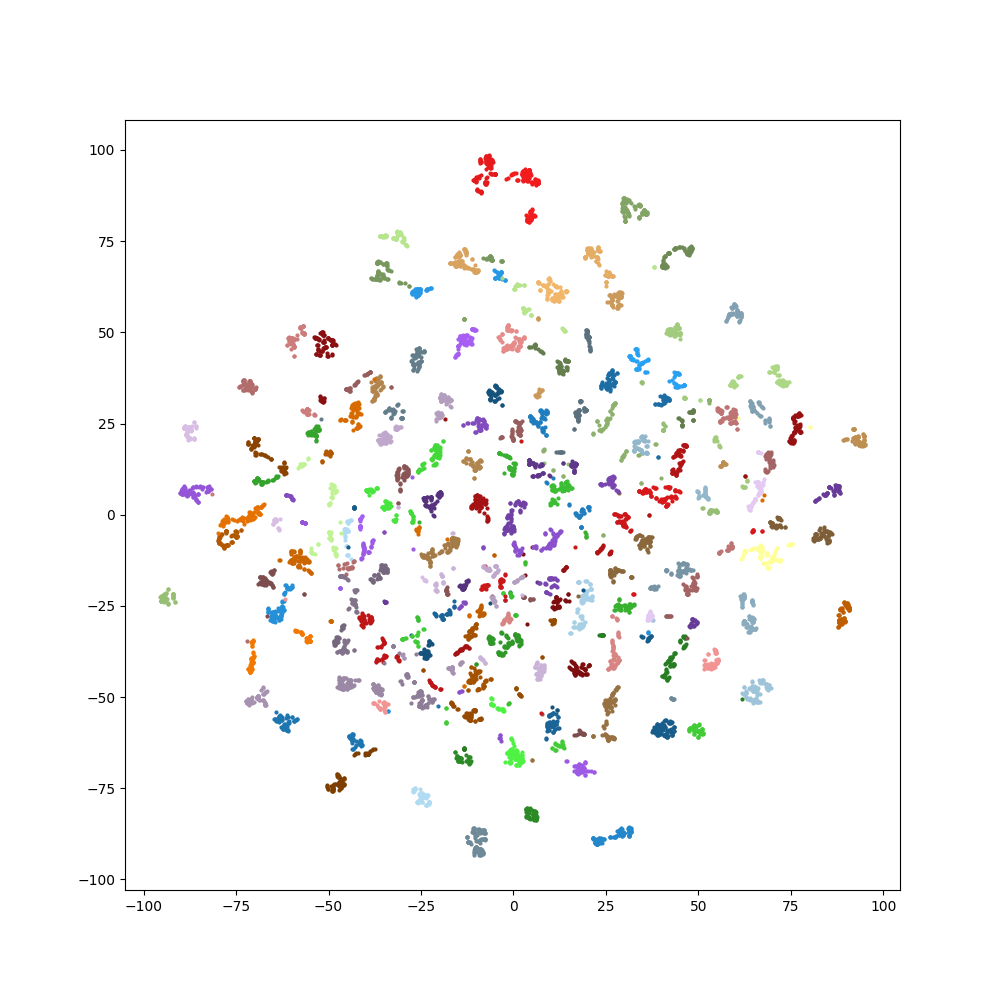}\hspace{4pt}}}
    \subfigure[\Surprise perplexity: 40 (10 subj.) \label{fig:clustering:surprise_10_perp_40}]{{\hspace{4pt}\includegraphics[trim= 92pt 81pt 73pt 89pt, clip, height=1.40in, width=0.45\linewidth]{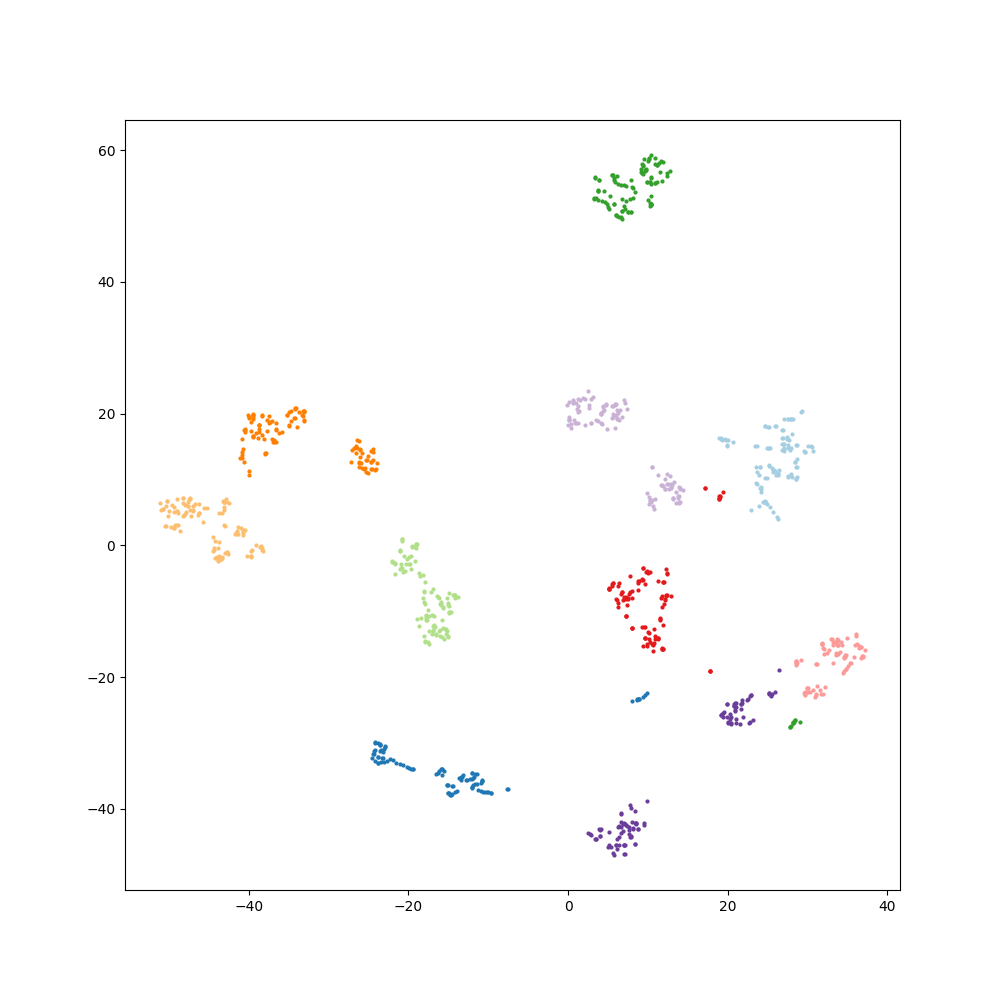}\hspace{4pt}}}
    
    \subfigure[\Surprise perplexity: 50 (full) \label{fig:clustering:surprise_full_perp_50}]{{\hspace{4pt}\includegraphics[trim= 92pt 81pt 73pt 89pt, clip, height=1.40in, width=0.45\linewidth]{figures/clustering/tsne/Surprise_wasserstein_perplexity-50.png}\hspace{4pt}}}
    \subfigure[\Surprise perplexity: 50 (10 subj.) \label{fig:clustering:surprise_10_perp_50}]{{\hspace{4pt}\includegraphics[trim= 92pt 81pt 73pt 89pt, clip, height=1.40in, width=0.45\linewidth]{figures/clustering/tsne/Surprise_wasserstein_perplexity-50_10subjects.png}\hspace{4pt}}}
    
    \subfigure[\Surprise perplexity: 100 (full) \label{fig:clustering:surprise_full_perp_100}]{{\hspace{4pt}\includegraphics[trim= 92pt 81pt 73pt 89pt, clip, height=1.40in, width=0.45\linewidth]{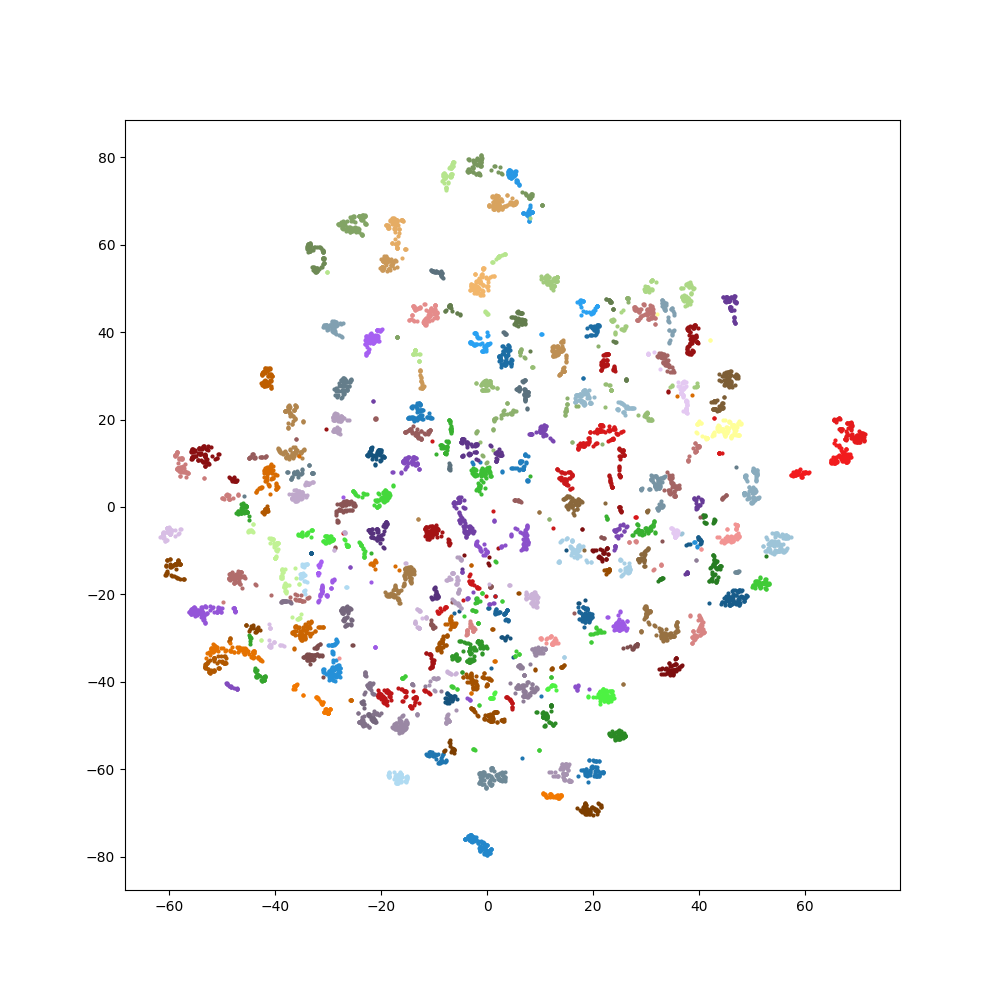}\hspace{4pt}}}
    \subfigure[\Surprise perplexity: 100 (10 subj.) \label{fig:clustering:surprise_10_perp_100}]{{\hspace{4pt}\includegraphics[trim= 92pt 81pt 73pt 89pt, clip, height=1.40in, width=0.45\linewidth]{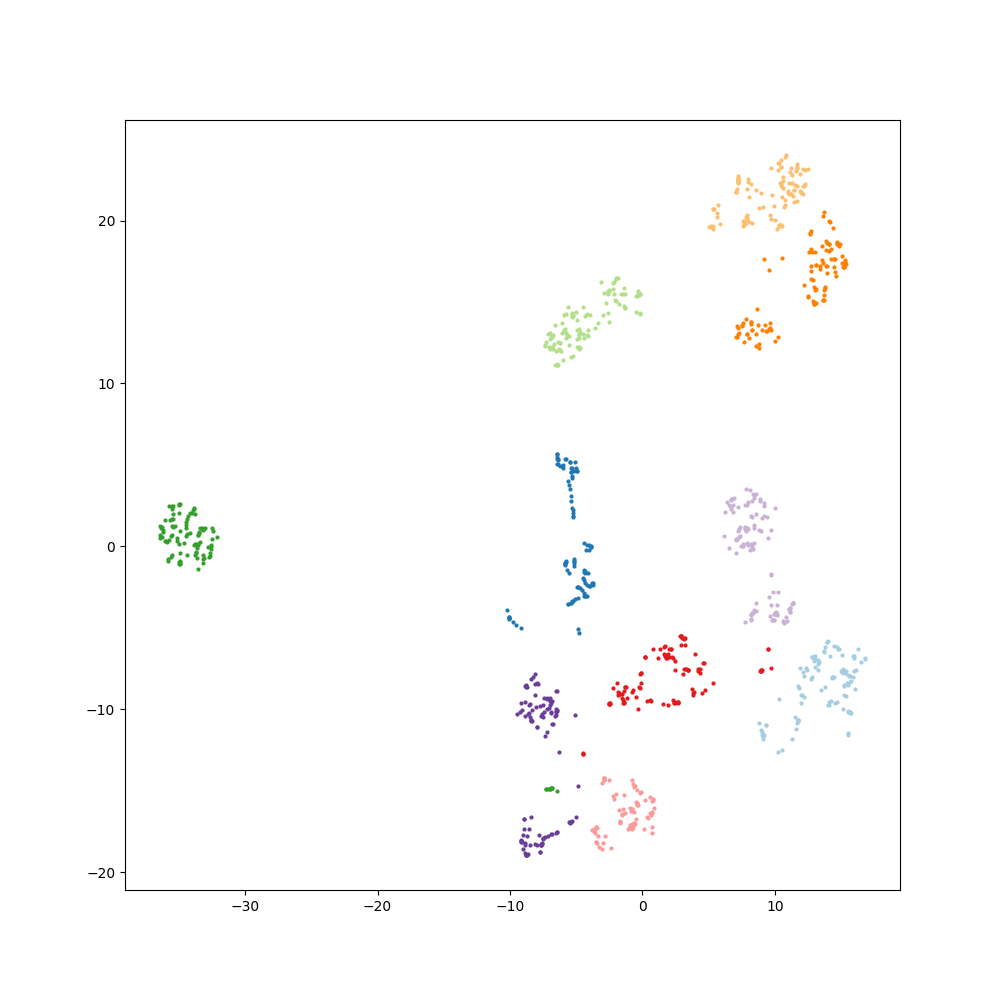}\hspace{4pt}}}
    
    \caption{t-SNE clustering of individual topological data for \Surprise emotion at different perplexities.}
    \label{fig:clustering:surprise}
\end{figure}

%%%%%%%%%%%%%%%%%%%%%%%%%%%%%%%%%%%%%%%%%%%%%%%%%%%%%%%%%%%%%%%%%%%%%%%%%%%%%%%%%%%%%%%%%%%%%%%%%
%%%%%%%%%%%%%%%%%%%%%%%%%%%%%%%%%%%%%%%%%%%%% UMAP %%%%%%%%%%%%%%%%%%%%%%%%%%%%%%%%%%%%%%%%%%%%%%
%%%%%%%%%%%%%%%%%%%%%%%%%%%%%%%%%%%%%%%%%%%%%%%%%%%%%%%%%%%%%%%%%%%%%%%%%%%%%%%%%%%%%%%%%%%%%%%%%

% \Anger

\begin{figure}[!b]
    \centering
    \subfigure[\Anger perplexity: 30 (full) \label{fig:umap:angry_full_perp_30}]{{\hspace{4pt}\includegraphics[trim= 92pt 81pt 73pt 89pt, clip, height=1.40in, width=0.45\linewidth]{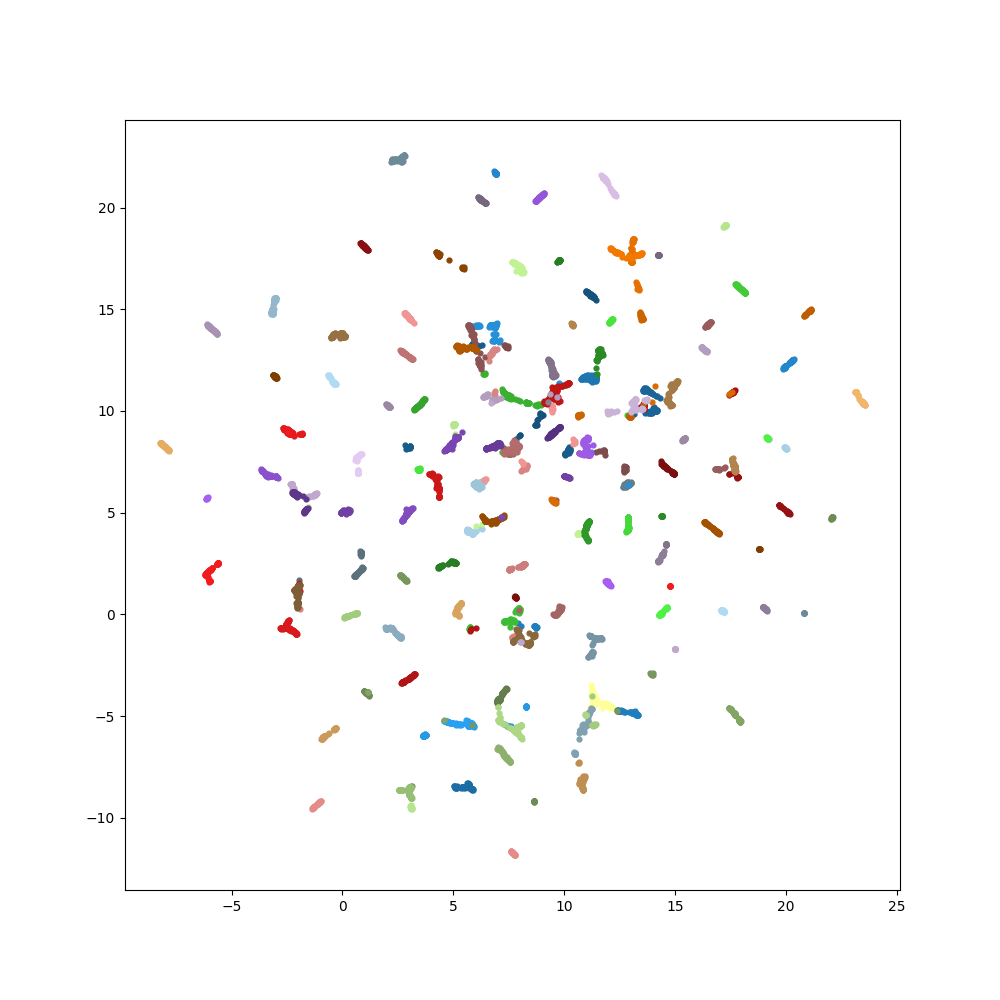}\hspace{4pt}}}
    \subfigure[\Anger perplexity: 30 (10 subj.) \label{fig:umap:angry_10_perp_30}]{{\hspace{4pt}\includegraphics[trim= 92pt 81pt 73pt 89pt, clip, height=1.40in, width=0.45\linewidth]{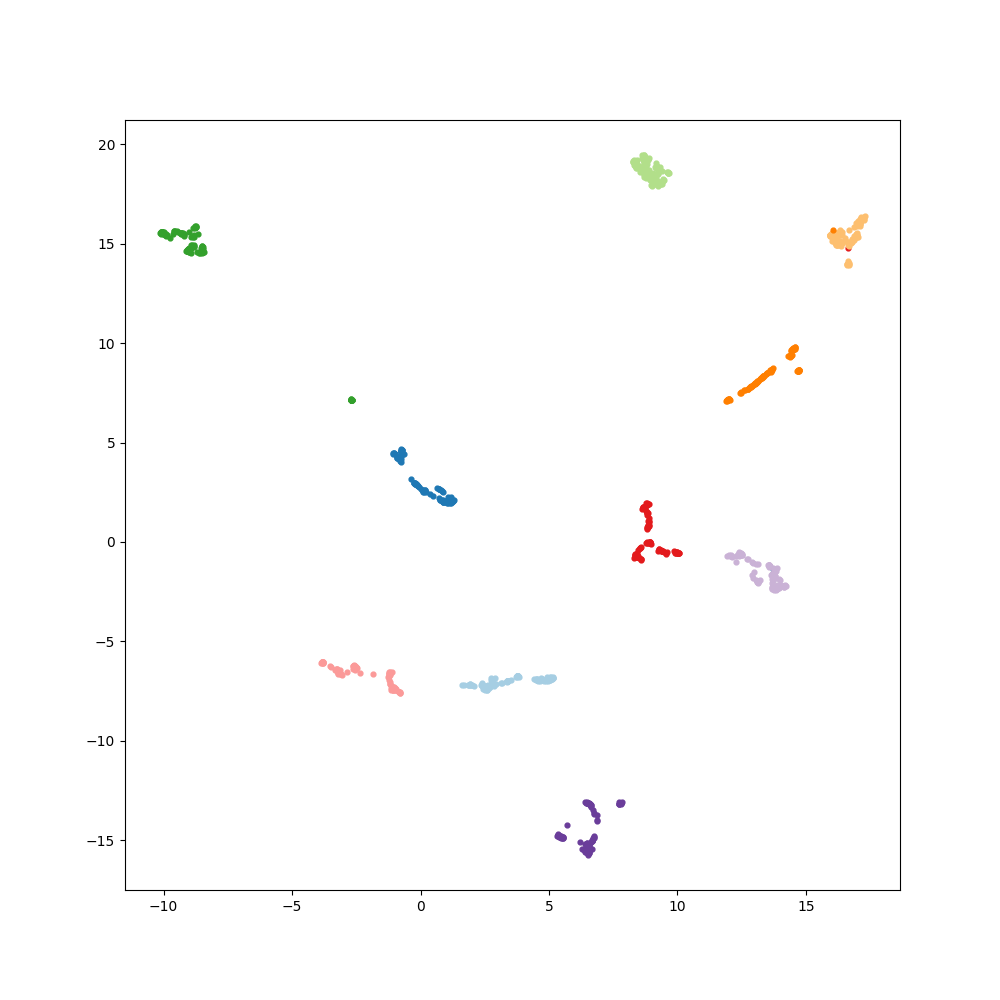}\hspace{4pt}}}
    
    \subfigure[\Anger perplexity: 40 (full) \label{fig:umap:angry_full_perp_40}]{{\hspace{4pt}\includegraphics[trim= 92pt 81pt 73pt 89pt, clip, height=1.40in, width=0.45\linewidth]{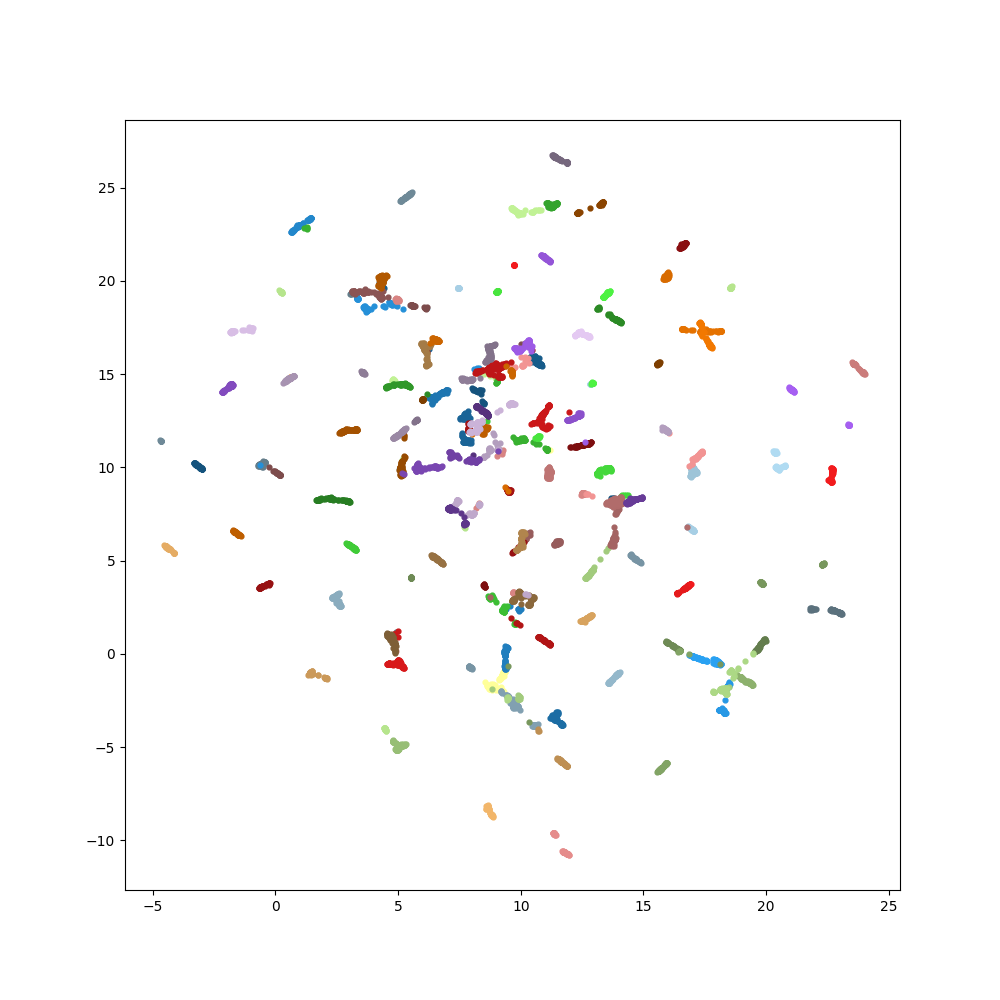}\hspace{4pt}}}
    \subfigure[\Anger (10 subj.)  Perplexity=40 \label{fig:umap:angry_10_perp_40}]{{\hspace{4pt}\includegraphics[trim= 92pt 81pt 73pt 89pt, clip, height=1.40in, width=0.45\linewidth]{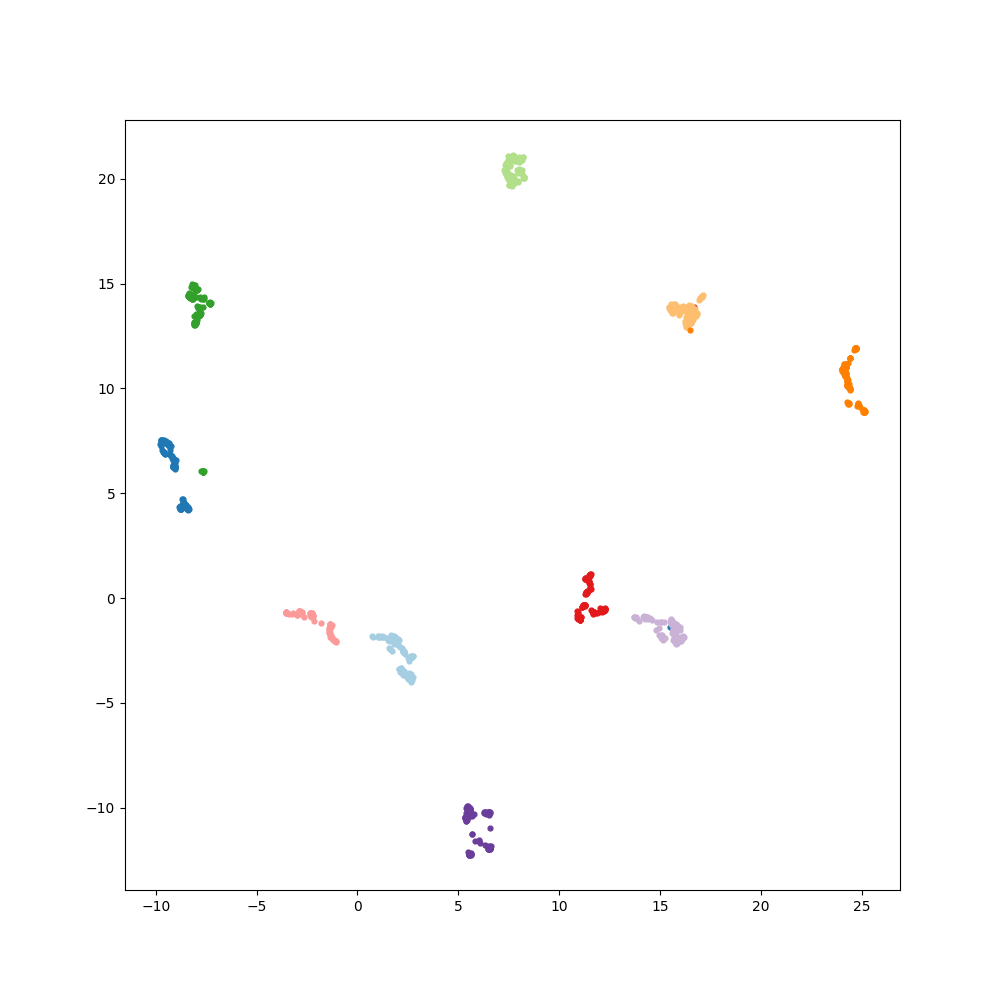}\hspace{4pt}}}
    
    \subfigure[\Anger perplexity: 50 (full) \label{fig:umap:angry_full_perp_50}]{{\hspace{4pt}\includegraphics[trim= 92pt 81pt 73pt 89pt, clip, height=1.40in, width=0.45\linewidth]{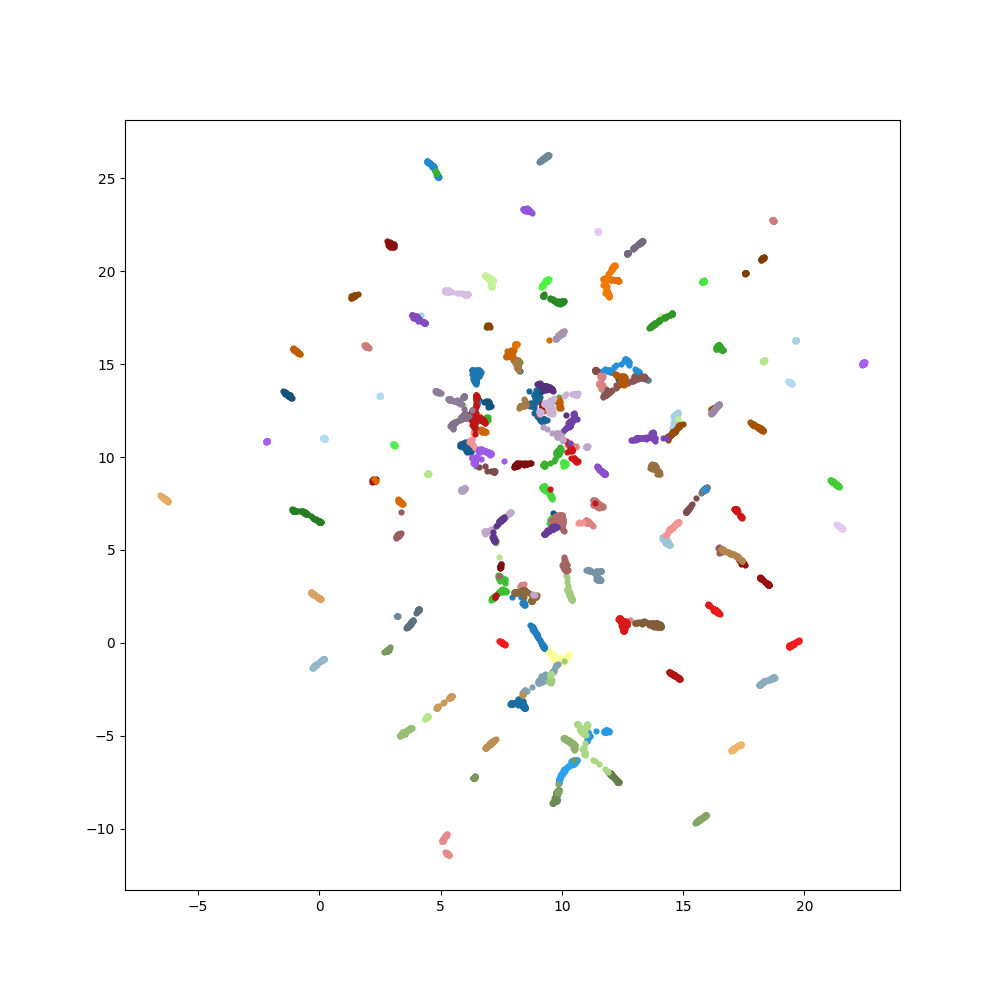}\hspace{4pt}}}
    \subfigure[\Anger perplexity: 50 (10 subj.) \label{fig:umap:angry_10_perp_50}]{{\hspace{4pt}\includegraphics[trim= 92pt 81pt 73pt 89pt, clip, height=1.40in, width=0.45\linewidth]{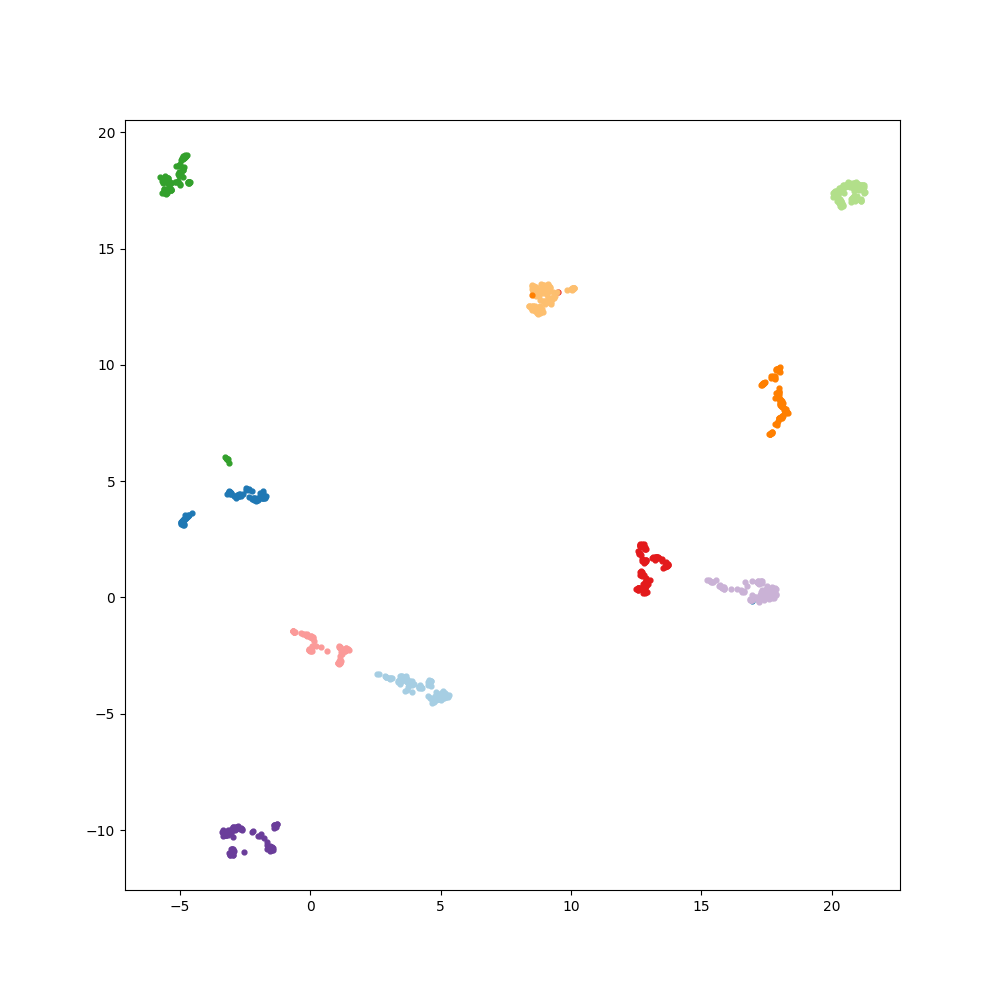}\hspace{4pt}}}
    
    \subfigure[\Anger perplexity: 100 (full) \label{fig:umap:angry_full_perp_100}]{{\hspace{4pt}\includegraphics[trim= 92pt 81pt 73pt 89pt, clip, height=1.40in, width=0.45\linewidth]{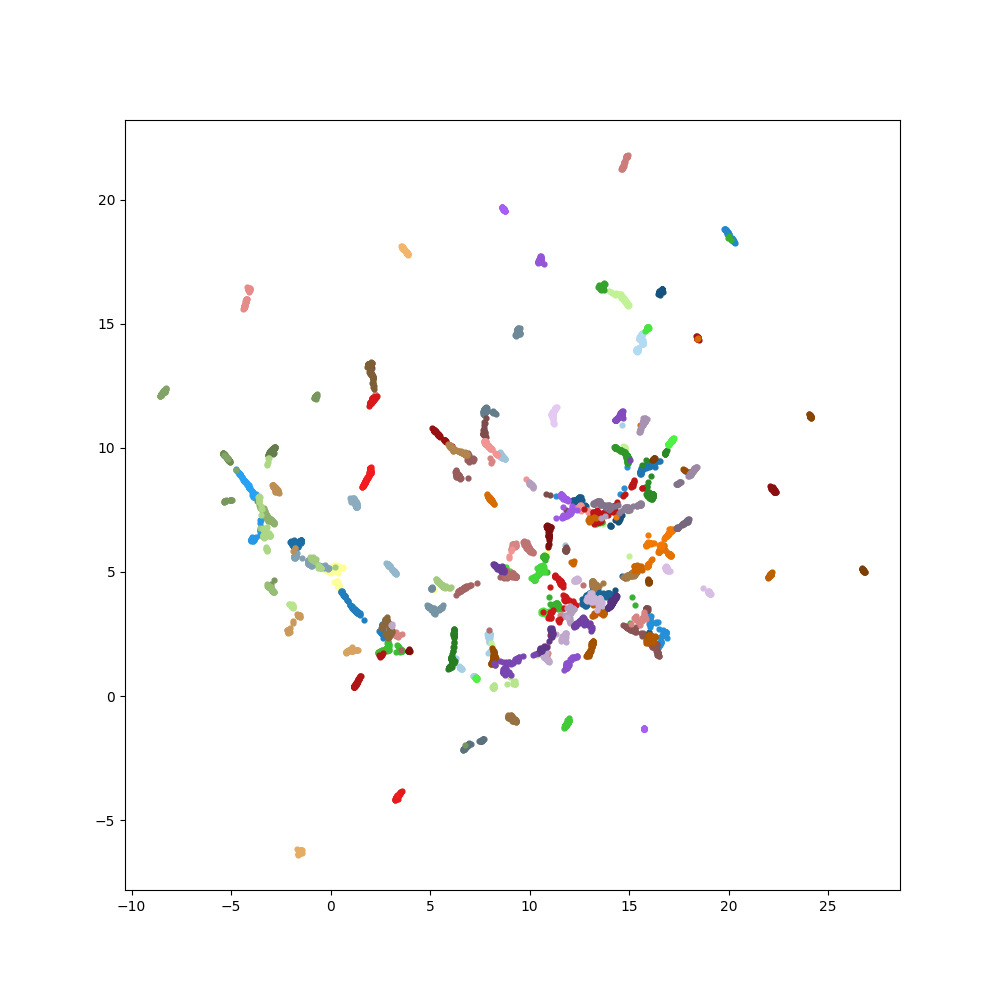}\hspace{4pt}}}
    \subfigure[\Anger perplexity: 100 (10 subj.) \label{fig:umap:angry_10_perp_100}]{{\hspace{4pt}\includegraphics[trim= 92pt 81pt 73pt 89pt, clip, height=1.40in, width=0.45\linewidth]{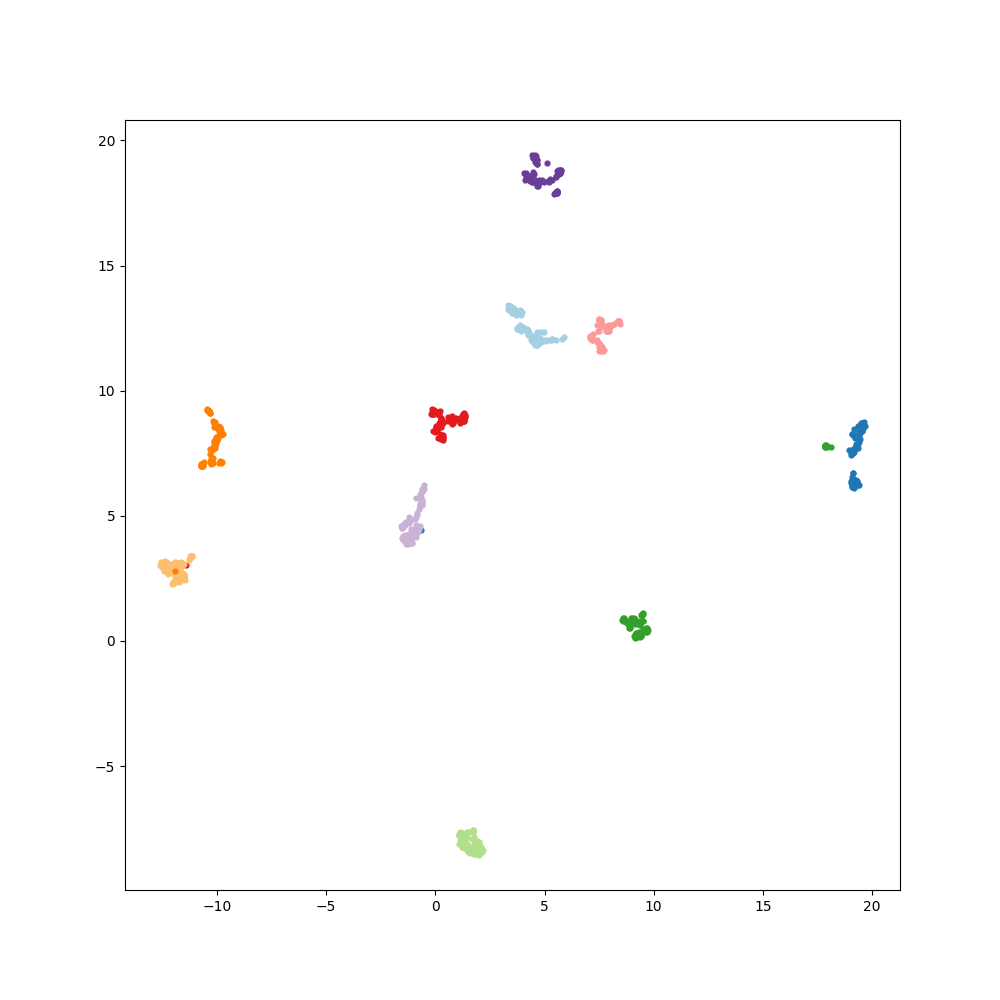}\hspace{4pt}}}
    
    \caption{UMAP clustering of individual topological data for \Anger emotion at different perplexities.}
    \label{fig:umap:angry}
\end{figure}

% \Disgust

\begin{figure}[!b]
    \centering
    \subfigure[\Disgust perplexity: 30 (full) \label{fig:umap:disgust_full_perp_30}]{{\hspace{4pt}\includegraphics[trim= 92pt 81pt 73pt 89pt, clip, height=1.40in, width=0.45\linewidth]{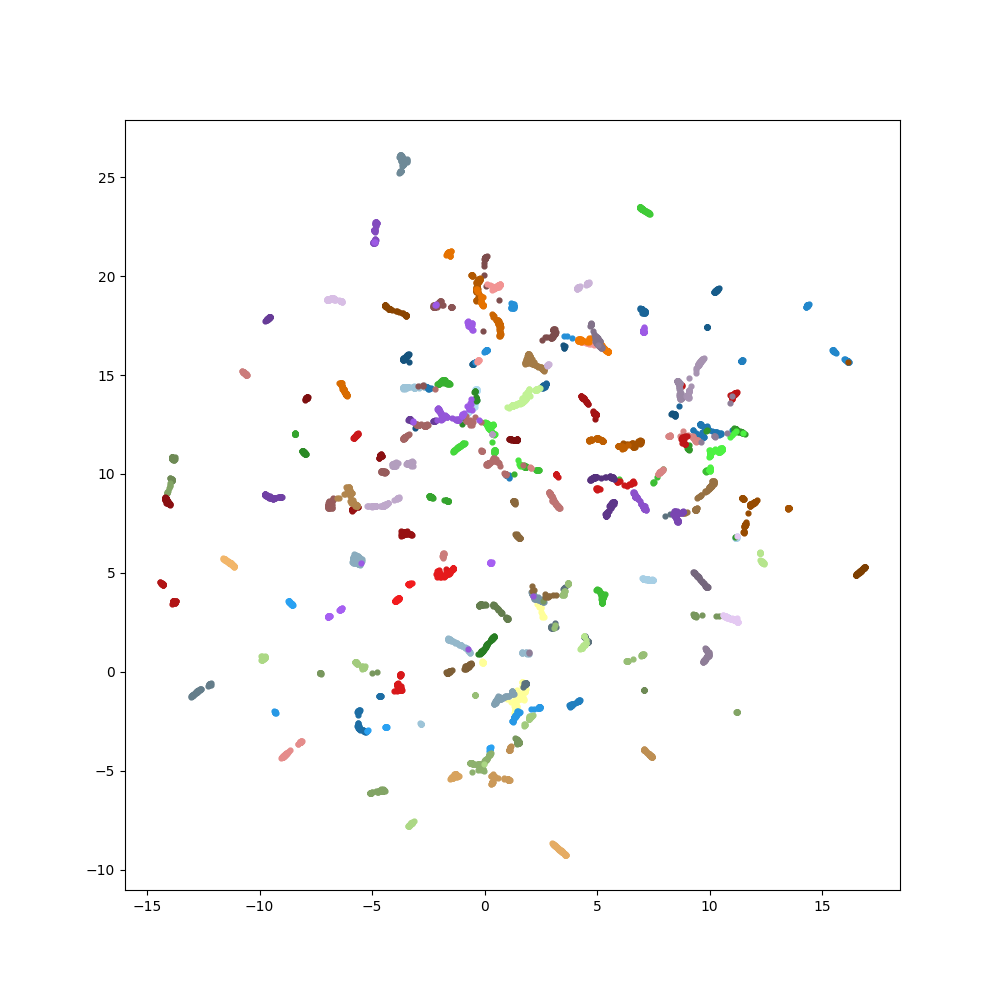}\hspace{4pt}}}
    \subfigure[\Disgust perplexity: 30 (10 subj.) \label{fig:umap:disgust_10_perp_30}]{{\hspace{4pt}\includegraphics[trim= 92pt 81pt 73pt 89pt, clip, height=1.40in, width=0.45\linewidth]{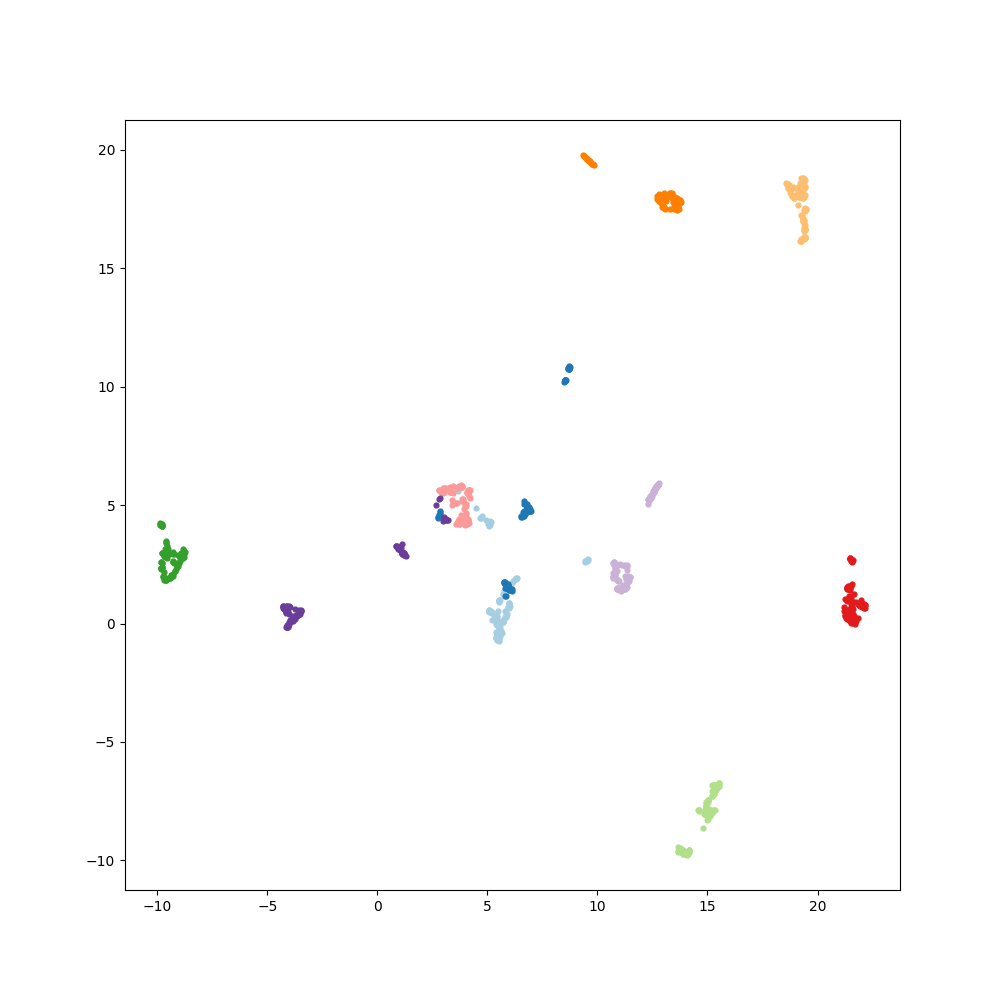}\hspace{4pt}}}
    
    \subfigure[\Disgust perplexity: 40 (full) \label{fig:umap:disgust_full_perp_40}]{{\hspace{4pt}\includegraphics[trim= 92pt 81pt 73pt 89pt, clip, height=1.40in, width=0.45\linewidth]{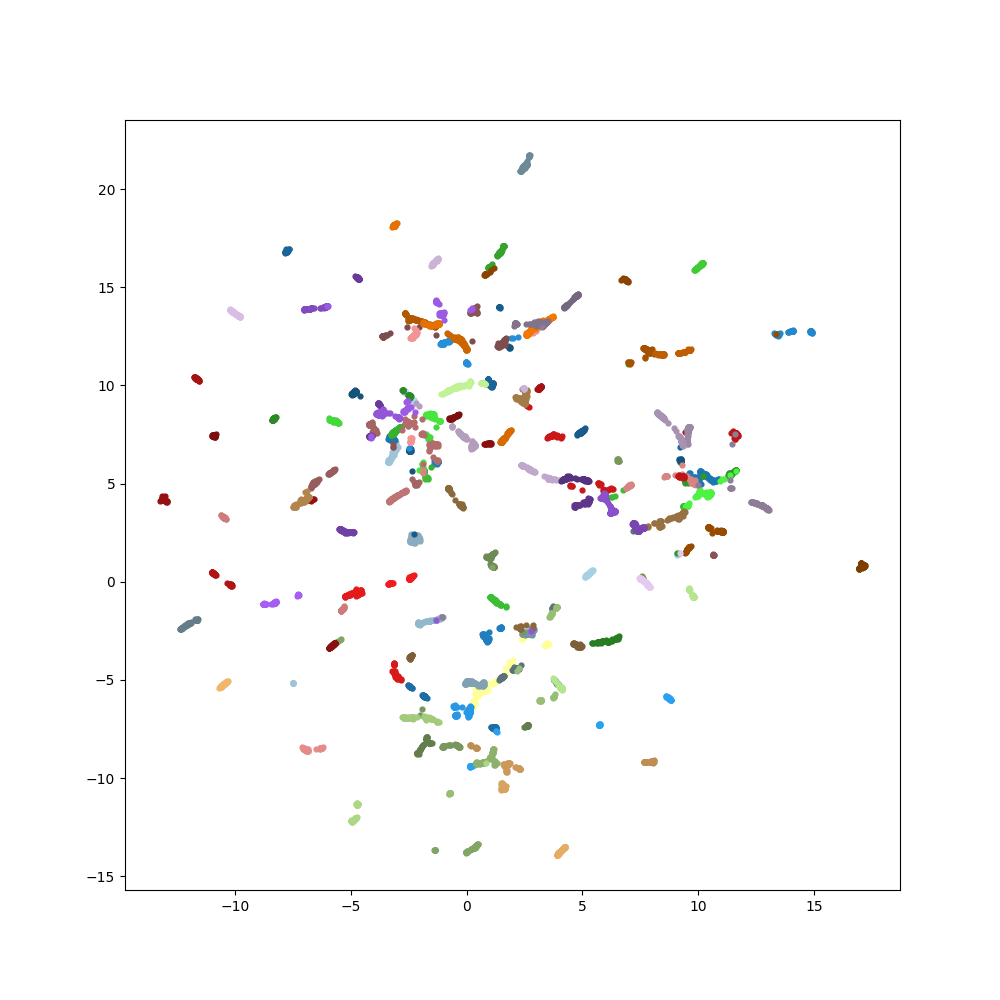}\hspace{4pt}}}
    \subfigure[\Disgust perplexity: 40 (10 subj.) \label{fig:umap:disgust_10_perp_40}]{{\hspace{4pt}\includegraphics[trim= 92pt 81pt 73pt 89pt, clip, height=1.40in, width=0.45\linewidth]{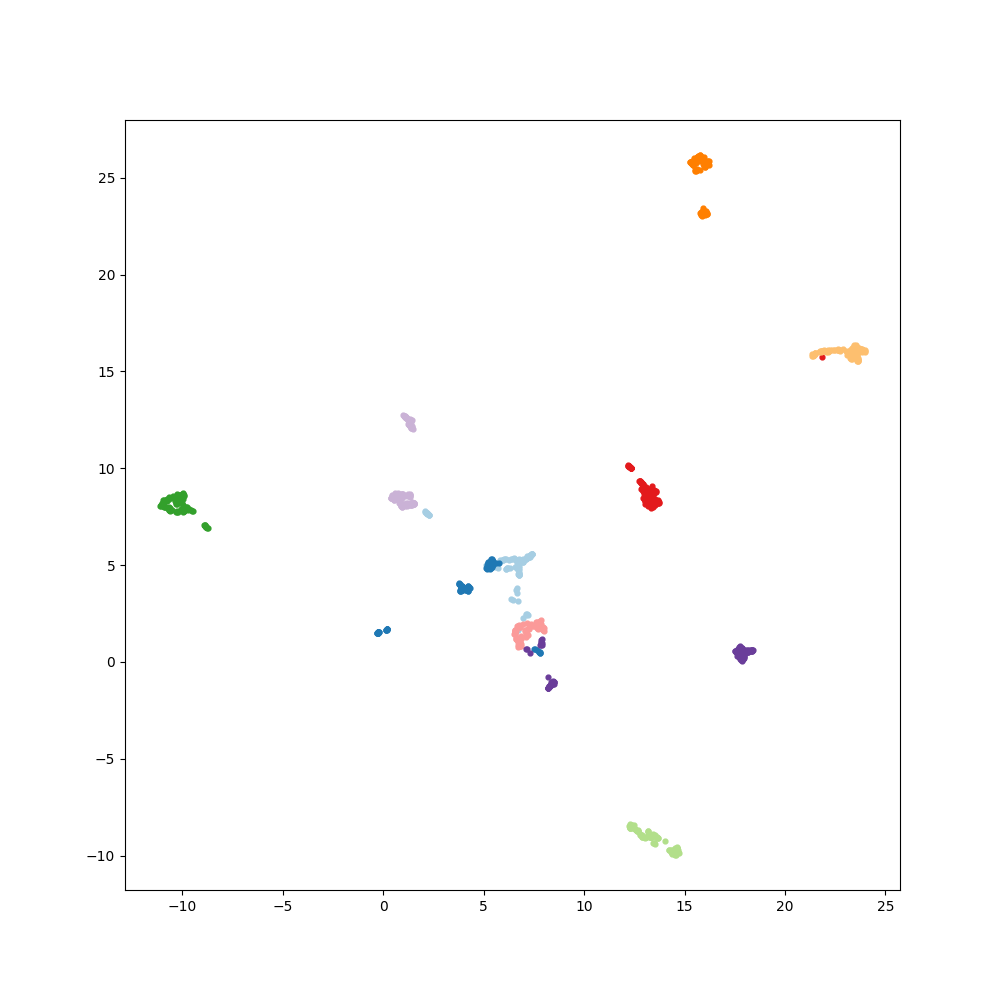}\hspace{4pt}}}
    
    \subfigure[\Disgust perplexity: 50 (full) \label{fig:umap:disgust_full_perp_50}]{{\hspace{4pt}\includegraphics[trim= 92pt 81pt 73pt 89pt, clip, height=1.40in, width=0.45\linewidth]{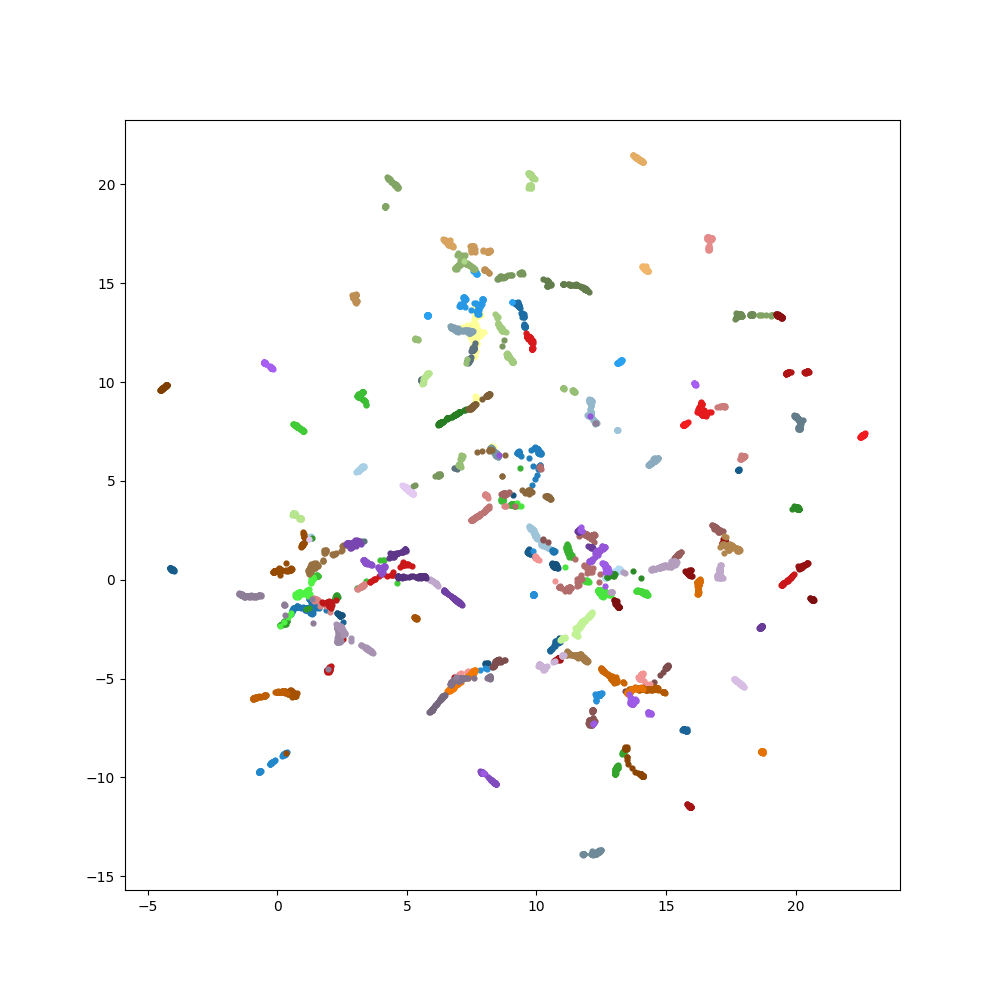}\hspace{4pt}}}
    \subfigure[\Disgust perplexity: 50 (10 subj.) \label{fig:umap:disgust_10_perp_50}]{{\hspace{4pt}\includegraphics[trim= 92pt 81pt 73pt 89pt, clip, height=1.40in, width=0.45\linewidth]{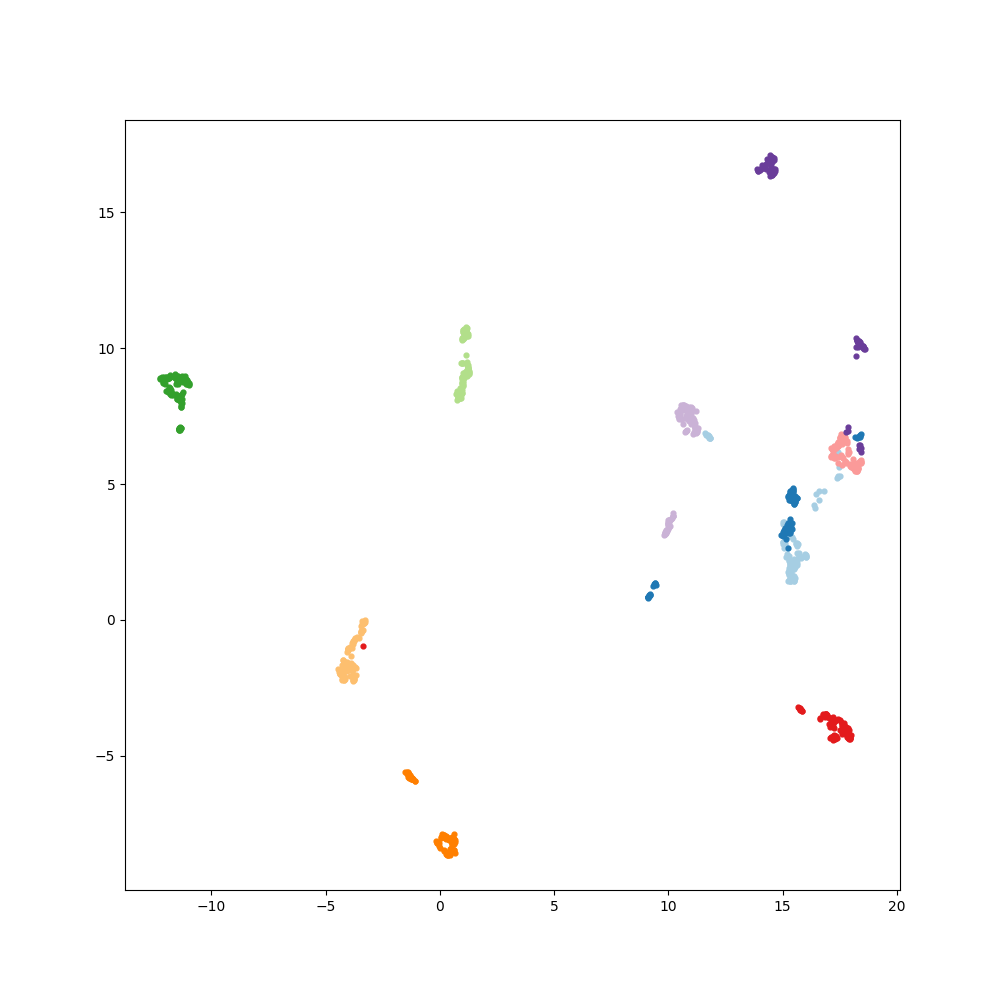}\hspace{4pt}}}
    
    \subfigure[\Disgust perplexity: 100 (full) \label{fig:umap:disgust_full_perp_100}]{{\hspace{4pt}\includegraphics[trim= 92pt 81pt 73pt 89pt, clip, height=1.40in, width=0.45\linewidth]{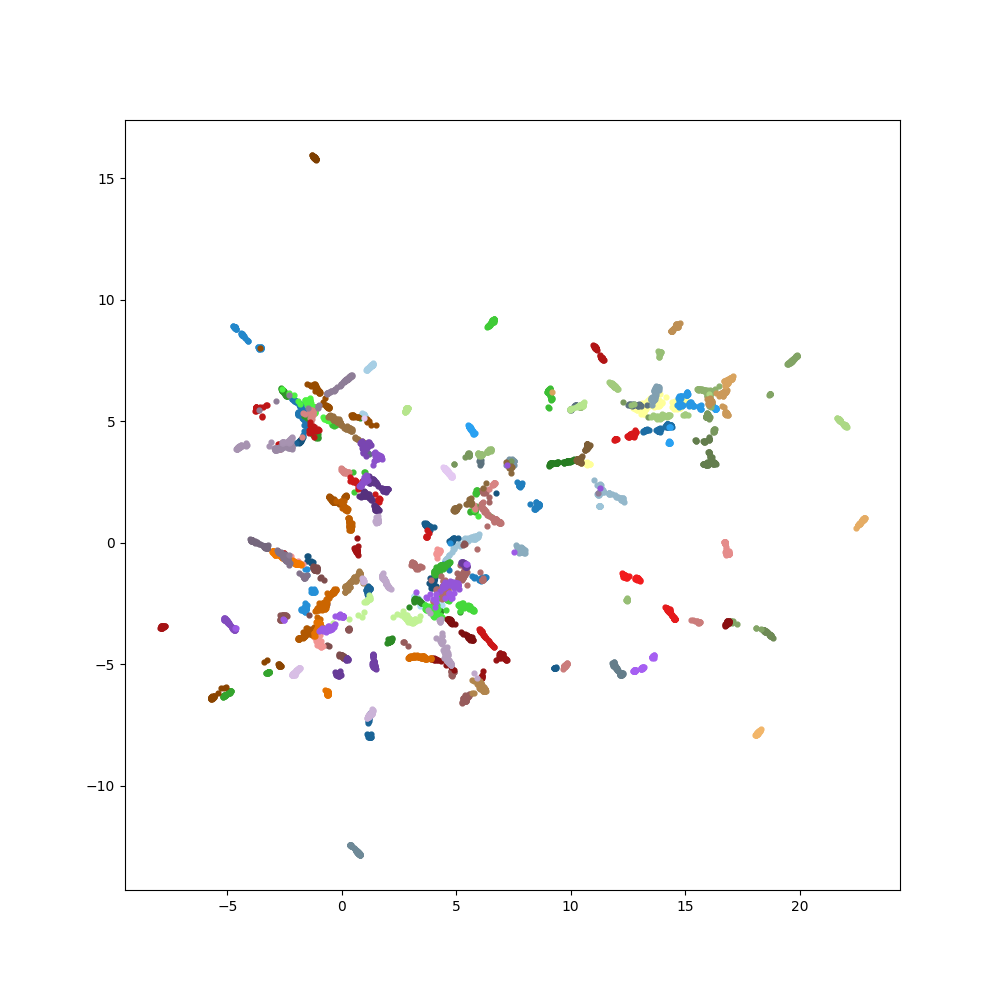}\hspace{4pt}}}
    \subfigure[\Disgust perplexity: 100 (10 subj.) \label{fig:umap:disgust_10_perp_100}]{{\hspace{4pt}\includegraphics[trim= 92pt 81pt 73pt 89pt, clip, height=1.40in, width=0.45\linewidth]{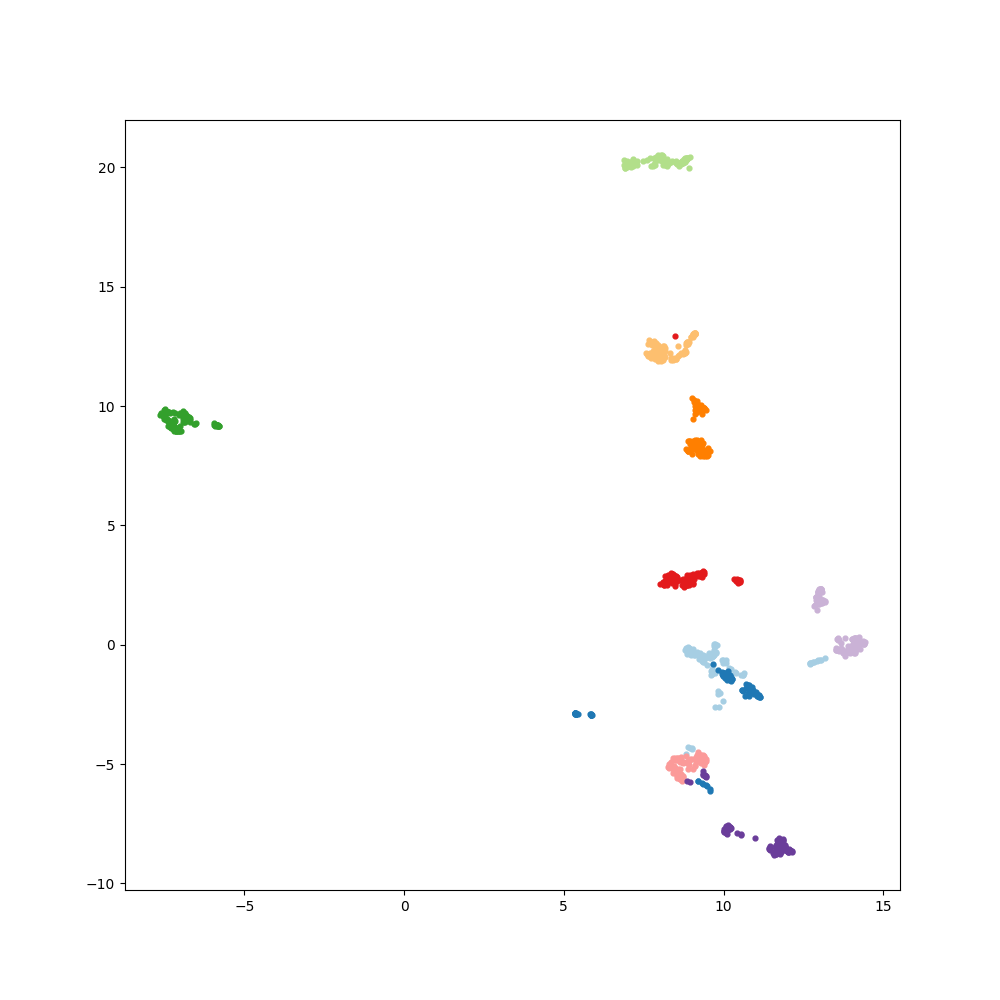}\hspace{4pt}}}
    
    \caption{UMAP clustering of individual topological data for \Disgust emotion at different perplexities.}
    \label{fig:umap:disgust}
\end{figure}

% \Fear

\begin{figure}[!b]
    \centering
    \subfigure[\Fear perplexity: 30 (full) \label{fig:umap:fear_full_perp_30}]{{\hspace{4pt}\includegraphics[trim= 92pt 81pt 73pt 89pt, clip, height=1.40in, width=0.45\linewidth]{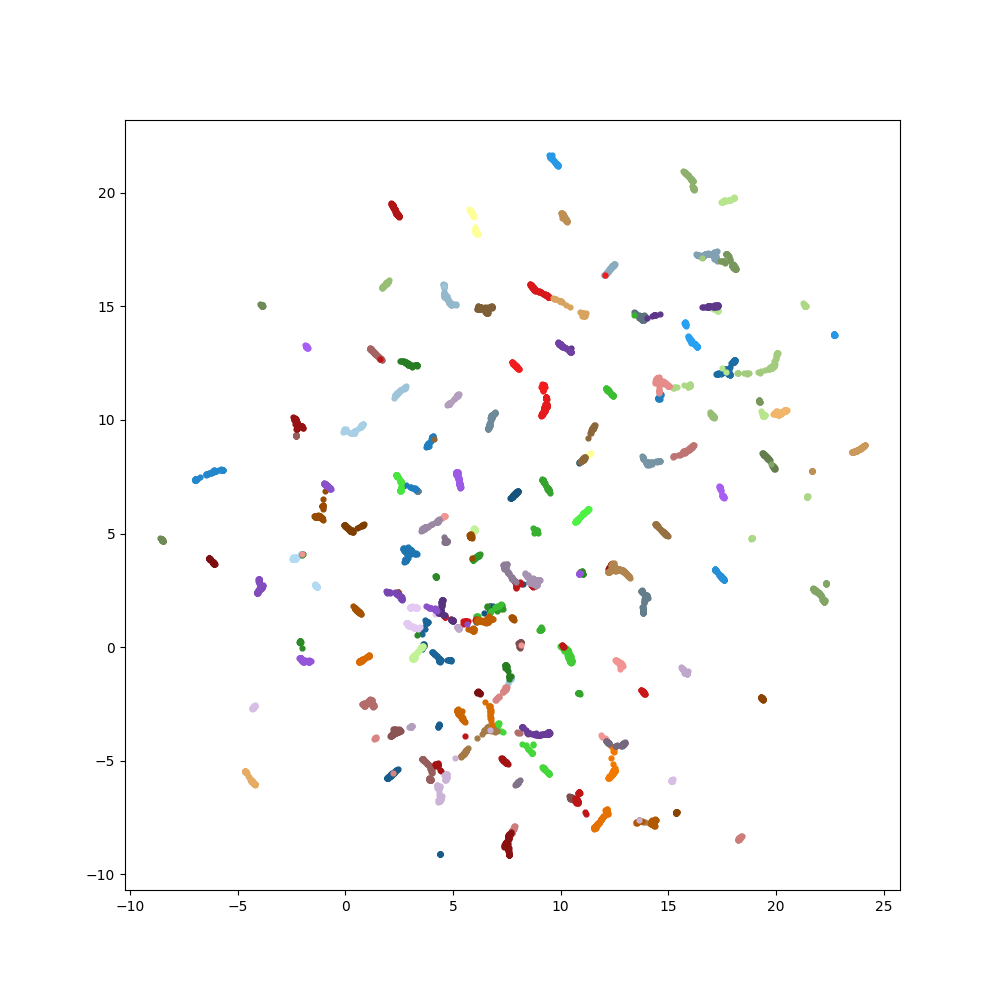}\hspace{4pt}}}
    \subfigure[\Fear perplexity: 30 (10 subj.) \label{fig:umap:fear_10_perp_30}]{{\hspace{4pt}\includegraphics[trim= 92pt 81pt 73pt 89pt, clip, height=1.40in, width=0.45\linewidth]{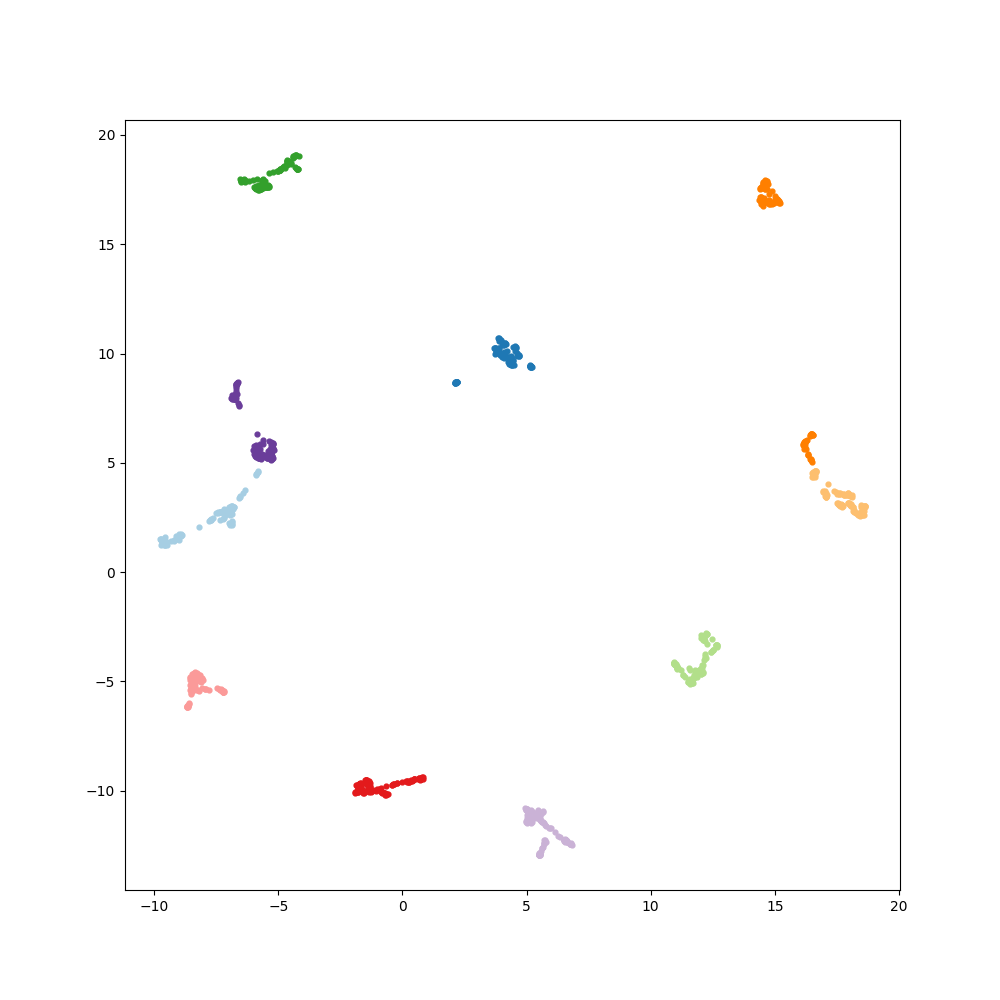}\hspace{4pt}}}
    
    \subfigure[\Fear perplexity: 40 (full) \label{fig:umap:fear_full_perp_40}]{{\hspace{4pt}\includegraphics[trim= 92pt 81pt 73pt 89pt, clip, height=1.40in, width=0.45\linewidth]{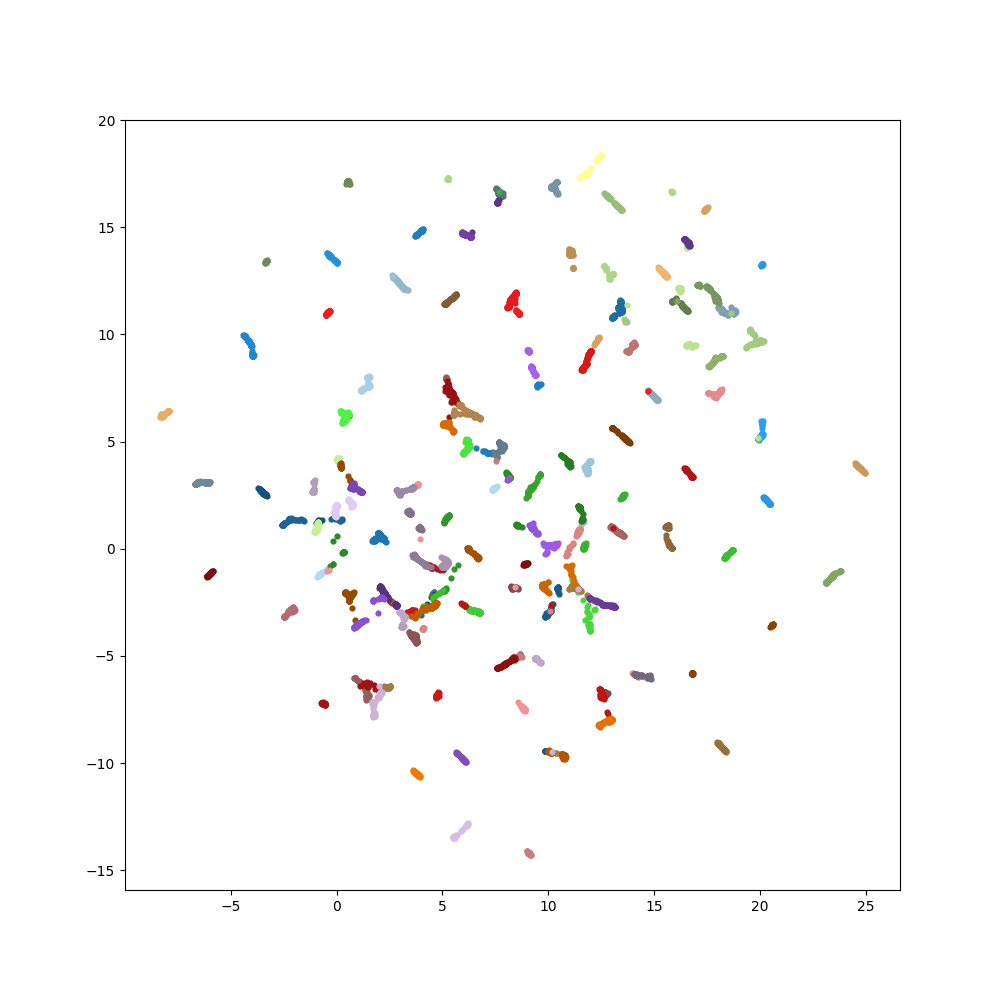}\hspace{4pt}}}
    \subfigure[\Fear perplexity: 40 (10 subj.) \label{fig:umap:fear_10_perp_40}]{{\hspace{4pt}\includegraphics[trim= 92pt 81pt 73pt 89pt, clip, height=1.40in, width=0.45\linewidth]{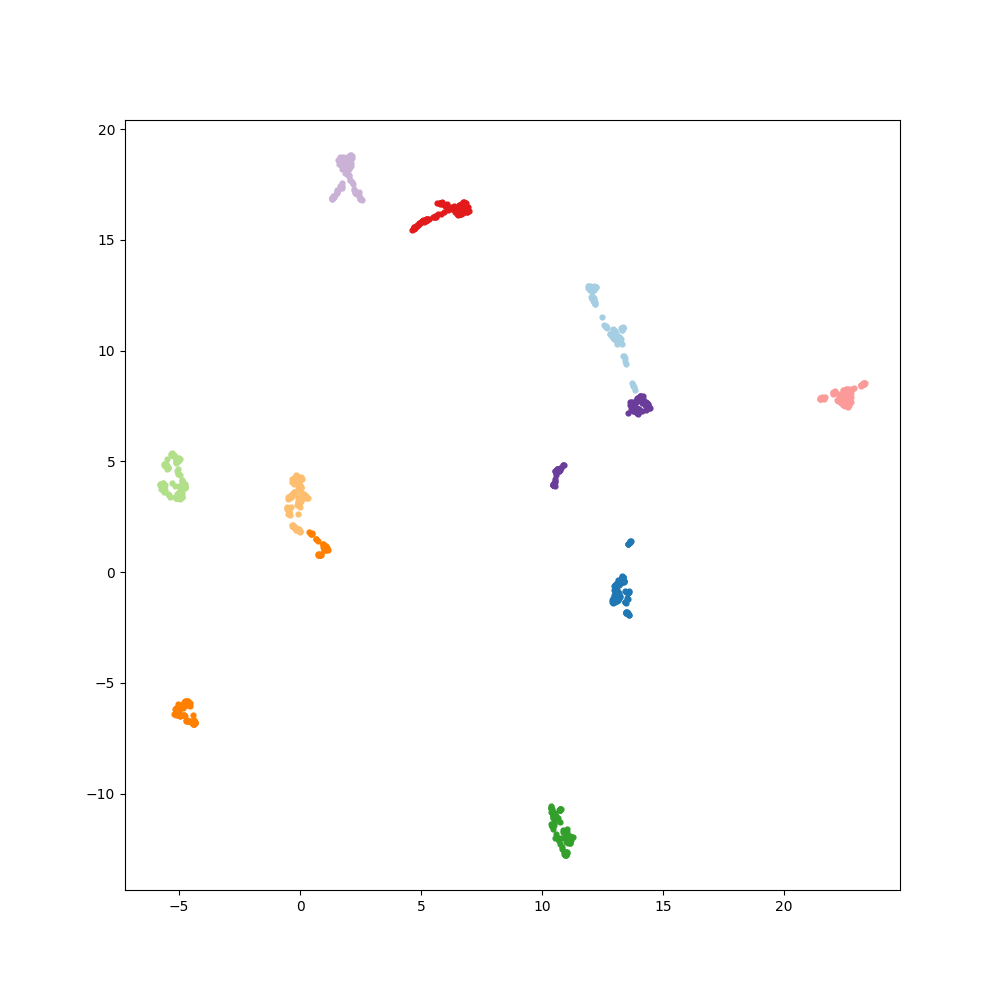}\hspace{4pt}}}
    
    \subfigure[\Fear perplexity: 50 (full) \label{fig:umap:fear_full_perp_50}]{{\hspace{4pt}\includegraphics[trim= 92pt 81pt 73pt 89pt, clip, height=1.40in, width=0.45\linewidth]{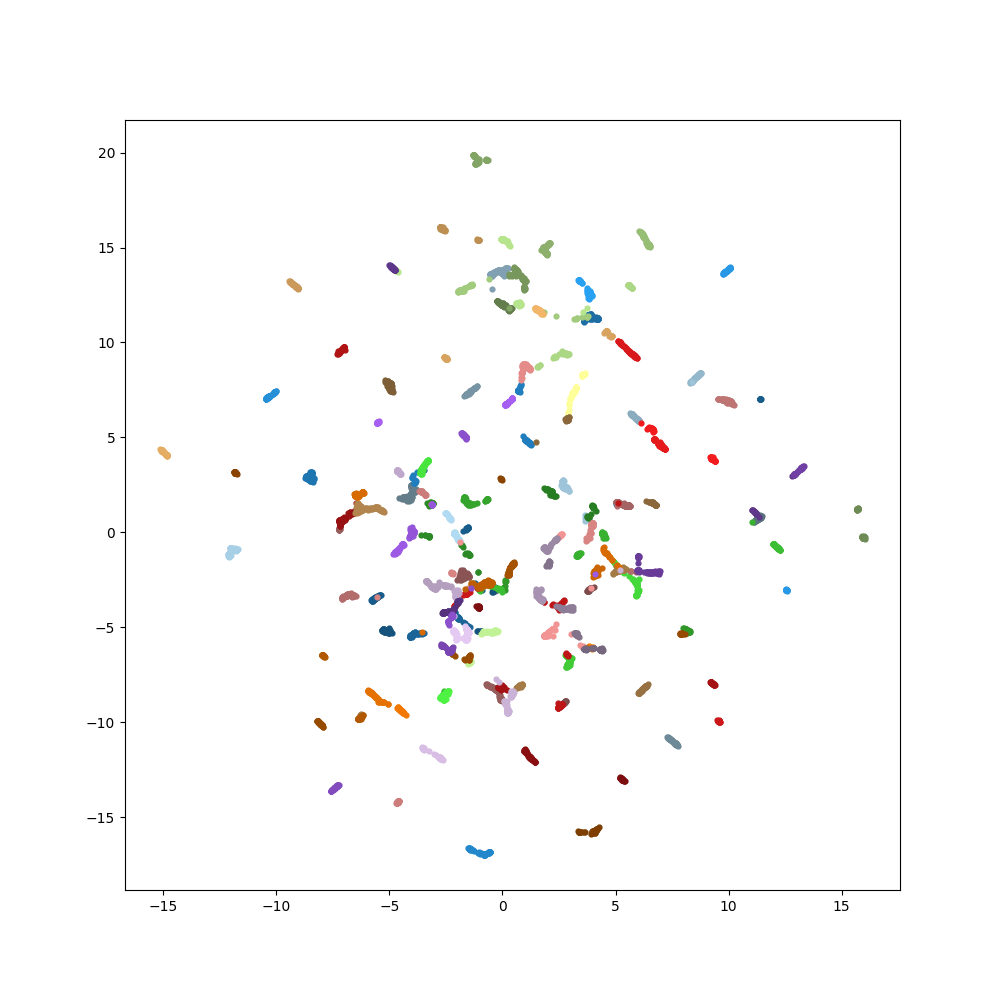}\hspace{4pt}}}
    \subfigure[\Fear perplexity: 50 (10 subj.) \label{fig:umap:fear_10_perp_50}]{{\hspace{4pt}\includegraphics[trim= 92pt 81pt 73pt 89pt, clip, height=1.40in, width=0.45\linewidth]{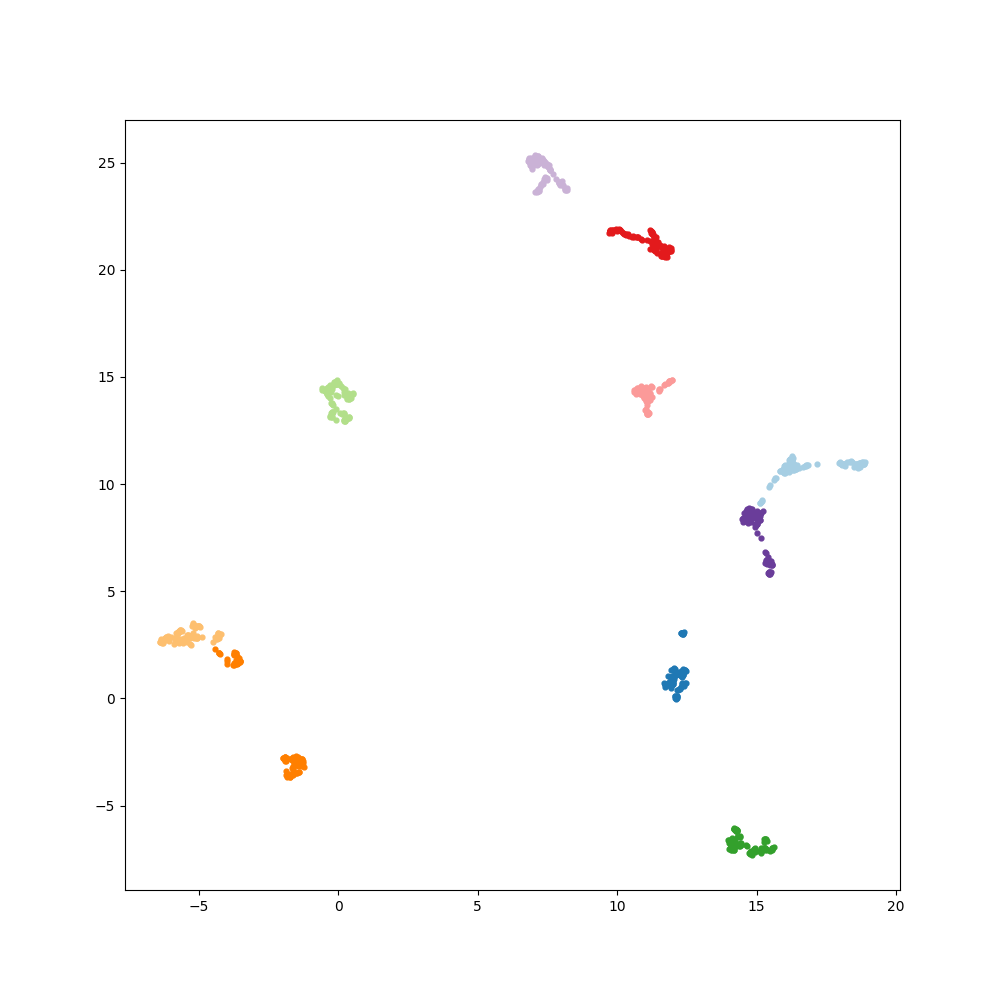}\hspace{4pt}}}
    
    \subfigure[\Fear perplexity: 100 (full) \label{fig:umap:fear_full_perp_100}]{{\hspace{4pt}\includegraphics[trim= 92pt 81pt 73pt 89pt, clip, height=1.40in, width=0.45\linewidth]{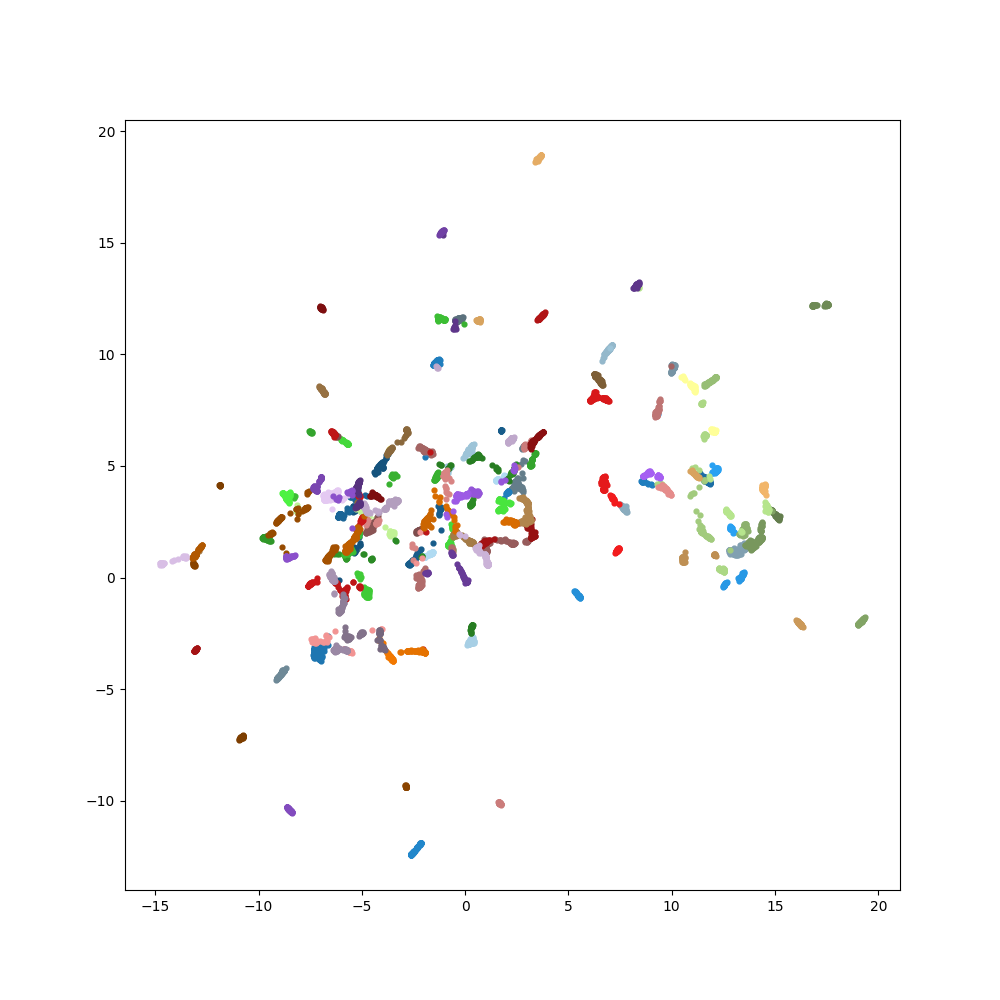}\hspace{4pt}}}
    \subfigure[\Fear perplexity: 100 (10 subj.) \label{fig:umap:fear_10_perp_100}]{{\hspace{4pt}\includegraphics[trim= 92pt 81pt 73pt 89pt, clip, height=1.40in, width=0.45\linewidth]{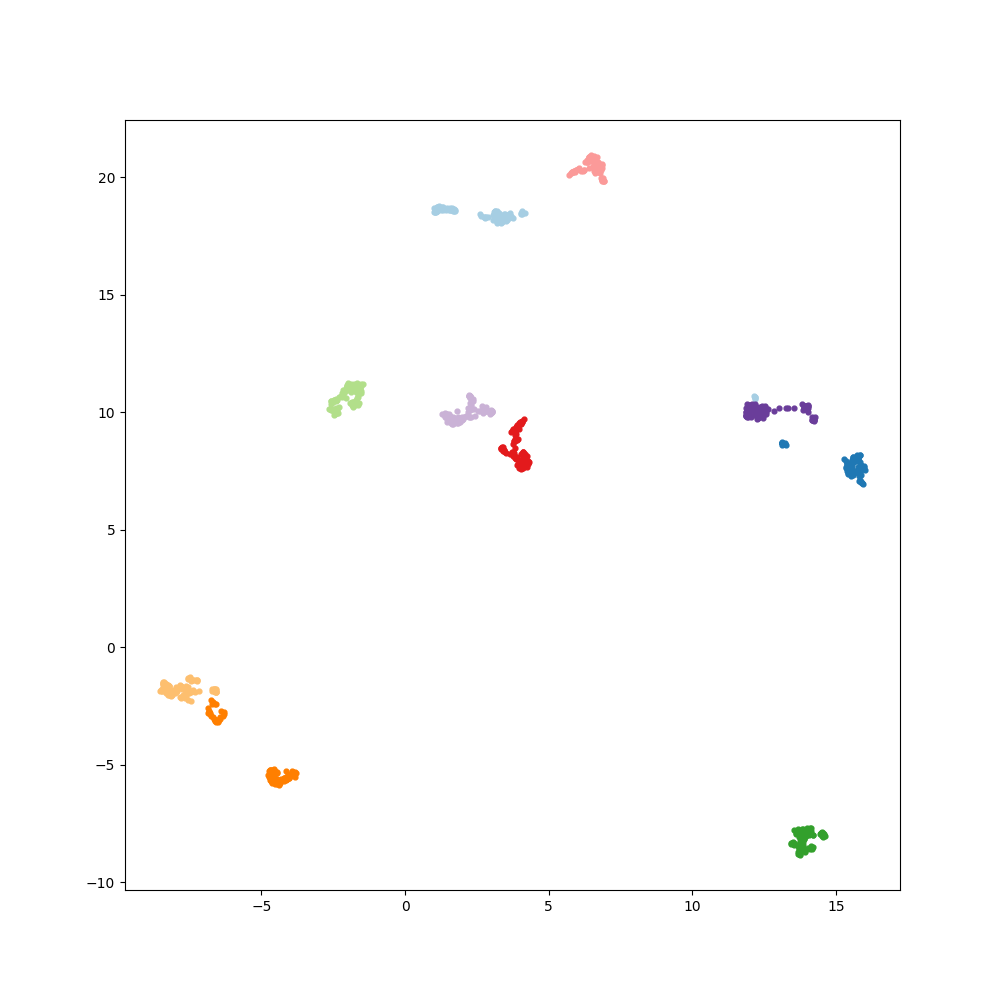}\hspace{4pt}}}
    
    \caption{UMAP clustering of individual topological data for \Fear emotion at different perplexities.}
    \label{fig:umap:fear}
\end{figure}

% Happy

\begin{figure}[!b]
    \centering
    \subfigure[\Happiness perp.: 30 (full) \label{fig:umap:happy_full_perp_30}]{{\hspace{4pt}\includegraphics[trim= 92pt 81pt 73pt 89pt, clip, height=1.40in, width=0.45\linewidth]{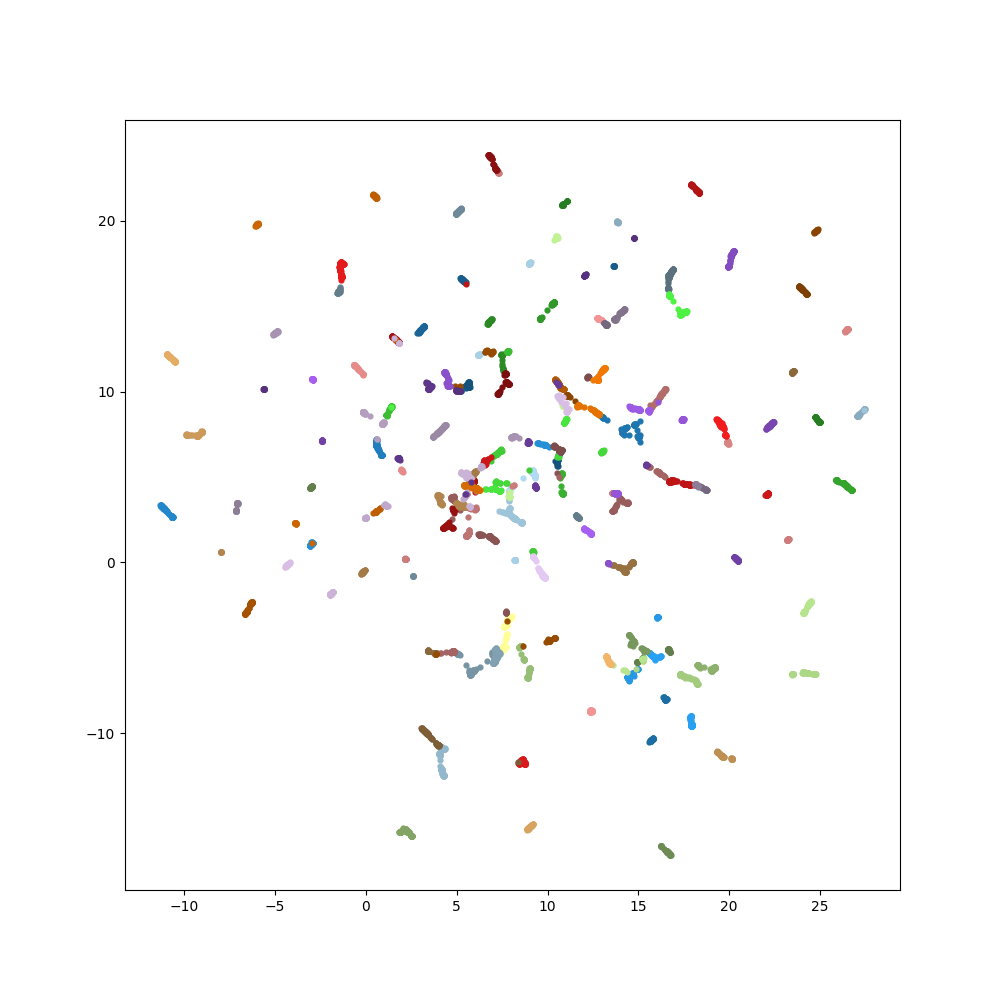}\hspace{4pt}}}
    \subfigure[\Happiness perp.: 30 (10 subj.) \label{fig:umap:happy_10_perp_30}]{{\hspace{4pt}\includegraphics[trim= 92pt 81pt 73pt 89pt, clip, height=1.40in, width=0.45\linewidth]{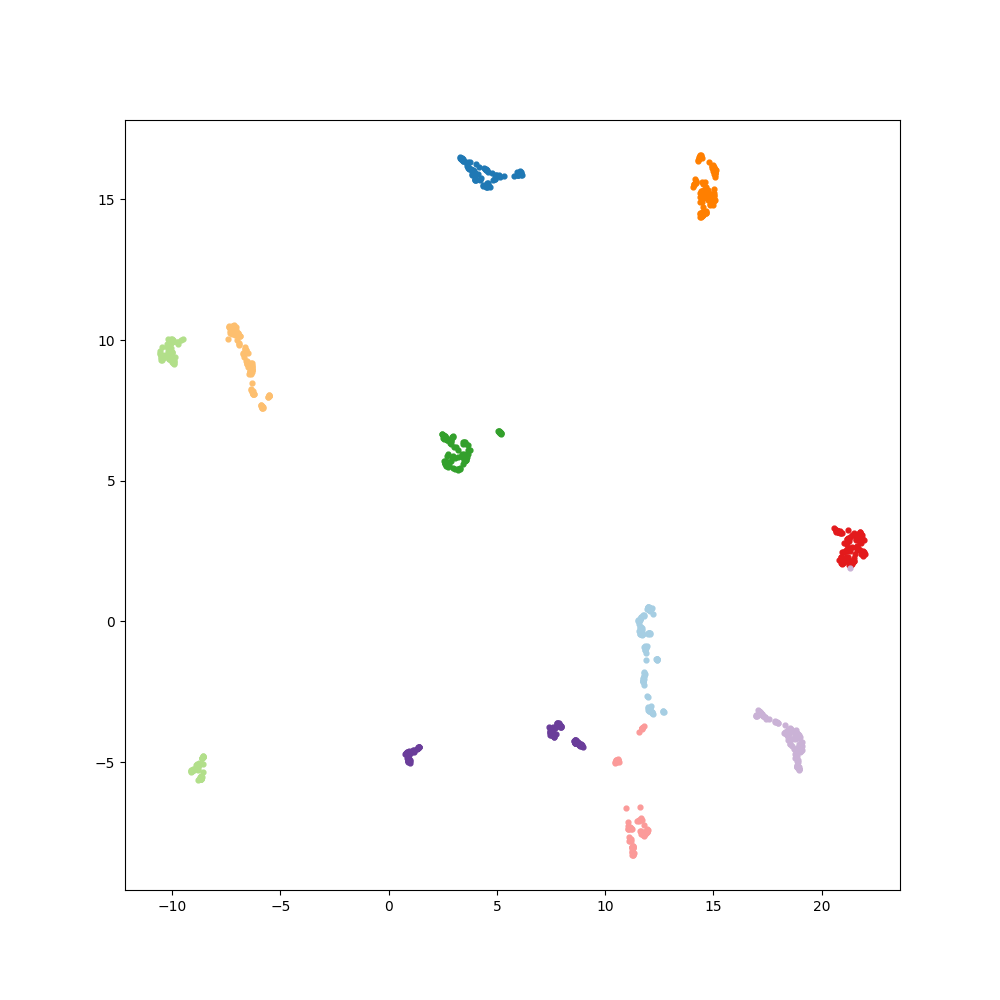}\hspace{4pt}}}
    
    \subfigure[\Happiness perp.: 40 (full) \label{fig:umap:happy_full_perp_40}]{{\hspace{4pt}\includegraphics[trim= 92pt 81pt 73pt 89pt, clip, height=1.40in, width=0.45\linewidth]{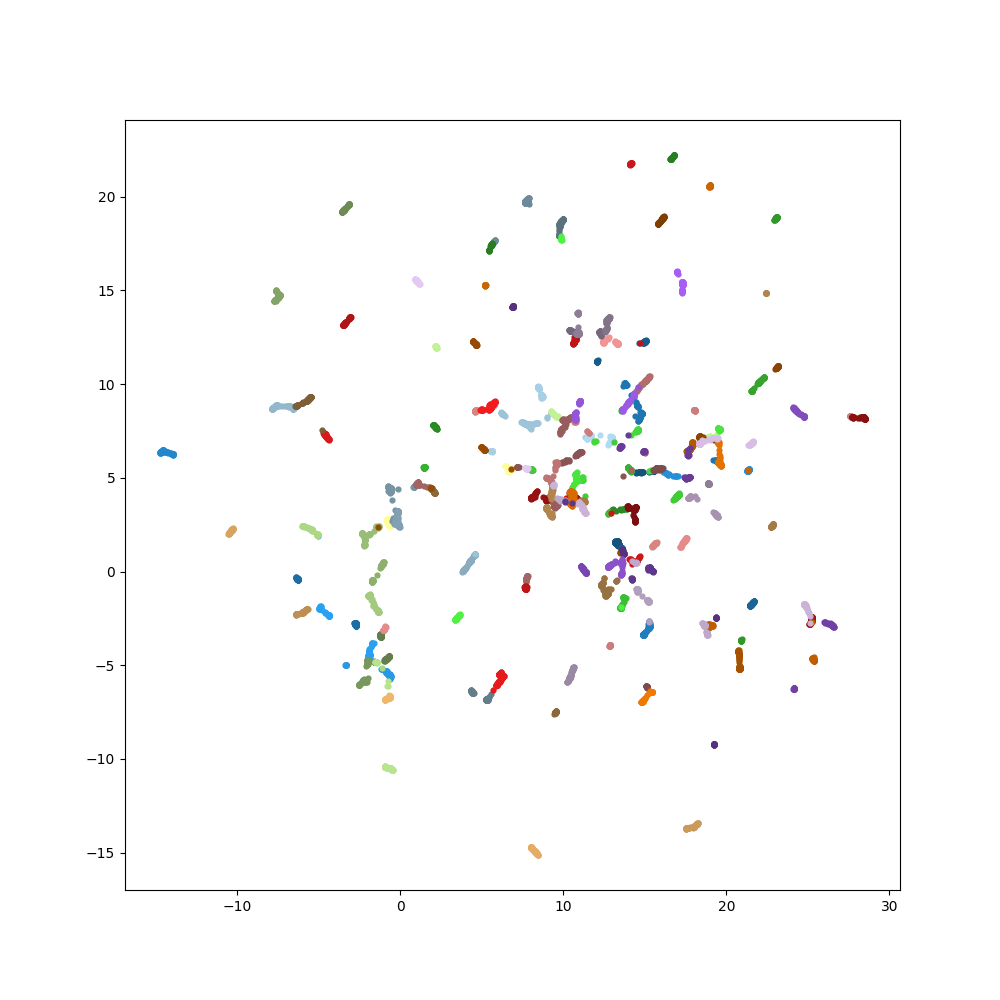}\hspace{4pt}}}
    \subfigure[\Happiness perp.: 40 (10 subj.) \label{fig:umap:happy_10_perp_40}]{{\hspace{4pt}\includegraphics[trim= 92pt 81pt 73pt 89pt, clip, height=1.40in, width=0.45\linewidth]{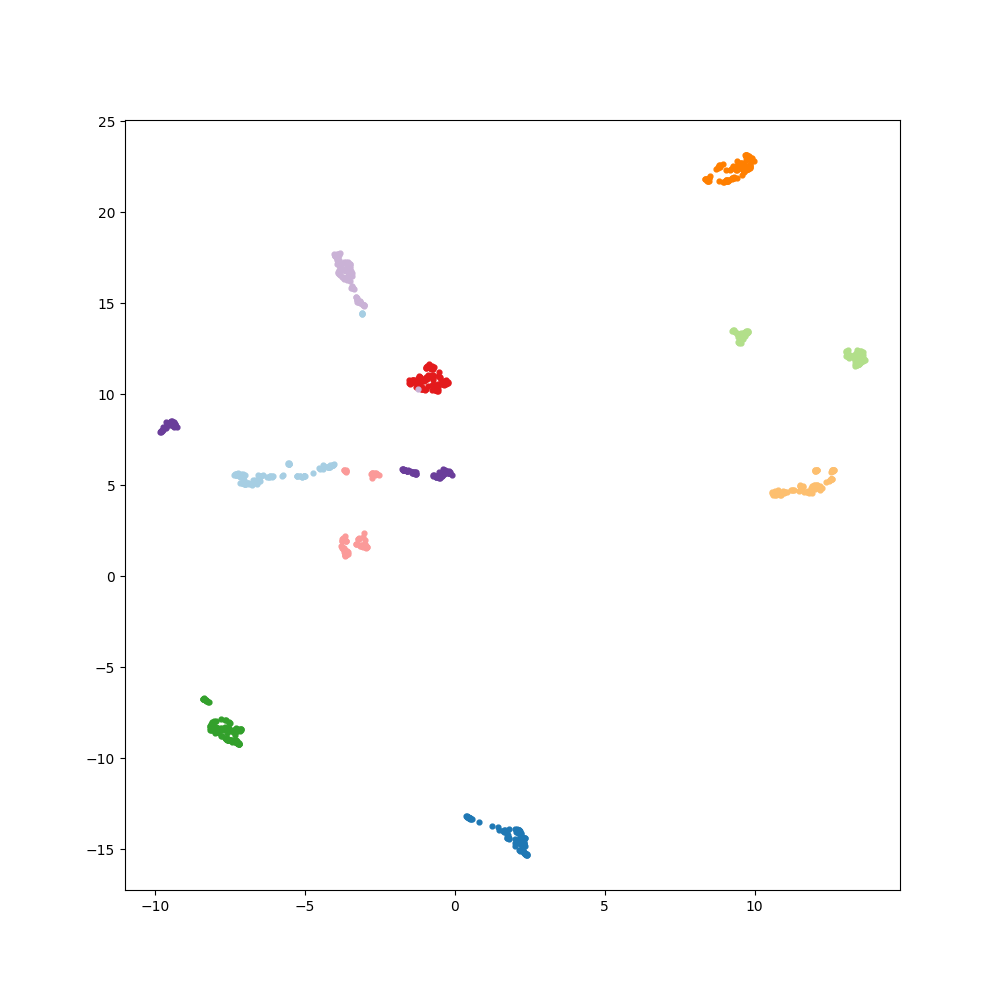}\hspace{4pt}}}
    
    \subfigure[\Happiness perp.: 50 (full) \label{fig:umap:happy_full_perp_50}]{{\hspace{4pt}\includegraphics[trim= 92pt 81pt 73pt 89pt, clip, height=1.40in, width=0.45\linewidth]{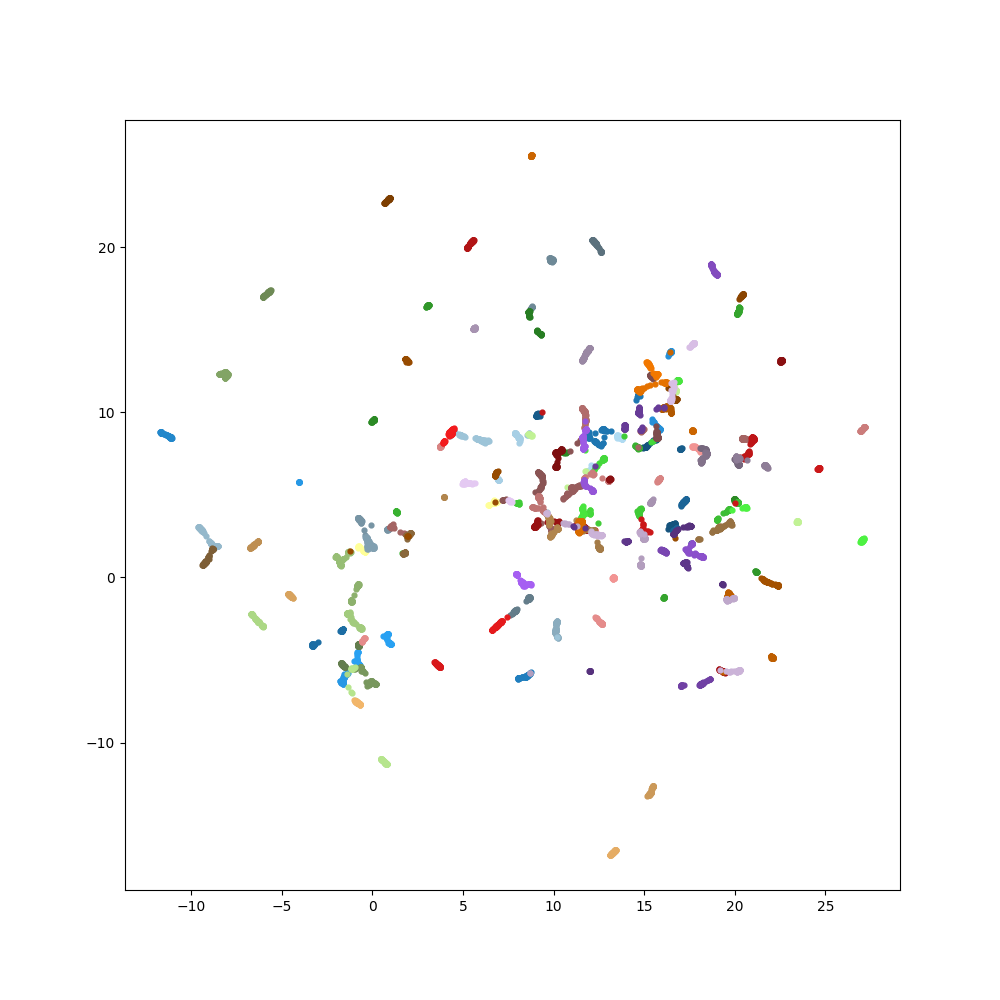}\hspace{4pt}}}
    \subfigure[\Happiness perp.: 50 (10 subj.) \label{fig:umap:happy_10_perp_50}]{{\hspace{4pt}\includegraphics[trim= 92pt 81pt 73pt 89pt, clip, height=1.40in, width=0.45\linewidth]{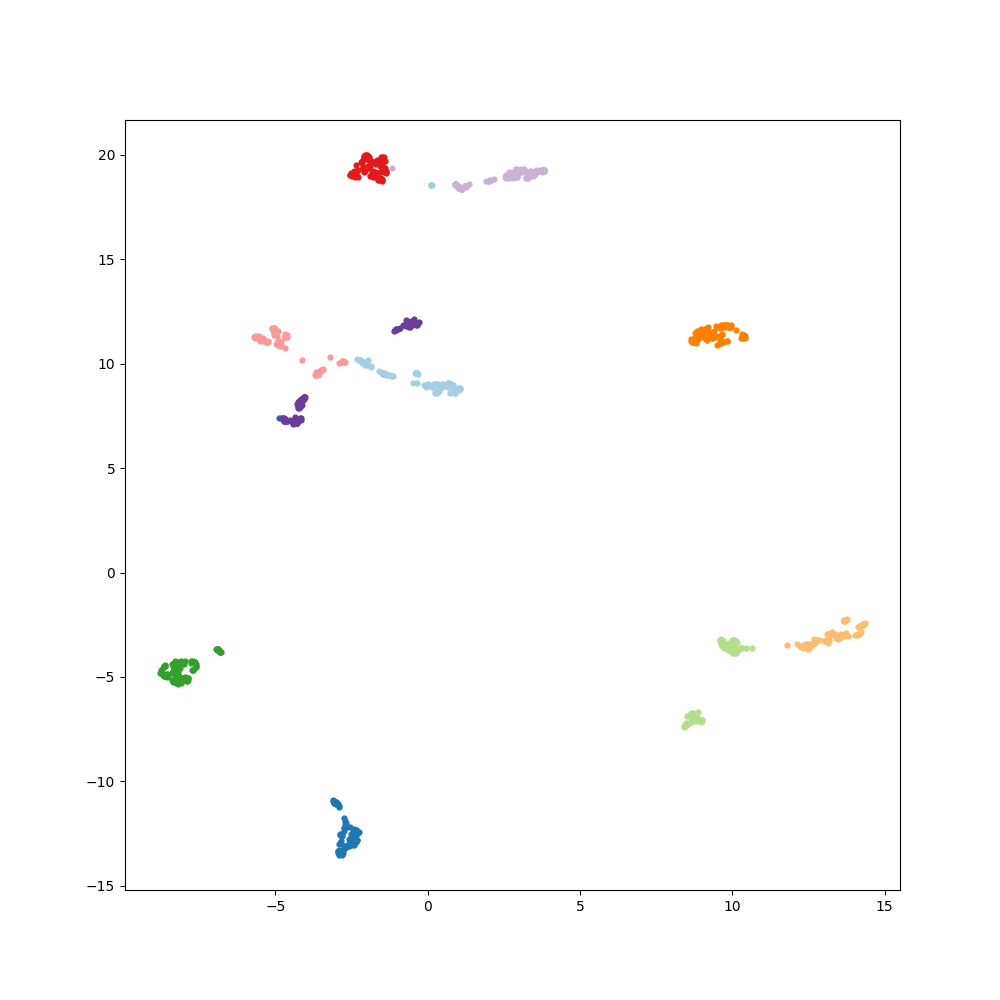}\hspace{4pt}}}
    
    \subfigure[\Happiness perp.: 100 (full) \label{fig:umap:happy_full_perp_100}]{{\hspace{4pt}\includegraphics[trim= 92pt 81pt 73pt 89pt, clip, height=1.40in, width=0.45\linewidth]{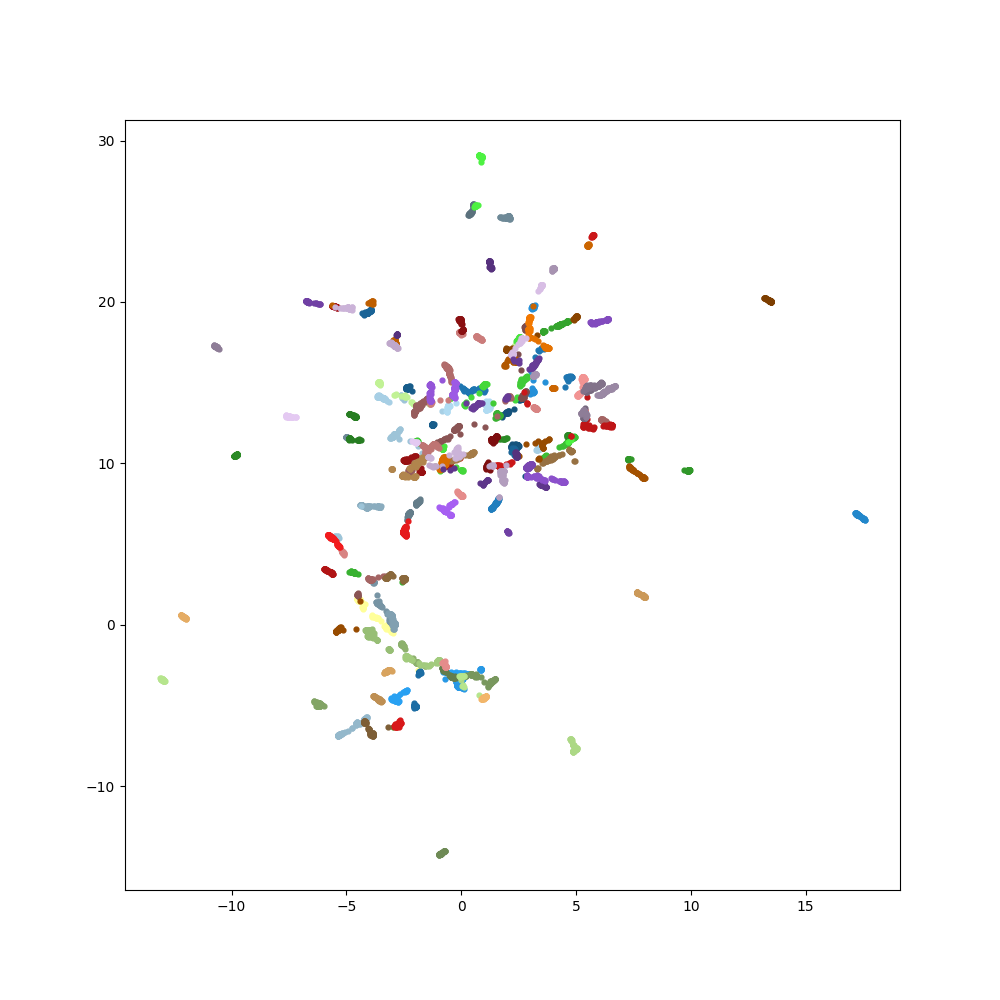}\hspace{4pt}}}
    \subfigure[\Happiness perp.: 100 (10 subj.) \label{fig:umap:happy_10_perp_100}]{{\hspace{4pt}\includegraphics[trim= 92pt 81pt 73pt 89pt, clip, height=1.40in, width=0.45\linewidth]{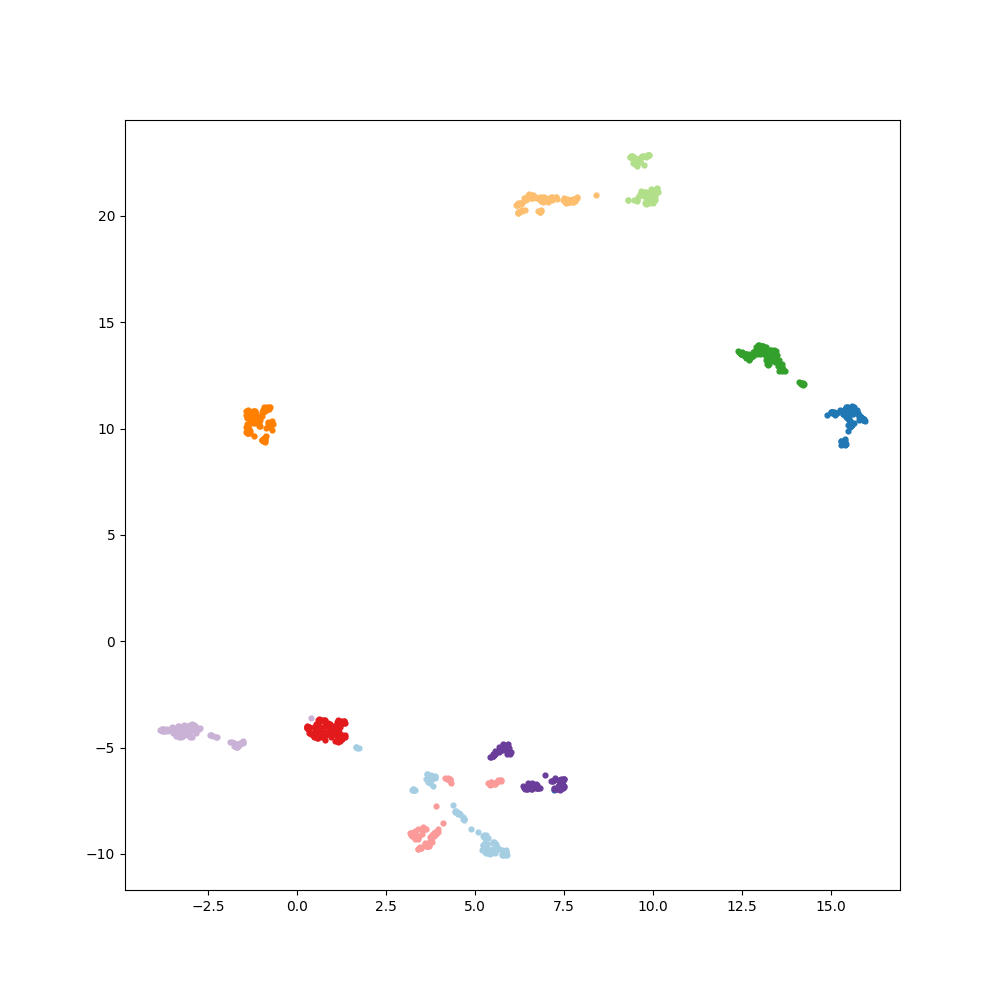}\hspace{4pt}}}
    
    \caption{UMAP clustering of individual topological data for \Happiness emotion at different perplexities.}
    \label{fig:umap:happy}
\end{figure}

% \Sadness

\begin{figure}[!b]
    \centering
    \subfigure[\Sadness perplexity: 30 (full) \label{fig:umap:sad_full_perp_30}]{{\hspace{4pt}\includegraphics[trim= 92pt 81pt 73pt 89pt, clip, height=1.40in, width=0.45\linewidth]{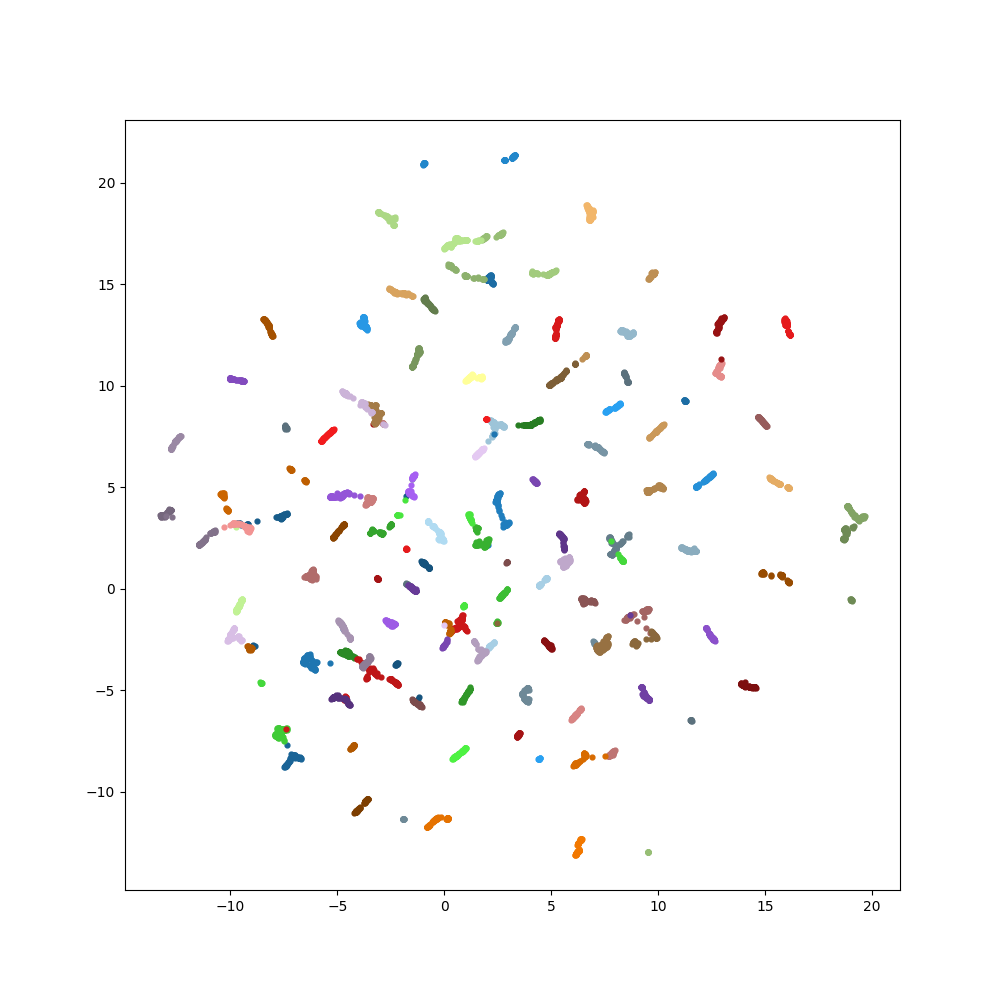}\hspace{4pt}}}
    \subfigure[\Sadness perplexity: 30 (10 subj.) \label{fig:umap:sad_10_perp_30}]{{\hspace{4pt}\includegraphics[trim= 92pt 81pt 73pt 89pt, clip, height=1.40in, width=0.45\linewidth]{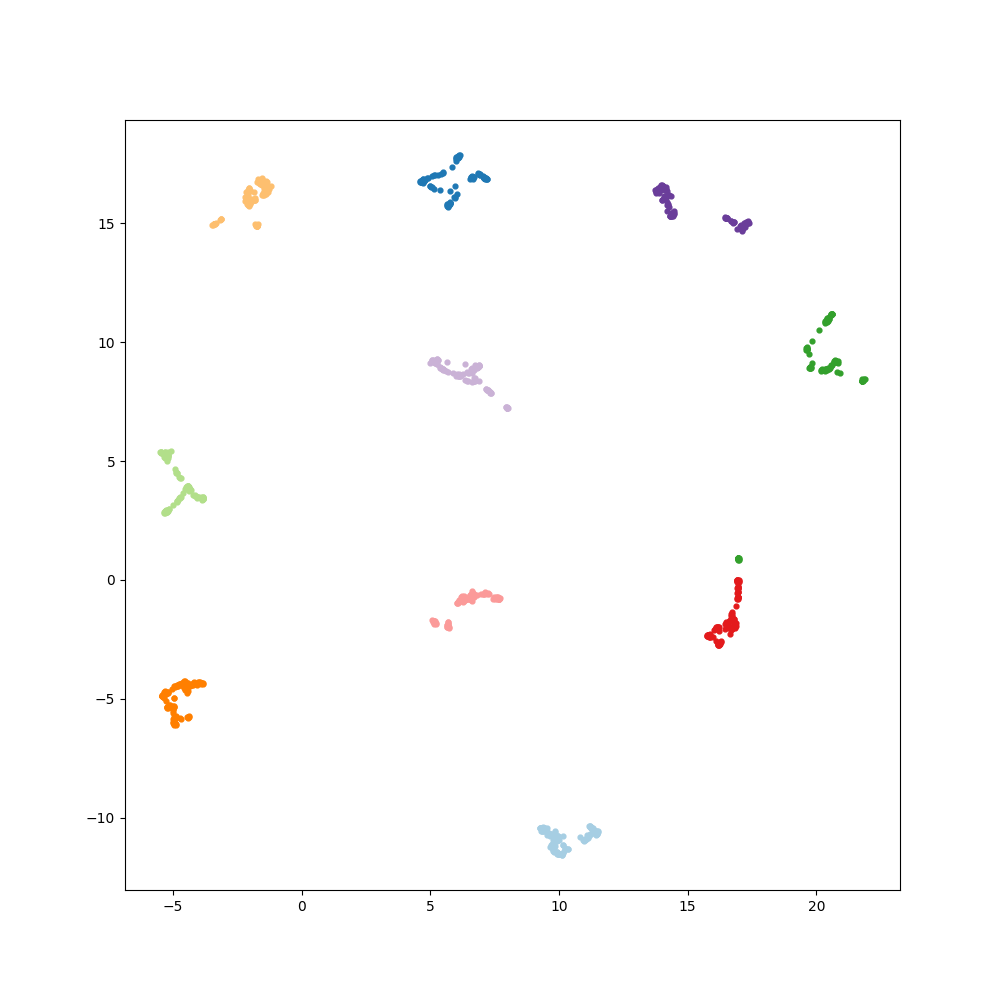}\hspace{4pt}}}
    
    \subfigure[\Sadness perplexity: 40 (full) \label{fig:umap:sad_full_perp_40}]{{\hspace{4pt}\includegraphics[trim= 92pt 81pt 73pt 89pt, clip, height=1.40in, width=0.45\linewidth]{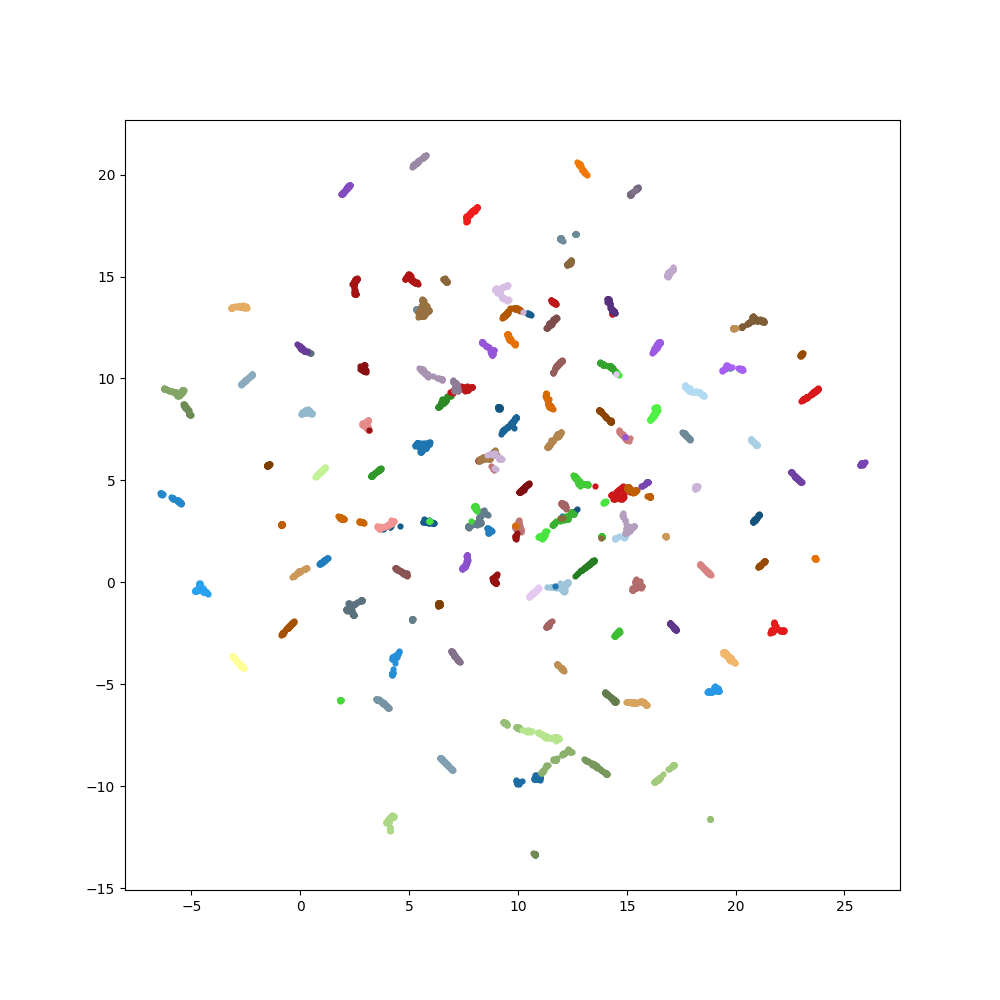}\hspace{4pt}}}
    \subfigure[\Sadness perplexity: 40 (10 subj.) \label{fig:umap:sad_10_perp_40}]{{\hspace{4pt}\includegraphics[trim= 92pt 81pt 73pt 89pt, clip, height=1.40in, width=0.45\linewidth]{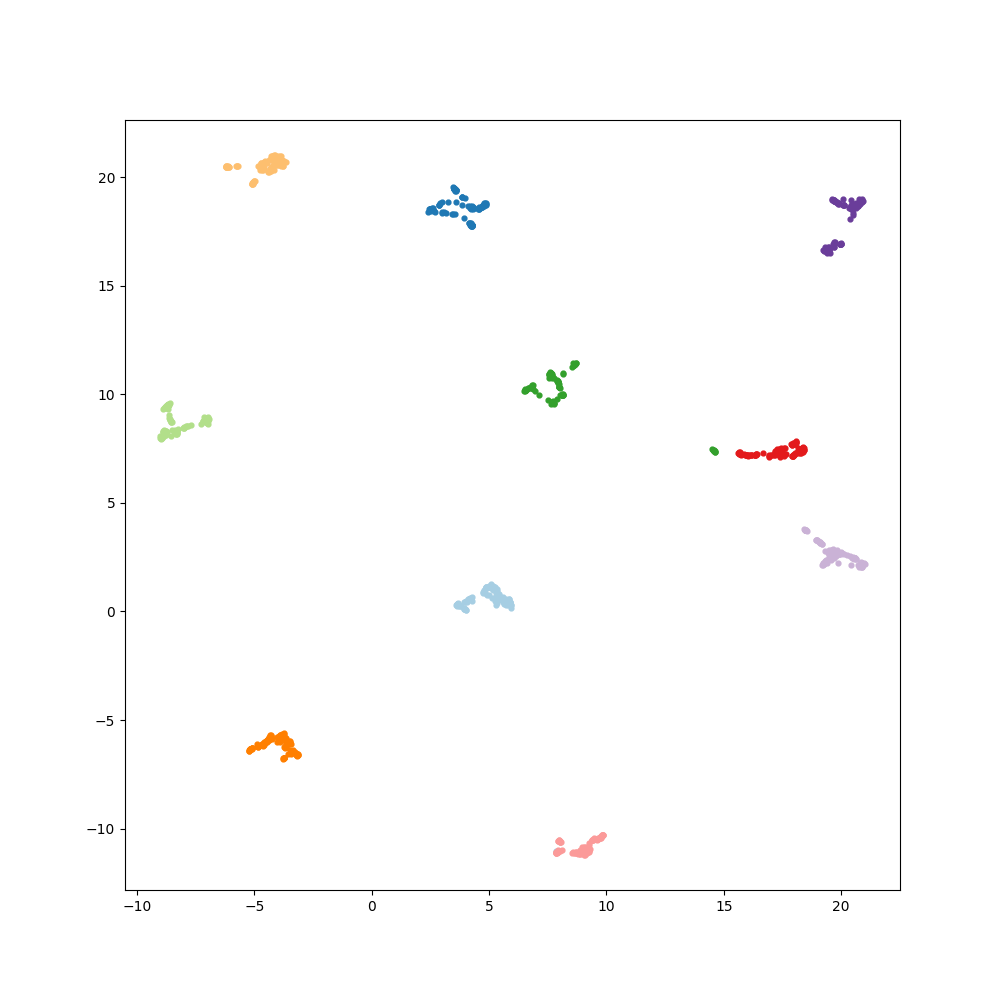}\hspace{4pt}}}
    
    \subfigure[\Sadness perplexity: 50 (full) \label{fig:umap:sad_full_perp_50}]{{\hspace{4pt}\includegraphics[trim= 92pt 81pt 73pt 89pt, clip, height=1.40in, width=0.45\linewidth]{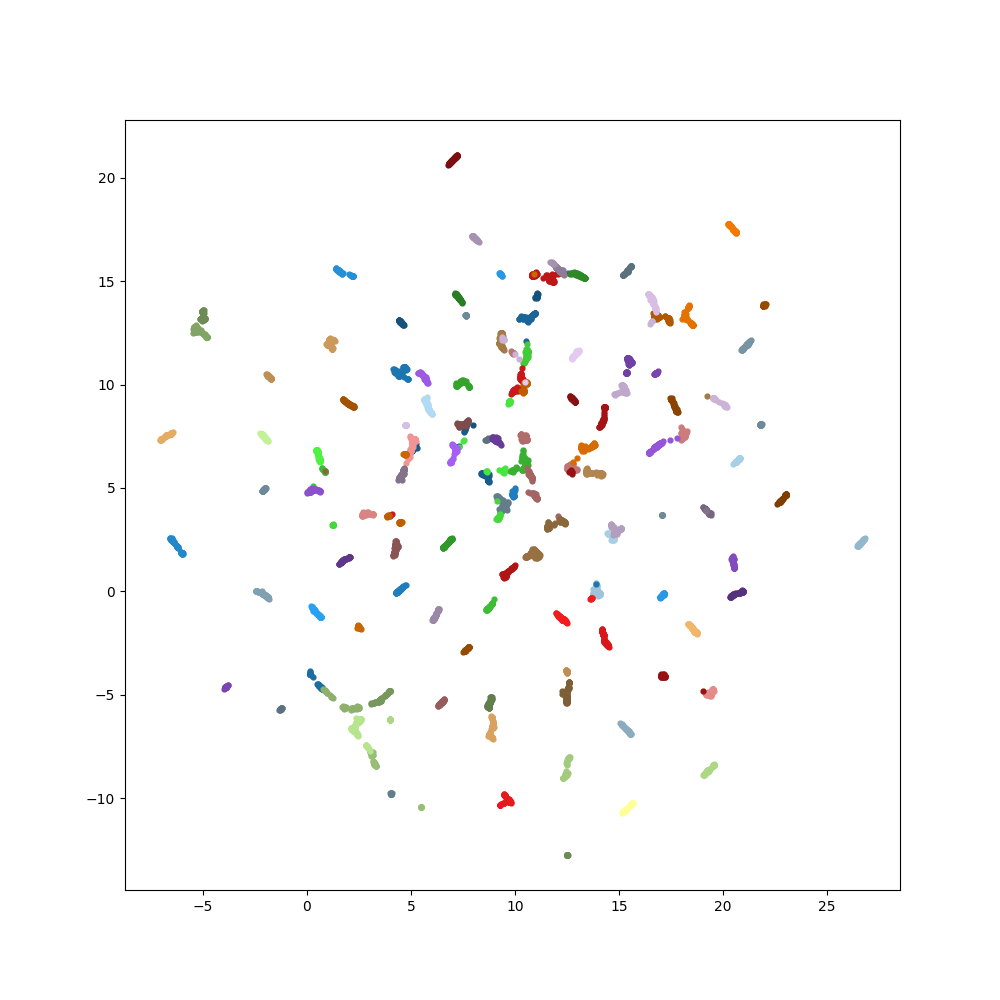}\hspace{4pt}}}
    \subfigure[\Sadness perplexity: 50 (10 subj.) \label{fig:umap:sad_10_perp_50}]{{\hspace{4pt}\includegraphics[trim= 92pt 81pt 73pt 89pt, clip, height=1.40in, width=0.45\linewidth]{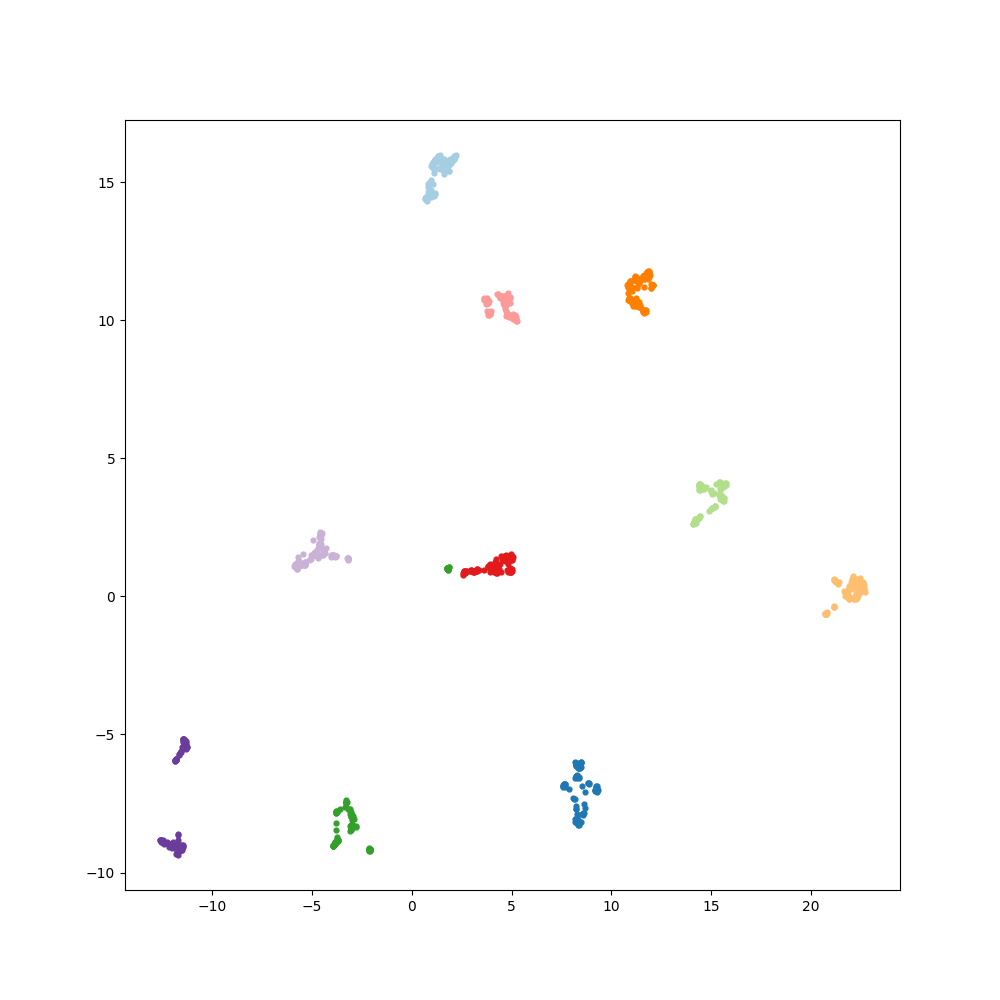}\hspace{4pt}}}
    
    \subfigure[\Sadness perplexity: 100 (full) \label{fig:umap:sad_full_perp_100}]{{\hspace{4pt}\includegraphics[trim= 92pt 81pt 73pt 89pt, clip, height=1.40in, width=0.45\linewidth]{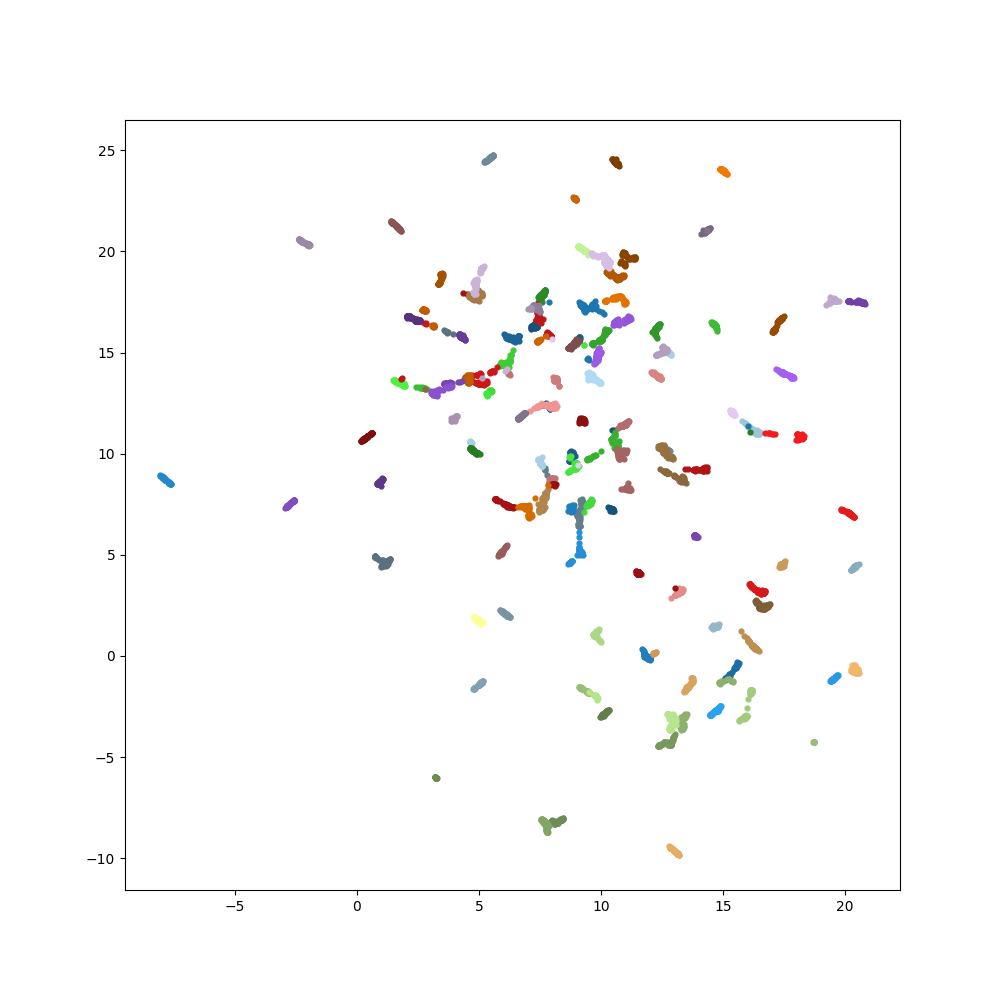}\hspace{4pt}}}
    \subfigure[\Sadness perplexity: 100 (10 subj.) \label{fig:umap:sad_10_perp_100}]{{\hspace{4pt}\includegraphics[trim= 92pt 81pt 73pt 89pt, clip, height=1.40in, width=0.45\linewidth]{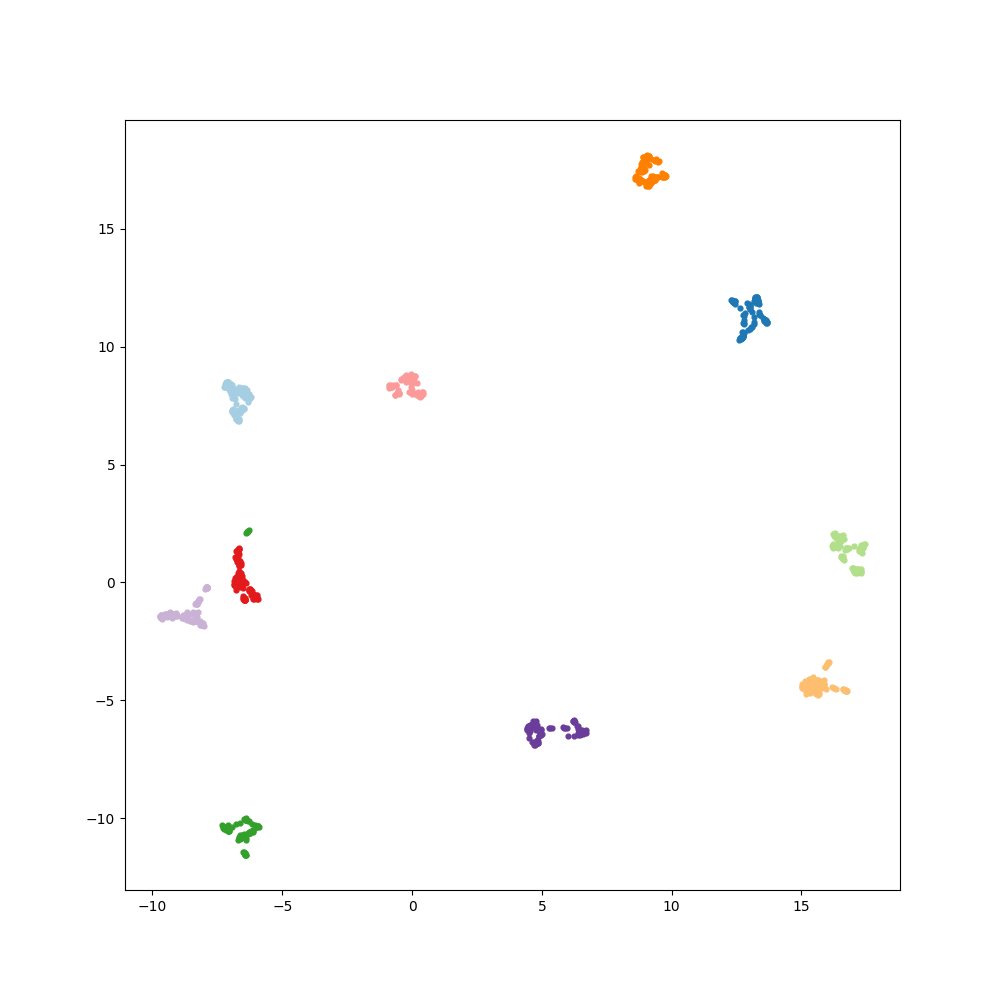}\hspace{4pt}}}
    
    \caption{UMAP clustering of individual topological data for \Sadness emotion at different perplexities.}
    \label{fig:umap:sad}
\end{figure}

% \Surprise

\begin{figure}[!b]
    \centering
    \subfigure[\Surprise perplexity: 30 (full) \label{fig:umap:surprise_full_perp_30}]{{\hspace{4pt}\includegraphics[trim= 92pt 81pt 73pt 89pt, clip, height=1.40in, width=0.45\linewidth]{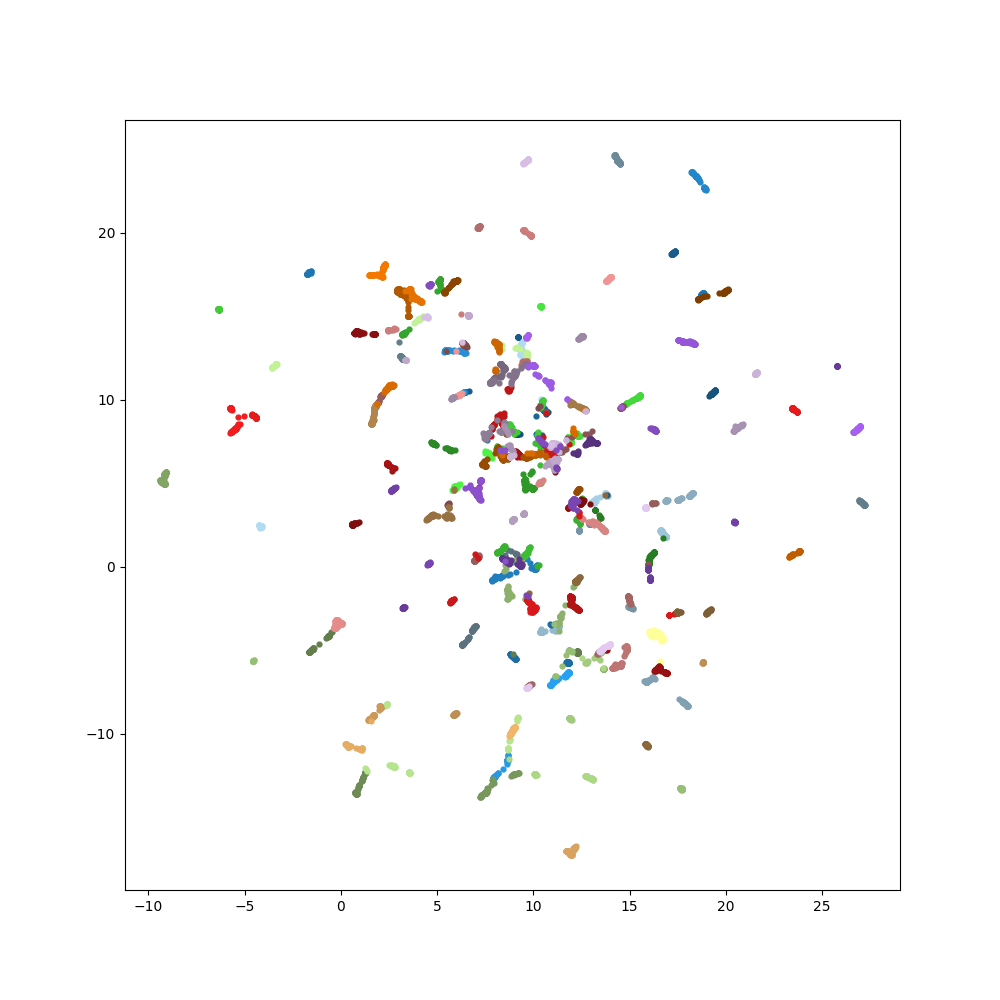}\hspace{4pt}}}
    \subfigure[\Surprise perplexity: 30 (10 subj.) \label{fig:umap:surprise_10_perp_30}]{{\hspace{4pt}\includegraphics[trim= 92pt 81pt 73pt 89pt, clip, height=1.40in, width=0.45\linewidth]{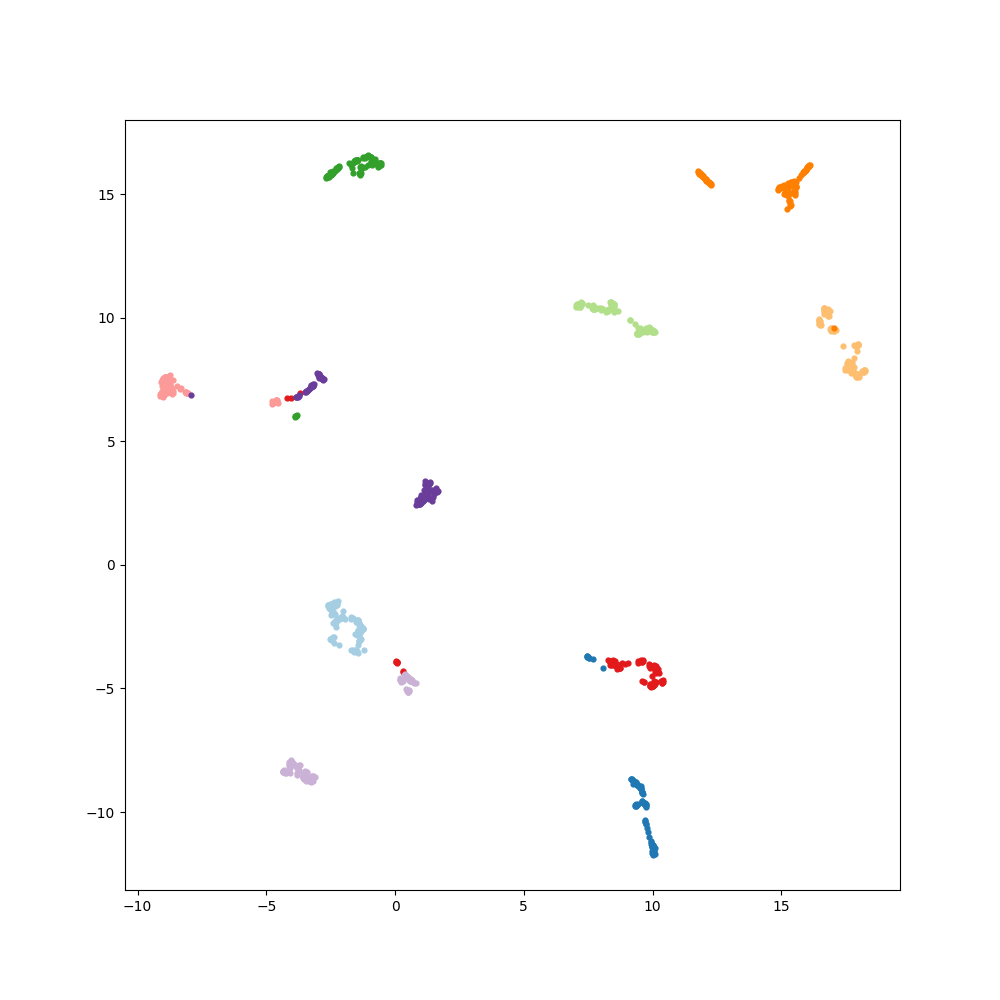}\hspace{4pt}}}
    
    \subfigure[\Surprise perplexity: 40 (full) \label{fig:umap:surprise_full_perp_40}]{{\hspace{4pt}\includegraphics[trim= 92pt 81pt 73pt 89pt, clip, height=1.40in, width=0.45\linewidth]{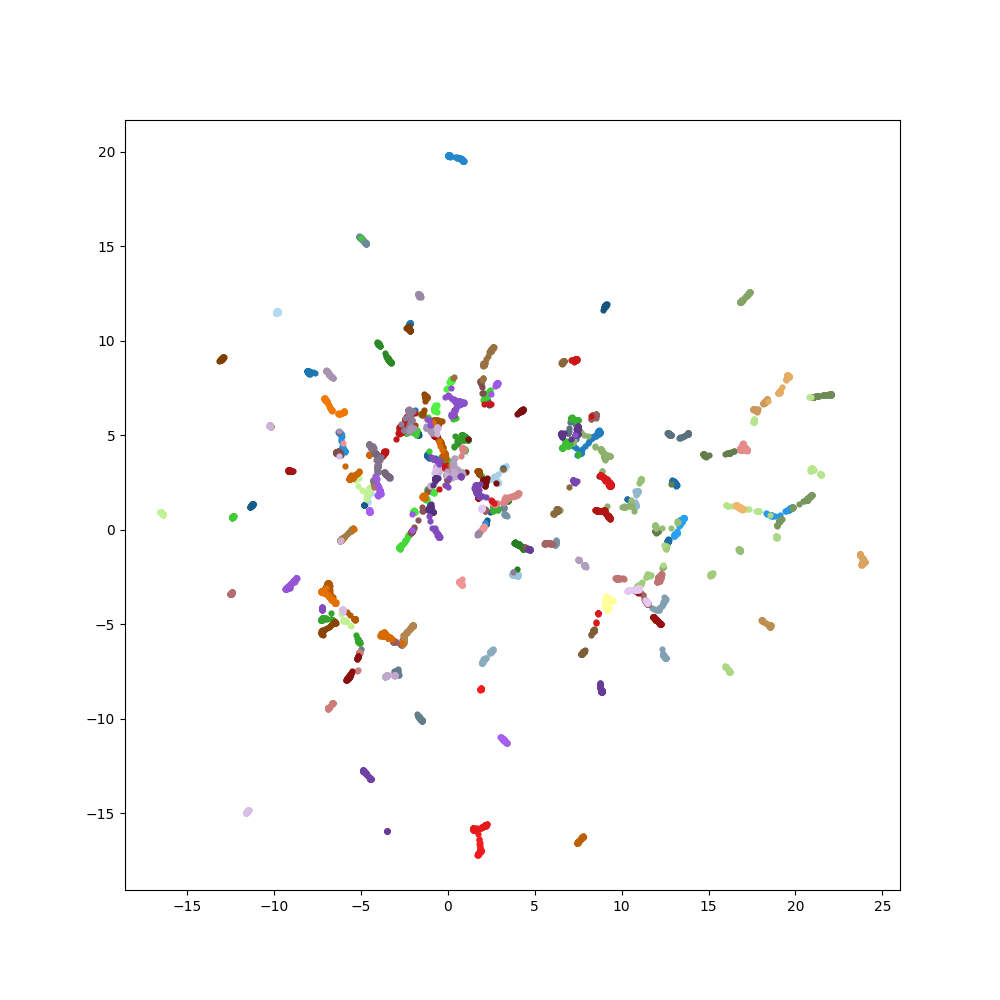}\hspace{4pt}}}
    \subfigure[\Surprise perplexity: 40 (10 subj.) \label{fig:umap:surprise_10_perp_40}]{{\hspace{4pt}\includegraphics[trim= 92pt 81pt 73pt 89pt, clip, height=1.40in, width=0.45\linewidth]{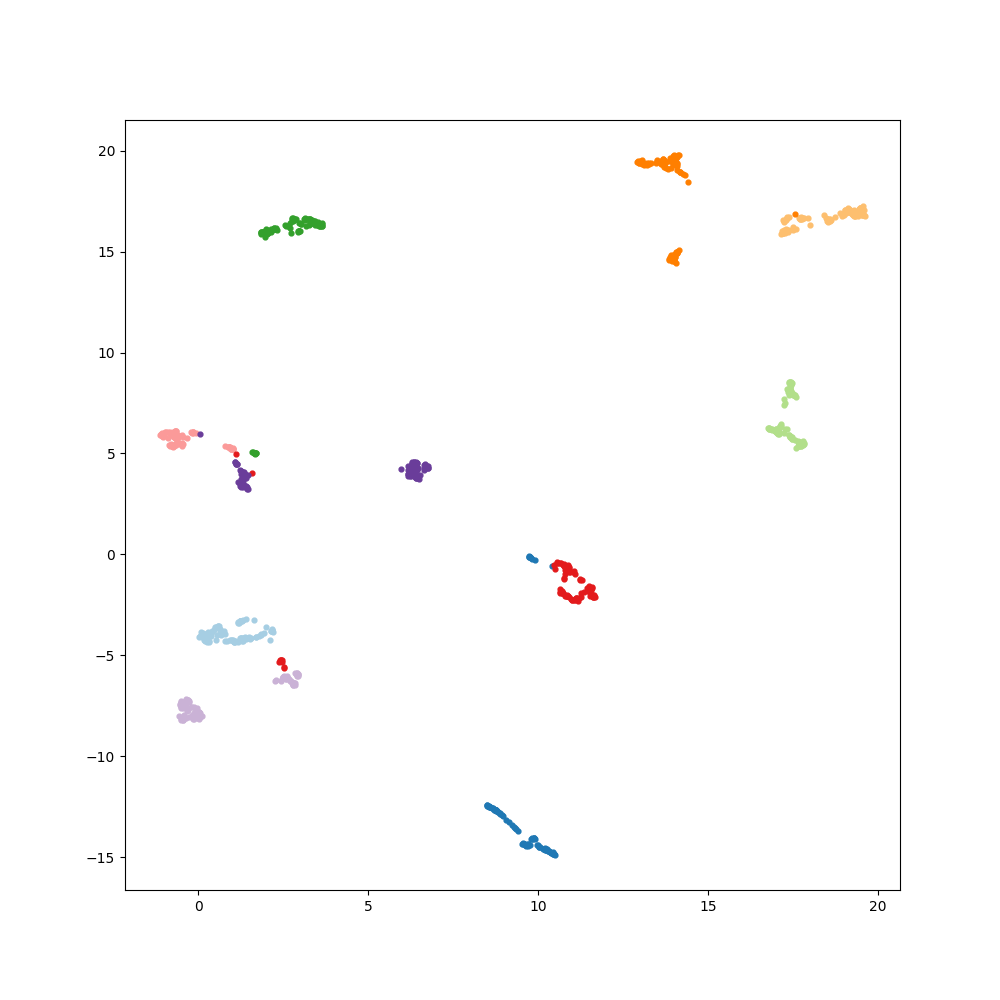}\hspace{4pt}}}
    
    \subfigure[\Surprise perplexity: 50 (full) \label{fig:umap:surprise_full_perp_50}]{{\hspace{4pt}\includegraphics[trim= 92pt 81pt 73pt 89pt, clip, height=1.40in, width=0.45\linewidth]{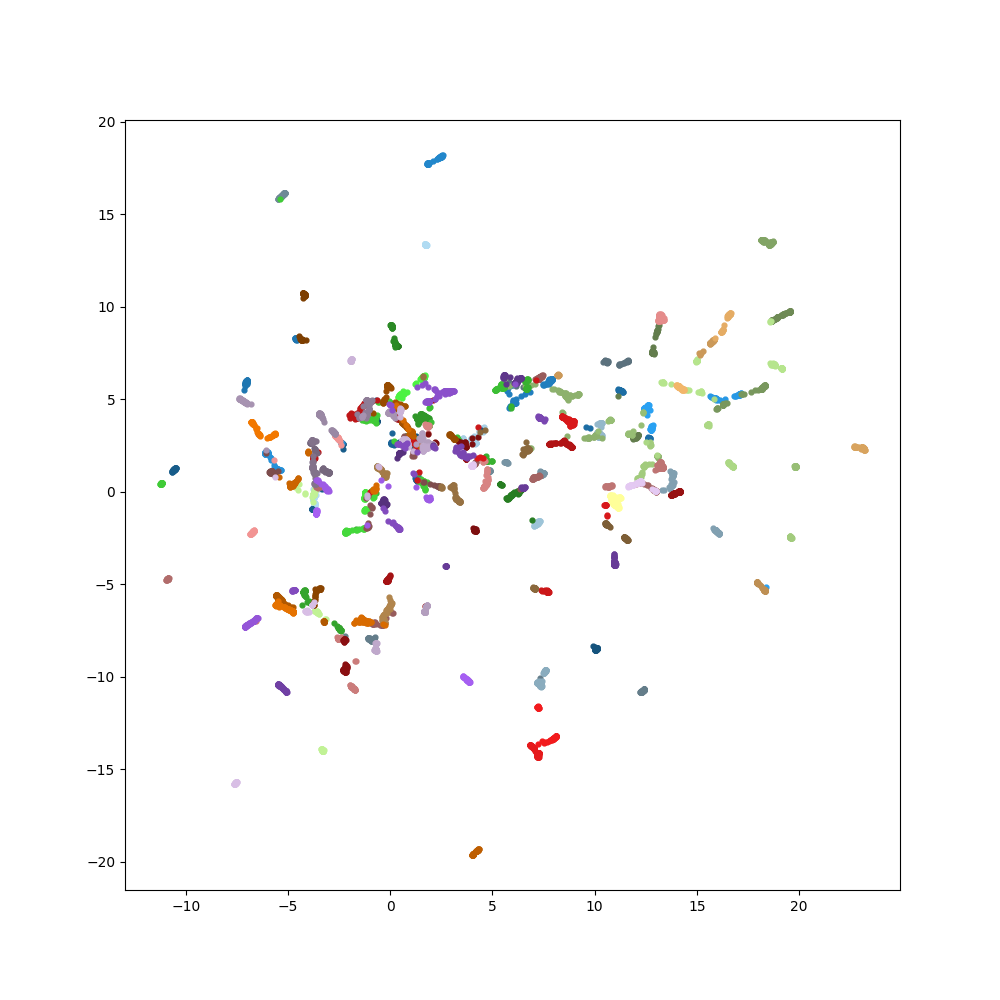}\hspace{4pt}}}
    \subfigure[\Surprise perplexity: 50 (10 subj.) \label{fig:umap:surprise_10_perp_50}]{{\hspace{4pt}\includegraphics[trim= 92pt 81pt 73pt 89pt, clip, height=1.40in, width=0.45\linewidth]{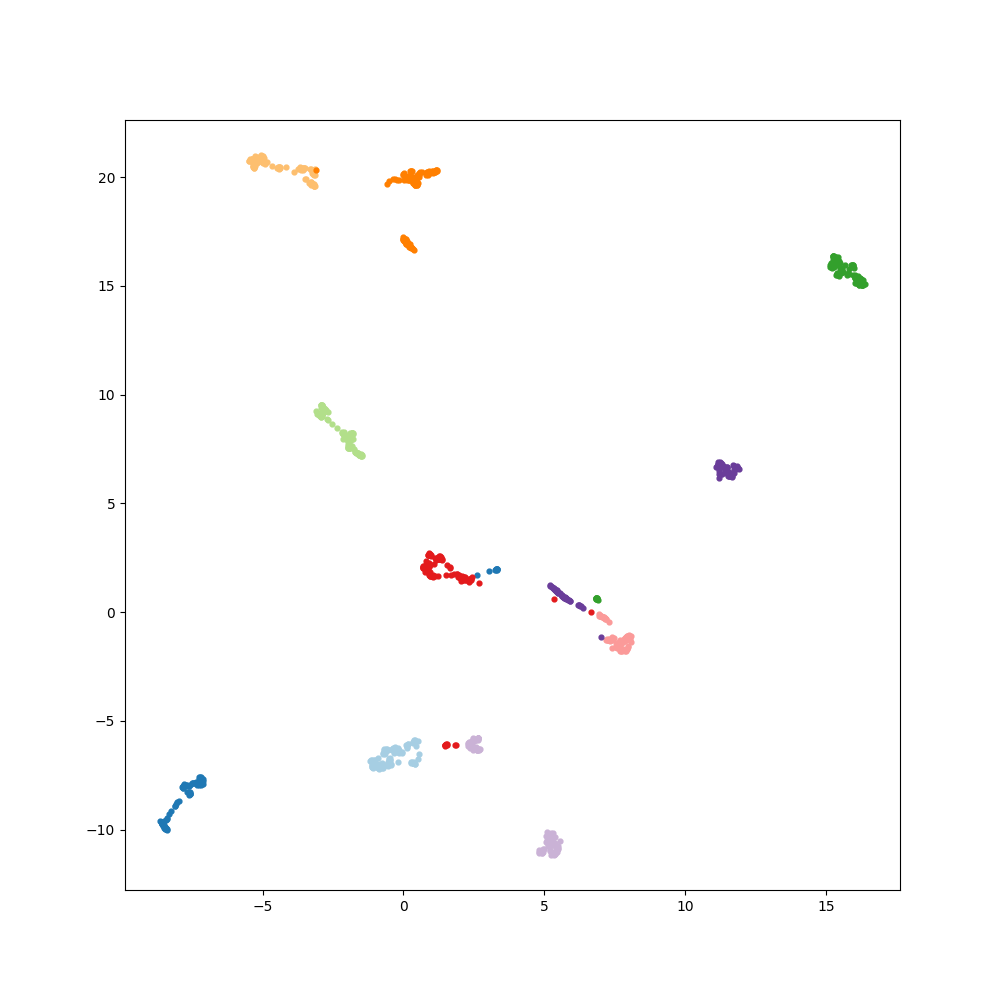}\hspace{4pt}}}
    
    \subfigure[\Surprise perplexity: 100 (full) \label{fig:umap:surprise_full_perp_100}]{{\hspace{4pt}\includegraphics[trim= 92pt 81pt 73pt 89pt, clip, height=1.40in, width=0.45\linewidth]{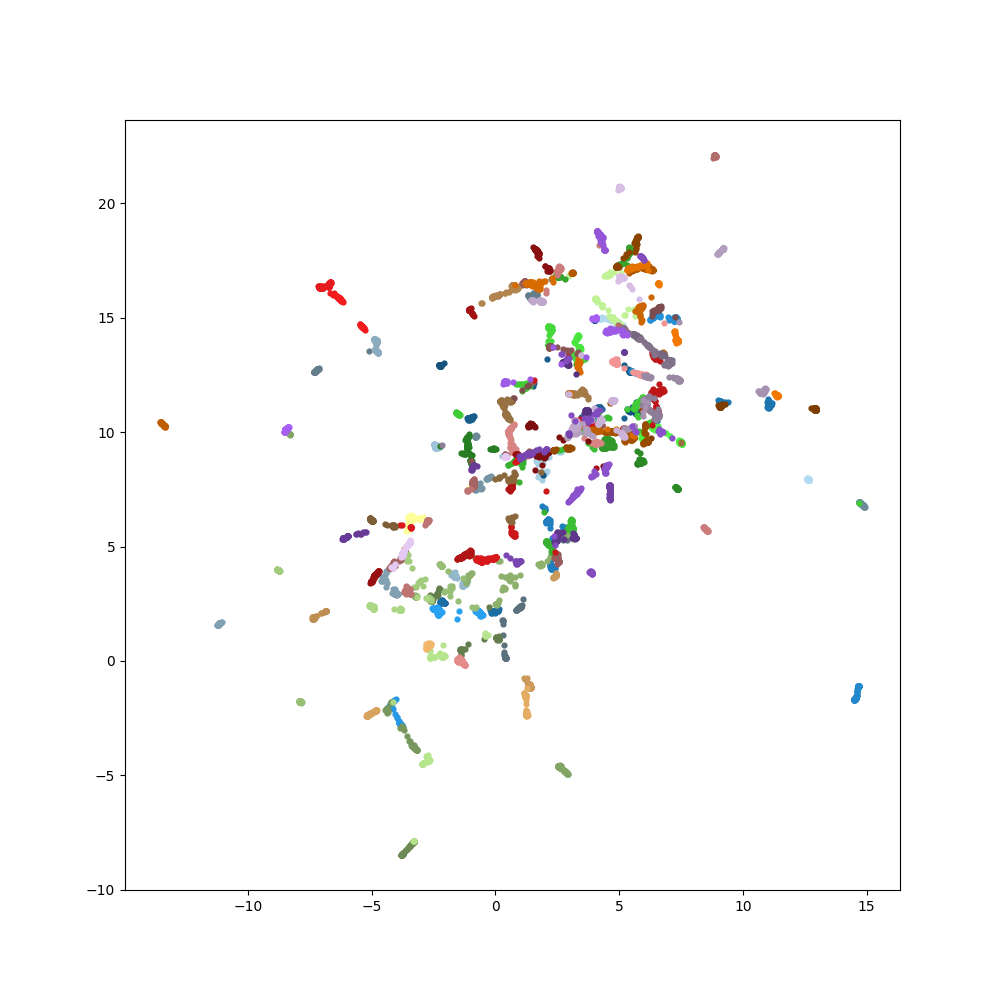}\hspace{4pt}}}
    \subfigure[\Surprise perplexity: 100 (10 subj.) \label{fig:umap:surprise_10_perp_100}]{{\hspace{4pt}\includegraphics[trim= 92pt 81pt 73pt 89pt, clip, height=1.40in, width=0.45\linewidth]{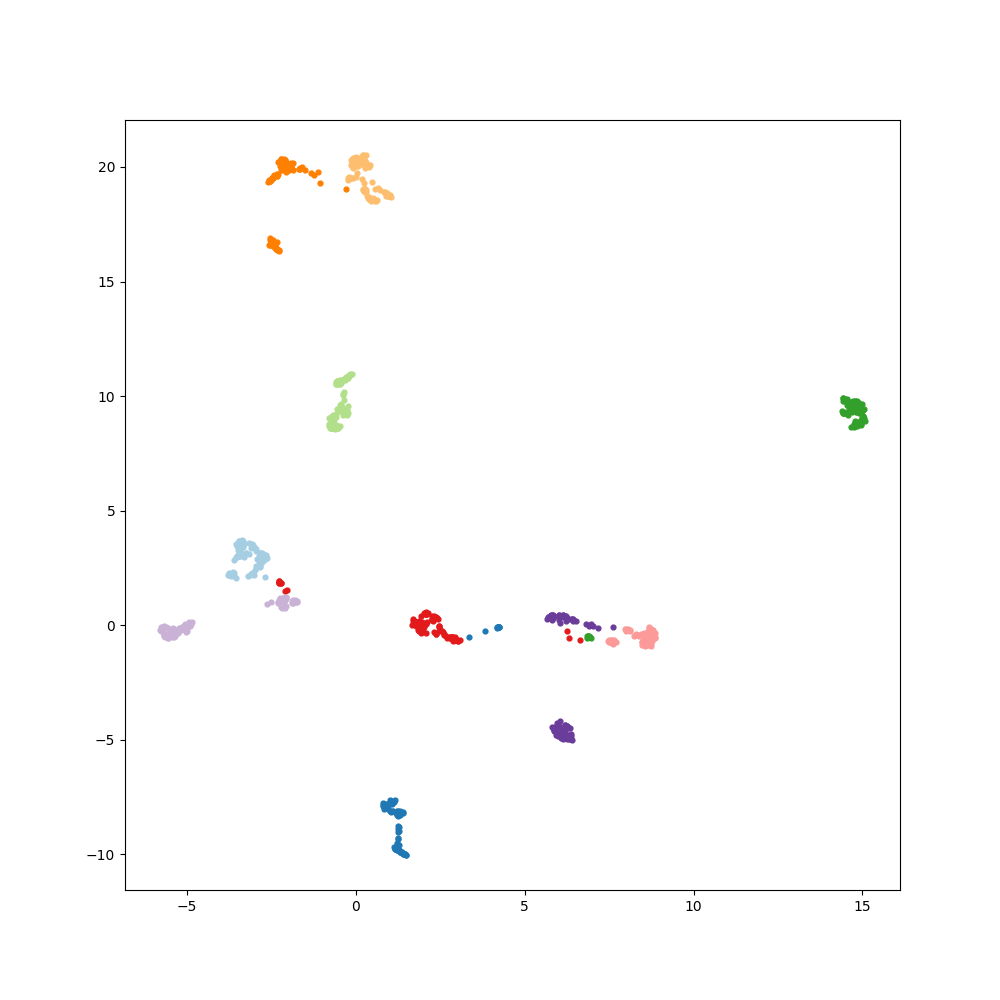}\hspace{4pt}}}
    
    \caption{UMAP clustering of individual topological data for \Surprise emotion at different perplexities.}
    \label{fig:umap:surprise}
\end{figure}

\end{document}